\documentclass[11pt]{article}
\usepackage[utf8]{inputenc}
\usepackage[letterpaper, portrait, margin=1in]{geometry}
\usepackage[T1]{fontenc}

\usepackage{color,amsmath,amsfonts,amsthm,mathrsfs,fancyhdr,verbatim,framed,xspace,mdframed,tabularx,booktabs,enumitem,comment,float,mathtools}
\usepackage{algorithm}
\usepackage{algpseudocode}
\usepackage{hyperref}
\usepackage[dvipsnames]{xcolor}
\definecolor{linkblue}{HTML}{0072BB}
\usepackage{graphicx}
\hypersetup{
    colorlinks=true,
    linkcolor=blue,
    citecolor=teal,
    hypertexnames=false
}
\makeatletter
\providecommand*{\toclevel@algorithm}{0}
\makeatother

\usepackage{fec}
\newcommand{\compact}{2}

\ifcase\compact\relax
  \usepackage{mathpazo}
\or
  \usepackage{newtxtext,newtxmath}
\or
  \usepackage[tt=false]{libertine}
  \usepackage[libertine]{newtxmath}
  \usepackage[varqu]{zi4}
\else
  \usepackage{mathpazo}
\fi
\usepackage{etoolbox}
\makeatletter
\AddToHookNext{cmd/@maketitle/before}{%
  \def\@makefnmark{\hbox{\@textsuperscript{\normalfont\@thefnmark}}}}
\makeatother

\newtheorem{theorem}{Theorem}[section]
\newtheorem{lemma}[theorem]{Lemma}
\newtheorem{proposition}[theorem]{Proposition}
\newtheorem{corollary}[theorem]{Corollary}

\newtheorem{fact}[theorem]{Fact}

\newtheorem*{restated}{Restatement}
\theoremstyle{definition}
\newtheorem{definition}[theorem]{Definition}

\newtheorem*{inassumption*}{Quantum Input Assumptions}
\theoremstyle{remark}

\theoremstyle{plain}
\usepackage{braket}
\usepackage{longtable} 
\usepackage{tikz}
\usetikzlibrary{arrows.meta,calc,decorations.pathreplacing,positioning,shapes.geometric}

\usepackage{authblk}

\newcommand{\proofstep}[1]{\par\smallskip\noindent\textit{#1}\enspace\ignorespaces}

\title{Quantum Speedups for Integer Linear Programming}

\author{Brandon Augustino\thanks{brandon.augustino@jpmchase.com}}
\author{Shouvanik Chakrabarti\thanks{shouvanik.chakrabarti@jpmchase.com}}
\affil{Global Technology Applied Research, JPMorganChase, New York, NY 10001, USA}

\date{}
\begin{document}
\maketitle
{\renewcommand{\thefootnote}{}%
 \footnotetext{\copyright~2026 JPMorganChase All rights reserved.}}

\begin{abstract}
We give a quantum algorithm for integer linear programming
(ILP), which, given a rational matrix $A$ with $n$ columns and a rational vector
$b$, asks to find $x\in\Z{n}$ with $Ax\leq b$, or show that none exists.
For input length $L$ and any fixed failure probability, it uses at most
$2^{O(n)}(\log(2n))^{n}\operatorname{poly}(L)$ gates.  To our knowledge,
it is the first quantum algorithm in the literature to beat the best
previously published classical worst-case bound for general ILP,
$2^{O(n)}(\log(2n))^{4n}\operatorname{poly}(L)$, due to Reis and
Rothvoss (FOCS 2023). Our algorithm searches the tree of their
subspace-flatness recursion, using lattice sparsification to give each
branch a short address, reversibly decodable without listing the others.
We also improve the classical bound to
$2^{O(n)}(\log(2n))^{2n}\operatorname{poly}(L)$: a rank-weighted variant
of Dadush's Gaussian hyperplane descent, which needs no stable
filtration, constructs subspaces along which the recursion branches only
$O(\log^2(2n))$ ways per dimension it removes, compared with
$O(\log^4(2n))$ for Reis and Rothvoss. The quantum speedup is thus
quadratic up to $2^{O(n)}$ factors, and holds instance by instance.  With
high probability, the quantum gate count is at most
$2^{O(n)}\operatorname{poly}(L)$ times the square root of the time of the
address-ordered depth-first classical reference run of the recursion,
with our improved flatness routine and the same random setup.  For
bounded equation form $Ax=b$, $0\leq x\leq u$, with positive integer
bounds $u_i$, a quantum algorithm uses at most
$2^{O(n)}(\prod_i\log_2(u_i+1))^{1/2}\operatorname{poly}(L)$ gates for
feasibility and optimization, accelerating the algorithm of Dadush,
Eisenbrand, and Rothvoss (Mathematical Programming, 2025).
\end{abstract}

\clearpage
\tableofcontents
\clearpage

\section{Introduction}
\label{sec:introduction}

Given a rational matrix $A$ with $n$ columns and a rational vector $b$,
\emph{integer linear programming} (ILP) asks
\begin{equation}\tag{ILP}\label{eq:ilp}
    \text{find}~x \in \Z{n}~\text{such that}~Ax\leq b,~\text{or show that none exists}.
\end{equation}
Integer programs model discrete decisions in operations
research, computer science, finance, and economics
\cite{schrijver1998theory,grotschel2012geometric,wolsey1999integer,conforti2014integer}.
In many applications we want the best solution, not just some solution.
Given an objective vector $c\in\mathbb{Q}^n$, \emph{integer linear
optimization} (ILO) asks to compute
\begin{equation}\tag{ILO}
    \max\{c^\top x : Ax\leq b,\ x\in\mathbb{Z}^n\},
    \label{eq:ilo}
\end{equation}
or to report that the feasible set is empty or that the objective is
unbounded above.  Deciding feasibility is NP-complete, and optimization
is NP-hard \cite{papadimitriou1981complexity,schrijver1998theory}.  We
write $L$ for the encoding length of the rational input.

Discrete optimization, and integer programming in particular, has been a
central target of quantum algorithms since Grover's
algorithm~\cite{grover1996fast}; see
\cite[Section~IV.A]{abbas2024challenges} for a recent survey.  This
algorithm finds one of $N$ candidates with $O(\sqrt N)$ calls to an
oracle that recognizes solutions, rather than the $\Theta(N)$ calls
needed classically, a quadratic speedup that is optimal for unstructured
search~\cite{bennett1997strengths}.  Yet no previous quantum algorithm
improves on the best classical worst-case bound for general ILP.

Larger speedups have come from structure in the problem.  Adiabatic
optimization~\cite{farhi2000quantum} and the quantum approximate
optimization algorithm (QAOA)~\cite{farhi2014quantum} model the objective
using a Hamiltonian whose ground states encode optimal solutions.
Adiabatic optimization interpolates slowly to this Hamiltonian from one
with an easily prepared ground state, at a speed limited by the spectral
gaps along the way.  The QAOA alternates a chosen number of
evolutions under the objective and under a mixing Hamiltonian.
Neither has a known worst-case bound for general ILP, and constraints
require penalties or mixers that preserve feasibility.  Short-path
methods combine quantum dynamics with amplitude
amplification~\cite{brassard2002quantum} and, given suitable spectral
properties, provably beat exhaustive search by more than a quadratic
factor~\cite{hastings2018shortPath,dalzell2022mind}.
Feasibility-preserving variants do the same over Markov-chain
search~\cite{chakrabarti2024generalized}, including on Gomory's group
relaxation~\cite{gomory1965relation,gomory1969some,augustino2026group}.
Decoded quantum interferometry exploits algebraic structure instead.  For
approximating optimal polynomial intersection over finite fields, it is
superpolynomially faster than known classical algorithms, because the
structure of this problem makes its decoding step
easy~\cite{jordan2025dqi}.  General ILP is not known to have the spectral
or algebraic structure that these methods exploit.

Without such structure, quantum speedups for ILP can potentially come from
accelerating a classical algorithm, but the quantum bound is then only
as good as that algorithm.  In the simplest case, the classical algorithm tests
every point of a box $[-M,M]^n$ that contains a solution (if one
exists), and its quantum version makes $O((2M+1)^{n/2})$ calls to a
membership oracle for the feasible set.  A box with
$M=2^{\operatorname{poly}(L)}$ always
suffices~\cite[Corollary~17.1b]{schrijver1998theory}, but even for $n=1$
all solutions can have magnitude $2^{\Omega(L)}$.  For fixed $n$ this
approach is then exponential in $L$, while classical algorithms run in
polynomial time.

Quantum variants of backtracking, branch-and-bound, and dynamic
programming
\cite{montanaro2018backtracking,ambainis2017quantum,montanaro2020branchAndBound,chakrabarti2022universal,doi:10.1137/1.9781611975482.107}
start from classical algorithms that can be much faster than testing
every point of the box.  For tree search, the cost falls to
roughly the square root of the tree size.  Montanaro's quantum
backtracking algorithm decides whether a bounded-degree tree with $T$
vertices and depth $h$ has a marked vertex with about $\sqrt T$ queries,
up to factors polynomial in $h$ \cite{montanaro2018backtracking}.
Ambainis and Kokainis replace $T$ by the number of vertices that
depth-first search visits before it first reaches a marked one
\cite{ambainis2017quantum}.  None of these methods, however, is known to
accelerate the best classical algorithms for ILP.

The best classical worst-case bounds for ILP come from geometric
recursions.  These descend from Lenstra's algorithm
\cite{lenstra1983integer}, which runs in time $2^{O(n^2)}\operatorname{poly}(L)$
\cite{reis2023subspace}. His approach exploits flatness: either the
polyhedron $K=\{x\in\mathbb R^n:Ax\leq b\}$ contains a lattice point, or it
is thin in some lattice direction, so that only a few parallel lattice
hyperplanes meet it.  Branching on these hyperplanes leaves
lower-dimensional subproblems, which are solved recursively.
Kannan~\cite{kannan1987minkowski} improved the dependence on $n$ to
$n^{O(n)}$ by branching on lattice subspaces rather than single
directions: he projects $K$ onto a subspace along which it is thin,
enumerates the projected lattice points, and recurses on the fiber above
each one.  Dadush~\cite{dadush2012integer} improved the bound to
$2^{O(n)}n^n$, enumerating these points with $M$-ellipsoid coverings, and
Reis and Rothvoss~\cite{reis2023subspace} to
$2^{O(n)}(\log(2n))^{4n}\operatorname{poly}(L)$, by finding much flatter
subspaces.

These recursions are natural candidates for quantum tree search, but
they do not fit it as they stand.  Quantum tree search needs local
access: given a node and an index $j$, it must compute the $j$th child
without generating the others.  The recursions produce the children of
a node only as a list, by enumerating lattice points.  To compute one
child, a circuit would have to regenerate this list at every query, at a
cost proportional to the number of branches, which is exactly the factor
that quantum search is meant to save.

Our main result is a quantum algorithm whose worst-case bound for general ILP
beats the best known classical one. It quantizes the recursion of
\cite{dadush2012integer,reis2023subspace} by giving the branches short
addresses that can be decoded reversibly, so each child is computed on
its own without listing the others. Constructing these addresses is a
key technical step to obtain the instance-by-instance comparison.  With high probability, the gate count on each instance is at most
$2^{O(n)}\operatorname{poly}(L)$ times the square root of the time of the
classical reference defined below: a depth-first run of the classical
recursion that uses the same random setup and decodes each node's
addresses in a fixed order.

The classical and quantum algorithms call the same flatness routine, so a
better routine improves both.  To make the quantum algorithm as fast as
we can, we therefore also improve the classical one.  We construct
subspaces witnessing flatness with factor $O(\log^2(2n))$ instead of
$O(\log^4(2n))$.  The classical recursion then runs in time
\[
    2^{O(n)}(\log(2n))^{2n}\operatorname{poly}(L),
\]
and the quantum algorithm uses
\[
    2^{O(n)}(\log(2n))^{n}\operatorname{poly}(L)
\]
gates.  For general ILP, this is a quadratic speedup up to $2^{O(n)}$
factors.

Problems in bounded equation form,
\begin{equation}\label{e:bounded_eq}
    \text{find}~x \in \Z{n}~\text{such that}~Ax= b~\text{and}~0 \leq x \leq u,~\text{or show that none exists},    
\end{equation}
with positive integer bounds $u_i$, behave differently.  The
approximate-to-exact framework of Dadush, Eisenbrand, and Rothvoss
(DER)~\cite{dadush2025approximate} solves them in time
$2^{O(n)}\Psi(u)\operatorname{poly}(L)$, where
$\Psi(u):=\prod_i\log_2(u_i+1)$.  When every $u_i$ is at most $n^{O(1)}$,
this is at most $2^{O(n)}(\log(2n))^n\operatorname{poly}(L)$, our
quantum bound for general ILP, so on these instances our worst-case
bound is no better than the classical one.  We therefore quantize DER
as well, and obtain $2^{O(n)}\Psi(u)^{1/2}\operatorname{poly}(L)$ gates.
Table~\ref{tab:bounds} collects these bounds.

\begin{table}[t]
\caption{Exact algorithms for integer programs in dimension $n$, up to
$2^{O(n)}\operatorname{poly}(L,\log(1/\delta))$ factors.}
\label{tab:bounds}
\centering
\footnotesize
\begin{tabular}{@{}l@{\hspace{14pt}}l@{\hspace{4pt}}l@{\hspace{14pt}}l@{\hspace{4pt}}l@{}}
\toprule
Problem & \multicolumn{2}{l}{Classical time} & \multicolumn{2}{l}{Quantum gates} \\
\midrule
General, $Ax\leq b$ & $(\log(2n))^{4n}$ & \cite{reis2023subspace} & & \\
 & $(\log(2n))^{2n}$ & Cor.~\ref{cor:classical-ip} & $(\log(2n))^{n}$ & Cor.~\ref{cor:worst-case-ip} \\
Bounded, $Ax=b$, $0\leq x\leq u$ & $\prod_i\log_2(u_i+1)$ & \cite{dadush2025approximate}
 & $\bigl(\prod_i\log_2(u_i+1)\bigr)^{1/2}$ & Cor.~\ref{cor:bounded-equation-form} \\
\bottomrule
\end{tabular}
\end{table}

\subsection{Constructive flatness and improved classical ILP}
\label{subsec:flatness-proof-idea}

The recursion in~\cite{dadush2012integer,reis2023subspace} branches on the lattice points of a projection of $K$, so
its running time depends on how thin $K$ is along some subspace.
Thinness is measured by the covering radius $\mu=\mu(\Lambda,K)$ of a
body $K$ in the span of a lattice $\Lambda$, the least $\mu\geq0$ such
that every translate of $\mu K$ within $\operatorname{span}\Lambda$ meets
$\Lambda$; if $\mu\leq1$, then $K$ contains a lattice point.  When $K$
is the Euclidean unit ball, $\mu$ is the largest distance from a point of
$\operatorname{span}\Lambda$ to $\Lambda$.  Each projection onto a lattice
subspace, one spanned by vectors of the dual lattice, gives a lower bound
on $\mu$.  If every translate of $\alpha K$ meets $\Lambda$, then
every translate of the projection of $\alpha K$ meets the projected
lattice, so the projection has at least the volume of a cell of the
projected lattice.  Hence a projection of $\alpha K$ with smaller volume
shows that $\mu>\alpha$.   

A subspace whose bound is within a factor $r$ of $\mu$ leaves at most
$2^{O(d)}r^d$ branches, where $d$ is its dimension: shrinking $K$ toward
an interior point by the right factor preserves feasibility and leaves at
most that many lattice points in the projection of the shrunken body.  The recursion therefore runs in time
$2^{O(n)}r^n\operatorname{poly}(L)$, and the question is how small $r$ can
be.  Kannan and Lov\'asz
showed that $r=n$ always suffices and that a simplex requires
$r=\Omega(\log n)$ \cite{kannan1988covering}.  The subspace flatness
conjecture \cite{dadush2012integer}, \cite[Conjecture~1]{reis2023subspace} asserts that
$r=O(\log(2n))$ always suffices, which would be optimal, and Dadush
showed that finding such subspaces in time
$(\log(2n))^{O(n)}\operatorname{poly}(L)$ would solve ILP within the same
time \cite{dadush2012integer}.
Reis and Rothvoss proved $r=O(\log^3(2n))$
\cite{reis2023subspace}, and Letwin, Reis, and Rothvoss recently improved
this to $r=O(\log^{3/2}(2n))$ \cite{letwin2026strongly} using optimal
mean-width and mean-norm estimates for convex bodies
\cite{bizeul2026optimal,paouris2026optimal}.  These results show that a
good subspace exists, but the recursion must also find one, and the
algorithm of Reis and Rothvoss finds one with $r=O(\log^4(2n))$. 

We show how to construct a subspace with $r=O(\log^2(2n))$, within
$\sqrt{\log(2n)}$ of the best existential bound.  The existential proofs
split the lattice along a filtration whose successive quotients are
stable (meaning that no sublattice of a quotient is denser than the
quotient itself) and add up a covering radius for each quotient.  The
Reis--Rothvoss algorithm computes its filtration with Dadush's Gaussian
hyperplane descent \cite{dadush2019approximating}.  This descent
guarantees stability only up to a factor proportional to the reverse Minkowski
threshold $\xi$, the scale above which a lattice without dense
sublattices has small Gaussian mass.  Their covering bound for the
quotients pays $\xi$ again, and with $\xi=O(\log(2n))$ this gives
$O(\log^4(2n))$, one factor $\xi$ above their existential bound
\cite[Theorems~50 and~55]{reis2023subspace}.  Our algorithm is a
rank-weighted variant of Dadush's descent~\cite{dadush2019approximating}.  It needs no stable
filtration and pays $\xi$ only once, so the estimates
in~\cite{reis2023subspace} already give $O(\log^3(2n))$.

The construction works in stages.  A stage holds a lattice of rank $m$,
initially $\Lambda$ of rank $k$, finds a primitive sublattice of rank $q$ whose dual
has small Gaussian mass at a scale proportional to $q/m$, and continues
with the projection orthogonal to that sublattice.  The covering radius
of $\Lambda$ is at most the sum of the covering contributions of these
sublattices \cite[Lemmas~40--41]{reis2023subspace}, and small dual
Gaussian mass bounds each contribution.  By Poisson summation, a Gaussian
vector at that scale is then nearly uniform modulo the sublattice, so a
symmetric body that holds more than three quarters of the Gaussian's
mass covers the span of the sublattice, after doubling, with its lattice translates.

The point is that the scale charges each stage in proportion to the
fraction of the rank it removes.  If every stage removes rank one, the
stages are charged $1/k,1/(k-1),\ldots,1$, and in general the fractions
sum to at most $1+\log k$ however the ranks split, whereas a fixed scale
would charge each of up to $k$ stages in full.

To find such a sublattice we adapt Dadush's descent.  It samples a dual
lattice vector from a discrete Gaussian distribution and, if the sample
is nonzero, cuts the current lattice by the hyperplane orthogonal to it,
lowering the rank by one.  We keep the original rank $m$ and determinant
scale fixed during a descent, let the scale of the mass test shrink in
proportion to the remaining rank $q$, and stop when zero samples are
frequent.  An exact discrete Gaussian sample is zero with probability
equal to the reciprocal of the dual's Gaussian mass, so frequent zeros
indicate small mass with high confidence, up to a constant gap, without
enumerating the dual.
The witness that a descent can succeed is a sublattice $N$ of minimal
rank-penalized determinant, whose dual has no sublattice that is dense at
the descent's scale.  While
the current lattice contains $N$ and the tests are accurate, reverse
Minkowski and a Gaussian projection inequality show that each cut keeps
$N$ with constant probability, and the descent stops by the time it
reaches $N$, so $2^{O(m)}$ attempts suffice for constant success
probability.  The algorithm never computes $N$,
and its output need not equal $N$ or be stable.

Each stage first puts its projected body in approximately isotropic
position.  The stage's covering bound is then a product of two factors:
the reverse Minkowski threshold $\xi$, through the smoothing scale, and
the expected gauge of a standard Gaussian vector in the symmetric core of
the body.  Two recent estimates each save a further $\sqrt{\log(2n)}$
factor.  Letwin, Reis, and Rothvoss prove
$\xi=O(\sqrt{\log(2n)})$ \cite{letwin2026strongly}, and the
Bizeul--Klartag mean-norm bound \cite{bizeul2025distances}, now sharp
\cite[Theorem~1.4]{bizeul2026optimal}, bounds the gauge by
$O(\sqrt{\log(2n)})$.  Each stage's covering contribution is then at
most $O(\log(2n))$ times the fraction of the rank it removes, times the
normalized determinant-volume ratio $R_W$ of its subspace $W$, defined
in~\eqref{eq:flatness-interface}, and the harmonic sum gives
$O(\log^2(2n))$.

Dadush's filtration, combined with the covering theorem
\cite[Theorem~1.2]{letwin2026strongly} that removes the second $\xi$,
also reaches $O(\log^2(2n))$, but only after
\cite[Lemma~4.1]{letwin2026strongly} is adapted to approximate stability.  Without these recent results,
the filtration route stays at $O(\log^4(2n))$.  Either way one factor
$\xi$ remains, and it is the gap to the existential bound.  With the
hidden minimizer $N$ in place of the descent's output, and the covering
theorem of Letwin, Reis, and Rothvoss in place of our Gaussian-mass
certificate, the same chain of stages reproves their existential
$O(\log^{3/2}(2n))$ bound without a stable filtration.
Our construction loses this $\xi$ because it certifies covering through
Gaussian mass, which we can test by sampling.

\begin{theorem}[Constructive subspace flatness]
\label{thm:constructive-flatness}
For $k\geq1$, let $\Lambda\subseteq\mathbb R^k$ be a rational full-rank lattice and
$K\subseteq\mathbb R^k$ a full-dimensional projection of a rational
polytope (its \emph{rational lift}).
Put $\Phi(k):=C_\Phi\lceil\log_2(2k)\rceil^2$ for a universal integer $C_\Phi$.  A randomized
algorithm running in time $2^{O(k)}\operatorname{poly}(L,\log(1/\delta))$
finds, with probability at least $1-\delta$, a nonzero lattice subspace
$W$ of dimension $d$ and an estimate $R$ such that
\begin{equation}\label{eq:flatness-interface}
    \mu(\Lambda,K)\leq\Phi(k)R_W,
    \qquad R_W\leq R\leq C_RR_W,
    \qquad R_W^d:=\frac{\det(\Pi_W\Lambda)}{\operatorname{Vol}_d(\Pi_WK)},
\end{equation}
where $C_R$ is a universal integer, $\Pi_W$ is the
orthogonal projection onto $W$, and lattice subspaces are spanned by
dual lattice vectors.
\end{theorem}
The algorithm returns the stage subspace with the largest estimated
determinant-volume ratio.  Section~\ref{subsec:subspace-flatness}
sketches the proof, and Appendix~\ref{app:flatness-proofs} completes it.

The classical integer programming recursion improves accordingly, and we measure our
quantum speedup against this improved bound.
\begin{corollary}[Classical integer programming]
\label{cor:classical-ip}
The recursion \cite{dadush2012integer,reis2023subspace}, with
Theorem~\ref{thm:constructive-flatness} as its flatness routine, runs in
time
\begin{equation}
    2^{O(n)}\Phi(n)^n\operatorname{poly}(L,\log(1/\delta))
    =2^{O(n)}(\log(2n))^{2n}\operatorname{poly}(L,\log(1/\delta))
    \label{eq:classical-bound}
\end{equation}
on every random tape and solves~\eqref{eq:ilp} with probability at least
$1-\delta$.
\end{corollary}

\subsection{Quantum speedups for the improved recursion}
\label{subsec:quantum-entry}

\begin{figure}[t]
\centering
\definecolor{kfa}{rgb}{0.642,0.642,0.654}
\definecolor{kfb}{rgb}{0.506,0.506,0.518}
\definecolor{kfc}{rgb}{0.500,0.500,0.512}
\definecolor{kfd}{rgb}{0.899,0.899,0.911}
\definecolor{kfe}{rgb}{0.505,0.505,0.517}
\definecolor{kff}{rgb}{0.621,0.621,0.633}
\definecolor{kfg}{rgb}{0.516,0.516,0.528}
\definecolor{kfh}{rgb}{0.700,0.700,0.712}
\definecolor{kfi}{rgb}{0.500,0.500,0.512}
\definecolor{tfa}{rgb}{0.692,0.692,1.000}
\definecolor{tfb}{rgb}{0.604,0.604,1.000}
\definecolor{tfc}{rgb}{0.600,0.600,1.000}
\definecolor{tfd}{rgb}{0.860,0.860,1.000}
\definecolor{tfe}{rgb}{0.603,0.603,1.000}
\definecolor{tff}{rgb}{0.679,0.679,1.000}
\definecolor{tfg}{rgb}{0.610,0.610,1.000}
\definecolor{tfh}{rgb}{0.730,0.730,1.000}
\definecolor{tfi}{rgb}{0.600,0.600,1.000}
\begin{tikzpicture}[line join=round,line cap=round,
        floor/.style={fill=gray!5,draw=gray!45,thin},
        shK/.style={fill=gray!17,draw=gray!55,thin},
        shT/.style={fill=blue!13,draw=blue!55!black,semithick},
        gridpt/.style={gray!50},
        xpt/.style={teal!85!green!90!black},
        drop/.style={densely dashed,teal!85!green!70,thin},
        kback/.style={fill opacity=0.6},
        kbackedge/.style={gray!60,densely dashed,thin},
        kfront/.style={fill opacity=0.42},
        kedge/.style={gray!70!black,thick},
        tback/.style={fill opacity=0.45},
        tbackedge/.style={blue!50,densely dashed,very thin},
        tfront/.style={fill opacity=0.4},
        tedge/.style={blue!65!black,semithick},
        fiber/.style={teal!85!green!80!black,very thick},
        bluedot/.style={fill=blue!75!black},
        graydot/.style={fill=white,draw=gray!70!black,thin},
        lab/.style={inner sep=1.5pt},
        leader/.style={blue!65!black,very thin},
        key/.style={anchor=west,font=\footnotesize}]
    \path[floor] (0.956,1.832) -- (-4.027,1.127) -- (-2.594,-0.682) -- (2.388,0.023) -- cycle;
    \path[shK] (0.200,-0.099) -- (1.311,0.394) -- (0.300,0.975) -- (-0.364,1.283) -- (-2.459,0.619) -- (-2.661,0.311) -- (-2.398,0.009) -- (-0.242,-0.093) -- cycle;
    \path[shT] (0.025,0.046) -- (0.858,0.416) -- (0.099,0.852) -- (-0.399,1.083) -- (-1.969,0.584) -- (-2.121,0.354) -- (-1.924,0.127) -- (-0.307,0.051) -- cycle;
    \foreach \p in {(0.699,1.614),(0.934,1.318),(1.169,1.021),(1.404,0.724),(1.638,0.428),(1.873,0.131),(-0.003,1.515),(0.232,1.218),(0.467,0.922),(0.702,0.625),(0.937,0.329),(1.171,0.032),(-0.704,1.416),(-0.470,1.119),(0.470,-0.067),(-1.406,1.317),(-1.171,1.020),(-0.232,-0.166),(-2.108,1.217),(-1.873,0.921),(-1.169,0.031),(-0.934,-0.266),(-2.810,1.118),(-2.575,0.821),(-2.340,0.525),(-2.105,0.228),(-1.870,-0.068),(-1.636,-0.365),(-3.511,1.019),(-3.277,0.722),(-3.042,0.426),(-2.807,0.129),(-2.572,-0.168),(-2.337,-0.464)}
        \fill[gridpt] \p circle (1.05pt);
    \draw[drop] (-0.235,1.892) -- (-0.235,0.823);
    \draw[drop] (-0.937,2.625) -- (-0.937,0.723);
    \draw[drop] (-1.638,3.616) -- (-1.638,0.624);
    \draw[drop] (0.000,1.721) -- (0.000,0.526);
    \draw[drop] (-0.702,1.898) -- (-0.702,0.427);
    \draw[drop] (-1.404,2.899) -- (-1.404,0.328);
    \draw[drop] (0.235,1.568) -- (0.235,0.229);
    \draw[drop] (-0.467,1.665) -- (-0.467,0.130);
    \foreach \p in {(-0.235,0.823),(-0.937,0.723),(-1.638,0.624),(0.000,0.526),(-0.702,0.427),(-1.404,0.328),(0.235,0.229),(-0.467,0.130)}
        \fill[xpt] \p circle (1.7pt);
    \path[kback,fill=kfa] (0.300,4.944) -- (1.311,2.927) -- (1.086,1.692) -- (-0.101,2.065) -- (-0.364,2.859) -- cycle;
    \path[kback,fill=kfg] (-0.101,2.065) -- (-0.606,1.760) -- (-2.661,4.692) -- (-2.459,5.015) -- (-0.364,2.859) -- cycle;
    \path[kback,fill=kfh] (-0.101,2.065) -- (1.086,1.692) -- (-0.228,1.500) -- (-0.606,1.760) -- cycle;
    \path[kback,fill=kfi] (-2.459,5.015) -- (-0.305,5.733) -- (0.300,4.944) -- (-0.364,2.859) -- cycle;
    \draw[kbackedge] (1.086,1.692) -- (-0.101,2.065);
    \draw[kbackedge] (-0.364,2.859) -- (-0.101,2.065);
    \draw[kbackedge] (0.300,4.944) -- (-0.364,2.859);
    \draw[kbackedge] (-0.606,1.760) -- (-0.101,2.065);
    \draw[kbackedge] (-2.459,5.015) -- (-0.364,2.859);
    \draw[fiber] (-0.235,2.042) -- (-0.235,2.375);
    \draw[fiber] (-0.235,4.549) -- (-0.235,5.052);
    \draw[fiber] (-0.937,2.737) -- (-0.937,3.195);
    \draw[fiber] (-0.937,4.613) -- (-0.937,5.013);
    \draw[fiber] (-1.638,3.738) -- (-1.638,4.186);
    \draw[fiber] (-1.638,4.677) -- (-1.638,4.779);
    \draw[fiber] (0.000,2.076) -- (0.000,2.204);
    \path[tback,fill=tfa] (0.099,4.615) -- (0.858,3.103) -- (0.689,2.177) -- (-0.201,2.456) -- (-0.399,3.052) -- cycle;
    \path[tback,fill=tfg] (-0.201,2.456) -- (-0.580,2.228) -- (-2.121,4.427) -- (-1.969,4.669) -- (-0.399,3.052) -- cycle;
    \path[tback,fill=tfh] (-0.201,2.456) -- (0.689,2.177) -- (-0.297,2.033) -- (-0.580,2.228) -- cycle;
    \path[tback,fill=tfi] (-1.969,4.669) -- (-0.354,5.207) -- (0.099,4.615) -- (-0.399,3.052) -- cycle;
    \draw[tbackedge] (0.689,2.177) -- (-0.201,2.456);
    \draw[tbackedge] (-0.399,3.052) -- (-0.201,2.456);
    \draw[tbackedge] (0.099,4.615) -- (-0.399,3.052);
    \draw[tbackedge] (-0.580,2.228) -- (-0.201,2.456);
    \draw[tbackedge] (-1.969,4.669) -- (-0.399,3.052);
    \draw[fiber] (-0.235,1.892) -- (-0.235,2.042);
    \draw[fiber] (-0.235,2.375) -- (-0.235,4.549);
    \draw[fiber] (-0.235,5.052) -- (-0.235,5.402);
    \foreach \p in {(-0.235,2.979),(-0.235,3.650),(-0.235,4.321)} \path[bluedot] \p circle (1.5pt);
    \draw[fiber] (-0.937,2.625) -- (-0.937,2.737);
    \draw[fiber] (-0.937,3.195) -- (-0.937,4.613);
    \draw[fiber] (-0.937,5.013) -- (-0.937,5.466);
    \foreach \p in {(-0.937,3.551),(-0.937,4.221)} \path[bluedot] \p circle (1.5pt);
    \draw[fiber] (-1.638,3.616) -- (-1.638,3.738);
    \draw[fiber] (-1.638,4.186) -- (-1.638,4.677);
    \draw[fiber] (-1.638,4.779) -- (-1.638,5.249);
    \draw[fiber] (0.000,1.721) -- (0.000,2.076);
    \draw[fiber] (0.000,2.204) -- (0.000,4.534);
    \draw[fiber] (0.000,4.745) -- (0.000,5.088);
    \foreach \p in {(0.000,2.683),(0.000,3.353),(0.000,4.024)} \path[bluedot] \p circle (1.5pt);
    \draw[fiber] (-0.702,1.980) -- (-0.702,2.402);
    \draw[fiber] (-0.702,2.468) -- (-0.702,4.887);
    \draw[fiber] (-0.702,5.091) -- (-0.702,5.340);
    \foreach \p in {(-0.702,2.583),(-0.702,3.254),(-0.702,3.925),(-0.702,4.595)} \path[bluedot] \p circle (1.5pt);
    \draw[fiber] (-1.404,2.983) -- (-1.404,3.403);
    \draw[fiber] (-1.404,3.471) -- (-1.404,4.605);
    \draw[fiber] (-1.404,4.858) -- (-1.404,5.058);
    \foreach \p in {(-1.404,3.826),(-1.404,4.496)} \path[bluedot] \p circle (1.5pt);
    \draw[fiber] (0.235,1.813) -- (0.235,2.111);
    \draw[fiber] (0.235,3.175) -- (0.235,3.948);
    \draw[fiber] (0.235,4.345) -- (0.235,4.502);
    \foreach \p in {(0.235,3.727)} \path[bluedot] \p circle (1.5pt);
    \draw[fiber] (-0.467,1.827) -- (-0.467,2.150);
    \draw[fiber] (-0.467,2.387) -- (-0.467,4.696);
    \foreach \p in {(-0.467,2.958),(-0.467,3.628),(-0.467,4.299)} \path[bluedot] \p circle (1.5pt);
    \path[tfront,fill=tfb] (-0.307,4.672) -- (0.025,4.145) -- (0.858,3.103) -- (0.099,4.615) -- (-0.354,5.207) -- cycle;
    \path[tfront,fill=tfc] (-0.545,2.436) -- (-0.297,2.033) -- (0.689,2.177) -- (0.858,3.103) -- (0.025,4.145) -- cycle;
    \path[tfront,fill=tfd] (-1.924,4.286) -- (-0.307,4.672) -- (-0.354,5.207) -- (-1.969,4.669) -- (-2.121,4.427) -- cycle;
    \path[tfront,fill=tfe] (-2.121,4.427) -- (-0.580,2.228) -- (-0.297,2.033) -- (-0.545,2.436) -- (-1.924,4.286) -- cycle;
    \path[tfront,fill=tff] (-1.924,4.286) -- (-0.545,2.436) -- (0.025,4.145) -- (-0.307,4.672) -- cycle;
    \draw[tedge] (0.858,3.103) -- (0.099,4.615);
    \draw[tedge] (0.689,2.177) -- (0.858,3.103);
    \draw[tedge] (0.025,4.145) -- (-0.307,4.672);
    \draw[tedge] (0.858,3.103) -- (0.025,4.145);
    \draw[tedge] (0.099,4.615) -- (-0.354,5.207);
    \draw[tedge] (-0.354,5.207) -- (-0.307,4.672);
    \draw[tedge] (-0.297,2.033) -- (-0.545,2.436);
    \draw[tedge] (0.689,2.177) -- (-0.297,2.033);
    \draw[tedge] (-0.545,2.436) -- (0.025,4.145);
    \draw[tedge] (-1.924,4.286) -- (-0.307,4.672);
    \draw[tedge] (-0.354,5.207) -- (-1.969,4.669);
    \draw[tedge] (-2.121,4.427) -- (-1.969,4.669);
    \draw[tedge] (-1.924,4.286) -- (-2.121,4.427);
    \draw[tedge] (-2.121,4.427) -- (-0.580,2.228);
    \draw[tedge] (-0.297,2.033) -- (-0.580,2.228);
    \draw[tedge] (-0.545,2.436) -- (-1.924,4.286);
    \draw[fiber] (0.000,4.534) -- (0.000,4.745);
    \draw[fiber] (-0.702,2.402) -- (-0.702,2.468);
    \draw[fiber] (-0.702,4.887) -- (-0.702,5.091);
    \draw[fiber] (-1.404,3.403) -- (-1.404,3.471);
    \draw[fiber] (-1.404,4.605) -- (-1.404,4.858);
    \draw[fiber] (0.235,2.111) -- (0.235,3.175);
    \draw[fiber] (0.235,3.948) -- (0.235,4.345);
    \draw[fiber] (-0.467,2.150) -- (-0.467,2.387);
    \draw[fiber] (-0.467,4.696) -- (-0.467,5.149);
    \path[kfront,fill=kfb] (-0.242,5.019) -- (0.200,4.317) -- (1.311,2.927) -- (0.300,4.944) -- (-0.305,5.733) -- cycle;
    \path[kfront,fill=kfc] (-0.559,2.038) -- (-0.228,1.500) -- (1.086,1.692) -- (1.311,2.927) -- (0.200,4.317) -- cycle;
    \path[kfront,fill=kfd] (-2.398,4.504) -- (-0.242,5.019) -- (-0.305,5.733) -- (-2.459,5.015) -- (-2.661,4.692) -- cycle;
    \path[kfront,fill=kfe] (-2.661,4.692) -- (-0.606,1.760) -- (-0.228,1.500) -- (-0.559,2.038) -- (-2.398,4.504) -- cycle;
    \path[kfront,fill=kff] (-2.398,4.504) -- (-0.559,2.038) -- (0.200,4.317) -- (-0.242,5.019) -- cycle;
    \draw[kedge] (1.311,2.927) -- (0.300,4.944);
    \draw[kedge] (1.086,1.692) -- (1.311,2.927);
    \draw[kedge] (0.200,4.317) -- (-0.242,5.019);
    \draw[kedge] (1.311,2.927) -- (0.200,4.317);
    \draw[kedge] (0.300,4.944) -- (-0.305,5.733);
    \draw[kedge] (-0.305,5.733) -- (-0.242,5.019);
    \draw[kedge] (-0.228,1.500) -- (-0.559,2.038);
    \draw[kedge] (1.086,1.692) -- (-0.228,1.500);
    \draw[kedge] (-0.559,2.038) -- (0.200,4.317);
    \draw[kedge] (-2.398,4.504) -- (-0.242,5.019);
    \draw[kedge] (-0.305,5.733) -- (-2.459,5.015);
    \draw[kedge] (-2.661,4.692) -- (-2.459,5.015);
    \draw[kedge] (-2.398,4.504) -- (-2.661,4.692);
    \draw[kedge] (-2.661,4.692) -- (-0.606,1.760);
    \draw[kedge] (-0.228,1.500) -- (-0.606,1.760);
    \draw[kedge] (-0.559,2.038) -- (-2.398,4.504);
    \draw[drop] (-0.702,1.980) -- (-0.702,1.898);
    \draw[drop] (-1.404,2.983) -- (-1.404,2.899);
    \draw[drop] (0.235,1.813) -- (0.235,1.568);
    \draw[drop] (-0.467,1.827) -- (-0.467,1.665);
    \node[lab,anchor=west,text=gray!70!black] at (0.978,4.040) {$K$};
    \draw[leader] (-1.635,4.159) -- (-2.487,3.809) node[lab,anchor=east,text=blue!65!black] {$\widetilde K$};
    \node[lab,anchor=east,text=gray!70!black] at (-4.077,1.127) {$W$};
    \draw[fiber] (3.088,3.456) -- (3.088,3.796);
    \node[key] at (3.388,3.626) {child $K\cap\Pi_W^{-1}(y)$ for $y\in X$};
    \path[bluedot] (3.088,3.076) circle (1.5pt);
    \node[key] at (3.388,3.076) {lattice point of $\widetilde K$};
    \fill[xpt] (3.088,2.526) circle (1.7pt);
    \node[key] at (3.388,2.526) {$X=K_W\cap\Pi_W(x_0+\Lambda)$};
    \path[shT] (2.958,1.876) rectangle (3.218,2.076);
    \node[key] at (3.388,1.976) {$K_W=\Pi_W\widetilde K$};
    \fill[gridpt] (3.088,1.426) circle (1.05pt);
    \node[key] at (3.388,1.426) {$\Pi_W(x_0+\Lambda)$};
\end{tikzpicture}
\caption{A node of the classical recursion in dimension $k=3$, branching on a
lattice subspace $W$ of dimension $d=2$; here $x_0+\Lambda=\mathbb Z^3$ and $W$
is horizontal, drawn below the body for clarity.  The recursion shrinks the
polytope $K$ about its center to $\widetilde K$, which keeps the node feasible
if it was, and projects $\widetilde K$ to $K_W=\Pi_W\widetilde K$.  Each point
$y$ of the branch set $X$ indexes one child, the fiber $K\cap\Pi_W^{-1}(y)$ of
the unshrunken body, here a vertical segment (teal).  Every lattice point of
$\widetilde K$ (blue dots) lies on one of these fibers.  A $d$-dimensional
projection removes $d$ dimensions at once, at the price of up to
$\kappa_k(d)$ branches.}
\label{fig:classical-projection}
\end{figure}
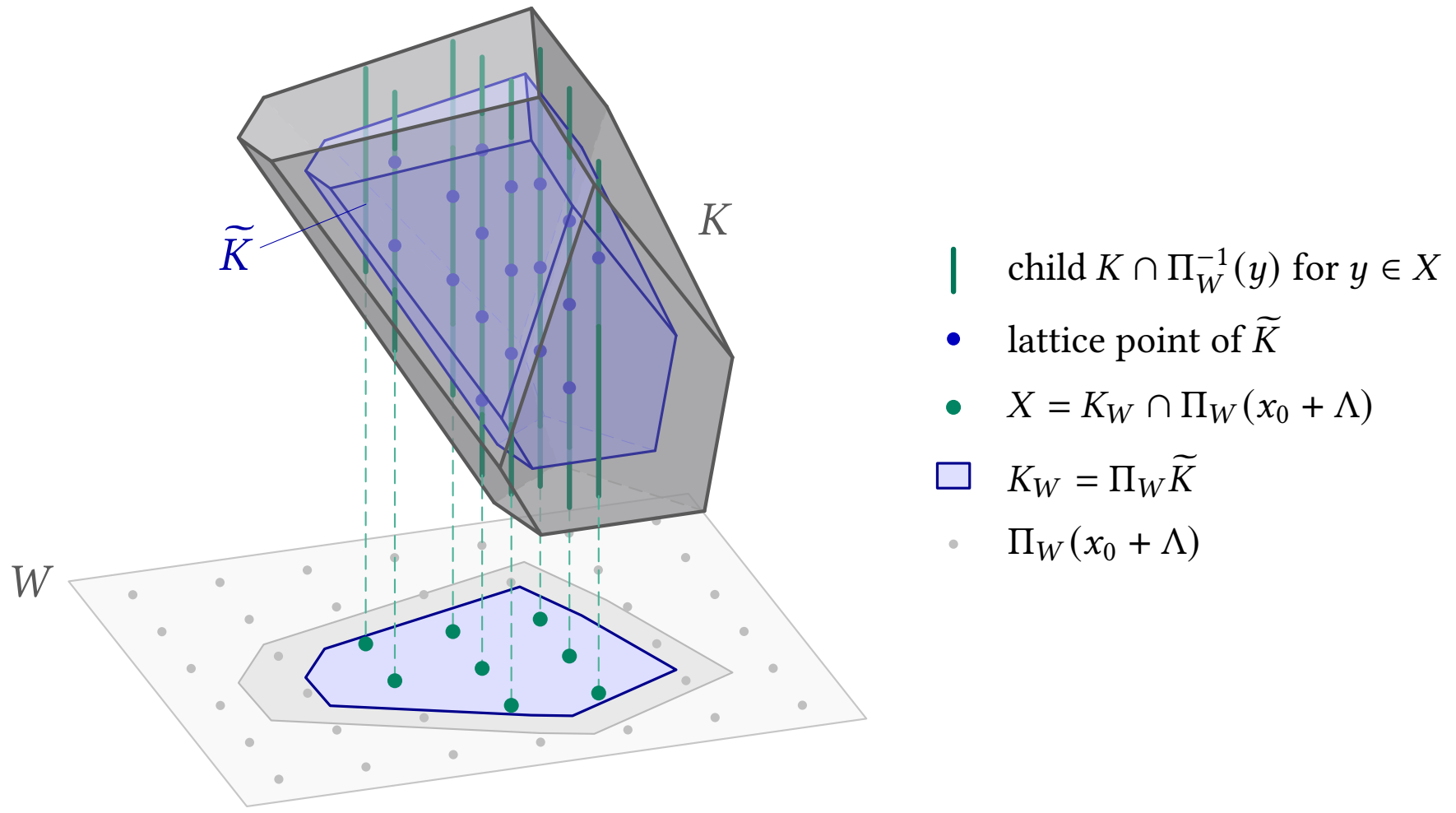

Our main quantum algorithm searches the tree of the classical recursion of~\cite{dadush2012integer,reis2023subspace}. The recursion decides
whether $K$ meets $x_0+\Lambda$ for a rational convex body $K$, made compact by standard bounds
(Section~\ref{subsec:classical-recursion}). At a node of dimension $k$, it
finds a lattice subspace $W$ of dimension $d\geq1$ and a shrunken body
$\widetilde K$ that preserves feasibility, and branches on
\begin{equation*}
    X:=\Pi_W\widetilde K\cap\Pi_W(x_0+\Lambda),
    \qquad |X|\leq\kappa_k(d):=(C_{\mathrm{br}}\Phi(k))^d,
\end{equation*}
where $C_{\mathrm{br}}$ is a universal integer. The child of $y\in X$ is
the slice of $K$ above $y$, of dimension at most $k-d$
(Figure~\ref{fig:classical-projection}).

This bound counts the branches, but as noted above, tree search must
also index them.  Regenerating the list $X$ at every query costs up to
$|X|$, and replaying the enumeration up to its $j$th output still costs
$\Omega(j)$.  Nor can the enumeration serve as the tree.
Tree search pays polynomially in the depth, but the Voronoi traversal
\cite{dadush2011enumerative} can be as deep as $|X|$, since collinear
points in a thin body form a path.  Coordinate-wise quantum enumeration
trees \cite{aono2018quantum} are shallow, but their intermediate levels
consist of lattice points in projections of the body, which can far outnumber
$X$; no $2^{O(d)}\kappa_k(d)$ bound on their size is known, and the
standard $d^{O(d)}$ analysis, as in Kannan's algorithm
\cite{kannan1987minkowski}, gives only $n^{O(n)}$ overall.

We instead give each branch a short address in two layers
(Figure~\ref{fig:two-layer-address}).  The first
layer is an ellipsoid of a cover of the projected body by $2^{O(d)}$
translates of an $M$-ellipsoid, each containing at most
$2^{O(d)}\kappa_k(d)$ projected lattice points \cite{dadush2013near}.  The
second picks one point inside the ellipsoid by a linear hash.  Writing
the lattice points as $\Pi_Wx_0+Bz$ with $z\in\mathbb Z^d$, the hash
\begin{equation*}
    r_a(\Pi_Wx_0+Bz)=a^\top z\bmod p,
\end{equation*}
for a prime $p$ and a nonzero vector $a\in\mathbb F_p^d$, splits the
projected lattice into $p$ buckets, each an explicit affine lattice.  If
the hash is injective on the lattice points of the ellipsoid, each bucket
holds at most one of them, and one exact closest-vector computation from
the ellipsoid's center finds that point or shows that there is none
\cite{micciancio2010deterministic,dadush2012integer}.  The residue is then
an address, decoded without listing the other points.  Random sublattices
of this kind, combined with closest-vector computations, are standard
tools of lattice sparsification
\cite{khot2003hardness,dadush2013sparse,dadush2014closest}.  The
difficulty is certifying, without listing an ellipsoid's $\nu$ points,
that the hash is injective on them with modulus at most
$2^{O(d)}(\nu+1)$.  To follow an
individual run, the address range must scale with the points actually
present, which can be far fewer than $\kappa_k(d)$; otherwise empty
residues would dominate the search.

One random setup must serve all $2^{\operatorname{poly}(L)}$ possible
nodes, so a naive union bound over unchecked hashes would call for
exponentially large moduli.  We certify each hash instead.  The hash is
injective exactly when the kernel of its linear part contains no nonzero
difference of two lattice points of the ellipsoid.  We cannot list these
differences, and the lattice points of the difference body can be
arbitrarily more numerous, even when the ellipsoid contains a single
lattice point.  For an ellipsoid containing at least one lattice point,
the parallelogram identity instead places every difference in a set built
from midpoints, with at most $2^{d+1}\nu-1$ elements, and exact closest-
and shortest-vector computations decide whether the kernel avoids its
nonzero points without listing them.  Since $\nu$ is unknown, we try
primes from successive dyadic intervals.  Each pass of the test certifies
injectivity, and with high probability the search stops at a modulus of at
most $2^{O(d)}\nu$.  A point that lies in several cover ellipsoids keeps
only its address in the first of them, its \emph{canonical} address.

Drawing all random choices once, into a setup $\omega$, and
replacing each address range by a balanced binary tree turns the
recursion into one fixed binary tree.
Quantum tree search then needs about the square root of the time
$T_{\rm C}(I,\omega)$ of our classical reference, a depth-first run of the
recursion on the same $\omega$ that decodes each node's addresses in order
and stops at its first verified point.  Early termination matters here.
On a feasible instance the reference may stop long before it has explored
the whole tree, and the tree search of Ambainis and Kokainis pays only
for the part of the tree that depth-first search visits
\cite{ambainis2017quantum}.  On a good setup, $T_{\rm C}(I,\omega)$ is at
most $2^{O(n)}\operatorname{poly}(L,\log(1/\delta))$ times the sum, over
the nodes the reference visits, of one plus the number of lattice points
in their cover ellipsoids, counted with multiplicity, and on infeasible
instances it is at least this sum.

\begin{theorem}[Instance-wise quantum speedup for ILP]
\label{thm:main}
\label{thm:instance-wise}
There is a quantum algorithm that draws a random setup $\omega$, shared with the
classical reference defined above, such that,
for every integer-feasibility instance $I=(K,x_0+\Lambda)$ in dimension
$n$, where $K$ is a polyhedron and $x_0+\Lambda$ an affine lattice, the
following holds.  Let $T_{\rm C}(I,\omega)$ be the reference's running
time in bit operations.  With probability at least $1-\delta$, both
algorithms are correct and the quantum algorithm uses at most
\begin{equation}
    2^{O(n)}\operatorname{poly}\!\left(L,\log\frac1\delta\right)
    \sqrt{T_{\rm C}(I,\omega)}
    \label{eq:instance-wise-bound}
\end{equation}
gates.  It uses at most
$2^{O(n)}\Phi(n)^{n/2}\operatorname{poly}(L,\log(1/\delta))$ gates on every input and setup.
\end{theorem}
Section~\ref{sec:quantum-ip} proves the theorem.  Each local query of
the tree search replays node processing along a root-to-vertex path, at
a cost of $2^{O(n)}\operatorname{poly}(L,\log(1/\delta))$ gates, and this
replay is part of the single-exponential factor
in~\eqref{eq:instance-wise-bound}. The
worst-case bound alone admits a much simpler proof, which amplifies the success
probability of a single random root-to-leaf path of the recursion and
needs neither certified hashes nor the address tree; that route gives no
instance-wise comparison.  Both routes extend to optimization (including
the unbounded case) and to mixed-integer feasibility.

Since $r=\Omega(\log n)$ is necessary \cite{kannan1988covering}, no
worst-case factor $r$ brings the classical recursion's bound
$2^{O(n)}r^n\operatorname{poly}(L)$ below $2^{O(n)}(\log(2n))^n$, a
bound our quantum algorithm already attains in the worst case.  Since
$\log\Phi(n)^n=\Theta(n\log\log(4n))$, the factor $2^{O(n)}$ costs only
$O(1/\log\log(4n))$ in the exponent: the worst-case gate bound is
$\Phi(n)^{n(1/2+O(1/\log\log(4n)))}\operatorname{poly}(L,\log(1/\delta))$.
A proof
of the subspace flatness conjecture whose subspaces can be found as in
Theorem~\ref{thm:constructive-flatness} would give
$2^{O(n)}(\log(2n))^n\operatorname{poly}(L)$ classically, matching our
quantum bound, and $2^{O(n)}(\log(2n))^{n/2}\operatorname{poly}(L)$
quantumly. For bounded equation form~\eqref{e:bounded_eq}, DER tests explicitly
indexed cells in order, so tree search over the cell indices uses
$2^{O(n)}\Psi(u)^{1/2}\operatorname{poly}(L)$ gates, with the same
instance-wise comparison. 

The rest of this paper is organized as follows. Section~\ref{sec:preliminaries} fixes the input and computation model and
collects the lattice and convex-geometric facts we use.
Section~\ref{sec:classical-flatness} outlines the proof of
Theorem~\ref{thm:constructive-flatness} and compares the construction
with the filtration route.  Section~\ref{sec:search} constructs the
addresses that give tree search local access.
Section~\ref{sec:quantum-ip} builds the classical recursion, turns it
into one searchable binary tree, outlines the proof of Theorem~\ref{thm:main}, and derives
the worst-case, optimization, and bounded-equation-form results, and
Section~\ref{sec:discussion} discusses limits and open problems.  The
appendices prove the indexable-cover lemma
(Appendix~\ref{app:indexable-cover-proof}), implement the classical
geometric subroutines in finite precision
(Appendix~\ref{app:finite-geometric-primitives}), give the recursive
encoding of nodes, the circuit precision, and the standard reductions to
optimization and mixed-integer feasibility
(Appendix~\ref{app:recursion-details}), verify the DER cells and prove the
bounded-equation-form results (Appendix~\ref{app:der-cell-test}), extend
the comparison to priority-driven cover enumeration
(Appendix~\ref{subsec:priority-cover}), and implement the quantum
recursion with its deferred proofs
(Appendix~\ref{app:finite-quantum-recursion}).

The main text
gives the idea of every proof. The key appendices are
Appendix~\ref{app:flatness-proofs}, which completes the proof of
Theorem~\ref{thm:constructive-flatness}, and
Appendix~\ref{app:hash-proofs}, which proves the hashing results of
Section~\ref{sec:sparsification}.  The others make the proofs finite,
capped, and reversible, extend the comparison, or give standard
reductions, and can be skipped on a first reading;
Appendix~\ref{subsec:random-path} proves the worst-case quantum bound
without the address tree.

\section{Preliminaries}
\label{sec:preliminaries}

We fix the input and cost model and the lattice notation, including the
exact closest-vector solver, and then record the determinant and Gaussian
estimates used in the proof of Theorem~\ref{thm:constructive-flatness}.

\subsection{Input and computation model}
\label{subsec:input-model}

An instance of integer feasibility is a pair $I=(K,x_0+\Lambda)$ of a
rational polyhedron $K\subseteq\mathbb R^n$ and a rational full-rank affine
lattice $x_0+\Lambda=x_0+B\mathbb Z^n$, with $x_0\in\mathbb Q^n$ and a
nonsingular $B\in\mathbb Q^{n\times n}$.  The task is to find a point of
$K\cap(x_0+\Lambda)$ or to report that there is none.
Problem~\eqref{eq:ilo} optimizes over $K=\{x:Ax\leq b\}$ and
$\mathbb Z^n$; Corollaries~\ref{cor:mixed-integer-linear-feasibility}
and~\ref{cor:integer-linear-optimization} reduce mixed-integer feasibility
and optimization to the feasibility algorithm.  The encoding length $L$
counts the bits of a binary encoding of all dimensions and of the
numerators and denominators of all rational data, so $n\leq L$.  We write
$\delta\in(0,1/2)$ for a failure probability and measure classical running
time in bit operations.

The bodies met by the recursion are projections and slices of polyhedra,
so we present every convex set by a \emph{rational lift}
\begin{equation*}
    D=\{x\in\mathbb R^d:A_Dx+F_Dw\leq b_D\text{ for some }w\in\mathbb R^{d'}\}
\end{equation*}
with rational $A_D$, $F_D$, and $b_D$.  The encoding length of $D$ is that
of $(A_D,F_D,b_D)$, and a polyhedron is its own lift with $d'=0$.  Rational
linear programming over the lift decides membership of rational points
and maximizes rational linear functions over $D$: it reports that $D$ is
empty or that the objective is unbounded above on $D$ when either occurs,
and otherwise returns an exact rational maximizer.  If $D$ is nonempty, a
Farkas certificate separates a rational point outside $D$ from $D$ by a
rational hyperplane.  All these operations are exact and take polynomial
time \cite{grotschel2012geometric,schrijver1998theory}.

Lifts are closed under
the operations of the recursion.  For the orthogonal projection $\Pi_W$
onto a rational subspace $W$, the projection $\Pi_WD$ and the fiber
$D\cap\Pi_W^{-1}(y)$ over a rational point $y$ are lifts obtained by
appending variables and linear equalities, and a shrunken body
$c+\theta(D-c)$ with rational $c$ and $\theta>0$ is a lift obtained by an
affine change of variables.  In each case the encoding length grows only
polynomially.

The recursion works in the coefficient coordinates of the root lattice.
Standard small-solution bounds give a box $[-M,M]^n$ with
$M\leq2^{\operatorname{poly}(L)}$ that preserves
feasibility~\eqref{eq:root-compactification}.  After each restriction to a
hyperplane or fiber, Lemma~\ref{lem:rank-preserving-frank-tardos} replaces
the affine subspace of the node by one of the same dimension that has the
same integer points in the box and a $\operatorname{poly}(L)$-bit
description.  Nodes and returned witnesses therefore have
$\operatorname{poly}(L)$-bit encodings.  Auxiliary data computed by
randomized routines, such as projection parameters and cover descriptions,
have $\operatorname{poly}(L,\log(1/\delta))$ bits, and every subroutine is
charged for the actual encoding length of its input.
Appendix~\ref{app:recursion-details} gives the details.

A circuit spends a fixed, finite number of gates on every input, so every
classical routine that we run inside one is \emph{capped}.  Its loops,
registers, and random tape have explicit bounds, fixed in advance as
functions of its input length; it raises a failure flag and halts on
overflow or timeout; and it is padded to its cap.  With its random tape
fixed, a capped routine is a deterministic Boolean computation, and keeping
the tape and the computation history until uncomputation gives reversible
circuits for the routine and its inverse with polynomial overhead
\cite{bennett1989time}.  Appendix~\ref{app:finite-geometric-primitives}
gives capped implementations of the geometric subroutines we use.

Quantum gate counts are over a fixed finite universal gate set closed under
inverses and include all arithmetic.  We assume no quantum random-access
memory.  The input, the shared setup $\omega$ of random choices
(Section~\ref{subsec:fixed-tree}), and all work data are held in ordinary
registers, and every access to them is a circuit whose gates are counted.
Non-native phase rotations are synthesized to an explicit error budget
(Appendix~\ref{subsec:phase-synthesis}).  Every returned point is verified
exactly in rational arithmetic, so no approximation error can make an
algorithm output an invalid witness.

\subsection{Lattice geometry and closest vectors}
\label{subsec:lattice-preliminaries}

This subsection fixes the lattice notation, defines the covering radius
by which we measure flatness, and records the exact closest-vector
solver that decodes the addresses of Section~\ref{sec:sparsification}.
A \emph{lattice} of rank $r$ is a set $\Lambda=B\mathbb Z^r$, where the
columns of $B\in\mathbb R^{m\times r}$, a \emph{basis} of $\Lambda$, are
linearly independent.  It is \emph{full-rank} in a linear subspace $V$ if
$\operatorname{span}\Lambda=V$, and \emph{rational} if it has a rational
basis.  Its determinant $\det\Lambda:=\det(B^\top B)^{1/2}$ does not depend
on the basis.  An \emph{affine lattice} is a translate $x_0+\Lambda$.  The
\emph{dual lattice}
\begin{equation*}
    \Lambda^*:=\{y\in\operatorname{span}\Lambda:\langle y,v\rangle\in\mathbb Z
       \text{ for every }v\in\Lambda\}
\end{equation*}
has basis $B(B^\top B)^{-1}$ and determinant $1/\det\Lambda$
\cite[Definitions~3.6--3.7]{dadush2019approximating}; applying the basis
formula twice gives $(\Lambda^*)^*=\Lambda$.

A linear subspace $W$ is a \emph{lattice subspace} of $\Lambda$ if it is
spanned by vectors of $\Lambda$, that is, if
$W=\operatorname{span}(\Lambda\cap W)$
\cite[Definition~3.8]{dadush2019approximating}.  For a subspace
$W\subseteq\operatorname{span}\Lambda$, the projection $\Pi_W\Lambda$ is a
lattice if and only if $W$ is a lattice subspace of $\Lambda^*$
\cite[Lemma~3.9]{dadush2019approximating}, and the flat subspaces of Theorem~\ref{thm:constructive-flatness} are of this
kind.  If $W$ is a lattice subspace of $\Lambda^*$ of dimension $d$, then
$\Lambda\cap W^\perp$ is a sublattice of rank $r-d$, primitive in the sense
of Section~\ref{subsec:normalized-determinants}, and $\Pi_W\Lambda$ is a
lattice of rank $d$ with
\begin{equation*}
    (\Pi_W\Lambda)^*=\Lambda^*\cap W
\end{equation*}
\cite[Lemmas~3.9--3.10]{dadush2019approximating}, applied to the lattice
subspace $W^\perp\cap\operatorname{span}\Lambda$ of $\Lambda$. Note that $\Pi_W\Lambda$ is
the quotient $\Lambda/(\Lambda\cap W^\perp)$, so
$\det\Lambda=\det(\Lambda\cap W^\perp)\det(\Pi_W\Lambda)$.

If $\Lambda$ is
rational, then so is every lattice subspace $W$ of $\Lambda^*$, the matrix
of $\Pi_W$ is rational, and lattice normal forms compute bases of
$\Pi_W\Lambda$ and $\Lambda\cap W^\perp$ in polynomial time
\cite{schrijver1998theory}.  For an affine lattice,
$\Pi_W(x_0+\Lambda)=\Pi_Wx_0+\Pi_W\Lambda$ is an affine lattice in $W$, and
the recursion branches on its points in a projected body.

For a lattice $\Lambda$ and a convex body
$K\subseteq\operatorname{span}\Lambda$, compact with nonempty interior
relative to $\operatorname{span}\Lambda$, the \emph{covering radius}
\begin{equation*}
    \mu(\Lambda,K):=\min\{t\geq0:\Lambda+tK=\operatorname{span}\Lambda\}
\end{equation*}
is the least $t$ such that every translate of $tK$ within
$\operatorname{span}\Lambda$ meets $\Lambda$ \cite{kannan1988covering}.  It
does not change when $K$ is translated or when one invertible linear map
is applied to both arguments, and $\mu(\Lambda,sK)=\mu(\Lambda,K)/s$ for
$s>0$.  If $\mu(\Lambda,K)\leq1$, then $K$ meets every affine lattice
$x_0+\Lambda$ with $x_0\in\operatorname{span}\Lambda$.  Projection does not
increase the covering radius: if $W\subseteq\operatorname{span}\Lambda$ is a
lattice subspace of $\Lambda^*$, then
$\mu(\Pi_W\Lambda,\Pi_WK)\leq\mu(\Lambda,K)$, because
$\Lambda+tK=\operatorname{span}\Lambda$ implies
$\Pi_W\Lambda+t\,\Pi_WK=W$.

For a compact set
$D\subseteq\operatorname{span}\Lambda$, the largest number of lattice
points in a translate of $D$,
\begin{equation}
    \nu_{\max}(\Lambda,D):=
    \max_{t\in\operatorname{span}\Lambda}|(D+t)\cap\Lambda|
    \label{eq:G-parameter}
\end{equation}
is finite because $D$ is bounded and $\Lambda$ is discrete.  It bounds the
number of points of every affine lattice $x_0+\Lambda$ with
$x_0\in\operatorname{span}\Lambda$ in $D$, since
$D\cap(x_0+\Lambda)=x_0+((D-x_0)\cap\Lambda)$.  Both $\mu$ and
$\nu_{\max}$ are invariant under translating the body, so for a body in a
translate $a+\operatorname{span}\Lambda$, as at the nodes of the recursion,
we evaluate them on the body minus $a$.

Closest vectors are taken in ellipsoidal norms.  Let $x_0+B\mathbb Z^d$ be
a rational full-rank affine lattice in $\mathbb R^d$, let $Q$ be a rational
positive definite $d\times d$ matrix, and let $t\in\mathbb Q^d$ be a
target.  An \emph{exact closest vector} is a point
\begin{equation}
    y^\star\in\operatorname*{arg\,min}_{y\in x_0+B\mathbb Z^d}
       (y-t)^\top Q(y-t),
    \label{eq:cvp-definition}
\end{equation}
and an exact shortest vector minimizes $y^\top Qy$ over
$B\mathbb Z^d\setminus\{0\}$.  Moving the target to $t-x_0$ reduces the
affine case to the linear one.  The Voronoi-cell algorithm of Micciancio
and Voulgaris solves both problems deterministically in
$2^{O(d)}\operatorname{poly}(L)$ time and space, where $L$ is the encoding
length of $(x_0,B,Q,t)$ \cite{micciancio2010deterministic},
\cite[Theorem~5.2.2]{dadush2012integer}.  The optimal values are rational,
so comparisons with rational thresholds are exact, and Bennett simulation
turns the solver into reversible circuits with
$2^{O(d)}\operatorname{poly}(L)$ gates.  Section~\ref{sec:sparsification}
decodes each address of a branch with one such computation.

\subsection{Normalized determinants and duality}
\label{subsec:normalized-determinants}

The proof of Lemma~\ref{lem:rank-weighted-dual-density} compares
sublattices of different ranks and passes between a lattice and its
dual, and the filtration route of
Section~\ref{subsec:filtration-route} is phrased in terms of stability.
This subsection collects the definitions both use.
Duals and orthogonal complements are taken inside the span of the lattice.
A sublattice $P\subseteq\Lambda$ is primitive if
$P=\Lambda\cap\operatorname{span}P$.  Its saturation
$\Lambda\cap\operatorname{span}P$ has the same rank and no larger
determinant.  For primitive $P$, the quotient lattice is
$\Lambda/P:=\Pi_{(\operatorname{span}P)^\perp}\Lambda$, with
$\det\Lambda=\det P\det(\Lambda/P)$
\cite[Lemmas~3.9--3.11]{dadush2019approximating}.

Determinants of different-rank sublattices cannot be compared directly.
For a nonzero lattice $\Lambda$, define its \emph{normalized determinant} and
minimum normalized determinant by
\begin{equation*}
    \operatorname{nd}(\Lambda):=\det(\Lambda)^{1/\operatorname{rank}\Lambda},
    \qquad
    \operatorname{nd}_{\min}(\Lambda):=
        \min_{0\neq W\subseteq\Lambda}\operatorname{nd}(W).
\end{equation*}
The normalized determinant measures a length scale, so smaller values mean
denser sublattices.  We use the convention $\det\{0\}=1$.

A lattice $\Lambda$ is \emph{stable} if every nonzero sublattice
$W\subseteq\Lambda$ satisfies $\operatorname{nd}(W)\geq\operatorname{nd}(\Lambda)$.
Stability is invariant under scaling.  After rescaling by
$1/\operatorname{nd}(\Lambda)$ to determinant one, it says that every
nonzero sublattice has determinant at least one.  Determinant one alone
is not enough: $\tfrac12\mathbb Ze_1\oplus2\mathbb Ze_2$ has
determinant one but contains the rank-one sublattice
$\tfrac12\mathbb Ze_1$, whose normalized determinant is $1/2$.

If $N_2$ is a primitive
rank-$j$ sublattice of the dual of a rank-$s$ lattice $N$, then its
annihilator $N_1:=N\cap(\operatorname{span}N_2)^\perp$ has rank $s-j$ and
\begin{equation}
    \det N_2=\frac{\det N_1}{\det N}.
    \label{eq:primitive-determinantal-duality}
\end{equation}
By this determinantal duality, lower bounds on determinants in $N$ become lower bounds on determinants
in $N^*$ after dividing by $\det N$
\cite[Lemma~3.14]{dadush2019approximating}.

\subsection{Gaussian mass and convex-body estimates}
\label{subsec:gaussian-geometric-preliminaries}

Our flatness construction certifies covering through Gaussian mass,
which can be estimated by sampling, and compares that mass with volume
through estimates for convex bodies in near-isotropic position.  This
subsection collects both kinds of facts.
For a convex body $D$ containing the origin in its relative interior,
its gauge on $V=\operatorname{span}D$ is
\begin{equation*}
 \|x\|_D:=\inf\{t>0:x\in tD\},\qquad x\in V.
\end{equation*}
The gauge is convex and positively homogeneous; it is a norm when
$D=-D$, but need not be symmetric otherwise.
Here and below, a symmetric body means an origin-symmetric body.
The symbol $G$ denotes a standard centered Gaussian with identity
covariance on the Euclidean space currently under consideration,
including on a specified subspace when appropriate.
We use $C_{\mathrm{iso}}$ for the universal constant in
Fact~\ref{fact:isotropic-mean-norm} below.
For a discrete set $A$, write
$\rho(A):=\sum_{x\in A}e^{-\pi\|x\|_2^2}$.
The centered discrete Gaussian $D_{\Lambda,s}$ on a lattice $\Lambda$, with parameter
$s>0$, gives $v\in\Lambda$ probability
\begin{equation*}
    \Pr_{X\sim D_{\Lambda,s}}[X=v]
    =\frac{e^{-\pi\|v\|_2^2/s^2}}{\rho(\Lambda/s)}.
\end{equation*}
In particular, zero has probability $1/\rho(\Lambda/s)$.  This identity lets
us estimate a Gaussian mass by counting zero samples, rather than summing
over an infinite lattice.
Our descent samples from $D_{\mathcal S^*,1/\sigma}$, whose zero probability
is $1/\rho(\sigma\mathcal S^*)$.  For a fixed lattice $\mathcal S$,
increasing $\sigma$ narrows this distribution and makes zero more likely.
The descent, however, decreases $\sigma$ as the rank falls, and so
requires small dual mass at progressively smaller geometric scales.

Minkowski's first theorem says that a symmetric convex body
$C\subseteq\operatorname{span}\Lambda$ of volume greater than $2^d\det\Lambda$
contains a nonzero vector of a rank-$d$ lattice $\Lambda$
\cite{schrijver1998theory}.  Reverse Minkowski,
Fact~\ref{fact:reverse-minkowski}, controls the collective contribution of
short vectors from determinants of \emph{all} sublattices, not just the
determinant of $\Lambda$ itself.

The two estimates behind our bound,
Facts~\ref{fact:reverse-minkowski} and~\ref{fact:isotropic-mean-norm},
are stated in Section~\ref{sec:flatness-core}, where they are used; we
check here that they follow directly from the cited results.  If
$\operatorname{nd}_{\min}(\Lambda)\geq\xi(m)$, every sublattice of
$\xi(m)^{-1}\Lambda$ has determinant at least one, so this lattice is
weakly stable in the sense of Letwin, Reis, and Rothvoss.  Their spectral
bound \cite[Corollary~4.4 and proof of Theorem~1.7]{letwin2026strongly},
applied in $\operatorname{span}\Lambda$ with parameter $s=1/\xi(m)$ and
$r=\operatorname{rank}\Lambda\leq m$, gives
\begin{equation*}
    \rho(\Lambda)\leq\bigl(1-e^{-\xi(m)^2/(2e)}\bigr)^{-r}
    \leq\Bigl(1-\frac1{3r}\Bigr)^{-r}\leq\frac32,
\end{equation*}
which is Fact~\ref{fact:reverse-minkowski}.
Fact~\ref{fact:isotropic-mean-norm} is the sharp form of the
Bizeul--Klartag mean-norm estimate \cite{bizeul2025distances}, recorded in
\cite[Theorem~1.4]{bizeul2026optimal} using the newer dimension-free
bound on $\kappa_m$.  For $m\geq2$ that theorem states
$\mathbb E\|\theta\|_K\leq C\sqrt{\log m/m}$ for a uniform unit vector
$\theta$, which gives the fact because
$\mathbb E\|G\|_K=\mathbb E\|G\|_2\,\mathbb E\|\theta\|_K\leq\sqrt m\,
\mathbb E\|\theta\|_K$.  For $m=1$ the body is $[-\sqrt3,\sqrt3]$.
Both estimates are recent.  The earlier reverse Minkowski threshold
$O(\log(2m))$ \cite[Definition~2.12 and Theorem~2.13]{dadush2019approximating}
and Pisier's $\ell$-position give the weaker variant of
Appendix~\ref{subsec:published-variant}.

The connection to hyperplane descent is a Gaussian projection inequality.
If $N\subseteq\Lambda$ is primitive, then
$\Pi_{\operatorname{span}N}(\Lambda^*)=N^*$, and
\begin{equation}
    \Pr_{v\sim D_{\Lambda^*,1/\sigma}}[v\perp N]
    \geq\frac1{\rho(\sigma N^*)}.
    \label{eq:gaussian-projection-preservation}
\end{equation}
The Gaussian mass of a lattice is at most the mass of its projection times
that of its kernel, and dividing by the total mass gives the displayed
probability bound
\cite[Lemma~3.31]{dadush2019approximating}.
Thus a small Gaussian mass on $N^*$ makes a random dual direction likely to
be orthogonal to $N$, and the corresponding hyperplane cut then keeps $N$
inside the current lattice.

Each stage of the proof of Theorem~\ref{thm:constructive-flatness} uses
the following bounds.

\begin{lemma}[Approximately isotropic cores]
\label{lem:isotropic-core}
Let $K\subseteq\mathbb R^m$ be a convex body whose uniform distribution has
barycenter $b$ and covariance $\Sigma$.  Let $\widetilde x$ satisfy
$\|b-\widetilde x\|_{K-K}\leq1/(2(m+1))$, put
$C:=(K-\widetilde x)\cap(\widetilde x-K)$, and let $T$ be invertible with
$\frac12I_m\preceq T\Sigma T^\top\preceq2I_m$.  Then
\begin{equation}
    \operatorname{Vol}_m(TK)^{1/m}\leq\sqrt{4\pi e},
    \qquad
    \mathbb E\|G\|_{TC}\leq4\sqrt2\,C_{\mathrm{iso}}\sqrt{\log(2m)}.
    \label{eq:isotropic-core}
\end{equation}
\end{lemma}

\begin{proof}
The volume bound compares the uniform distribution on the body with the
Gaussian of the same mean and covariance, and the norm bound reduces to
Fact~\ref{fact:isotropic-mean-norm}.
Put $K':=T(K-\widetilde x)$, so that $TC=K'\cap(-K')$, and let
$b':=T(b-\widetilde x)$ and $\Sigma':=T\Sigma T^\top$ be the barycenter
and covariance of $K'$.  Let $u$ be the uniform density on $K'$ and
$\varphi$ the Gaussian density with mean $b'$ and covariance $\Sigma'$.
Since $u$ and $\varphi$ have the same first two moments, nonnegativity of
relative entropy gives
\begin{equation*}
    0\leq\int u\log\frac u\varphi
    =\frac12\log\bigl((2\pi e)^m\det\Sigma'\bigr)
       -\log\operatorname{Vol}_m(K').
\end{equation*}
With $\det\Sigma'\leq2^m$ and $\operatorname{Vol}_m(TK)=\operatorname{Vol}_m(K')$,
this proves the volume bound.

For the norm bound, $\|x\|_{TC}=\max\{\|x\|_{K'},\|-x\|_{K'}\}$, so
$\mathbb E\|G\|_{TC}\leq2\,\mathbb E\|G\|_{K'}$.  The barycenter
inequality $-(K'-b')\subseteq m(K'-b')$ \cite[Section~5]{reis2023subspace}
gives $K'-K'\subseteq(m+1)(K'-b')$.  The hypothesis on $\widetilde x$ is
invariant under $T$, so $-b'\in\frac12(K'-b')$; hence
$\frac12(K'-b')\subseteq K'$ and $\|x\|_{K'}\leq2\|x\|_{K'-b'}$.  Finally,
$K_0:=\Sigma'^{-1/2}(K'-b')$ is isotropic, and $\Sigma'^{-1/2}G$ is a
centered Gaussian with covariance $\Sigma'^{-1}\preceq2I_m$.  Writing
$\sqrt2G$ in law as $\Sigma'^{-1/2}G+H$ with an independent centered
Gaussian $H$, Jensen's inequality and
Fact~\ref{fact:isotropic-mean-norm} give
\begin{equation*}
    \mathbb E\|G\|_{K'-b'}
    =\mathbb E\|\Sigma'^{-1/2}G\|_{K_0}
    \leq\sqrt2\,\mathbb E\|G\|_{K_0}
    \leq\sqrt2\,C_{\mathrm{iso}}\sqrt{\log(2m)}.
\end{equation*}
The three factors $2$, $2$, and $\sqrt2$ give~\eqref{eq:isotropic-core}.
\end{proof}

\section{Constructive subspace flatness}
\label{sec:classical-flatness}
\label{sec:flatness-core}
\label{app:rank-weighted-flatness}

This section outlines the proof of Theorem~\ref{thm:constructive-flatness},
whose full proof is in Appendix~\ref{app:finite-geometric-primitives}.
Section~\ref{subsec:subspace-flatness} sets up a stage-by-stage covering
argument in which smoothness, a Gaussian-mass condition that can be
tested by sampling, takes the place of stability, and ends with the
outline of the proof.  Section~\ref{subsec:rank-weighted-descent} states
the three lemmas behind this argument and proves the first two;
Appendix~\ref{app:finite-geometric-primitives} proves the third.
Appendix~\ref{subsec:published-variant} derives an $O(\log^3(2k))$ variant
from the estimates of Reis and Rothvoss, and
Section~\ref{subsec:filtration-route} compares our route with their
filtration route.  Section~\ref{subsec:classical-recursion} places the
theorem in the classical recursion.

\subsection{Subspace flatness}
\label{subsec:subspace-flatness}

The proof of Theorem~\ref{thm:constructive-flatness} splits the lattice
into pieces, one stage at a time.  A stage works with a lattice
$\mathcal L$ of rank $m$, initially $\Lambda$.  It finds a primitive
sublattice $\mathcal S\subseteq\mathcal L$ of rank $q$, bounds the
covering radius of $\mathcal S$ in the section of the body through
$\operatorname{span}\mathcal S$, and continues with the projection of
$\mathcal L$ orthogonal to $\mathcal S$.  The covering radius of
$\Lambda$ is at most the sum of these bounds
\cite[Lemmas~40--41]{reis2023subspace}.

Reis and Rothvoss make each piece stable.  We ask only for what the
covering bound uses: each piece should be \emph{smooth}, meaning that its
dual has small Gaussian mass at a given scale (defined below).  Our key
change is to weight this scale by rank.  We require smoothness at a scale
proportional to $q/m$, so each stage is charged in proportion to the
fraction of the rank that it removes.  Over the at most $k$ stages, these
fractions sum to at most $1+\log k$ however the ranks split, as in the
proof of \cite[Lemma~6.3]{dadush2019approximating}, whereas a fixed scale
would charge every stage in full.
Lemmas~\ref{lem:gaussian-covering-certificate}--\ref{lem:rank-weighted-smooth-sublattice}
in Section~\ref{subsec:rank-weighted-descent} turn smoothness into a
covering bound, show that some sublattice is smooth at this scale, and
find one.

In the geometric statements of this section, ambient dimensions and the
ranks of input lattices are positive unless we say otherwise.  A residual
quotient of rank zero ends the construction.

The finite proof in Appendix~\ref{app:finite-geometric-primitives} also
uses the following data.  Rational linear programming finds an inner
simplex and an enclosing coordinate box for $K$.  As in the proof of
Lemma~\ref{lem:rational-indexed-cover}, these supply rational inner
and outer radius data of polynomial bit length.  Following
\cite{kannan1988covering}, write
$\mu_{\mathrm{KL}}(\Lambda,K):=\max_W R_W$, with the maximum over
nonzero lattice subspaces of $\Lambda^*$.

We use $\Lambda$ for an ambient lattice or the fixed input to a lattice
subroutine, $\mathcal L$ for a residual quotient lattice in the flatness
construction, and $\mathcal S$ for a sublattice being analyzed or
constructed.  We reserve $\Gamma$ for the projected lattice used to index
branches.  These roles separate the input, which stays fixed during a
descent attempt, from the sublattice, which changes after each cut.

For $\sigma>0$, call a lattice $\mathcal S$ \emph{$\sigma$-smooth} if
$\rho(\sigma\mathcal S^*)\leq3/2$; equivalently, $\sigma$ is at least the
smoothing parameter $\eta_{1/2}(\mathcal S)$ of Micciancio and Regev
\cite{micciancio2007worst}.  By Poisson summation, the Gaussian with
density proportional to $e^{-\pi\|x\|_2^2/\sigma^2}$ then has density at
most $3/2$ times uniform modulo $\mathcal S$.  Unlike stability,
smoothness can be tested by sampling with high confidence, up to a constant gap: a discrete Gaussian sample from
$\mathcal S^*$ with parameter $1/\sigma$ is zero with probability exactly
$1/\rho(\sigma\mathcal S^*)$.

The covering bound below is the product of two quantities: the scale at
which a piece is smooth, and the Gaussian mean norm of the body.  The
following estimates control them.

\begin{fact}[{Reverse Minkowski \cite{letwin2026strongly}}]
\label{fact:reverse-minkowski}
Let $\xi(m):=\lceil\sqrt{2e\log(3m)}\,\rceil$.  Then
\begin{equation}
    \operatorname{nd}_{\min}(\Lambda)\geq\xi(m)
    \quad\Longrightarrow\quad\rho(\Lambda)\leq3/2
    \qquad(\operatorname{rank}\Lambda\leq m).
    \label{eq:reverse-minkowski-threshold}
\end{equation}
\end{fact}

\begin{fact}[{Bizeul--Klartag \cite{bizeul2025distances}, \cite[Theorem~1.4]{bizeul2026optimal}}]
\label{fact:isotropic-mean-norm}
There is a universal $C_{\mathrm{iso}}$ such that every convex body
$K\subseteq\mathbb R^m$ whose uniform distribution has mean zero and
covariance $I_m$ satisfies
$\mathbb E\|G\|_K\leq C_{\mathrm{iso}}\sqrt{\log(2m)}$.
\end{fact}

We can now outline the proof of Theorem~\ref{thm:constructive-flatness};
Appendix~\ref{app:finite-geometric-primitives} gives the full proof.  At a
stage with residual subspace $W$ of dimension $m$, put
$\mathcal L:=\Pi_W\Lambda$, $K_{\mathcal L}:=\Pi_WK$, and
$C:=(K_{\mathcal L}-\widetilde x)\cap(\widetilde x-K_{\mathcal L})$ for an
estimated barycenter $\widetilde x$.  A rational map $T$ putting
$K_{\mathcal L}$ in approximately isotropic position gives
$\operatorname{Vol}_m(TK_{\mathcal L})^{1/m}=O(1)$ and
$\mathbb E\|G\|_{TC}=O(\sqrt{\log(2m)})$
(Fact~\ref{fact:isotropic-mean-norm} and Lemma~\ref{lem:isotropic-core}).
We apply Lemma~\ref{lem:rank-weighted-smooth-sublattice} to $T\mathcal L$ and
pull the output back to a rank-$q$ sublattice $\mathcal S$.  As covering
radii are invariant under $T$ and
$\operatorname{nd}(T\mathcal L)=R_W\operatorname{Vol}_m(TK_{\mathcal L})^{1/m}$,
Lemma~\ref{lem:gaussian-covering-certificate} gives
\begin{equation*}
    \mu(\mathcal S,C\cap\operatorname{span}\mathcal S)
    \leq O(\xi(m))\frac qm\operatorname{nd}(T\mathcal L)\,
       \mathbb E\|G\|_{TC}
    \leq O(\log(2m))\frac qmR_W.
\end{equation*}

The next stage works in $W\cap(\operatorname{span}\mathcal S)^\perp$.  For
stages $W_i$ of dimensions $m_i$, summing over the stages and using
$\sum_i(m_i-m_{i+1})/m_i\leq1+\log k$ gives
\begin{equation*}
    \mu(\Lambda,K)\leq O(\log(2k))\sum_i\frac{m_i-m_{i+1}}{m_i}R_{W_i}
       \leq O(\log^2(2k))\max_iR_{W_i},
\end{equation*}
and the algorithm returns the $W_i$ with the largest estimate of
$R_{W_i}$.

The same chain also gives the existential bound.  With the sublattice $N$ of
Lemma~\ref{lem:rank-weighted-dual-density} and the covering theorem
\cite[Theorem~1.2]{letwin2026strongly} in place of $\mathcal S$ and
Lemma~\ref{lem:gaussian-covering-certificate}, $O(\xi(m))$ becomes $O(1)$,
recovering the existential $O(\log^{3/2}(2k))$ bound of Letwin, Reis, and
Rothvoss without a stable filtration (Section~\ref{subsec:filtration-route}); the
additional $\sqrt{\log(2k)}$ in the constructive bound comes from our
Gaussian-mass certificate.

The full proof in Appendix~\ref{app:finite-geometric-primitives} follows
this outline.  It adds what the outline leaves out: rational position
maps, the barycenter, covariance, and volume estimates, encoding bounds
for the quotient lattices, and the allocation of failure probabilities.

\subsection{Rank-weighted Gaussian descent}
\label{subsec:rank-weighted-descent}

This subsection states the three lemmas announced in
Section~\ref{subsec:subspace-flatness} in order and proves the first two;
Appendix~\ref{app:finite-geometric-primitives} proves the third and Theorem~\ref{thm:constructive-flatness}.  Small dual Gaussian mass implies
a covering bound in any symmetric convex body
(Lemma~\ref{lem:gaussian-covering-certificate}).  A rank-penalized
determinant provides a hidden sublattice compatible with the
rank-dependent scale (Lemma~\ref{lem:rank-weighted-dual-density}),
without a canonical filtration.  A descent with a zero-frequency stopping
test then constructs a sublattice satisfying that mass bound
(Lemma~\ref{lem:rank-weighted-smooth-sublattice}).

\begin{lemma}[Gaussian-mass covering certificate]
\label{lem:gaussian-covering-certificate}
Let $\mathcal S$ have rank $q$, and let $C$ be a symmetric convex body in
its span.  If $\mathcal S$ is $\sigma$-smooth, then
\begin{equation}
    \mu(\mathcal S,C)\leq\frac{8\sigma}{\sqrt{2\pi}}\,\mathbb E\|G\|_C.
    \label{eq:gaussian-mass-covering}
\end{equation}
If $C$ is a full-dimensional symmetric body in a larger Euclidean space,
the same bound holds for $C\cap\operatorname{span}\mathcal S$ with the
ambient Gaussian on the right.
\end{lemma}

\begin{proof}
Let $\pi$ be the quotient map onto the torus
$\operatorname{span}\mathcal S/\mathcal S$, with normalized Haar measure
$\lambda$, and put $Y:=\sigma G/\sqrt{2\pi}$.  By Poisson summation, the
density of $\pi(Y)$ relative to $\lambda$ is at most
$\rho(\sigma\mathcal S^*)\leq3/2$.  For
$r>4\sigma\,\mathbb E\|G\|_C/\sqrt{2\pi}$, Markov's inequality gives
$\Pr[Y\in rC]>3/4$, so $\lambda(\pi(rC))>1/2$.  For every torus element $t$,
symmetry gives $t-\pi(rC)$ the same measure, so it meets $\pi(rC)$ and
$t\in\pi(2rC)$.  Hence $\mathcal S+2rC=\operatorname{span}\mathcal S$, and
letting $r$ decrease proves~\eqref{eq:gaussian-mass-covering}.  For the
ambient assertion, the projection of a standard Gaussian onto
$\operatorname{span}\mathcal S$ is standard there, and Jensen's
inequality, applied conditionally on this projection, bounds its gauge in
$C\cap\operatorname{span}\mathcal S$ by $\mathbb E\|G\|_C$.
\end{proof}

Every lattice $\Lambda$ of rank $m$ has a sublattice that is smooth at the
rank-weighted scale.  Put $g:=\operatorname{nd}(\Lambda)$ and let $N$ minimize the rank-penalized
determinant $\mathcal D(N):=\det N/((3g/m)^ss!)$, where
$s=\operatorname{rank}N$, over all primitive sublattices of $\Lambda$,
including $\{0\}$.  Determinantal duality turns the minimality of $N$
into a lower bound on the normalized determinant of every sublattice
of $N^*$.  The weight
$3g/m$ contributes the factor $m$ in the resulting bound, while the factorial keeps
$N$ nonzero, since $\mathcal D(\Lambda)\leq(e/3)^m<1=\mathcal D(\{0\})$,
at the cost of a factor $s$.

\begin{lemma}[A rank-weighted dual-density bound]
\label{lem:rank-weighted-dual-density}
Every lattice $\Lambda$ of rank $m\geq1$, with
$g=\operatorname{nd}(\Lambda)$, has a nonzero primitive sublattice $N$ of
rank $s$ satisfying
\begin{equation}
    \operatorname{nd}_{\min}(N^*)\geq\frac{m}{3sg}.
    \label{eq:rank-weighted-dual-density}
\end{equation}
\end{lemma}

\begin{proof}
Put $a:=3g/m$.  Among the primitive sublattices of $\Lambda$, including
zero, choose $N$ minimizing the rank-penalized determinant
\begin{equation*}
    \mathcal D(N):=\frac{\det N}{a^r r!},
    \qquad r=\operatorname{rank}N.
\end{equation*}
At each rank, only finitely many primitive sublattices have determinant
below any fixed bound, so the minimum is attained.  With $\det\{0\}=0!=1$,
\begin{equation*}
    \mathcal D(\Lambda)=\frac{(m/3)^m}{m!}
       \leq(e/3)^m<1=\mathcal D(\{0\}).
\end{equation*}
Thus $s:=\operatorname{rank}N\geq1$.

Fix a primitive rank-$j$ sublattice $N_2\subseteq N^*$ and its annihilator
$N_1:=N\cap(\operatorname{span}N_2)^\perp$, of rank $s-j$.  Since $N$ is
primitive in $\Lambda$, so is $N_1$, and minimality gives
$\mathcal D(N_1)\geq\mathcal D(N)$.  Determinantal
duality~\eqref{eq:primitive-determinantal-duality} therefore yields
\begin{equation*}
    \det N_2=\frac{\det N_1}{\det N}
       \geq a^{-j}\frac{(s-j)!}{s!}
       \geq(as)^{-j}.
\end{equation*}
For a nonprimitive $N_2$, saturation in $N^*$ decreases its determinant
without changing rank, so the same bound holds.  Taking the minimum over
all nonzero $N_2$ gives
$\operatorname{nd}_{\min}(N^*)\geq1/(as)=m/(3sg)$.
\end{proof}

By Fact~\ref{fact:reverse-minkowski}, $N$ is $\sigma$-smooth for
$\sigma=3\xi(m)(s/m)g$, the rank-weighted scale for rank $s$.

To find a smooth sublattice, we adapt Dadush's Gaussian hyperplane
descent \cite[Section~4]{dadush2019approximating}.
If $v\in\mathcal S^*$ is
nonzero, then $\mathcal S\cap v^\perp$ is a primitive sublattice of rank one less
than $\mathcal S$.  A primitive sublattice $N\subseteq\mathcal S$ survives this cut exactly
when $v\perp N$, and by~\eqref{eq:gaussian-projection-preservation} a small
Gaussian mass on $N^*$ makes this event likely.
Dadush's dense-sublattice algorithm samples and cuts in the same way.
It seeks a small normalized
determinant, recomputes the sampling scale from the current lattice, and
retains a recursive candidate when its normalized determinant improves.
We instead keep the original rank $m$ and determinant
scale $g$ fixed during an attempt, let the scale shrink with the remaining
rank $q$, and stop when the current dual has small Gaussian mass.  This
gives a covering bound proportional to $q/m$, even when $q$ is small.

\begin{lemma}[Rank-weighted smooth sublattice]
\label{lem:rank-weighted-smooth-sublattice}
For a rank-$m\geq1$ lattice $\Lambda$ with a rational basis of encoding length $L$, a randomized algorithm running in
time $2^{O(m)}\operatorname{poly}(L,\log(1/\delta))$ returns, with
probability at least $1-\delta$, a nonzero primitive sublattice
$\mathcal S\subseteq\Lambda$ of rank $q$ and a scale $\sigma_q>0$
satisfying
\begin{equation}
    \rho(\sigma_q\mathcal S^*)\leq3/2
    \quad(\text{$\mathcal S$ is $\sigma_q$-smooth}),
    \qquad
    \sigma_q\leq16\xi(m)\frac qm\operatorname{nd}(\Lambda).
    \label{eq:rank-weighted-smoothness}
\end{equation}
\end{lemma}
The algorithm descends from $\Lambda$ by hyperplane cuts, and the
minimizing sublattice $N$ of Lemma~\ref{lem:rank-weighted-dual-density}
is never computed; it serves only as a witness that the descent can
succeed.  An attempt (a full descent from the input lattice) starts with
$\mathcal S=\Lambda$.  At rank $q$, the algorithm tests whether the current
lattice $\mathcal S$ is $\sigma_q$-smooth, for $\sigma_q$ proportional to
$\xi(m)(q/m)g$, by sampling its dual.  It returns the lattice if the estimated frequency of zero
samples meets the acceptance threshold. Otherwise it draws one more dual
sample: if that sample is zero, the attempt fails, and if it is
nonzero, the algorithm cuts by the hyperplane orthogonal to it.  Since $\rho(\sigma_q\mathcal S^*)\leq3/2$ is equivalent
to zero probability at least $2/3$, frequent zero samples recognize a
$\sigma_q$-smooth output without enumerating $\mathcal S^*$, and the
acceptance threshold $3/4$ leaves room for sampling and estimation error.

The invariant behind the analysis is that the current lattice still
contains $N$.  While it does, we have $q\geq s$, and the scale gives
$\operatorname{nd}_{\min}(\sigma_qN^*)\geq2\xi(m)$, so
$\rho(\sigma_qN^*)\leq17/16$ and $N$ is smooth with room to spare.  If an
accurate test continues, the cut then keeps $N$ with probability greater
than $1/8$, and once $\mathcal S=N$ every accurate test accepts.  The
descent never needs to know which cuts keep $N$, and its output need not
be $N$.  If all tests were accurate, an attempt would therefore succeed
with probability at least $8^{-m}$.  The proof in
Appendix~\ref{app:finite-geometric-primitives} removes this assumption by
coupling the descent with a corrected process whose tests are always
accurate, so that $O(8^m\log(1/\delta))$ attempts suffice.  It also runs
the discrete Gaussian sampler at a fixed constant accuracy, maps long
samples to zero, and bounds the encoding of every intermediate basis.

\subsection{Comparison with the filtration route}
\label{subsec:filtration-route}

This subsection counts how often each route pays the reverse Minkowski
threshold.  With the recent estimates, both routes pay it once, and this
single payment accounts for the gap between the constructive factor
$O(\log^2(2k))$ and the existential factor $O(\log^{3/2}(2k))$.

Reis and Rothvoss obtain their constructive bound from a filtration
computed by Dadush's algorithm \cite[Theorem~6.4]{dadush2019approximating},
which runs in $2^{O(k)}$ time up to polynomial factors and, with probability at least $1-2^{-k}$, returns a
$3C_\eta(k)$-stable filtration.  Here $C_\eta(k)$ is the reverse Minkowski
threshold in the form of Fact~\ref{fact:reverse-minkowski}
\cite[Definition~2.12]{dadush2019approximating}.  Their covering bound for
the resulting quotients uses the threshold once more
\cite[Lemma~27 and Proposition~46]{reis2023subspace}.  With the earlier
threshold $O(\log(2k))$, the two uses contribute $O(\log^2(2k))$, and the
total factor is $O(\log^4(2k))$ \cite[Theorem~55]{reis2023subspace}.  The
variant of Appendix~\ref{subsec:published-variant} pays the threshold once and gives $O(\log^3(2k))$ from the
same estimates.

The covering theorem of Letwin, Reis, and Rothvoss for lattices with
weakly stable duals \cite[Theorem~1.2]{letwin2026strongly} removes the
second use of the threshold.  The computed filtrations are only
$3C_\eta(k)$-stable, so we first merge their quotients and then adapt
\cite[Lemma~4.1]{letwin2026strongly} to approximate stability.

In a $\gamma$-stable filtration in the sense
of \cite[Definition~6.1]{dadush2019approximating}, the quotients $L_j$
have increasing normalized determinants $r_j$ and satisfy
$\operatorname{nd}_{\min}(L_j)\geq r_j/\gamma$ and
$\operatorname{nd}_{\min}(L_j^*)\geq1/(\gamma r_j)$.  Merge consecutive
quotients greedily, starting a new group at the first quotient whose
normalized determinant exceeds twice that of the current group's first
quotient.  Denote the groups by $I_h$ and their merged quotients by $M_h$.  Put
$n_j:=\operatorname{rank}L_j$, $d_h:=\sum_{j\in I_h}n_j$, and
\begin{equation*}
    R_h:=\operatorname{nd}(M_h)
        =\left(\prod_{j\in I_h}r_j^{n_j}\right)^{1/d_h},
    \qquad
    a_h:=r_{\min I_h}.
\end{equation*}
For $j\in I_h$ we have $a_h\leq r_j\leq2a_h$, and hence
$a_h\leq R_h\leq2a_h$.  Along the induced filtration of $M_h$, the
determinant of a sublattice is at least the product of the determinants
of its projected pieces, which are sublattices of the original
quotients.  Applying the same argument to the reversed dual filtration,
whose quotients are the $L_j^*$, gives
\begin{align*}
    \operatorname{nd}_{\min}(M_h)
       &\geq\min_{j\in I_h}\frac{r_j}{\gamma}
        \geq\frac{R_h}{2\gamma},\\
    \operatorname{nd}_{\min}(M_h^*)
       &\geq\min_{j\in I_h}\frac1{\gamma r_j}
        \geq\frac1{2\gamma R_h}.
\end{align*}
Thus the merged filtration is $2\gamma$-stable.  Since
$a_{h+1}>2a_h$, its normalized determinants satisfy
$R_h<R_{h+1}$ and $R_h\leq R_{h+2}/2$.

The proof of
\cite[Lemma~4.1]{letwin2026strongly}
uses its filtration only through these separation inequalities and,
in the $\gamma=1$ case, weak stability of $2R_hM_h^*$.
Here $2\gamma R_hM_h^*$ is weakly stable, so replacing $2R_h$ by
$2\gamma R_h$ in that proof loses a factor $\gamma$.  The recursion of
\cite[Corollary~1.3]{letwin2026strongly}, run with these computed
filtrations, approximately isotropic positions, and estimated volumes, then
gives a constructive factor $O(\gamma\sqrt{\log(2k)}\log(2k))$.  By
Fact~\ref{fact:reverse-minkowski}, $\gamma=3C_\eta(k)=O(\sqrt{\log(2k)})$,
so this route also gives $O(\log^2(2k))$.  Both routes pay the threshold
once.  Removing it from this route, for example by computing
$O(1)$-stable filtrations, would give the existential factor
$O(\log^{3/2}(2k))$.

Our chain needs no filtration even for the existential bound.  At each
stage, apply Lemma~\ref{lem:rank-weighted-dual-density} to $T\mathcal L$
instead of running the descent, and let $N\subseteq\mathcal L$ be the
pullback by $T^{-1}$ of the sublattice it provides, of rank $s$.  Put
$c:=3s\operatorname{nd}(T\mathcal L)/m$.  Every sublattice of
$c(TN)^*=(c^{-1}TN)^*$ then has determinant at least one, so this dual is
weakly stable.  Theorem~1.2 of \cite{letwin2026strongly}, applied to
$c^{-1}TN$ in $\operatorname{span}(TN)$, and the Jensen step at the end of
the proof of Lemma~\ref{lem:gaussian-covering-certificate} give
\begin{equation*}
    \mu(N,C\cap\operatorname{span}N)
    \leq\sqrt{\frac{2e}{\pi}}\,c\,\mathbb E\|G\|_{TC}
    =\sqrt{\frac{2e}{\pi}}\,\frac{3s}{m}\operatorname{nd}(T\mathcal L)\,
       \mathbb E\|G\|_{TC}.
\end{equation*}
This is the first line of the per-stage bound~\eqref{eq:rank-weighted-cover}
from the proof in Appendix~\ref{app:finite-geometric-primitives}, with $q$
replaced by $s$ and $C_4\xi(m)$ by a constant.  The same composition then
gives the stage sum~\eqref{eq:rank-weighted-sum} with $\log(2m_i)$ replaced by
$\sqrt{\log(2m_i)}$, and the harmonic sum gives
$\mu(\Lambda,K)\leq O(\log^{3/2}(2k))\max_iR_{W_i}(\Lambda,K)$.  This is
the existential bound of \cite[Corollary~1.3]{letwin2026strongly}, proved
without a filtration.

The existential display uses weak stability of $c(TN)^*$ directly.
Fact~\ref{fact:reverse-minkowski} instead gives the Gaussian-mass
certificate $\rho(\xi(m)c(TN)^*)\leq3/2$, and
Lemma~\ref{lem:gaussian-covering-certificate} then pays $\xi(m)$ in the
covering estimate.  Our algorithm constructs a possibly different
sublattice and certifies its covering bound in the same way, through its
dual Gaussian mass.  The extra factor therefore comes from using the
Gaussian-mass certificate at all.  Sampling and estimation errors affect
only universal constants, and even an exact Gaussian-mass test would
leave this use of reverse Minkowski in the present analysis unchanged.

Our analysis pays $\xi(m)$ for this sufficient certificate but does not
show that certifying smoothness, or constructing a good subspace,
requires this loss.  Suppose that a
$2^{O(m)}\operatorname{poly}(L)$-time test, given the current sublattice
$\mathcal S$ of rank $q$, either certifies that every nonzero sublattice of
$\mathcal S^*$ has normalized determinant at least a constant multiple of
$m/(3q\operatorname{nd}(T\mathcal L))$, or returns a sublattice of
$\mathcal S^*$ whose annihilator in $\mathcal S$ lowers the potential
$\mathcal D$ from the proof of Lemma~\ref{lem:rank-weighted-dual-density}.
Such annihilators have smaller rank and keep $\mathcal D$ below
$\mathcal D(\{0\})=1$, so the test certifies after fewer than $m$ cuts, and
the certified case gives the display above with $s$ replaced by $q$, up to
that constant.  This would give $O(\log^{3/2}(2k))$ constructively.

\section{Local lattice-point access by sparsification}
\label{sec:search}
\label{app:local-access}

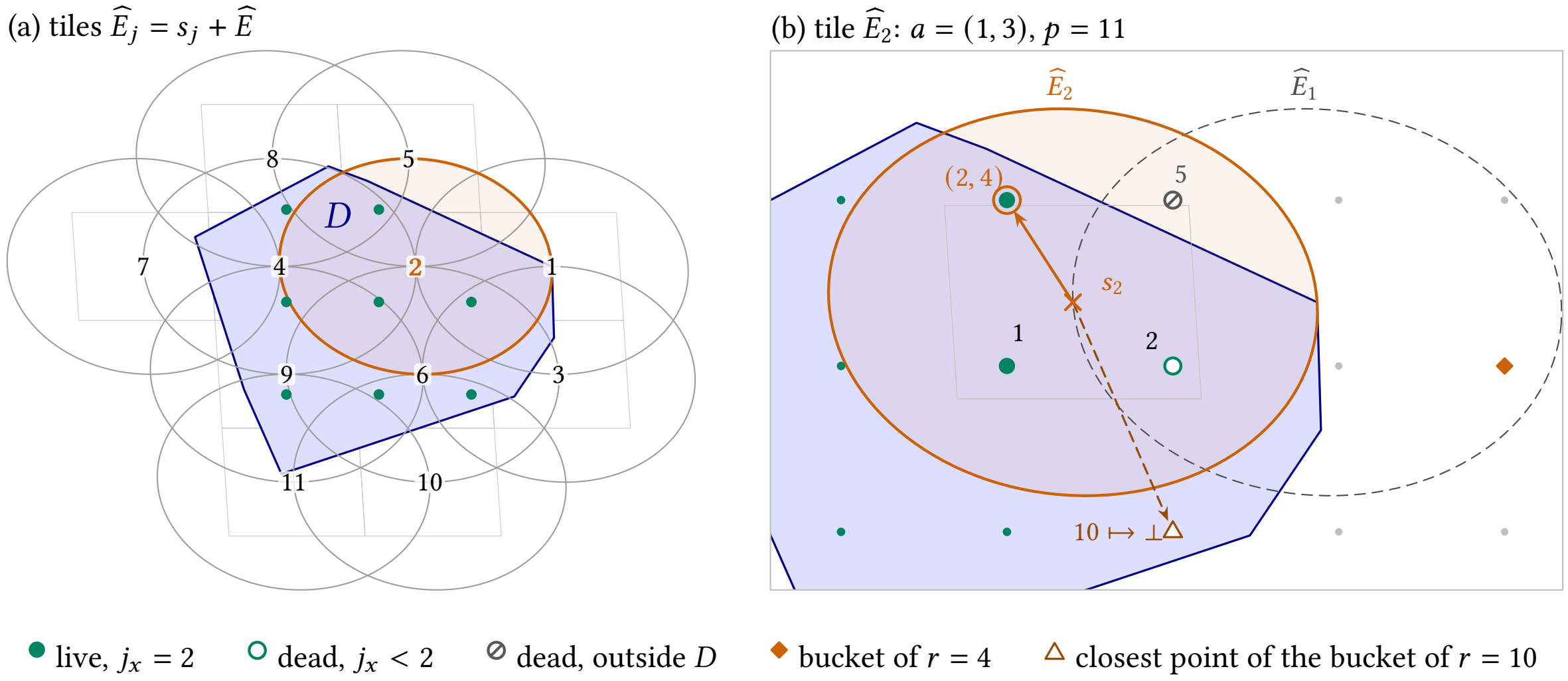
\begin{figure}[t]
\centering
\definecolor{acc}{rgb}{0.80,0.38,0.02}
\begin{tikzpicture}[line join=round,line cap=round,
        shT/.style={fill=blue!13,draw=blue!55!black,semithick},
        xpt/.style={teal!85!green!90!black},
        gridpt/.style={gray!50},
        para/.style={gray!45,very thin},
        tile/.style={gray!75,thin},
        hitile/.style={acc,thick,fill=acc,fill opacity=0.07},
        earlytile/.style={gray!60!black,densely dashed,thin},
        frame/.style={gray!50,thin},
        num/.style={font=\scriptsize,inner sep=0.6pt,fill=white,fill opacity=0.8,text opacity=1,rounded corners=1pt},
        hinum/.style={text=acc,font=\scriptsize\bfseries},
        center/.style={acc,thick},
        livept/.style={circle,fill=teal!85!green!90!black,inner sep=0pt,minimum size=4.6pt},
        earlypt/.style={circle,draw=teal!85!green!90!black,fill=white,thick,inner sep=0pt,minimum size=4.6pt},
        outpt/.style={circle,draw=gray!70!black,fill=white,thick,inner sep=0pt,minimum size=4.6pt,
            path picture={\draw[gray!70!black,thick] (path picture bounding box.south west) -- (path picture bounding box.north east);}},
        retpt/.style={circle,draw=acc,thick,inner sep=0pt,minimum size=7.6pt},
        bucket/.style={diamond,draw=acc,fill=acc,semithick,inner sep=0pt,minimum size=4.6pt},
        botpt/.style={regular polygon,regular polygon sides=3,draw=acc!75!black,fill=white,semithick,inner sep=0pt,minimum size=6pt},
        cvparrow/.style={acc,thick,-{Stealth[length=4.5pt]},shorten >=3.6pt},
        botarrow/.style={acc!75!black,semithick,densely dashed,-{Stealth[length=4pt]},shorten >=3.3pt},
        rlab/.style={font=\scriptsize,inner sep=0.5pt},
        lab/.style={inner sep=1.5pt},
        ptitle/.style={font=\footnotesize,inner sep=0pt},
        keym/.style={font=\footnotesize,inner sep=1pt}]
    \draw[para] (6.071,3.761) -- (4.715,3.761) -- (4.784,2.687) -- (6.141,2.687) -- cycle;
    \draw[para] (4.715,3.761) -- (3.358,3.761) -- (3.428,2.687) -- (4.784,2.687) -- cycle;
    \draw[para] (6.141,2.687) -- (4.784,2.687) -- (4.854,1.612) -- (6.211,1.612) -- cycle;
    \draw[para] (3.358,3.761) -- (2.002,3.761) -- (2.071,2.687) -- (3.428,2.687) -- cycle;
    \draw[para] (4.645,4.836) -- (3.288,4.836) -- (3.358,3.761) -- (4.715,3.761) -- cycle;
    \draw[para] (4.784,2.687) -- (3.428,2.687) -- (3.498,1.612) -- (4.854,1.612) -- cycle;
    \draw[para] (2.002,3.761) -- (0.645,3.761) -- (0.715,2.687) -- (2.071,2.687) -- cycle;
    \draw[para] (3.288,4.836) -- (1.932,4.836) -- (2.002,3.761) -- (3.358,3.761) -- cycle;
    \draw[para] (3.428,2.687) -- (2.071,2.687) -- (2.141,1.612) -- (3.498,1.612) -- cycle;
    \draw[para] (4.854,1.612) -- (3.498,1.612) -- (3.567,0.537) -- (4.924,0.537) -- cycle;
    \draw[para] (3.498,1.612) -- (2.141,1.612) -- (2.211,0.537) -- (3.567,0.537) -- cycle;
    \path[shT] (3.202,4.221) -- (1.872,3.517) -- (2.360,1.998) -- (2.733,1.156) -- (5.054,1.928) -- (5.449,2.513) -- (5.428,3.224) -- (3.590,4.076) -- cycle;
    \draw[tile] (5.428,3.224) ellipse [x radius=1.361, y radius=1.071, rotate=-6.13];
    \draw[tile] (5.497,2.149) ellipse [x radius=1.361, y radius=1.071, rotate=-6.13];
    \draw[tile] (2.715,3.224) ellipse [x radius=1.361, y radius=1.071, rotate=-6.13];
    \draw[tile] (4.002,4.299) ellipse [x radius=1.361, y radius=1.071, rotate=-6.13];
    \draw[tile] (4.141,2.149) ellipse [x radius=1.361, y radius=1.071, rotate=-6.13];
    \draw[tile] (1.358,3.224) ellipse [x radius=1.361, y radius=1.071, rotate=-6.13];
    \draw[tile] (2.645,4.299) ellipse [x radius=1.361, y radius=1.071, rotate=-6.13];
    \draw[tile] (2.785,2.149) ellipse [x radius=1.361, y radius=1.071, rotate=-6.13];
    \draw[tile] (4.211,1.075) ellipse [x radius=1.361, y radius=1.071, rotate=-6.13];
    \draw[tile] (2.854,1.075) ellipse [x radius=1.361, y radius=1.071, rotate=-6.13];
    \draw[hitile] (4.071,3.224) ellipse [x radius=1.361, y radius=1.071, rotate=-6.13];
    \foreach \p in {(2.782,1.948),(2.782,2.870),(2.782,3.791),(3.704,1.948),(3.704,2.870),(3.704,3.791),(4.626,1.948),(4.626,2.870)}
        \fill[xpt] \p circle (1.5pt);
    \node[num] at (5.428,3.224) {1};
    \node[num,hinum] at (4.071,3.224) {2};
    \node[num] at (5.497,2.149) {3};
    \node[num] at (2.715,3.224) {4};
    \node[num] at (4.002,4.299) {5};
    \node[num] at (4.141,2.149) {6};
    \node[num] at (1.358,3.224) {7};
    \node[num] at (2.645,4.299) {8};
    \node[num] at (2.785,2.149) {9};
    \node[num] at (4.211,1.075) {10};
    \node[num] at (2.854,1.075) {11};
    \node[lab,text=blue!55!black] at (3.301,3.737) {$D$};
    \node[ptitle,anchor=south west] at (0.000,5.423) {(a) tiles $\widehat E_j=s_j+\widehat E$};
    \begin{scope}
    \clip (7.606,0.000) rectangle (15.500,5.373);
    \path[shT] (9.061,4.655) -- (6.676,3.392) -- (7.552,0.669) -- (8.220,-0.841) -- (12.383,0.543) -- (13.091,1.593) -- (13.053,2.867) -- (9.757,4.396) -- cycle;
    \draw[para] (11.774,3.831) -- (9.341,3.831) -- (9.467,1.904) -- (11.899,1.904) -- cycle;
    \draw[earlytile] (13.053,2.867) ellipse [x radius=2.441, y radius=1.921, rotate=-6.13];
    \draw[hitile] (10.620,2.867) ellipse [x radius=2.441, y radius=1.921, rotate=-6.13];
    \foreach \p in {(13.268,0.579),(13.268,2.232),(13.268,3.885),(14.921,0.579),(14.921,2.232),(14.921,3.885)}
        \fill[gridpt] \p circle (1.05pt);
    \foreach \p in {(8.309,0.579),(8.309,2.232),(8.309,3.885),(9.962,0.579),(11.615,0.579)}
        \fill[xpt] \p circle (1.2pt);
    \end{scope}
    \draw[frame] (7.606,0.000) rectangle (15.500,5.373);
    \foreach \p in {(14.921,2.232)}
        \node[bucket] at \p {};
    \draw[botarrow] (10.620,2.867) -- (11.615,0.579);
    \node[botpt] at (11.615,0.579) {};
    \draw[cvparrow] (10.620,2.867) -- (9.962,3.885);
    \draw[center] (10.545,2.792) -- (10.695,2.942) (10.545,2.942) -- (10.695,2.792);
    \node[livept] at (9.962,2.232) {};
    \node[livept] at (9.962,3.885) {};
    \node[earlypt] at (11.615,2.232) {};
    \node[outpt] at (11.615,3.885) {};
    \node[retpt] at (9.962,3.885) {};
    \node[rlab,text=acc] at (9.636,4.097) {$(2,4)$};
    \node[rlab] at (10.075,2.564) {$1$};
    \node[rlab] at (11.406,2.486) {$2$};
    \node[rlab,text=gray!70!black] at (11.697,4.144) {$5$};
    \node[rlab,text=acc!75!black] at (11.075,0.579) {$10\mapsto\bot$};
    \node[rlab,text=acc] at (11.018,3.028) {$s_{2}$};
    \node[rlab,text=gray!60!black] at (12.928,5.045) {$\widehat E_{1}$};
    \node[rlab,text=acc] at (10.495,5.045) {$\widehat E_{2}$};
    \node[ptitle,anchor=south west] at (7.606,5.423) {(b) tile $\widehat E_{2}$: $a=(1,3)$, $p=11$};
    \matrix[anchor=north,column sep=2pt,inner sep=0pt] at (7.750,-0.500) {
        \node[livept] {}; & \node[keym] {live, $j_x=2$}; &[12pt] \node[earlypt] {}; & \node[keym] {dead, $j_x<2$}; &[12pt] \node[outpt] {}; & \node[keym] {dead, outside $D$}; &[12pt] \node[bucket] {}; & \node[keym] {bucket of $r=4$}; &[12pt] \node[botpt] {}; & \node[keym] {closest point of the bucket of $r=10$}; \\
    };
\end{tikzpicture}
\caption{The two-layer address of Section~\ref{sec:search} for $d=2$,
$x_0+\Lambda=\mathbb Z^2$ and $B=I$, so that $r_a(z)=a^{\top}z\bmod p$.  The
body $D$ is the projected body $K_W=\Pi_W\widetilde K$ of
Figure~\ref{fig:classical-projection}, and its lattice points (teal) are the
branch set $X$.  (a)~The first layer: translates $\widehat E_j=s_j+\widehat E$
of one ellipse, centered on a tiling of the plane by translates of a
parallelogram cell (gray) whose closure lies in $\widehat E$, so each closed
cell lies in its tile.  A tile is kept when its closed cell meets
$D$, and the kept tiles are numbered at their centers by breadth-first search
through face-adjacent cells from tile~$1$, whose center is the point $a_0$ of
$D$.  (b)~The second layer in the tile $\widehat E_{2}$: its
$\nu_{2}=4$ lattice points carry distinct residues under the
certified hash with $a=(1,3)$ and $p=11$.  To decode
$(2,4)$, one bucket CVP from $s_{2}$ finds the point of the bucket
$\{z:r_a(z)=4\}$ (filled diamonds) closest to $s_{2}$ in
$\lVert\cdot\rVert_{\widehat Q}$; it lies in $\widehat E_{2}\cap D$, so it
is returned, and since $j_x=2$ the address is live.  Three kinds of dead
address are illustrated: $(2,10)$ decodes to $\bot$ because its
closest bucket point (triangle) lies outside $\widehat E_{2}$; $(2,5)$ decodes
to $\bot$ because its point (gray, struck through) lies outside $D$; and $(2,2)$ decodes to a
point $x$ (hollow teal) that also lies in the earlier tile
$\widehat E_{1}$, so $j_x=1$, the address is not canonical, and the
canonical address of $x$ is $(1,r_{a_{1}}(x))=(1,2)$,
where tile~$1$ has $a_{1}=(1,3)$ and $p_{1}=7$.}
\label{fig:two-layer-address}
\end{figure}

Tree search on the recursion must compute each child of a node, a point
of the set $X$ of Section~\ref{subsec:quantum-entry}, by a reversible
circuit without listing the others.  We give each point an address with
two layers (Figure~\ref{fig:two-layer-address}): an ellipsoid of a small
cover of the body, and a residue that
picks one point inside that ellipsoid.  We work with a compact
full-dimensional convex body $D\subseteq\mathbb R^d$, given by a rational
lift (Appendix~\ref{app:recursion-details}), and a full-rank affine
lattice $x_0+\Lambda=x_0+B\mathbb Z^d$; the recursion uses
$D=\Pi_W\widetilde K$.

Section~\ref{subsec:indexable-cover} builds the first layer, an indexed
cover of $D$ by translates of one ellipsoid.
Section~\ref{sec:sparsification} builds the second: it makes each hash
bucket an explicit affine lattice, certifies one hash that is injective
on all lattice points of an ellipsoid, attaches one such hash to each
tile of the cover, and states the resulting access interface,
Theorem~\ref{thm:coherent-access}, whose proof is completed in
Appendix~\ref{app:finite-quantum-recursion}.  The hashing and
geometric-indexing results of this section assume $d\ge 1$; the proof
of Theorem~\ref{thm:coherent-access} treats the case $d=0$ separately.

For a lattice $\Lambda$, $\nu_{\max}(\Lambda,D)$ is the largest number of
lattice points in a translate of $D$, as in~\eqref{eq:G-parameter}.  For a
rational positive definite matrix $Q$, write
$\lVert x\rVert_Q:=\sqrt{x^\top Qx}$ and
$E(t,Q):=\{x:\lVert x-t\rVert_Q\leq1\}$.

\subsection{Coherently indexable \texorpdfstring{$M$}{M}-ellipsoid covers}
\label{subsec:indexable-cover}

The first layer of an address selects one ellipsoid from a small cover of
$D$.  An \emph{$M$-ellipsoid} of $D$ is an ellipsoid $E$ such that
$2^{O(d)}$ translates of $E$ cover $D$ and $2^{O(d)}$ translates of $D$
cover $E$ \cite{dadush2011enumerative}; the first property gives a small
cover, and the second bounds the lattice points in each translate of $E$
by $2^{O(d)}\nu_{\max}(\Lambda,D)$.  To index the cover, we tile space by
translates of a parallelepiped $P$ whose closure lies in a rational
rounding $\widehat E$ of $E$, keep
the tiles that meet $D$, and order them by a breadth-first search through
adjacent tiles, which the circuit recomputes.  Deterministic tests certify each accepted candidate for $E$, and amplification
and a union bound yield one seed that works for all bodies of bounded encoding length.

\begin{lemma}[Indexable $M$-ellipsoid cover]
\label{lem:rational-indexed-cover}
There are universal integers $C_{\rm cov},C_{\rm pt}\geq2$ with the
following property.  Let $L'\geq1$ and $\tau\in(0,1/2)$.  With probability
at least $1-\tau$, a uniformly random seed of
$\operatorname{poly}(L',\log(1/\tau))$ bits is good for all compact
full-dimensional convex bodies $D\subseteq\mathbb R^d$ of encoding length
at most $L'$ at once: for each such $D$, a reversible circuit of size
$2^{O(d)}\operatorname{poly}(L',\log(1/\tau))$ maps each
$j\in\{1,\ldots,C_{\rm cov}^d\}$ to a center $s_j$ and a validity bit such
that, for an ellipsoid $\widehat E=E(0,\widehat Q)$ depending on $D$,
\begin{equation}
    D\subseteq\bigcup_{j\text{ valid}}(s_j+\widehat E)
    \label{eq:rational-indexed-cover}
\end{equation}
and $\nu_{\max}(\Lambda,\widehat E)\leq C_{\rm pt}^d\nu_{\max}(\Lambda,D)$ for every lattice
$\Lambda$.
\end{lemma}

During setup we draw one seed and keep it fixed for every cover query in
that run; we suppress it in the notation.  In the recursion,
$L'=\operatorname{poly}(L,\log(1/\delta))$ bounds the encoding length of every projected body.
The lemma lower-bounds the probability that the drawn seed is good, and the
event that it is not is charged to the setup failure budget ($\tau/2$ in
the coherent-access construction below).  The algorithm never tests the seed:
every cover it accepts is certified whatever the seed, so goodness matters
only for completion.  Appendix~\ref{app:indexable-cover-proof}
gives the proof, building on \cite[Algorithms~4.4--4.5]{dadush2012integer}.

\subsection{Certified hash--CVP addresses}
\label{sec:sparsification}

Inside an ellipsoid we address points by lattice sparsification
\cite{khot2003hardness,dadush2013sparse,dadush2014closest}.  For a prime
$p$ and a nonzero vector $a\in\mathbb F_p^d$, which we call a
\emph{normal}, the hash $r_a(x_0+Bz):=a^\top z\bmod p$ splits
$x_0+\Lambda$ into $p$ buckets, each an explicit affine lattice.  If $r_a$ is injective on the $\nu$
lattice points of an ellipsoid, each bucket has at most one point in the
ellipsoid, and exact CVP finds it as the bucket's point closest to the
center in the ellipsoid's norm; the residue is then an address.  The
instance-wise bound needs two guarantees.  A certificate that $r_a$ is
injective on all $\nu$ points makes addresses decode bijectively on a
fixed tape, and the bound $p\leq2^{O(d)}(\nu+1)$ keeps empty buckets from
dominating the search.

The hash is injective on the lattice points of an ellipsoid exactly
when its kernel contains no nonzero difference of two of them.  We cannot
list these differences, and the obvious superset, the integer points of
the difference body, can be arbitrarily larger even when $\nu=1$ (see the
example in Section~\ref{subsec:certified-global-hash}).  We use midpoints
instead.  After Section~\ref{subsec:bucket-hash} makes each bucket an
explicit affine lattice that exact CVP can search,
Section~\ref{subsec:certified-global-hash} applies the parallelogram
identity to enclose all differences in a set $S$, sorted by the parity
class of their midpoints, with at most $2^{d+1}\nu-1$ elements
(Lemma~\ref{lem:ellipsoid-difference-enlargement}).

The set $S$ is never listed.  Given the $2^d$ parity-class distances of
the lemma, a further
$2^d-1$ CVP calls and one SVP call decide exactly whether the kernel of a
hash avoids $S\setminus\{0\}$, and once $p>2^{d+2}\nu$ a random normal
passes with probability greater than $3/4$
(Proposition~\ref{prop:certified-global-hash}).  Since $\nu$ is unknown,
we try primes of doubling size from a \emph{prime table}, one prime in
each interval $(2^{g+2},2^{g+3})$ for $g=0,\ldots,g_{\mathrm{top}}$
(Appendix~\ref{app:finite-quantum-recursion}), drawn once by sampling
with exact primality tests; the table is \emph{valid} if no interval is
missing.  Every pass certifies injectivity, and with high probability
the first pass has $p\leq2^{O(d)}\nu$.
Section~\ref{subsec:access-theorem} attaches one certified hash to each
cover ellipsoid and states the resulting access theorem.

\subsubsection{Hashing into decodable affine lattices}
\label{subsec:bucket-hash}

Decoding a residue means finding the point of its bucket closest to an
ellipsoid center, so exact CVP needs a basis of each bucket.  Fix a prime
$p$ and a nonzero normal $a\in\mathbb F_p^d$, and recall the hash
\begin{equation*}
    r_a(x_0+Bz):=a^\top z\bmod p.
\end{equation*}

Every bucket of this linear hash is an explicit affine lattice.  Let $i_a$
be the least index with $a_{i_a}\neq0$, and write $a_{i_a}^{-1}$ for its
inverse modulo $p$.  Using representatives in $\{0,\ldots,p-1\}$, set
\begin{equation*}
    v_i=e_i-(a_ia_{i_a}^{-1}\bmod p)e_{i_a}\quad(i\neq i_a),
    \qquad
    v_{i_a}=pe_{i_a},
\end{equation*}
and let $V_a$ have columns $v_1,\ldots,v_d$.  For $r\in\mathbb F_p$, set
$z_{a,r}=(ra_{i_a}^{-1}\bmod p)e_{i_a}$.

\begin{lemma}[Explicit bucket basis]
\label{lem:bucket-basis}
The bucket of residue $r$ is the affine lattice
\begin{equation}
    \{x_0+Bz:a^\top z=r\bmod p\}
    =x_0+Bz_{a,r}+BV_a\mathbb Z^d.
    \label{eq:bucket-affine-lattice}
\end{equation}
Moreover, $|\det V_a|=p$, and its shift and basis can be computed reversibly
using $\operatorname{poly}(d,\log p,L)$ gates.
\end{lemma}

\begin{proof}
Every column of $V_a$ lies in the kernel of $z\mapsto a^\top z\bmod p$.
This map is onto, so its kernel has index $p$ in $\mathbb Z^d$.  Since
$|\det V_a|=p$, the column lattice of $V_a$ is exactly that kernel.
Finally, $a^\top z_{a,r}=r\bmod p$, which proves
\eqref{eq:bucket-affine-lattice}.  The construction uses only modular
arithmetic on $O(\log p)$-bit registers.
\end{proof}

This explicit bucket basis is classical
\cite[Claim~2.15]{stephensdavidowitz2016discrete}; it makes each residue
class an input to exact CVP.  Section~\ref{subsec:certified-global-hash}
certifies a hash for which no class contains two points of the ellipsoid.

\subsubsection{Certifying one globally injective hash}
\label{subsec:certified-global-hash}

We want one hash that is injective on all lattice points of an
ellipsoid, and we want to check this exactly before the hash is used.
The check needs an enlargement of the actual differences that is small
and that CVP can access without listing it.  The size of
$(E-E)\cap\Lambda$ is not controlled by the number of lattice points in $E$ for
shifted ellipsoids (see the example below), so we track the parity class of each
possible midpoint instead.
Lemma~\ref{lem:ellipsoid-difference-enlargement} builds and bounds this
enlargement; Proposition~\ref{prop:certified-global-hash} turns it into an
exact test and a randomized search.

\begin{lemma}[An accessible enlargement of ellipsoid differences]
\label{lem:ellipsoid-difference-enlargement}
Let $E=E(t,Q)$ and $X_E:=E\cap(x_0+B\mathbb Z^d)$, with $\nu:=|X_E|\geq1$.
In coefficient coordinates put
\begin{align*}
    t_B&:=B^{-1}(t-x_0),& Q_B&:=B^\top QB,\\
    q(z)&:=(z-t_B)^\top Q_B(z-t_B),&
    Z&:=\{z\in\mathbb Z^d:q(z)\leq1\}.
\end{align*}
For $e\in\{0,1\}^d$, compute the exact rational squared distance
\begin{align}
    q^\star_e&:=\min_{m\in e/2+\mathbb Z^d}q(m),\notag\\
    S_e&:=\left\{v\in e+2\mathbb Z^d:
        q^\star_e+\frac14v^\top Q_Bv\leq1\right\},\notag\\
    S&:=\bigcup_e S_e.
    \label{eq:parity-difference-enlargement}
\end{align}
Then
\begin{equation}
    Z-Z\subseteq S,\qquad
    |S|\leq2^{d+1}\nu-1,
    \qquad |X_E-X_E|\leq2^{d+1}\nu-1.
    \label{eq:ellipsoid-difference-bound}
\end{equation}
The description of $S$ requires $2^d$ exact ellipsoidal CVP calls and
$2^{O(d)}\operatorname{poly}(L)$ bits and operations, without enumerating
$X_E$ or $S$.
\end{lemma}

For $z,w\in Z$, the midpoint $(z+w)/2$ lies in $e/2+\mathbb Z^d$ with
$e\equiv z-w\pmod2$, and the parallelogram identity puts $z-w$ in $S_e$.
Conversely, if $m_e$ attains $q^\star_e$ and $v\in S_e$, one of the
integer vectors $m_e\pm v/2$ lies in $Z$, so $|S_e|\leq2\nu$.  The proof
is in Appendix~\ref{app:finite-quantum-recursion}.

\begin{proposition}[A checked globally injective ellipsoid hash]
\label{prop:certified-global-hash}
For the nonempty ellipsoid intersection above and
$\tau\in(0,1/2)$, a capped randomized algorithm uses
$2^{O(d)}\operatorname{poly}(L,\log(1/\tau))$ time and space and returns
either failure or an odd prime $p$ and nonzero normal
$a\in\mathbb F_p^d$.  Every returned pair is certified to make
$z\mapsto a^\top z\bmod p$ injective on $Z$.  With probability at least
$1-\tau$ it returns such a pair with $p\leq C2^d\nu$, for a universal
constant $C$.  The test for a proposed pair is deterministic and uses
$2^d-1$ exact ellipsoidal CVP calls and one SVP call in dimension $d$.
Neither the construction nor the test enumerates $X_E$ or $S$.
\end{proposition}
The test works in the kernel lattice $\mathcal K_a=V_a\mathbb Z^d$ of the
hash, which has the explicit basis of Lemma~\ref{lem:bucket-basis}.  A
kernel vector $v$ of parity $e$ lies in $S_e$ exactly when
$q^\star_e+(1/4)v^\top Q_Bv\leq1$, so it suffices to minimize
$v^\top Q_Bv$ over the kernel vectors of each parity.  Since $p$ is odd,
these vectors form a coset of $2\mathcal K_a$; writing $v=2w$, the
comparison becomes $q^\star_e+w^\top Q_Bw\leq1$ with $w$ ranging over a
coset of $\mathcal K_a$.  For $e\neq0$ this coset avoids zero, and the
minimum is a CVP; for $e=0$ the coset is $\mathcal K_a$ itself and we need
a shortest nonzero vector of $\mathcal K_a$, which is an SVP.  Passing all
$2^d$ comparisons
certifies injectivity because $Z-Z\subseteq S$, although a rejection
need not exhibit an actual collision.

Once $p>2^{d+2}\nu$, no nonzero
$v\in S$ is divisible by $p$.  Indeed, if $v=pw\in S_e$, the odd
multiples $\pm w,\pm3w,\ldots,\pm pw$ keep the parity and do not
increase the norm, so they would put $p+1>2\nu$ vectors into $S_e$.  A
uniform nonzero normal therefore annihilates each such $v$ with
probability less than $1/p$, and a union bound over the at most
$2^d\nu-1$ nonzero antipodal pairs in $S$ keeps the rejection probability below
$1/4$.

Since $\nu$ is unknown, the algorithm climbs a ladder of primes of
doubling size and tries $O(\log(1/\tau))$ normals at each level.  Every
pass certifies injectivity, whatever its level, and with high
probability the search stops by the first level whose primes exceed
$2^{d+2}\nu$, where $p\leq C2^d\nu$.  The ladder, the caps on levels and
trials, and the split of the failure probability are in
Appendix~\ref{app:finite-quantum-recursion}, where the proof is
completed.

The parity enlargement matters for shifted ellipsoids.  Take
$\Lambda=\mathbb Z^2$, $x_0=0$, and an integer $M\geq1$, and let
\begin{equation*}
    E_M=\left\{(x,y):2x^2+32M^2(y-1/4)^2\leq2M^2+1\right\}.
\end{equation*}
Its sole integer point is $(0,0)$, but $(E_M-E_M)\cap\mathbb Z^2$
has $4M+1$ points, whereas the parity enlargement
\eqref{eq:parity-difference-enlargement} is only
$\{(0,0),(1,0),(-1,0)\}$.  Replacing the actual differences by the
ordinary difference body would therefore give an arbitrarily worse
modulus bound.

Enumeration-free sampling is already classical: the CVP reductions of
\cite[Proposition~3.2, Lemma~3.3, and Theorem~3.5]{stephensdavidowitz2016discrete}
sample and approximately count lattice points without listing them.
The deterministic sparsifier of \cite[Section~5]{dadush2013sparse} avoids
enumerating the original, potentially large, set by enumerating bounded
local sets in successive sublattices.  Its intermediate projection tests
use enumeration access to preserve injectivity, and its final scalar
restriction carries a different guarantee, of sparsification and metric
preservation.  Parity-coset CVP is also classical
\cite[Theorem~5.2.5]{dadush2012integer}.  The construction here combines
the parity enlargement of
Lemma~\ref{lem:ellipsoid-difference-enlargement}, whose size is controlled
by $\nu$, with an exact test that the hash kernel avoids its nonzero
points, and together they certify injectivity on all points.  The instance-wise comparison relies on this
certificate, which leaves the exponent of the IP bound unchanged.

\subsubsection{Addresses for the lattice points of a body}
\label{subsec:access-theorem}

Once a hash for an ellipsoid passes the test, the residue $r=r_a(x)$ is
a unique address for each lattice point $x$ of the ellipsoid, and one
bucket CVP from the ellipsoid's center decodes every occupied residue.
A tree search must name every child of a node by an address it can
decode locally, so we extend these addresses to all lattice points of a
body.  Each nonempty cover ellipsoid gets its own certified hash.
Because cover ellipsoids overlap, a point can be reached from several of
them, and we keep only its address in the first.  Each point then has
exactly one canonical address, and with high probability the number of
addresses is at most $2^{O(d)}$ times the \emph{occupancy} of the cover,
one plus the number of lattice points in its ellipsoids, counted with
multiplicity.  We fix the tiles, certify one hash per tile, define the
addresses, and collect their properties in
Theorem~\ref{thm:certified-geometric-indexing}.

Fix a rational compact full-dimensional lifted body $D$, a rational affine
lattice $x_0+\Lambda=x_0+B\mathbb Z^d$, and the indexed cover of
Lemma~\ref{lem:rational-indexed-cover}.  Pad the slots to
$q_D:=2^{\lceil\log_2C_{\mathrm{cov}}^d\rceil}<2C_{\mathrm{cov}}^d$,
with the added slots invalid.  Let $V\subseteq[q_D]$ be the valid slots,
with centers $s_j$, and write
\begin{equation}
    \widehat E_j:=s_j+\widehat E,
    \qquad
    \nu_j:=|\widehat E_j\cap(x_0+\Lambda)|,
    \qquad j\in V.
    \label{eq:tile-counts}
\end{equation}
We call the cover ellipsoid $\widehat E_j$ a \emph{tile}.  Conditional on
a good cover seed, every point of $D$ lies in a tile.

The hash moduli come from a ladder~\eqref{eq:prime-ladder} of primes
$p_g$ with $2^{g+2}<p_g<2^{g+3}$, one per scale $g$; its generation and
validity are in Appendix~\ref{app:finite-quantum-recursion}.
For each valid tile, test emptiness by exact CVP.  For an empty tile, set
$P_j=p_0$ and $a_j=e_1$, with every residue dead.  For a nonempty tile,
run the certificate test of Proposition~\ref{prop:certified-global-hash}
on proposed normals at increasing prime scales, and retain the first
passing pair $(P_j,a_j)$.  If a nonempty tile has no passing pair before
the cap, setup fails.  Fix universal integers $C_{\mathrm{hash}}\geq32$ and
$C_{\mathrm{addr}}\geq4C_{\mathrm{hash}}C_{\mathrm{cov}}$ large enough
that, on the simultaneous small-modulus event,
\begin{equation}
    \nu_j\leq P_j\leq C_{\mathrm{hash}}^d(\nu_j+1).
    \label{eq:tile-prime}
\end{equation}
The lower bound follows from certified injectivity alone, so it also
holds for accepted overshoots; it need not be strict.

The \emph{raw address set} is
\begin{equation}
    \mathcal A:=\{(j,r):j\in V,\ r\in\mathbb F_{P_j}\}\cup\{\dagger\},
    \qquad
    |\mathcal A|=1+\sum_{j\in V}P_j,
    \label{eq:output-sensitive-addresses}
\end{equation}
ordered lexicographically with the always-dead sentinel $\dagger$ last.
To decode $(j,r)$, one bucket CVP from $s_j$ gives the closest point $x$;
return $x$ if $x\in\widehat E_j\cap D$, and $\bot$ otherwise.
For $x\in D\cap(x_0+\Lambda)$ put
\begin{equation*}
    j_x:=\min\{j\in V:x\in\widehat E_j\}.
\end{equation*}
An address decoding to $x\neq\bot$ is \emph{canonical}, or \emph{live},
exactly when $j=j_x$, and the inverse address of $x$ is
\begin{equation}
    c(x):=(j_x,r_{a_{j_x}}(x)).
    \label{eq:certified-cover-inverse}
\end{equation}
All other addresses are dead.

\begin{theorem}[Certified geometric indexing]
\label{thm:certified-geometric-indexing}
Given a successfully certified cover of $D$ and $\tau\in(0,1/2)$, the setup
above has a capped
$2^{O(d)}\operatorname{poly}(L,\log(1/\tau))$ implementation such that:
\begin{enumerate}
    \item On every completed setup, live decoding is a bijection with
    $X:=D\cap(x_0+\Lambda)$, with inverse
    \eqref{eq:certified-cover-inverse}.  Hash correctness is certified,
    not a further probabilistic completeness assumption.
    \item With probability at least $1-\tau$, setup completes and
    \eqref{eq:tile-prime} holds for every valid tile.  On this event,
    \begin{equation}
        1+\sum_{j\in V}\nu_j
        \leq|\mathcal A|
        \leq C_{\mathrm{addr}}^d
             \Bigl(1+\sum_{j\in V}\nu_j\Bigr).
        \label{eq:address-count}
    \end{equation}
    \item Given the setup or its retained random tape, fixed-length
    reversible circuits compute $|\mathcal A|$, convert raw indices to
    addresses and back, decode and test canonicity, and recover $c(x)$
    from $x$.  Each uses
    $2^{O(d)}\operatorname{poly}(L,\log(1/\tau))$ gates and qubits and
    clears its private work.  All failures and off-domain inputs have
    capped default outputs.
\end{enumerate}
\end{theorem}
Item~1 is deterministic.  In a nonempty tile the certified hash is
injective on the tile's lattice points, so each point of the tile is the
unique point of its residue bucket inside the tile, and one exact bucket
CVP from the tile center recovers it.  A point may lie in several tiles;
the first-tile rule keeps only its address in tile $j_x$, so every point
of $X$ has exactly one live address, namely~\eqref{eq:certified-cover-inverse}.
For item~2, Proposition~\ref{prop:certified-global-hash} is applied to
each nonempty tile with a failure budget of order $\tau/q_D$, and a union
bound runs over the tiles, not over the points.  The proofs of items~2
and~3 are in Appendix~\ref{app:finite-quantum-recursion}.  The raw range
is sparse, so tree search uses its known size and the partial bijection
rather than dense rank/select access to $X$.

The following theorem packages the cover, the certified hashes, and
these addresses into the interface that the quantum search uses.

\begin{theorem}[Local access by sparsification]
\label{thm:coherent-access}
Let $D\subseteq\mathbb R^d$ be a compact full-dimensional convex body given
by a rational lift, let $x_0+\Lambda$ be a rational full-rank affine
lattice, and let $\kappa\geq \nu_{\max}(\Lambda,D)$ be an integer and
$\tau\in(0,1/2)$.  For each valid slot $j$ of
Lemma~\ref{lem:rational-indexed-cover}, let $\nu_j$ be the number of points
of $x_0+\Lambda$ in the cover ellipsoid $s_j+\widehat E$, and put
$\chi:=1+\sum_j\nu_j$.  Capped randomized preprocessing returns failure or
an ordered raw domain $\mathcal A$ of size at most $2^{O(d)}\kappa$, a
decoder, and a test for canonical addresses.  Every completed setup maps the
canonical addresses bijectively onto $D\cap(x_0+\Lambda)$:
cover validity and hash correctness are certified.  With probability at
least $1-\tau$, setup completes and $|\mathcal A|\leq2^{O(d)}\chi$.  Setup
and each query cost $2^{O(d)}\operatorname{poly}(L,\log\kappa,\log(1/\tau))$
time.  The decoders and the test have reversible circuits of the same size
that return all work registers to zero.
\end{theorem}
The proof combines the cover of Lemma~\ref{lem:rational-indexed-cover}
with one certified hash per tile, as in
Theorem~\ref{thm:certified-geometric-indexing}.  Setup first runs the
cover construction and fails if it produces no valid slot.  The
deterministic covering tests of Appendix~\ref{app:indexable-cover-proof}
certify every accepted cover, so the random seed affects
only whether setup completes, not whether the cover is correct.  By the
lemma, every tile then holds at most $\lceil C_{\rm pt}^d\kappa\rceil$
lattice points, so the prime ladder can be truncated after
$O(d+\log\kappa)$ levels while still containing the sufficient prime
scale of Proposition~\ref{prop:certified-global-hash} for every tile.
The truncation also bounds every accepted prime, including overshoots,
by $2^{O(d)}\kappa$, which gives $|\mathcal A|\leq2^{O(d)}\kappa$ on
every completed setup.  A union bound over the cover seed, the prime
ladder, and the tiles gives completion with a small modulus in every
tile, and the address count \eqref{eq:address-count} of
Theorem~\ref{thm:certified-geometric-indexing} then yields
$|\mathcal A|\leq2^{O(d)}\chi$.
Correctness needs none of these probabilities: certified coverage,
certified injectivity, exact bucket CVP, and the first-tile rule make
the canonical addresses decode bijectively on every completed setup.
The same theorem supplies the reversible circuits.
Appendix~\ref{app:finite-quantum-recursion} gives the full proof,
including the split of the failure budget, the trivial case $d=0$, and
the caps on preprocessing.

The address bound tracks $\chi$, the occupancy of the cover, rather than
$|D\cap(x_0+\Lambda)|$: addresses that decode outside $D$, or to a point
of an earlier ellipsoid, are dead.  On a good setup
(Section~\ref{sec:instance-wise}), the classical reference decodes at most
$2^{O(n)}$ times the total occupancy of the nodes it visits, and decodes every address only on exhaustive runs, such as
infeasible instances; full cover enumeration
(Corollary~\ref{cor:cover-enumeration}) pays the full occupancy at every
node it selects.  Section~\ref{sec:instance-wise} compares the quantum
search with the reference's actual run, and
Corollary~\ref{cor:cover-enumeration} compares it with cover enumeration
under admissible priorities.  On every tape, $2^{O(d)}\kappa_k(d)$ bounds
each node's address count.

\section{The quantum integer programming algorithm}
\label{sec:quantum-ip}
\label{sec:ip}
\label{sec:instance-wise}
\label{app:instance-wise}

This section places Theorem~\ref{thm:constructive-flatness} in the
classical recursion (Section~\ref{subsec:classical-recursion}), turns the
recursion into one tree that a quantum algorithm can search, using the
addresses of Theorem~\ref{thm:coherent-access}, and proves
Theorem~\ref{thm:main} by searching it
(Section~\ref{subsec:quantum-algorithm}).
Section~\ref{subsec:never-slower} runs the quantum algorithm alongside its
classical reference, Section~\ref{subsec:ip-consequences} derives the
worst-case consequences, and Section~\ref{subsec:bounded-equality} treats
bounded equation form.  Appendix~\ref{app:finite-quantum-recursion} gives
the formal definitions and statements behind
Section~\ref{subsec:quantum-algorithm}, with their proofs, and
Appendix~\ref{subsec:priority-cover} extends the comparison to
priority-driven cover enumeration.

\subsection{The classical recursion}
\label{subsec:classical-recursion}
\label{subsec:classical-complexity}

We now place Theorem~\ref{thm:constructive-flatness} in the recursion of
\cite{dadush2012integer,reis2023subspace}, which branches on the
projection of the body to a flat subspace.
A node is a rational compact\footnote{Standard small-solution bounds make the root body compact, and
rank-preserving Frank--Tardos compression keeps the description of every
recursive affine subspace polynomial in $L$
(Appendix~\ref{app:recursion-details}).}
convex body $K$ with a full-rank affine lattice $x_0+\Lambda$ in a
$k$-dimensional affine space.  If linear programming shows
$K=\varnothing$, the node is infeasible.  Otherwise the routine of
\cite[Lemma~60]{reis2023subspace} either rounds $K$, finding a rational
center $c_K$ and an ellipsoid $E$ with
$c_K+(k+1)^{-3/2}E\subseteq K\subseteq c_K+E$, or restricts the node,
without branching, to a hyperplane containing $K\cap(x_0+\Lambda)$.

At a rounded node, Theorem~\ref{thm:constructive-flatness} finds a lattice
subspace $W$ of dimension $d\geq1$ and an estimate $R$ with
$\mu(\Lambda,K)\leq\Phi(k)R$.  If $\Phi(k)R<1$, then $K$ meets every
lattice translate with room to spare, so shrinking it about $c_K$ keeps
it feasible.  For $\theta:=\min\{\Phi(k)R,1\}$, the body
$\widetilde K:=c_K+\theta(K-c_K)\subseteq K$ satisfies
$\mu(\Lambda,\widetilde K)=\mu(\Lambda,K)/\theta\leq1$ when $\theta<1$,
so in either case
\begin{equation}
	K\cap(x_0+\Lambda)\neq\varnothing
	\quad\Longleftrightarrow\quad
	\widetilde K\cap(x_0+\Lambda)\neq\varnothing.
	\label{eq:shrinkage-equivalence}
\end{equation}

The recursion branches on $X:=K_W\cap\Pi_W(x_0+\Lambda)$, where
$K_W:=\Pi_W\widetilde K$, and we put $\Gamma:=\Pi_W\Lambda$.  The number
of branches is bounded as follows; the shrinking factor $\theta$ and the
estimate $R$ cancel in the bound, so Theorem~\ref{thm:constructive-flatness}
enters the running time only through $\Phi(k)$.

\begin{lemma}[Branch bound]
\label{lem:branch-bound}
In the setting above, if
\eqref{eq:flatness-interface} holds, then
$|X|\leq\nu_{\max}(\Gamma,K_W)\leq(2C_R\Phi(k))^d$.
\end{lemma}

\begin{proof}
Dadush's lattice-point estimate \cite[Lemma~7.4.1]{dadush2012integer}
(see also \cite[Lemma~59]{reis2023subspace}) gives
\begin{equation*}
 \nu_{\max}(\Gamma,K_W)
 \leq2^d\max\{\mu(\Gamma,K_W)^d,1\}
			 \frac{\operatorname{Vol}_d(K_W)}{\det\Gamma}.
\end{equation*}
Projection does not increase covering radius, and scaling by $\theta$ gives
\begin{equation*}
 \mu(\Gamma,K_W)\leq\frac{\Phi(k)R}{\theta},
 \qquad
 \frac{\operatorname{Vol}_d(K_W)}{\det\Gamma}
	 =\left(\frac{\theta}{R_W}\right)^d
	 \leq\left(\frac{C_R\theta}{R}\right)^d.
\end{equation*}
Since $\theta=\min\{\Phi(k)R,1\}$, we have $\Phi(k)R/\theta\geq1$, so
$\max\{\mu(\Gamma,K_W)^d,1\}\leq(\Phi(k)R/\theta)^d$.  Multiplying by the
volume ratio and by $2^d$, the factors $\theta$ and $R$ cancel, and
$\nu_{\max}(\Gamma,K_W)\leq(2C_R\Phi(k))^d$.
\end{proof}

In the notation of Section~\ref{subsec:quantum-entry}, the lemma reads
\begin{equation}
	|X|\leq\nu_{\max}(\Gamma,K_W)
	\leq(C_{\rm br}\Phi(k))^d=\kappa_k(d),
	\qquad C_{\rm br}=2C_R.
	\label{eq:branch-bound}
\end{equation}

The children are the fibers $K\cap\Pi_W^{-1}(y)$ of $K$, not
of $\widetilde K$, so each node is the root body restricted to a
compressed affine subspace, and by~\eqref{eq:shrinkage-equivalence} a feasible node
has a feasible child; Figure~\ref{fig:classical-projection} shows a node
in three dimensions.

The recursion calls Theorem~\ref{thm:constructive-flatness} at every
rounded node.  Both the classical reference and the quantum algorithm of
Section~\ref{subsec:quantum-algorithm} need a version whose cost is bounded on
every random tape and whose outputs are checked before use, and the
following capped implementation provides it.

\begin{proposition}[Finite capped subspace flatness]
\label{prop:finite-reversible-flatness}
At every rounded rational residual instance of dimension $k\geq1$ and for every
$\beta\in(0,1/2)$, constructive subspace flatness has a capped randomized
bit-model implementation using
$2^{O(k)}\operatorname{poly}(L,\log(1/\beta))$ time, space, and random bits.  It writes a
rational output $(W,d,R)$ together with timeout and failure flags.  An exact
structural checker accepts only outputs with
$\operatorname{poly}(L,\log(1/\beta))$-bit encodings,
$1\leq d\leq k$, and projected and fiber lattice ranks $d$ and $k-d$.  With
probability at least $1-\beta$ over the random tape, the output is
\emph{geometrically valid}: it passes the checker and satisfies
\eqref{eq:flatness-interface}, and the
resulting $\widetilde K$ and $K_W$ satisfy \eqref{eq:shrinkage-equivalence}
and \eqref{eq:branch-bound}.
\end{proposition}
The implementation runs the routine of
Theorem~\ref{thm:constructive-flatness} with failure budget $\beta$.  The
proof of the theorem already fixes loop, norm, determinant, and bit caps,
so every random tape stays within the stated resources.  Before a
subspace is used, exact rational algebra checks the encoding, the
positivity of $R$, and the projected and fiber ranks, and a rejected
output becomes a failure leaf.  This check is structural and does not
certify~\eqref{eq:flatness-interface}.  That inequality holds with
probability at least $1-\beta$, and on this event the shrinking argument
above and Lemma~\ref{lem:branch-bound} give~\eqref{eq:shrinkage-equivalence}
and~\eqref{eq:branch-bound}.  Even a well-formed output that is not
geometrically valid lowers the dimension by at least one, and exact
verification of the leaves keeps it from producing an invalid accepted
point.  The proof is in Appendix~\ref{app:finite-geometric-primitives}, which
also pads the random tape, keeps the timeout and failure flags, and
applies Bennett simulation to obtain a reversible implementation with the
same resource bound.

Let $T(k)$ be the running time of the recursion on a node of dimension
$k$, after suppressing one common
$\operatorname{poly}(L,\log(1/\delta))$ factor, with $T(0)=1$.  Processing
a node, including the flatness routine of
Proposition~\ref{prop:finite-reversible-flatness}, takes $2^{O(k)}$ time,
and by \eqref{eq:branch-bound} the node has at most $\kappa_k(d)$
children, each of dimension at most $k-d$.  On a good setup, enumerating $X$ by the
addresses of Section~\ref{sec:sparsification} gives
\begin{equation}
    T(k)\leq2^{O(k)}+\max_{1\leq d\leq k}2^{O(d)}\kappa_k(d)T(k-d),
    \label{eq:classical-recurrence}
\end{equation}
which yields~\eqref{eq:classical-bound}.  The proof of
Corollary~\ref{cor:classical-ip}, in Appendix~\ref{app:reductions},
makes this precise.

The flatness routine is randomized, and a geometrically invalid output
can lose feasibility or enlarge a branch set.  The classical reference
of Section~\ref{subsec:reference} therefore draws all random choices
once and caps its node processings and address decodings, so that it is
correct on a good setup and runs within~\eqref{eq:classical-bound} on
every tape (Lemma~\ref{lem:classical-reference}); optimization follows by
the reduction in the proof of
Corollary~\ref{cor:integer-linear-optimization}.  The reference
enumerates $X$ through the addresses of Theorem~\ref{thm:coherent-access},
so it and the quantum algorithm process the same children in the same
order.

\subsection{The quantum algorithm}
\label{subsec:quantum-algorithm}
\label{subsec:universal-theorem}

The quantum algorithm searches the tree of the recursion of
Section~\ref{subsec:classical-recursion}.  We sketch the proof of
Theorem~\ref{thm:main}: we state the tree search, fix the tree, define
the classical reference, make the tree binary, and compare the search
with the reference.
Appendix~\ref{app:finite-quantum-recursion} gives the formal statements
and their proofs.

\proofstep{Quantum tree search.}
Quantum backtracking \cite{montanaro2018backtracking} decides whether a
tree has a marked leaf with a number of queries close to the square root
of the tree's size.  Ambainis and Kokainis \cite{ambainis2017quantum}
replace the size by the number of vertices that depth-first search
visits before the first marked leaf.  Both work in the local-access
model.  There, a rooted tree $\mathsf T$ is given by the encoding of its
root and by reversible circuits that compute, for a vertex $v$, its
depth, its number of children $\deg(v)$, its $i$th child for
$i\in[\deg(v)]$, and a label
$\operatorname{mark}(v)\in\{\textsc{marked},\textsc{unmarked},\textsc{internal}\}$.
The first two labels occur exactly at leaves.  Distinct vertices have
distinct encodings, of at most $b$ bits.  For a vertex $v$, the subtree
$\mathsf T(v)$ rooted at $v$ is given in the same model by the same
circuits, with depths shifted by $\operatorname{depth}(v)$.

Depth-first search visits the children of every vertex in increasing
order.  Let $T_{\mathrm{dfs}}(\mathsf T)$ be the number of vertices that
it visits up to and including the first marked vertex, or $|\mathsf T|$
if there is none.  Combining the detection algorithm of Ambainis and
Kokainis with a descent from the root turns detection into search.
Appendix~\ref{subsec:tree-search-proof} gives the proof.

\begin{theorem}[Quantum tree search with early termination]
\label{thm:tree-search}
Let $\mathsf{T}$ be a rooted tree in which every internal vertex has exactly
two children, given in the local-access model.  Let $\Delta,T_1\geq1$ be
integers and $\epsilon\in(0,1/2)$.  There are a universal effective constant
$c_{\mathrm{tree}}$ and a quantum algorithm with classical control and
intermediate measurements with the following properties.
\begin{enumerate}[label=(\alph*)]
    \item Every vertex that it returns has been checked to be marked.
    \item On every input, it applies the local-access circuits and their
    inverses at most
    \begin{equation}
        c_{\mathrm{tree}}\,\Delta^{5/2}\sqrt{T_1}\,
        \log^2\frac{4\Delta^2\log(2T_1)}{\epsilon}
        \label{eq:tree-search-worst}
    \end{equation}
    times.
    \item If $\Delta\geq\operatorname{depth}(\mathsf{T})$ and
    $T_1\geq|\mathsf{T}|$, then with probability at least $1-\epsilon$ it
    returns a marked vertex if $\mathsf{T}$ has one and
    \textnormal{\textsc{none}} otherwise, and it applies those circuits at
    most
    \begin{equation}
        c_{\mathrm{tree}}\,\Delta^{5/2}
        \sqrt{T_{\mathrm{dfs}}(\mathsf{T})}\,
        \log^2\frac{4\Delta^2\log(2T_1)}{\epsilon}
        \label{eq:tree-search-cost}
    \end{equation}
    times.
\end{enumerate}
Besides the work registers of those circuits, its other operations use
$\operatorname{poly}(b,\Delta,\log(T_1/\epsilon))$ qubits, and
$\operatorname{poly}(b,\Delta,\log(T_1/\epsilon))$ gates per application
of a local-access circuit.
\end{theorem}

As observed in \cite{montanaro2018backtracking}, a backtracking algorithm
with randomized heuristics fits this model once its random seed is fixed
in advance.

\proofstep{One setup for all nodes.}
\phantomsection
\label{subsec:fixed-tree}
A quantum walk queries each vertex many times in superposition, and it
must see the same children each time.  We therefore draw all random bits
of the recursion once, into a shared setup $\omega$.  It contains a
table of primes for the hash moduli, the cover seed of
Lemma~\ref{lem:rational-indexed-cover}, a flatness segment for
Proposition~\ref{prop:finite-reversible-flatness}, and a list of
proposed hash normals for every scale and dimension.  Processing a node
$H$ is then a capped deterministic computation
$\operatorname{Proc}_\omega(H)$ on its compressed affine subspace and
$\omega$ (Definition~\ref{def:node-processing}).  It runs the recursion
of Section~\ref{subsec:classical-recursion} with every random choice
read from $\omega$.  It certifies one hash per nonempty tile by trying
the stored normals at increasing prime scales.  The children of $H$ are
the compressed fibers of its canonical addresses, in address order.  The
resulting recursion tree $\mathcal T_\omega$ has at most $n$ internal
levels.

Which nodes occur depends on $\omega$, so we ask $\omega$ to work at
every node that can occur.  We call $\omega$ \emph{good} if every
flatness output is geometrically valid and every nonempty tile certifies
its hash with a modulus $P_j\leq C_{\mathrm{hash}}^d(\nu_j+1)$.
Compression bounds node encodings by
$L_{\mathrm{node}}=\operatorname{poly}(L)$, so at most
$2^{L_{\mathrm{node}}+1}$ nodes can occur.  A security parameter
$\lambda_{\mathrm{sec}}=\operatorname{poly}(L,\log(1/\delta))$ makes
each flatness call and each hash search fail with probability at most
$2^{-\lambda_{\mathrm{sec}}}$.  Given a valid prime table and a good
cover seed, a union bound over these nodes, not over their lattice
points, shows that a random setup is good with probability at least
$1-\delta/16$ (Lemma~\ref{lem:good-tapes}).  On a good setup,
$\mathcal T_\omega$ is the recursion tree of
Section~\ref{subsec:classical-recursion}, with children in
canonical-address order.

\proofstep{The classical reference.}
\phantomsection
\label{subsec:reference}
We compare the quantum algorithm with one classical implementation of the
recursion, Algorithm~\ref{alg:classical-reference}.  It runs on the same
setup $\omega$ and processes nodes depth first.  At an internal node it
decodes the addresses in increasing order, recurses on the child of each
canonical address, and stops at its first verified point.  An explicit
cap on its node processings and address decodings bounds its running
time on every tape.  Let $\operatorname{Dec}(I,\omega)$ be the number of
node processings and address decodings made by the same depth-first
exploration without this cap, \eqref{eq:decodings}, and
$T_{\mathrm C}(I,\omega)$ the reference's running time in bit operations.
On a good setup the cap is never reached, so $\operatorname{Dec}(I,\omega)$
counts exactly the operations of the reference, the reference is correct, and
\begin{equation*}
    \operatorname{Dec}(I,\omega)\leq T_{\mathrm C}(I,\omega)
    \leq2^{O(n)}\operatorname{poly}\!\left(L,\log\frac1\delta\right)
    \operatorname{Dec}(I,\omega).
\end{equation*}
Moreover, $\operatorname{Dec}(I,\omega)$ is at most $2^{O(n)}$ times the
sum, over the visited nodes, of one plus the number of lattice points in
their cover ellipsoids.  On infeasible instances it is at least this sum
over all nodes of $\mathcal T_\omega$.  On every tape,
$T_{\mathrm C}(I,\omega)\leq2^{O(n)}\Phi(n)^n\operatorname{poly}(L,\log(1/\delta))$
(Lemma~\ref{lem:classical-reference}).

\proofstep{The binary address tree.}
\phantomsection
\label{subsec:binary-address-tree}
Theorem~\ref{thm:tree-search} needs a binary tree of polynomial depth,
but a node of the recursion can have exponentially many addresses.  For
an integer $m\geq2$, the \emph{interval tree} $\mathcal I(m)$ has root
$[0,m)$.  An interval $[a,b)$ with $b-a\geq2$ has the children $[a,c)$
and $[c,b)$, where $c:=a+\lceil(b-a)/2\rceil$, and singletons are
leaves.  The tree has $m$ leaves, in the order $0,\ldots,m-1$ from left
to right, $m-1$ internal vertices, and depth $\lceil\log_2m\rceil$.
Every internal node $H$ of $\mathcal T_\omega$ has $|\mathcal A_H|>2$
addresses.  Replacing each address set by the leaves of
$\mathcal I(|\mathcal A_H|)$ turns $\mathcal T_\omega$ into one binary
tree $\mathsf T_\omega$ (Definition~\ref{def:binary-address-tree}).  A dead
address becomes an
unmarked leaf.  A canonical address is identified with the root of its
child: the root of the child's interval tree, or a single leaf if the
child is empty, failed, or a point.  The marked vertices are the point
leaves that pass exact verification.

A root-to-leaf path meets internal nodes whose projected dimensions
$d_H$ sum to at most $n$.  In the interval tree of each such node, it
uses at most $\lceil\log_2|\mathcal A_H|\rceil=O(d_H\log\log(4n))$
edges.  Hence $\mathsf T_\omega$ has depth at most
$\Delta_{\max}=O(n\log\log(4n))$.  A vertex is encoded by the canonical
addresses on the path to its node, together with an interval or leaf
tag.  Its local-access circuits replay $\operatorname{Proc}_\omega$
along this path, over at most $n$ levels, and then split the interval
or decode one address.  Each circuit uses
$2^{O(n)}\operatorname{poly}(L,\log(1/\delta))$ gates and returns its
work registers to zero (Lemmas~\ref{lem:address-tree-structure}
and~\ref{lem:address-tree-access}).

\proofstep{Comparison with the reference.}
Depth-first search on $\mathsf T_\omega$ meets the singletons of each
interval tree from left to right.  It explores the subtree below a
canonical singleton before it moves on.  It therefore meets the nodes of
$\mathcal T_\omega$ in the order in which the uncapped depth-first
exploration of the reference processes them, and it stops at the same
marked leaf.  At a visited internal node $H$
where the reference decodes $t_H$ addresses, it visits at most
$2t_H-1+\lceil\log_2|\mathcal A_H|\rceil$ vertices of the interval tree,
so the first inequality below holds on every setup.  On a good setup,
where the reference's cap is not reached, the second holds as well:
\begin{equation*}
    T_{\mathrm{dfs}}(\mathsf T_\omega)
    \leq(\Delta_{\max}+2)\operatorname{Dec}(I,\omega)
    \leq(\Delta_{\max}+2)T_{\mathrm C}(I,\omega).
\end{equation*}
On a good setup, $\mathsf T_\omega$ has a marked vertex exactly when the
instance is feasible, and it has at most
$T_{\max}=2^{O(n)}\Phi(n)^n$ vertices
(Lemma~\ref{lem:address-tree-structure}).

\proofstep{Proof of Theorem~\ref{thm:main}.}
In dimension zero, the algorithm checks the unique candidate exactly, so
we assume $n\geq1$.  The algorithm generates the prime table and the
cover seed, each with failure budget $\delta/32$, and draws the
recursion tape.  By Lemma~\ref{lem:good-tapes}, the setup is good with
probability at least $1-\delta/8$.  The algorithm applies
Theorem~\ref{thm:tree-search} to $\mathsf{T}_\omega$ with the circuits
of Lemma~\ref{lem:address-tree-access}, depth cap $\Delta_{\max}$, size
cap $T_{\max}$, and error $\delta/16$.  It verifies any returned point
exactly against the original input, which gives soundness on every tape.
On a good setup, Lemma~\ref{lem:address-tree-structure} shows that these
caps are valid and that $\mathsf{T}_\omega$ has a marked vertex exactly
when the instance is feasible.  Conditional on a good setup, the ideal
search fails with probability at most $\delta/16$, and gate synthesis
changes the law of its transcript by at most $\delta/16$.  Hence, with
conditional probability at least $1-\delta/8$, the search both answers
correctly and uses
$O(\Delta_{\max}^{5/2}\sqrt{T_{\mathrm{dfs}}(\mathsf{T}_\omega)})$
queries up to a polylogarithmic factor.  Since $\Delta_{\max}$ and
$\log T_{\max}$ are polynomially bounded and
$T_{\mathrm{dfs}}(\mathsf{T}_\omega)\leq(\Delta_{\max}+2)T_{\mathrm C}(I,\omega)$,
multiplying by the access cost
gives~\eqref{eq:instance-wise-bound} on the same event.  On every tape,
Theorem~\ref{thm:tree-search}(b) caps the queries with $\sqrt{T_{\max}}$
in place of $\sqrt{T_{\mathrm{dfs}}}$, and
$\sqrt{T_{\max}}=2^{O(n)}\Phi(n)^{n/2}$ gives the worst-case bound.  The
full proof, with the synthesis budget and the qubit count, is in
Appendix~\ref{app:finite-quantum-recursion}.

\subsection{Instance-wise speedups}
\label{subsec:never-slower}

To compare the two algorithms operation by operation, we measure both
in one unit.  Write $S$ for the number of bit operations of the common
initialization, which prepares the compactified root and the setup
$\omega$.  Let $\rho$
be the sequence of measurement outcomes of the quantum algorithm, and
let $Q(I,\omega;\rho)$ be the number of elementary operations it uses
once $\omega$ is fixed, where a native gate, a one-qubit preparation or
measurement, and a bit operation of the classical controller or of the
final verification each count as one.
Appendix~\ref{app:finite-quantum-recursion} shows that there are
$A=2^{O(n)}\operatorname{poly}(L,\log(1/\delta))\geq1$ and
$Q_{\max}=2^{O(n)}\Phi(n)^{n/2}\operatorname{poly}(L,\log(1/\delta))$
such that $S\leq A$ on every run, $Q(I,\omega;\rho)\leq Q_{\max}$ for
every setup and transcript, and, with probability at least $1-\delta$,
both algorithms are correct and
$Q(I,\omega;\rho)\leq A\sqrt{T_{\rm C}(I,\omega)}$.

\begin{corollary}[Never slower than the classical reference]
\label{cor:never-slower}
Let $\delta\in(0,1/2)$ and let $I$ be an input as in
Theorem~\ref{thm:main}. Perform the common initialization once, obtaining
the compactified root and one setup~$\omega$. Then run
Algorithm~\ref{alg:classical-reference} and the quantum algorithm of
Theorem~\ref{thm:main} on separate workspaces, both reading~$\omega$ but
neither writing to it. Alternate one classical bit operation with one
counted quantum operation. Stop at the first top-level final output of
either algorithm, breaking ties by a fixed rule. Let $H$ be the total number
of operations of this combined algorithm, including the setup. Let
$Q(I,\omega;\rho)$ be the cost the quantum computation would have if it ran
to completion with the same measurement outcomes~$\rho$; the proof shows
that $\rho$ has the distribution of a standalone transcript. Then the
following hold.
\begin{enumerate}
\item[(i)] For every setup and every transcript,
\begin{equation}\label{eq:never-slower-bound}
\begin{aligned}
H&\le S+2\min\{T_{\rm C}(I,\omega),\,Q(I,\omega;\rho)\}\\
 &\le S+2\min\{T_{\rm C}(I,\omega),\,Q_{\max}\}.
\end{aligned}
\end{equation}
\item[(ii)] With probability at least $1-\delta$, the output is correct and
\begin{equation*}
\begin{aligned}
H&\le S+2\min\bigl\{T_{\rm C}(I,\omega),\,A\sqrt{T_{\rm C}(I,\omega)}\bigr\}\\
 &\le S+2A\min\bigl\{T_{\rm C}(I,\omega),\,\sqrt{T_{\rm C}(I,\omega)}\bigr\}.
\end{aligned}
\end{equation*}
\item[(iii)] Every returned point is a verified feasible point of~$I$. On an
infeasible input, the combined algorithm always reports infeasibility. The
only possible error is a false report of infeasibility on a feasible input.
\end{enumerate}
\end{corollary}

The proof couples the combined run with standalone runs of the two
algorithms, so that the race stops within the first
$\min\{T_{\rm C},Q\}$ rounds.  Both algorithms read one setup, so no
union bound over two setups is needed: on the event of
Theorem~\ref{thm:main} that both are correct, whichever answers first is
correct.  The proof is in Appendix~\ref{app:finite-quantum-recursion}.

\subsection{Worst-case feasibility and optimization}
\label{subsec:ip-consequences}

Theorem~\ref{thm:main} decides integer feasibility.  Standard reductions
extend its worst-case bound to optimization and to mixed-integer
feasibility.

\begin{corollary}[Quantum integer programming]
\label{cor:worst-case-ip}
For every rational integer linear optimization instance $(A,b,c)$
in~\eqref{eq:ilo}, a quantum algorithm classifies the instance as
infeasible, unbounded, or finite with
probability at least $1-\delta$, and finds an optimizer if finite.  On
every input it uses at most
$2^{O(n)}\Phi(n)^{n/2}\operatorname{poly}(L,\log(1/\delta))$ gates.
Mixed-integer feasibility has the same bound, with $n$ integer variables.
\end{corollary}

The two assertions follow from the more precise
Corollaries~\ref{cor:mixed-integer-linear-feasibility}
and~\ref{cor:integer-linear-optimization}, which
Appendix~\ref{app:reductions} states and proves by standard reductions.
For mixed-integer feasibility, we project out the continuous variables,
keep the rational lift of the projection, and recover the continuous part
of a witness by linear programming.  For optimization, given a verified
integer feasible point, one linear program detects unboundedness, and
binary search over the objective value needs $\operatorname{poly}(L)$
feasibility calls.

The worst-case bounds above do not need the address tree.  Following one
random root-to-leaf path of the recursion and amplifying the probability
that it ends at a feasible point gives them directly, although this route
does not track the classical reference instance by instance.

\begin{proposition}[Worst-case bound by random recursion paths]
\label{prop:random-path-worst-case}
Let $I=(K,x_0+\Lambda)$ be an instance as in Theorem~\ref{thm:main} and
let $\delta\in(0,1/2)$.  If $K\cap(x_0+\Lambda)\neq\varnothing$, a quantum
algorithm returns one of its points with probability at least $1-\delta$.
It returns only exactly verified points and otherwise reports
infeasibility.  On every input it uses
$2^{O(n)}\Phi(n)^{n/2}\operatorname{poly}(L,\log(1/\delta))$ gates and
$2^{O(n)}\operatorname{poly}(L,\log(1/\delta))$ qubits.
\end{proposition}
The address-tree route fixes all randomness on $\omega$ and certifies
every hash, because a quantum walk must see the same children at every
query.  A single path is sampled afresh in each run, so it can draw new
coins at every rounded node and never certifies a hash.  At a rounded
node it picks a uniform cover slot, a uniform nonzero hash normal modulo
a prime fixed in advance for the pair $(k,d)$, and a uniform residue,
and decodes the child by one exact CVP in the bucket lattice of
Lemma~\ref{lem:bucket-basis}.  At a feasible node with a geometrically
valid flatness output, a fixed feasible child is reached when the slot's
tile contains it, the normal isolates it in that tile, and the residue is
its own, which happens with probability at least
$2^{-O(d)}\Phi(k)^{-d}$.  By induction on the dimension, one path ends at
a verified point with probability at least $2^{-O(n)}\Phi(n)^{-n}$ on a
feasible instance with a good cover seed and valid primes.  Amplitude
amplification with a random number of iterations below
$m=2^{O(n)}\Phi(n)^{n/2}$, as in the exponential search of Brassard,
H{\o}yer, Mosca, and Tapp~\cite{brassard2002quantum}, then finds such a
point.  The proof is in Appendix~\ref{subsec:random-path}.

\subsection{Bounded equation form}
\label{subsec:bounded-equality}
\label{app:bounded-equation}
\label{sec:der}

For $\{x\in\mathbb Z^n:Ax=b,\ 0\leq x\leq u\}$ with positive integer
bounds $u_i$, DER cuts each range $[0,u_i]$ at the distinct points
$0,u_i,2^{j-1},u_i-2^{j-1}$ for $0\leq j\leq\lceil\log_2u_i\rceil$, splitting the box into $2^{O(n)}\Psi(u)$ cells that stay in
$\prod_i[-1/2,u_i+1/2]$ when doubled about any of their points
\cite[Theorem~20 and Lemma~21]{dadush2025approximate}.  Given a good
cover seed, if a cell's slice of $\{x:Ax=b\}$ has an integer point,
approximate integer programming \cite{dadush2013sparse}
returns an integer point of the slice doubled about one of its points
(Lemma~\ref{lem:der-cell-test}, proved in
Appendix~\ref{app:der-cell-test}) with probability at least $1-\beta$ in
$2^{O(n)}\operatorname{poly}(L,\log(1/\beta))$ time.  This point lies in
$\prod_i[-1/2,u_i+1/2]$, so it is feasible by integrality.  Classically, testing the cells in order until one yields a
point, with the test's randomness fixed on a setup $\omega'$, takes time
$T_{\rm DER}(I,\omega')\leq2^{O(n)}\Psi(u)\operatorname{poly}(L,\log(1/\delta))$.

The DER reduction thus has the explicit search space that the geometric
recursion lacks: the cells have explicit mixed-radix indices and need no
addresses.  Amplitude amplification over all cells would improve the
worst-case bound but would not track the early termination of the
reference.  We instead put the cell indices in a balanced binary tree,
marked by the cell test, and apply Theorem~\ref{thm:tree-search}.  Depth-first
search follows the cells' classical order, so it stops where the
reference does (Theorem~\ref{thm:der-instance-wise}).  Running this
search and the algorithm of Theorem~\ref{thm:main} side by side gives the
minimum in Corollary~\ref{cor:bounded-equation-form}.

Write $I=(A,b,u)$, with rational $A,b$ and positive integers
$u_1,\ldots,u_n$, and put
\begin{equation}
    \mathcal P_I:=\{x:Ax=b,\ 0\leq x\leq u\},
    \qquad
    \Psi(u):=\prod_{i=1}^n\log_2(u_i+1).
    \label{eq:der-range-factor}
\end{equation}
The DER cells $\mathcal R_0,\ldots,\mathcal R_{N-1}$, with
$N\leq7^n\Psi(u)$, are listed in lexicographic order of their mixed-radix
indices, and $\mathcal R_j$ is computed from $j$ in polynomial time
(Appendix~\ref{app:der-cell-test}).

The classical DER reference tests the cells
$\mathcal R_0,\mathcal R_1,\ldots$ in order with the capped cell test of
Lemma~\ref{lem:der-cell-test} and returns the first verified point of
$\mathcal P_I\cap\mathbb Z^n$ that it finds; if no cell yields one, it
reports infeasibility.  Its shared setup holds one cover seed and one random tape for the
test, both reused in every cell; we write $\omega$ for this setup in
Theorem~\ref{thm:der-instance-wise}, and $\omega'$ when it runs alongside
the algorithm of Theorem~\ref{thm:main}.  The parameters are fixed in
Appendix~\ref{app:der-instance-wise}.

\begin{theorem}[Instance-wise DER speedup]
\label{thm:der-instance-wise}
For every bounded equation-form input $I$ and $\delta\in(0,1/2)$, there is a
quantum algorithm using the same setup $\omega$ as the DER reference such
that, with probability at least $1-\delta$, both are correct and its gate
count is at most
\begin{equation}
    2^{O(n)}\operatorname{poly}\!\left(L,\log\frac1\delta\right)
    \sqrt{T_{\mathrm{DER}}(I,\omega)}.
    \label{eq:der-instance-wise}
\end{equation}
On every input and tape the respective resource bounds are
\begin{align*}
     \text{classical:}\quad&2^{O(n)}\Psi(u)\operatorname{poly}(L,\log(1/\delta))
         \text{ bit operations},\\
     \text{quantum:}\quad&2^{O(n)}\Psi(u)^{1/2}\operatorname{poly}(L,\log(1/\delta))
         \text{ gates}.
\end{align*}
The
quantum algorithm uses
$2^{O(n)}\operatorname{poly}(L,\log(1/\delta))$ qubits.  Both return only
exactly verified points, and an infeasible instance never produces a point.
\end{theorem}
Once $\omega$ is fixed, the cell test is deterministic.  Handle $N=1$ by
running the single cell test directly; assume $N\geq2$ below.  The quantum
algorithm searches the interval tree $\mathcal I(N)$ of
Section~\ref{subsec:binary-address-tree} over the cell indices, with
leaf $j$ marked exactly when the test accepts $\mathcal R_j$.  Interval
splitting and mixed-radix decoding give the child circuits, and the
reversible cell test gives the mark, all at cost
$2^{O(n)}\operatorname{poly}(L,\log(1/\delta))$.  The leaves appear in the
reference's order, so every vertex that depth-first search visits before
the first marked leaf lies on the root path of a cell the reference
tests.  Hence $T_{\mathrm{dfs}}(\mathcal I(N))$ is at most
$\lceil\log_2N\rceil+1$ times $T_{\mathrm{DER}}(I,\omega)$, and
$\log N=\operatorname{poly}(L)$.  Theorem~\ref{thm:tree-search} with
$T_1=2N-1$ then gives~\eqref{eq:der-instance-wise}, and its every-input
cap together with~\eqref{eq:der-cell-count} gives the worst-case bound.
The proof, which also budgets the cover seed, the per-cell failure
parameter, and the synthesis of rotations, is in
Appendix~\ref{app:der-instance-wise}.

We can now accelerate whichever reference finishes first, without
deciding in advance which one is faster on the instance.

\begin{corollary}[Bounded equation form]
\label{cor:bounded-equation-form}
\label{cor:bounded-equation-portfolio}
For rational $A,b$, positive integers $u_1,\ldots,u_n$, and
$\delta\in(0,1/2)$, a quantum algorithm finds a verified point of
$\mathcal P_I\cap\mathbb Z^n$ if one exists, and reports infeasibility otherwise,
with probability at least $1-\delta$.  With the same probability, its gate
count is at most
\begin{equation}
    2^{O(n)}\operatorname{poly}\!\left(L,\log\frac1\delta\right)
    \sqrt{\min\{T_{\mathrm C}(I,\omega),
                     T_{\mathrm{DER}}(I,\omega')\}},
    \label{eq:ip-portfolio-instance-wise}
\end{equation}
where $T_{\mathrm C}(I,\omega)$ is the reference time of
Theorem~\ref{thm:main} on this instance with setup $\omega$, and
$T_{\mathrm{DER}}(I,\omega')$ is the reference time above with an
independent setup $\omega'$, each with error parameter $\delta/2$.  On every
input and tape, the algorithm uses at most
\begin{equation}
    2^{O(n)}\min\{\Phi(n)^{n/2},\Psi(u)^{1/2}\}
    \operatorname{poly}\!\left(L,\log\frac1\delta\right)
    \label{eq:bounded-equation-runtime}
\end{equation}
gates and $2^{O(n)}\operatorname{poly}(L,\log(1/\delta))$ qubits.  An
infeasible instance never produces a point.
\end{corollary}

\begin{proof}
Run the algorithms of Theorems~\ref{thm:main} and
\ref{thm:der-instance-wise} with error parameter $\delta/2$, interleaving
one elementary operation of each and stopping as soon as either halts.
Their classical setup costs are
$2^{O(n)}\operatorname{poly}(L,\log(1/\delta))$, which the overhead
in~\eqref{eq:ip-portfolio-instance-wise} absorbs because both reference
times are at least one.  By a union bound, with probability at least
$1-\delta$ both algorithms are correct and satisfy their instance-wise
bounds.  On this event the combined time is at most twice the smaller of
their times, which gives~\eqref{eq:ip-portfolio-instance-wise}.
Applied to the two deterministic caps, the same interleaving
gives~\eqref{eq:bounded-equation-runtime} on every tape.  The two
workspaces together have the stated size, and exact verification
preserves witness soundness.
\end{proof}

\begin{corollary}[Bounded equation-form optimization]
\label{cor:bounded-equation-optimization}
For rational $A,b,c$, positive integers $u_1,\ldots,u_n$, and
$\delta\in(0,1/2)$, a quantum algorithm maximizes $c^\top x$ over
$\{x\in\mathbb Z^n:Ax=b,\ 0\leq x\leq u\}$, or reports that this set
is empty, with probability at least $1-\delta$.  Including the objective
in the input length $L$, its every-tape bounds are
\begin{equation*}
  2^{O(n)}\Psi(u)^{1/2}\operatorname{poly}(L,\log(1/\delta))
  \quad\text{gates},\qquad
  2^{O(n)}\operatorname{poly}(L,\log(1/\delta))
  \quad\text{qubits}.
\end{equation*}
Every returned point is verified exactly.
\end{corollary}

\begin{proof}
Multiply $c$ by a positive common denominator to obtain
$\bar c\in\mathbb Z^n$ with the same maximizers, and put
\begin{equation*}
 B_{\mathrm{obj}}:=\sum_{i=1}^n|\bar c_i|u_i.
\end{equation*}
Every objective value $\bar c^\top x$ on the box belongs to
$[-B_{\mathrm{obj}},B_{\mathrm{obj}}]\cap\mathbb Z$, and
$\log(B_{\mathrm{obj}}+1)=\operatorname{poly}(L)$.
If $B_{\mathrm{obj}}=0$, then $\bar c=0$, because every $u_i$ is
positive, and one call to Theorem~\ref{thm:der-instance-wise} suffices.

Suppose $B_{\mathrm{obj}}\geq1$.  For an integer
$\gamma\in[-B_{\mathrm{obj}},B_{\mathrm{obj}}]$, consider the
bounded equation-form feasibility instance
\begin{equation*}
 Ax=b,\qquad \bar c^\top x-s=\gamma,\qquad
 0\leq x\leq u,\qquad 0\leq s\leq2B_{\mathrm{obj}},\qquad
 (x,s)\in\mathbb Z^{n+1}.
\end{equation*}
It is feasible if and only if an original feasible point satisfies
$\bar c^\top x\geq\gamma$.  Indeed, for such a point the integer
$s=\bar c^\top x-\gamma$ lies in $[0,2B_{\mathrm{obj}}]$; conversely,
its defining equation and $s\geq0$ imply the threshold inequality.
This instance has $n+1$ variables, positive integer upper bounds, and
encoding length polynomial in $L$, uniformly over these thresholds.
Its range factor is
\begin{equation*}
 \Psi(u_1,\ldots,u_n,2B_{\mathrm{obj}})=\Psi(u)\log_2(2B_{\mathrm{obj}}+1).
\end{equation*}
Thus Theorem~\ref{thm:der-instance-wise}, applied in dimension $n+1$,
uses at most
\begin{equation*}
 2^{O(n+1)}
 \sqrt{\Psi(u)\log_2(2B_{\mathrm{obj}}+1)}
 \operatorname{poly}(L,\log(J/\delta))
 =
 2^{O(n)}\Psi(u)^{1/2}\operatorname{poly}(L,\log(1/\delta))
\end{equation*}
gates per call, where
$J:=2+\lceil\log_2(2B_{\mathrm{obj}}+1)\rceil=\operatorname{poly}(L)$.

First call bounded equation-form feasibility on the original instance.
If it reports infeasibility, stop.  Otherwise retain its witness and
set $\ell=-B_{\mathrm{obj}}$ and $h=B_{\mathrm{obj}}$.  While $\ell<h$, set
$\gamma=\ell+\lceil(h-\ell)/2\rceil$ and test the augmented instance.
On a feasible answer set $\ell=\gamma$ and retain the returned
$x$; on an infeasible answer set $h=\gamma-1$.
There are at most $\lceil\log_2(2B_{\mathrm{obj}}+1)\rceil$ threshold
calls, hence fewer than $J$ calls in total.  Use fresh setups and error
at most $\delta/J$ for each call.  Conditional on any preceding history,
its error probability is at most $\delta/J$, so all calls are correct
with probability at least $1-\delta$.
On this event the optimum stays in $[\ell,h]$, and the retained
witness has objective at least $\ell$.  At termination $\ell=h$ is
the optimum, and that witness attains it.  Project away the slack
and verify the returned point and its objective exactly.
The box is bounded, so an unbounded objective is impossible.
The polynomial number of sequential calls preserves the gate bound;
their workspaces can be reused, giving the stated qubit bound.
The bounds hold on every tape by the feasibility algorithm's caps
and the fixed bound on the number and size of the calls.
\end{proof}

For $u_i\leq n^{O(1)}$, the gate bound for bounded equation form is
\begin{equation*}
    2^{O(n)}(\log(2n))^{n/2}\operatorname{poly}(L,\log(1/\delta)),
\end{equation*}
against DER's classical
\begin{equation*}
    2^{O(n)}(\log(2n))^n\operatorname{poly}(L).
\end{equation*}
The instance-wise comparison is stronger than a choice between these
worst-case formulas, because it follows whichever reference stops earlier
on each input.  It concerns only the two specified implementations.
\section{Discussion}
\label{sec:discussion}

Our speedup is quadratic up to $2^{O(n)}\operatorname{poly}(L)$ factors.
In the worst-case bounds the $2^{O(n)}$ factor changes the
coefficient of $n$ in the exponent of $\Phi(n)$ by only
$O(1/\log\log(4n))$, while the $\operatorname{poly}(L)$ factor
remains explicit.  Instance by instance, these factors matter when
$T_{\mathrm C}(I,\omega)$ is itself only $2^{O(n)}$.  Most of them (the
cover multiplicity, the address inflation, and the branch constant)
enlarge the tree and are already square-rooted.  The exception is
the replay: each local query recomputes node processing along its root
path, including flatness at the root, whereas the classical reference
processes each node once.  Carrying processed ancestor records in the vertex encoding, together
with an early-terminating tree search whose cost is weighted by node
costs, might give a bound near $\sqrt{T_{\mathrm C}(I,\omega)P}$,
where $P$ bounds both the classical cost of processing one node and
the classical cost of decoding one address.
Such a bound would also require gate-efficient coherent access to
the stored records and gate-efficient coherent address decoding
in our no-QRAM model.  We leave this open.

The shape of the recursion tree alone cannot guarantee a
more-than-quadratic advantage for arbitrary black-box leaf markings.
Consider a complete binary tree with $N$ leaves and a leaf predicate
available only by oracle queries, under the promise that either
no leaf or exactly one leaf is marked.  Any bounded-error quantum
algorithm distinguishing these cases makes $\Omega(\sqrt N)$
queries to that predicate~\cite{bennett1997strengths}, whereas
classical search requires $\Theta(N)$ queries in the worst case.
This is a query-model obstruction, not an ILP lower bound.
A larger advantage would need exploitable structure in the
markings, such as algebraic structure in the constraints
\cite{jordan2025dqi}, or a different algorithm.
We measure every quantum bound against the improved geometry.  Against the older
$2^{O(n)}(\log(2n))^{4n}\operatorname{poly}(L)$ baseline, our gate count
would look like a quartic speedup, crediting the search with what
the flatness routine gained.  The instance-wise comparison is with one
implementation of the recursion, the depth-first reference of
Section~\ref{subsec:reference} (Algorithm~\ref{alg:classical-reference}).

Improving the flatness guarantee improves both worst-case bounds
through $\Phi$: a constructive nondecreasing factor
$\Phi'$ with the interface of
Theorem~\ref{thm:constructive-flatness} would give
$2^{O(n)}\Phi'(n)^n\operatorname{poly}(L)$ classically and
$2^{O(n)}\Phi'(n)^{n/2}\operatorname{poly}(L)$ quantumly.

The natural place to look for such a $\Phi'$ is the one factor $\xi$ that
our construction loses by certifying covering through Gaussian mass
(Section~\ref{subsec:flatness-proof-idea}).
Consider a $2^{O(n)}\operatorname{poly}(L)$-time test that, for the
current sublattice and the stage's fixed $m$ and $g$, either certifies the
dual-density bound of Lemma~\ref{lem:rank-weighted-dual-density} up to a
constant factor or returns a dual sublattice whose annihilator lowers the
rank-penalized determinant.  Such a test would lower $\Phi(n)$ to
$O(\log^{3/2}(2n))$, giving $(\log(2n))^{3n/2}$ classically and
$(\log(2n))^{3n/4}$ quantumly.

Finally, the quantum algorithm uses $2^{O(n)}\operatorname{poly}(L,\log(1/\delta))$
qubits, matching classical space up to these factors; whether polynomially
many qubits suffice remains open.

\section*{AI Use Statement}
The authors used GPT-6.1 Sol and Opus 5.5 (accessed via GitHub Copilot) as LLM-based research assistants, to help explore
related work, check proof steps, and edit portions of this manuscript.
The human authors take responsibility for its mathematical claims, references,
and final text.

\section*{Disclaimer}
This paper was prepared for informational purposes by the Global Technology Applied Research center of JPMorgan Chase \& Co. This paper is not a product of the Research Department of JPMorgan Chase \& Co. or its affiliates. Neither JPMorgan Chase \& Co. nor any of its affiliates makes any explicit or implied representation or warranty and none of them accept any liability in connection with this paper, including, without limitation, with respect to the completeness, accuracy, or reliability of the information contained herein and the potential legal, compliance, tax, or accounting effects thereof. This document is not intended as investment research or investment advice, or as a recommendation, offer, or solicitation for the purchase or sale of any security, financial instrument, financial product or service, or to be used in any way for evaluating the merits of participating in any transaction.

\bibliographystyle{alpha}
\bibliography{references}

\clearpage
\appendix
\section{Proof of the coherently indexable cover theorem}
\label{app:indexable-cover-proof}

This appendix proves Lemma~\ref{lem:rational-indexed-cover}, the
indexable $M$-ellipsoid cover of Section~\ref{sec:search}.  The cover
supplies the first layer of the lattice-point addresses in
Section~\ref{sec:search}, and its seed is the cover seed in the shared
setup $\omega$ of the quantum algorithm in Section~\ref{sec:quantum-ip}.
Section~\ref{sec:classical-flatness} reuses the rational radius
computation from the first step of the proof, and the DER verifier of
Appendix~\ref{app:der-cell-test} runs its enumeration calls through the
cover.

The $M$-ellipsoid algorithms of Dadush \cite[Algorithms~4.4--4.5]{dadush2012integer} are
randomized and list the translates of their cover.  Local access needs a
single random seed that works for every body the recursion can produce
and a reversible circuit that maps an index $j$ to the $j$th
translate.  A union bound over the at most $2^{L'+1}$ bodies of encoding
length at most $L'$ supplies the seed.  A breadth-first traversal of a
rational grid of tiles, rerun inside the circuit, supplies the index
map.  We prove Lemma~\ref{lem:rational-indexed-cover} in the following
form, which makes its bit lengths explicit.

\begin{restated}[Lemma~\ref{lem:rational-indexed-cover} in full]
Let $L'\geq1$, $\tau\in(0,1/2)$, and
$L_\tau:=\operatorname{poly}(L',\log(1/\tau))$.  With probability at least
$1-\tau$, a uniform classical seed of $L_\tau$ bits is good: for every
compact full-dimensional convex body $D\subseteq\mathbb R^d$ given by a
rational lift of encoding length at most $L'$, a deterministic reversible
circuit of size $2^{O(d)}L_\tau$ with the seed hard-wired maps
$j\in[C_{\rm cov}^d]$ to a rational center $s_j$ and a validity bit such
that, for a fixed rational ellipsoid $\widehat E=E(0,\widehat Q)$,
\eqref{eq:rational-indexed-cover} holds and, for every lattice $\Lambda$,
\begin{equation}
    \nu_{\max}(\Lambda,\widehat E)\leq C_{\rm pt}^d\nu_{\max}(\Lambda,D).
    \label{eq:rational-cover-point-bound}
\end{equation}
All outputs have $L_\tau$ bits, and the circuit uses $2^{O(d)}L_\tau$
workspace.
\end{restated}

Throughout this appendix, $L'$ and $\tau$ denote the encoding-length
bound and the failure probability in Lemma~\ref{lem:rational-indexed-cover},
and $\operatorname{poly}(L)$ abbreviates
$\operatorname{poly}(L',\log(1/\tau))$.
In the proof, $E_0$ is the $M$-ellipsoid computed with the hard-wired seed,
$\widehat E=E(0,\widehat Q)$ is its rational rounding in
Lemma~\ref{lem:rational-indexed-cover}, and $\mathcal H$ is the set of selected
tile centers $s_j$.  We write the $2^{O(d)}$ gate count of
Lemma~\ref{lem:rational-indexed-cover} as
$C_{\rm idx}^d\operatorname{poly}(L)$, for a universal integer
$C_{\rm idx}\geq2$ fixed at the end of the proof.

The proof computes rational radii for $D$, fixes the seed, and makes the
M-Gen trials finite and certified.  It then rounds $E_0$ to
$\widehat E$, tiles space, bounds the numbers of tiles and lattice
points, and indexes the selected tiles by breadth-first search.

\begin{proof}[Proof of Lemma~\ref{lem:rational-indexed-cover}]
\emph{The case $d=0$.} Here $D=\mathbb R^0=\{0\}$.
Take $\widehat Q=I_0$, the empty identity matrix, and $s_1=0$.
There is exactly one valid slot, as $C_{\mathrm{cov}}^0=1$, and its
ellipsoid equals $D$.  Both the cover and the shift-invariant
lattice-point bound hold with equality.  No randomness is needed, and
the bit-length and circuit-size bounds are immediate.
For the rest of the proof, assume $d\geq1$.

\proofstep{Rational radii.}
Rational linear programming finds $d+1$ affinely independent rational points of $D$: once
a rational simplex of dimension $i<d$ has been found, optimize over $D$ a
rational basis of the directions orthogonal to its affine hull, in both
signs, to find a point outside that hull.  Full dimensionality
guarantees success.  Standard basic-solution bounds give all these
points $\operatorname{poly}(L)$ bits
\cite[Chapters~10 and~13]{schrijver1998theory}.  Let $a_D$ be their
rational barycenter.  The rational facet slacks of their simplex give an
explicit rational number $r_{\mathrm{in}}\geq2^{-\operatorname{poly}(L)}$
such that
\begin{equation}
    a_D+r_{\mathrm{in}}\mathbb{B}_2^d\subseteq D.
    \label{eq:rational-cover-inner-ball}
\end{equation}
Optimizing the coordinate directions over $D$ also gives a rational
$R_{\mathrm{out}}\leq2^{\operatorname{poly}(L)}$ with
$D\subseteq a_D+R_{\mathrm{out}}\mathbb{B}_2^d$.  Thus the centered symmetric lifted
polytope $D-D$ has explicit rational inner and outer radii.

\proofstep{Fixing the seed.}
At most $2^{L'+1}$ bodies have encoding length at most $L'$.  For each
such body $D$, the construction below reads a common uniformly random
classical seed and fails to produce~\eqref{eq:mgen-covering-guarantee}
with probability at most $\tau2^{-L'-1}$.  By a union bound, with
probability at least $1-\tau$ one seed succeeds for all these bodies
simultaneously; we call such a seed good and hard-wire it.  All
computations below are then deterministic.

\proofstep{Certified candidates.}
We use M-Gen and the covering certificates of
\cite[Algorithms~4.4--4.5 and Theorems~4.3.2--4.3.3]{dadush2012integer}.
A trial generates a centered rational candidate
\begin{equation*}
    E_0:=\{x:x^\top Q^{(0)}x\leq1\},
    \qquad Q^{(0)}\in\mathbb Q^{d\times d},
\end{equation*}
and then runs two capped covering traversals, one for $(D,E_0)$ and one
for $((D-D)^\circ,E_0^\circ)$.  Both bodies have rational lifted
descriptions.  Membership in $(D-D)^\circ$ is
\begin{equation*}
    \max_{x\in D}\langle y,x\rangle
       -\min_{x\in D}\langle y,x\rangle\leq1,
\end{equation*}
and linear-programming duality supplies its lift.  The logarithms of its
inner and outer radii are polynomially bounded by those of $D$.

For any rational body $A$ and rational ellipsoid $E$, the grid
construction below gives a cover of $A$ by translates of a fixed
dilation of $E$.  Its tile count is at most $c^dN(A,E)$ for a universal
$c$, without any $M$-ellipsoid assumption.  Conversely, completion with
$k$ tiles certifies $N(A,E)\leq c^dk$, since that fixed dilation is
covered by $c^d$ translates of $E$.  Each traversal stops once it would
exceed $C_{\mathrm{test}}^d$ tiles, for a sufficiently large universal
$C_{\mathrm{test}}$, and the candidate is rejected on that event.  If
both traversals complete, entropy duality, as used in
\cite[Theorem~4.3.3]{dadush2012integer}, certifies for a universal $C_M$
that
\begin{equation}
    N(D,E_0)\leq C_M^d,
    \qquad
    N(E_0,D)\leq C_M^d.
    \label{eq:mgen-covering-guarantee}
\end{equation}
The tests therefore never accept an incorrect candidate, even when its
random generation was unsuccessful.  For each fixed body $D$, before
the seed is fixed, candidate generation produces a candidate passing
both deterministic tests with probability bounded below by a universal
positive constant over its random tape, after enlarging the caps.  Bounded
dimensions are handled by ordinary rational ellipsoidal rounding.

\proofstep{Finite candidate generation.}
First round $D$ by a rational affine map.  M-Gen estimates its centroid
by $\widetilde c$ and samples the ideal tilt uniformly from the
dimension-scaled polar used in \cite[Algorithm~4.4]{dadush2012integer}:
\begin{align*}
    A_{\mathrm{sym}}&:=\operatorname{conv}\{D-\widetilde c,\widetilde c-D\},\\
    S_{\mathrm{tilt}}&:=dA_{\mathrm{sym}}^\circ
       =\{s:|\langle s,x\rangle|\leq d
                    \text{ for every }x\in D-\widetilde c\}.
\end{align*}
The convex hull and the factor $d$ belong to the sampling distribution;
the polar of the centered intersection is not used.  Rational linear
optimization in both signs decides membership in $S_{\mathrm{tilt}}$
exactly, and linear-programming duality gives a rational lift.

The centroid and covariance calls use
Lemma~\ref{lem:finite-bit-log-affine-moments}, with the inverse-polynomial
accuracies of
\cite[Theorem~4.3.2 and Corollaries~4.3.4--4.3.5]{dadush2012integer}
and a fixed small failure budget per trial.  The centroid call has $s=0$
and mean error at most $1/(3d)$ in its covariance norm, hence at most
$1/(d+1)$, as the cited M-Gen analysis requires.  Let $b$ be the exact
centroid.  The inclusion $-(D-b)\subseteq d(D-b)$
\cite[Section~5]{reis2023subspace} bounds each directional standard
deviation by $d$ times the corresponding support function of $D-b$.  A
successful estimate therefore lies in $b+\frac13(D-b)$, and $D$ has
polynomially bounded inner and outer radii about it in rounded
coordinates.  Reject unless $\widetilde c$ lies in the interior of $D$,
which rational linear programming checks.  For every retained tilt,
$|\langle s,x\rangle|\leq d$ on $D-\widetilde c$, so the tilted moment
call on this translated body uses $B=d$.  Check this bound exactly and
reject a trial that violates it.

We make the finite-tilt coupling quantitative.
The bounded dimensions have already been assigned to ordinary
rational ellipsoidal rounding, so assume $d\geq3$.
Conditional on the good centroid event, use a rational linear
rounding map $H$ for the symmetric body $S_{\mathrm{tilt}}$, with
\begin{equation*}
 \mathbb{B}_2^d\subseteq C_0:=HS_{\mathrm{tilt}}\subseteq q\mathbb{B}_2^d,
 \qquad q\geq1\text{ rational},\qquad q=\operatorname{poly}(d).
\end{equation*}
The map is linear, not translated, because $S_{\mathrm{tilt}}$ is
origin-symmetric.  For every $x\in D-\widetilde c$ its definition
and the inner-ball inclusion imply
\begin{equation*}
 \|H^{-\top}x\|_2
 =\sup_{\|v\|_2\leq1}|\langle H^{-1}v,x\rangle|
 \leq d.
\end{equation*}
Let $p_{\mathrm M}>0$ be a fixed effective lower bound on the ideal
candidate-generation success probability in the cited M-Gen
analysis, before spending its numerical failure budgets.
Fix a positive rational constant
$c_{\mathrm{tilt}}\leq\min\{1/1000,p_{\mathrm M}/8\}$.
Apply Lemma~\ref{lem:finite-grid-refinement} to $C_0$, with $\nu$ the
largest power of two not exceeding $c_{\mathrm{tilt}}/(2dq)$ and
mixing error $\theta=c_{\mathrm{tilt}}/(2d)$.
Using its $h,\alpha,\alpha'$ and
$\beta=\alpha+\alpha'+\sqrt d\,h$, we have
\begin{equation*}
 \beta\leq\frac{\nu}{4d}
       \leq\frac{c_{\mathrm{tilt}}}{8d^2q},\qquad
 C_0\subseteq U_h\subseteq(C_0)_{+\beta}
       \subseteq(1+\beta)C_0.
\end{equation*}
These choices have inverse-polynomial mesh and polynomial
mixing time.

For the analysis, let $Y$ be uniform on $U_h$ and
$Y_\star$ uniform on $C_0$.  A maximal coupling makes
$Y=Y_\star$ except with probability
\begin{equation*}
 1-\frac{\operatorname{Vol}(C_0)}{\operatorname{Vol}(U_h)}
 \leq(1+\beta)^d-1\leq2d\beta.
\end{equation*}
The cube center $a(Y)$ is uniform over the accepted cube centers.
Couple the actual walk center $Z$ to $a(Y)$ with error at most
$\theta$.
Choose the rational number
$\bar\beta:=\alpha+\alpha'+dh$, so that
$\beta\leq\bar\beta\leq2\beta$ for these parameters.
Return the tilt
$s:=H^{-1}Z/(1+\bar\beta)$.
Because $Z\in U_h\subseteq(1+\beta)C_0$, this tilt belongs to
$S_{\mathrm{tilt}}$; check that membership exactly and reject otherwise.
Set $s_\star:=H^{-1}Y_\star$, which is uniform on $S_{\mathrm{tilt}}$.
Outside an exceptional event of probability at most
\begin{equation*}
 2d\beta+\theta\leq\frac{3c_{\mathrm{tilt}}}{4d},
\end{equation*}
we have $Z=a(Y)$ and $Y=Y_\star$.  Since
$\|Z\|_2\leq q+1$ and $\|a(Y)-Y\|_2\leq\sqrt d\,h/2$,
\begin{equation*}
 \left\|\frac{Z}{1+\bar\beta}-Y_\star\right\|_2
 \leq\frac{\sqrt d\,h}{2}+2\beta(q+1)
 \leq\frac92q\beta
 \leq\frac{9c_{\mathrm{tilt}}}{16d^2}.
\end{equation*}
Consequently
\begin{equation*}
 \sup_{x\in D-\widetilde c}
 |\langle s-s_\star,x\rangle|
 \leq d\left\|\frac{Z}{1+\bar\beta}-Y_\star\right\|_2
 \leq\frac{9c_{\mathrm{tilt}}}{16d}
 <\frac1{100d}.
\end{equation*}
The ideal success event loses at most the displayed exceptional
probability, without any independence assumption.
Together with suitably small fixed centroid and moment failure
budgets, this leaves a fixed positive candidate-success probability.
The corresponding normalized densities differ by factors in
$[e^{-1/(50d)},e^{1/(50d)}]$.  Minimizing directional second moments
over their centers shows that the same bounds hold for their covariance
matrices in Loewner order.  Take relative covariance accuracy
$1/(100d)$ in the tilted moment call and form the candidate
$E_0=\{x:x^\top\widetilde\Sigma_s^{-1}x\leq d\}$.
It is a constant-factor approximation to the candidate of the source, so
the cited M-Gen success event survives discretization and moment
estimation once the universal covering caps are enlarged.  Candidate
generation therefore succeeds with constant probability, and the two
exact covering tests certify every accepted candidate.

\proofstep{Caps and amplification.}
High confidence comes from the independent capped trials below, not from
smaller numerical error inside one trial.  Exact positive-definiteness
checks, polynomial register caps, and deterministic per-trial time caps
reject all malformed or overlong trials, and these caps can be chosen to
lose only a fixed fraction of the success probability of a trial.  Each
trial, including its two covering tests and visited-list scans, costs
$2^{O(d)}\operatorname{poly}(L)$ time, space, and finite random bits.
Only candidate generation reads randomness, and it has a polynomial-time
cap; the covering tests are deterministic.  The random-tape length per
trial is therefore polynomial, despite the single-exponential
verification cost.

Run $O(L'+\log(1/\tau))$ independent trials and use the first candidate
that passes both tests.  Their constant success probability gives
failure at most $\tau2^{-L'-1}$ for each
fixed input body.  If all trials fail, output a fixed dummy
positive-definite matrix and only invalid slots.  All computations are
capped on every seed, and \eqref{eq:mgen-covering-guarantee} holds on a
good seed.  Retaining the trial histories for reversible simulation
still uses $2^{O(d)}\operatorname{poly}(L)$ gates and workspace.

\proofstep{Rational rounding.}
For a rational shape matrix, compute
\begin{equation*}
    Q^{(0)}=L_0\Delta_0L_0^\top,
    \qquad A_0:=L_0^{-\top}\Delta_0^{-1/2},
    \qquad E_0=A_0\mathbb{B}_2^d,
\end{equation*}
where $L_0$ is unit lower triangular and $\Delta_0$ positive diagonal.
Fraction-free elimination and Hadamard's minor bound give polynomial-bit
rational factors.  Bisection and exact squaring
approximate $\Delta_0^{-1/2}$ entrywise with relative error at most
$1/10$; put
\begin{equation*}
    \widetilde A_0:=L_0^{-\top}\widetilde\Delta_0^{-1/2},
    \qquad \widehat A_0:=\frac{10}{9}\widetilde A_0,
    \qquad \widehat Q:=\widehat A_0^{-\top}\widehat A_0^{-1}.
\end{equation*}
The matrix $A_0^{-1}(\widetilde A_0-A_0)$ is diagonal with these relative
errors, so
\begin{equation}
    \left\|A_0^{-1}
    (\widetilde A_0-A_0)\right\|_2\leq\frac1{10}.
    \label{eq:rational-ellipsoid-relative-error}
\end{equation}
The rounded matrices $\widehat A_0,\widehat A_0^{-1},\widehat Q$ are rational
with polynomial encoding, and
$\widehat E:=\widehat A_0\mathbb{B}_2^d=E(0,\widehat Q)$ satisfies
\begin{equation}
    E_0\subseteq\widehat E
    \subseteq\frac{11}{9}E_0.
    \label{eq:rational-ellipsoid-sandwich}
\end{equation}
The inclusions hold because the singular values of $A_0^{-1}\widehat A_0$ lie in $[1,11/9]$.

\proofstep{Tiles and the slot count.}
Let $m:=\lceil\sqrt d\rceil$ and
\begin{equation*}
    B_P:=\frac1m\widehat A_0,
    \qquad
    P:=B_P[-1,1)^d,
    \qquad
    \overline P:=B_P[-1,1]^d.
\end{equation*}
The matrix $B_P$ is rational and invertible, and
$\overline P\subseteq\widehat E$ because
$\|y\|_2/m\leq\sqrt d/m\leq1$ for $y\in[-1,1]^d$.  Moreover,
\begin{equation}
    \frac{\operatorname{Vol}_d(\widehat E)}
         {\operatorname{Vol}_d(P)}
    =\operatorname{Vol}_d(\mathbb{B}_2^d)\left(\frac m2\right)^d
    \leq(2\pi e)^{d/2}=:c_1^d.
    \label{eq:rational-tile-volume-ratio}
\end{equation}
The last inequality uses
$\operatorname{Vol}_d(\mathbb{B}_2^d)\leq(2\pi e/d)^{d/2}$ and
$m\leq2\sqrt d$.

Find a rational point $a_0\in D$ by linear programming and tile the ambient
space by the half-open sets
\begin{equation*}
    v+P,
    \qquad
    v\in\mathcal V:=a_0+2B_P\mathbb Z^d.
\end{equation*}
Select
\begin{equation*}
    \mathcal H:=
    \{v\in\mathcal V:(v+\overline P)\cap D\neq\emptyset\}.
\end{equation*}
For rational $v$, this predicate is exactly one rational linear
feasibility problem, formed by the lifted constraints for $D$ together
with $-\mathbf1\leq B_P^{-1}(x-v)\leq\mathbf1$.  Each test therefore
costs $\operatorname{poly}(L)$ gates, including on a common tile
boundary.  Every $x\in D$ belongs to a unique half-open tile $v+P$; that
$v$ lies in $\mathcal H$, and
\begin{equation*}
    x\in v+P\subseteq v+\widehat E.
\end{equation*}
Hence the translations in~\eqref{eq:rational-indexed-cover} are precisely the
rational tile centers $s_j=v\in\mathcal H$.

The tiles $v+P$ are disjoint, and the definition of $\mathcal H$ gives
$v+P\subseteq D+2\overline P$ for every $v\in\mathcal H$.  Cover $D$ by
$N(D,\widehat E)$ translates of $\widehat E$.  Each covering translate
contributes a subset of $\widehat E+2\overline P\subseteq3\widehat E$,
since $\overline P\subseteq\widehat E$, so
\begin{align*}
    |\mathcal H|\operatorname{Vol}_d(P)
    &\leq\operatorname{Vol}_d(D+2\overline P)\\
    &\leq N(D,\widehat E)3^d
        \operatorname{Vol}_d(\widehat E).
\end{align*}
Since $E_0\subseteq\widehat E$, we have
$N(D,\widehat E)\leq N(D,E_0)\leq C_M^d$, and
\eqref{eq:rational-tile-volume-ratio} yields
\begin{equation}
    |\mathcal H|\leq(3c_1C_M)^d.
    \label{eq:rational-cover-slot-count}
\end{equation}
Choose $C_{\mathrm{cov}}\geq3c_1C_M$.

\proofstep{Point bound.}
A maximal packing argument covers $cE_0$ by at most $(1+2c)^d$
translates of $E_0$ for every $c\geq1$.  Combining this fact at
$c=11/9$ with \eqref{eq:rational-ellipsoid-sandwich} and the cover of
$E_0$ by $C_M^d$ translates of $D$ gives
\begin{equation*}
    N(\widehat E,D)
    \leq\left(1+\frac{22}{9}\right)^d C_M^d.
\end{equation*}
Choose $C_{\mathrm{pt}}\geq\lceil(31/9)C_M\rceil$.  Covering
$\widehat E+x$ by these translates of $D$ and counting lattice points
proves \eqref{eq:rational-cover-point-bound} uniformly in $x$.

\proofstep{Indexing.}
Write each tile center uniquely as
\begin{equation*}
    v(z):=a_0+2B_Pz,
    \qquad z\in\mathbb Z^d,
\end{equation*}
and join two selected centers when their indices differ by a signed unit
vector.  This graph is connected.  For each selected center $v$, choose
$p\in D\cap(v+\overline P)$.  The segment $[a_0,p]\subseteq D$ meets a
sequence of selected tile closures from $v(0)+\overline P$ to
$v+\overline P$.  At a boundary crossing, including at the endpoint,
every incident tile contains the crossing point in its closure, so the
sequence can be refined to face-neighbor steps.  This is the grid connectivity used by
\cite[Theorems~4.2.1--4.2.2]{dadush2012integer}.

Run breadth-first search from $v(0)$, testing the $2d$ face-neighbors in
a fixed order and storing each newly selected center in a queue and a
visited list.  Exact closure-intersection tests decide selection, and a
linear scan of the visited list prevents repetitions.  The traversal
emits every selected center exactly once, in a deterministic order,
using
\begin{equation*}
    O\bigl(d|\mathcal H|^2\bigr)\operatorname{poly}(L)
\end{equation*}
operations and $|\mathcal H|\operatorname{poly}(L)$ space.  By
\eqref{eq:rational-cover-slot-count}, both are
$2^{O(d)}\operatorname{poly}(L)$.  Selected centers lie in
$D-\overline P$, so each coefficient vector has polynomial encoding.
This cover circuit needs no polynomial-space bound.

On input $j$, the circuit runs this deterministic traversal within its
fixed $C_{\mathrm{idx}}^d\operatorname{poly}(L)$ gate cap, copies the
$j$th center if it is emitted, and otherwise copies a fixed dummy center
with validity bit zero.  Once the traversal finishes, every remaining
slot of the padded index range is invalid, and
\eqref{eq:rational-cover-slot-count} guarantees that no selected center
is truncated.  After copying the output, the circuit reverses the
complete capped computation.  This compute--copy--uncompute circuit
works coherently on a superposition of indices.  Taking
$C_{\mathrm{idx}}$ large enough to dominate $C_{\mathrm{cov}}$ and the
fixed reversible-simulation overhead gives
$C_{\mathrm{idx}}^d\operatorname{poly}(L)$ gates and
$2^{O(d)}\operatorname{poly}(L)$ workspace per indexed call.
\end{proof}
\section{Finite implementation of classical geometric subroutines}
\label{app:finite-geometric-primitives}

The geometric routines we use are usually stated with real arithmetic and
loops of unbounded length.  A circuit needs a fixed finite number of bit
operations on every input, with every approximation error accounted for.
This appendix gives such implementations of the barycenter, moment,
$\ell$-position, and discrete Gaussian routines.  The constructive
flatness routine of Section~\ref{sec:classical-flatness} uses the barycenter, moment, and discrete Gaussian routines; the variant in Appendix~\ref{subsec:published-variant} uses the $\ell$-position routine.
The proof of the cover lemma in Appendix~\ref{app:indexable-cover-proof}
generates its $M$-ellipsoid candidates with the barycenter and moment
estimators, and the DER verifier of Appendix~\ref{app:der-cell-test}
uses the barycenter estimator.
Lemmas~\ref{lem:finite-bit-geometric-estimators}
and~\ref{lem:finite-bit-log-affine-moments} use a refined finite cube walk,
Lemma~\ref{lem:finite-bit-ell-position} optimizes a determinant against an
empirical Gaussian norm, and Lemma~\ref{lem:finite-bit-dgs} caps a discrete
Gaussian sampler by transcript coupling.  Each proof bounds the
approximation error and fixes all operation and register caps before
reversible simulation.  These are
classical randomized routines, and compiling them uses no coherent lattice
search.

\subsection{Finite realization of rank-weighted flatness}
\label{app:flatness-proofs}

\begin{proof}[Proof of Lemma~\ref{lem:rank-weighted-smooth-sublattice}]
We describe one attempt, check that accurate acceptance is valid, bound
the success probability through the hidden sublattice $N$, and treat
inaccurate tests, the sampler, and encoding.

Compute a rational $\widetilde g$ with
$\operatorname{nd}(\Lambda)\leq\widetilde g\leq2\operatorname{nd}(\Lambda)$ and an
integer $\widetilde\xi$ with $\xi(m)\leq\widetilde\xi\leq\xi(m)+1$, the ceiling
of a rational upper approximation of $\sqrt{2e\log(3m)}$ within $1/2$.
Like the original rank $m$, both stay fixed throughout an attempt.  At a current
primitive lattice $\mathcal S$ of rank $q\geq1$, set
\begin{equation}
    \sigma_q:=6\widetilde\xi\frac qm\widetilde g.
    \label{eq:rank-weighted-sigma}
\end{equation}
Thus each cut decreases $\sigma_q$ in proportion to the remaining rank.
Since $\xi(m)\geq3$, we have $\widetilde\xi\leq4\xi(m)/3$, so this
scale already satisfies the bound on $\sigma_q$
in~\eqref{eq:rank-weighted-smoothness}; it remains to make the output
$\sigma_q$-smooth.
Independent calls to Lemma~\ref{lem:finite-bit-dgs}, with sampler
parameter $1/\sigma_q$, sample $v\in\mathcal S^*$ within statistical
distance $1/128$ of the law proportional to
$e^{-\pi\sigma_q^2\|v\|_2^2}$.  This distance budget includes the norm
truncation described below.
Let
\begin{equation*}
    p_0(\mathcal S):=\frac1{\rho(\sigma_q\mathcal S^*)}.
\end{equation*}
Estimate the zero frequency using fresh samples.  If the estimate is at
least $3/4$, return $\mathcal S$.  Otherwise draw a fresh direction sample.  If it is
zero, fail this attempt; if it is nonzero, make it primitive in $\mathcal S^*$ and
replace $\mathcal S$ by $\mathcal S\cap v^\perp$.  A transition to rank zero also fails.

Call a test accurate if its estimate differs from $p_0(\mathcal S)$ by at
most $1/32$.  The sampler's zero probability differs from $p_0(\mathcal S)$
by at most $1/128$, and Hoeffding's inequality with $O(\log(1/\beta))$
samples estimates that biased probability to within $1/64$, except with
probability at most $\beta$.  Since $1/128+1/64<1/32$, each test is accurate
except with probability at most $\beta$.
If an accurate test accepts, then $p_0(\mathcal S)\geq23/32$, and hence
$\rho(\sigma_q\mathcal S^*)\leq32/23<3/2$, so the output is valid.  If it
continues, then $p_0(\mathcal S)\leq25/32$.

Next, let $N$ be the hidden rank-$s$ sublattice of
Lemma~\ref{lem:rank-weighted-dual-density}; it shows that, in the corrected
process defined below, whose tests are always accurate, an attempt accepts
before reaching rank zero with probability at least $8^{-m}$.
While $\mathcal S$ contains $N$, primitivity
gives $\Pi_{\operatorname{span}N}(\mathcal S^*)=N^*$, and
\eqref{eq:rank-weighted-dual-density} and \eqref{eq:rank-weighted-sigma}
imply, since $q\geq s$,
\begin{equation*}
    \operatorname{nd}_{\min}(\sigma_qN^*)
       \geq2\widetilde\xi\frac qs\geq2\xi(m).
\end{equation*}
Applying~\eqref{eq:reverse-minkowski-threshold} at half this scale and
then the scaling inequality $\rho(2A)-1\leq(\rho(A)-1)^4$ for a lattice
$A$, gives $\rho(\sigma_qN^*)-1\leq1/16$.
The Gaussian projection bound~\eqref{eq:gaussian-projection-preservation} gives
\begin{equation*}
    \Pr[v\perp N]\geq\frac1{\rho(\sigma_qN^*)}\geq\frac{16}{17}
\end{equation*}
for the ideal direction.  For the actual direction at a continuing
accurate test, therefore,
\begin{equation*}
    \Pr[v\neq0,\ v\perp N]
       \geq\frac{16}{17}-\frac{25}{32}-\frac1{128}>\frac18.
\end{equation*}
At rank $s$, retaining $N$ means $\mathcal S=N$, where every accurate test accepts.
The descent therefore need not know which cuts preserve $N$: the probability
that all required cuts do so is at least $8^{-m}$, and early accurate
acceptance is also a successful output.

To handle inaccurate tests, define a corrected process for the analysis
by replacing any estimate farther than $1/32$ from $p_0(\mathcal S)$ by
$p_0(\mathcal S)$ itself, without changing the fresh direction samples.  In this
corrected process every test has the required accuracy, so one attempt
succeeds with probability at least $8^{-m}$.  Run
$N_{\rm att}=\lceil8^m\log(4/\delta)\rceil$ independent attempts and give
each of the at most $mN_{\rm att}$ tests failure budget
$\delta/(4mN_{\rm att})$.  Couple the actual and
corrected processes until their first differing test.  Their current inputs
then agree, so the uniform conditional test guarantee and a union bound give
probability at most $\delta/4$ of any difference.  In the corrected process
the probability that all attempts fail is at most
$e^{-N_{\rm att}8^{-m}}\leq\delta/4$.
Each test uses $O(m+\log(1/\delta))$ samples.

The argument uses only a fixed small DGS bias, so the expensive DGS
primitive runs at constant accuracy, with no sampler confidence parameter
growing like $\log(1/\delta)$.  We apply
Lemma~\ref{lem:finite-bit-dgs} with its ambient parameter $m$ and a fixed
constant accuracy exponent, allowing room for truncation.  For a
sufficiently large universal $c$, standard centered Gaussian tails allow
us to map samples outside $\|v\|_2\leq cm/\sigma_1$ to zero while
retaining the stated bias.  All calls, including sampler setup, are
charged, at $2^{O(m)}\operatorname{poly}(L)$ each.
The frequency tests use the sampler's unconditional vector output,
including outputs mapped to zero after an internal timeout or truncation,
and never condition on an internal sampler flag.  The fixed bias of this
output is already included in the zero-probability and direction bounds
above.

The same norm cap bounds representation length on every branch.  Clear
denominators once by a positive integer that makes the scaled $\Lambda$
integral.  Each current $\mathcal S$ remains
primitive in $\Lambda$.  For a primitive $v\in\mathcal S^*$,
\begin{equation*}
    \det(\mathcal S\cap v^\perp)=\det(\mathcal S)\|v\|_2.
\end{equation*}
Making a direction primitive decreases its norm.  There are at most $m$
kernel steps, so the logarithm of every intermediate determinant is bounded
by an input-polynomial quantity.  LLL applied after each integer-kernel
operation then gives polynomial-bit bases, since the same scaling makes $\mathcal S$
integral; this is
the determinant-to-encoding argument of
\cite[Lemma~54]{reis2023subspace}.  Exact dual-membership, rank, and bit
checks reject malformed branches.  Fixed register, norm, and loop caps
therefore give the claimed all-branch resource bound.  Together with the
test and attempt bounds, this proves the lemma.
\end{proof}

We now prove Theorem~\ref{thm:constructive-flatness} along the outline
of Section~\ref{subsec:subspace-flatness}.

\begin{proof}[Proof of Theorem~\ref{thm:constructive-flatness}]
We apply the smooth-sublattice routine repeatedly to residual quotient
lattices.  At each stage we first choose a symmetric core and a position
in which the Gaussian covering bound can be compared with volume.  We then
remove the returned sublattice, add its covering contribution, and continue
on the lower-rank quotient.  The rank factor keeps the sum small
without requiring a large rank drop at any one stage.  Encoding and
confidence come last.

At a residual quotient lattice $\mathcal L$ of rank $m$, retain the projected
body $K_{\mathcal L}$ in lifted form; it inherits rational inner and outer
radius data.  Lemma~\ref{lem:finite-bit-geometric-estimators} computes
$\widetilde x$ with
$\|b(K_{\mathcal L})-\widetilde x\|_{K_{\mathcal L}-K_{\mathcal L}}\leq1/(2(m+1))$; let
$C=(K_{\mathcal L}-\widetilde x)\cap(\widetilde x-K_{\mathcal L})$.
Lemma~\ref{lem:finite-bit-log-affine-moments} with $s=0$ computes a
rational $\widetilde\Sigma$ with
$\frac9{10}\Sigma\preceq\widetilde\Sigma\preceq\frac{11}{10}\Sigma$, where
$\Sigma$ is the covariance of $K_{\mathcal L}$.  Factor
$\widetilde\Sigma=L_1\Delta_1L_1^\top$ with $L_1$ unit lower triangular and
$\Delta_1$ positive diagonal, and let $\widetilde\Delta$ be a rational
diagonal matrix whose entries approximate those of $\Delta_1^{-1/2}$ within
relative error $1/10$.  Then $T:=\widetilde\Delta L_1^{-1}$ is rational,
$T\widetilde\Sigma T^\top=\widetilde\Delta^2\Delta_1$ has diagonal entries
in $[0.81,1.21]$, and therefore
$\frac12I_m\preceq T\Sigma T^\top\preceq2I_m$.
Lemma~\ref{lem:isotropic-core} gives
\begin{equation*}
    \operatorname{Vol}_m(TK_{\mathcal L})^{1/m}\,\mathbb E\|G\|_{TC}
       \leq C_3\sqrt{\log(2m)}.
\end{equation*}
Exact structural, definiteness, and radius checks make these calls defined
on every tape.

Next, apply Lemma~\ref{lem:rank-weighted-smooth-sublattice} to $T\mathcal L$ and pull
its primitive rank-$q$ output back by $T^{-1}$, obtaining $\mathcal S\subseteq\mathcal L$.
Since $\operatorname{nd}(T\mathcal L)$ equals
$(\det\mathcal L/\operatorname{Vol}_mK_{\mathcal L})^{1/m}
\operatorname{Vol}_m(TK_{\mathcal L})^{1/m}$,
Lemma~\ref{lem:gaussian-covering-certificate}, linear invariance of
covering radii, and $\xi(m)=O(\sqrt{\log(2m)})$ yield
\begin{align}
    \mu(\mathcal S,C\cap\operatorname{span}\mathcal S)
    &\leq C_4\xi(m)\frac qm\operatorname{nd}(T\mathcal L)\,
       \mathbb E\|G\|_{TC}\notag\\
    &\leq C_5\log(2m)\frac qm
    \left(\frac{\det\mathcal L}{\operatorname{Vol}_mK_{\mathcal L}}\right)^{1/m}.
    \label{eq:rank-weighted-cover}
\end{align}
The restriction of the Gaussian norm to $\operatorname{span}(T\mathcal S)$ is at most
the ambient norm by the last assertion of
Lemma~\ref{lem:gaussian-covering-certificate}.  The fraction $q/m$ in this
bound replaces the requirement to remove half the rank.

To see why these contributions add, apply the covering-radius triangle
inequality to the centered body $K_{\mathcal L}-\widetilde x$.  Its section
through $\operatorname{span}\mathcal S$ contains
$C\cap\operatorname{span}\mathcal S$, and translating its projection
does not change the next quotient's covering radius.  Therefore
\begin{equation*}
    \mu(\mathcal L,K_{\mathcal L})
     \leq\mu(\mathcal S,C\cap\operatorname{span}\mathcal S)
         +\mu\bigl(\mathcal L/\mathcal S,
             \Pi_{(\operatorname{span}\mathcal S)^\perp}K_{\mathcal L}\bigr).
\end{equation*}
Each stage chooses its own center, so the chain needs no common center.

Start with $\Lambda_0=\{0\}$ and lift each $\mathcal S_i$ from the current quotient
to obtain a primitive chain
$\Lambda_0\subsetneq\Lambda_1\subsetneq\cdots\subsetneq\Lambda$.
Put $W_i=\operatorname{span}(\Lambda_i)^\perp$,
$\mathcal L_i=\Pi_{W_i}\Lambda$, $K_i=\Pi_{W_i}K$,
$m_i=\dim W_i$, and $q_i=m_i-m_{i+1}$.  There are at most $k$ stages.
The covering-radius triangle inequality
\cite[Lemmas~40--41]{reis2023subspace}, applied after translating each
quotient body to its chosen center, and \eqref{eq:rank-weighted-cover} give
\begin{equation}
    \mu(\Lambda,K)
       \leq C_5\sum_i\log(2m_i)\frac{q_i}{m_i}
              R_{W_i}(\Lambda,K).
    \label{eq:rank-weighted-sum}
\end{equation}
Here the full quotient determinant-volume ratio is exactly
$R_{W_i}(\Lambda,K)$; translations do not change covering radii or volumes.
Charging each stage by $q_i/m_i$ gives only a
harmonic sum, as in the proof of
\cite[Lemma~6.3]{dadush2019approximating}: for arbitrary positive rank
removals,
\begin{equation*}
    \sum_i\frac{q_i}{m_i}
       \leq\sum_i\sum_{j=m_{i+1}+1}^{m_i}\frac1j
       =\sum_{j=1}^k\frac1j\leq1+\log k.
\end{equation*}
Facts~\ref{fact:reverse-minkowski} and~\ref{fact:isotropic-mean-norm} each
contribute a factor $\sqrt{\log(2m)}$ to the per-stage estimate, and the
harmonic sum contributes the remaining logarithm.
For each stage with $m_i>0$, fix the rational coordinate chart of $W_i$
used below.  In it, $\mathcal L_i$ has a square rational full-rank basis,
so its determinant $D_i$ is a positive rational; let $V_i$ be the chart
volume of $K_i$.  The Euclidean determinant and volume carry the same
Jacobian factor, which cancels, so $R_{W_i}=(D_i/V_i)^{1/m_i}$.
Lemma~\ref{lem:finite-bit-geometric-estimators}, applied to the chart
image of $K_i$ with $\varepsilon=1/4$, returns a rational $\zeta_i>0$ with
$V_i/\zeta_i\in[3/4,5/4]$.  Put $A_i:=(D_i/((3/4)\zeta_i))^{1/m_i}$.
Since $A_i^{m_i}$ is rational, upward rational root isolation gives a
rational $R_i\in[A_i,(6/5)A_i]$ of polynomial encoding.  Then
\begin{equation*}
    R_{W_i}\leq A_i\leq R_i
       \leq\tfrac65\left(\tfrac53\right)^{1/m_i}R_{W_i}
       \leq2R_{W_i},
\end{equation*}
and we fix $C_R:=2$.  Return $W:=W_{i^*}$ and $R:=R_{i^*}$ for an index
$i^*$ maximizing $R_i$.  Then $R_{W_{i^*}}\leq R\leq2R_{W_{i^*}}$ and
$\max_iR_{W_i}\leq\max_iR_i=R\leq2R_{W_{i^*}}$, so the selection loses
the factor $C_R$ only once.  With $\ell:=\lceil\log_2(2k)\rceil$ we have
$\log(2m_i)\leq\ell$ and $1+\log k\leq\ell$, so
\eqref{eq:rank-weighted-sum} gives
\begin{equation*}
    \mu(\Lambda,K)\leq C_5\ell^2\max_iR_{W_i}\leq2C_5\ell^2R_{W_{i^*}}.
\end{equation*}
Hence \eqref{eq:flatness-interface} holds for every integer
$C_\Phi\geq\lceil2C_5\rceil$.  This constant is fixed after $C_R$ and
$C_5$ and before the algorithm runs, and it may be enlarged to absorb
other universal constants.  The volume calls use the confidence budget
$\delta/(8k)$ allocated below.

We finish with encoding and confidence, including the quotient sequence.
Work with rational coordinate charts of the residual subspaces; the
determinant-volume ratios and covering inequalities are invariant under
these charts.  Maintain each $\Lambda_i$ in the original coordinates rather
than recursively enlarging the encoding bound.  Let $b_j$ be the original
basis columns and choose a rational $F\geq\sum_j\|b_j\|_2$.
Fix rational $a\in\mathbb Q^k$ and
$0<r_{\mathrm{in}}\leq R_{\mathrm{out}}$, all of polynomial encoding
length in the original input, such that
\begin{equation*}
    a+r_{\mathrm{in}}B_2^k\subseteq K
    \subseteq a+R_{\mathrm{out}}B_2^k.
\end{equation*}
The initial simplex and enclosing box supply such data.  Put
\begin{equation}
    U_0=C_6(k+1)^2(1+F/r_{\mathrm{in}}),
    \qquad
    D_0=\max\{1,4R_{\mathrm{out}}U_0\}^{k}.
    \label{eq:rank-weighted-bit-cap}
\end{equation}
Since $\mu_{\mathrm{KL}}(\Lambda,K)\leq\mu(\Lambda,K)\leq F/r_{\mathrm{in}}$,
on the good event \eqref{eq:rank-weighted-cover} bounds each partial
covering radius by $U_0$.  Each symmetric centered quotient core lies in
$2R_{\mathrm{out}}B_2$, so the determinant-volume lower bound for covering
gives
\begin{equation*}
    \frac{\det\Lambda_{i+1}}{\det\Lambda_i}
    =\det(\mathcal S_i)\leq(4R_{\mathrm{out}}U_0)^{q_i}.
\end{equation*}
Since $\sum_i q_i=k$, every lifted determinant is at most $D_0$.
After one denominator clearing, all lifted lattices are integral.  LLL
therefore supplies bases whose encoding lengths are bounded by one fixed polynomial
in the original data, by \cite[Lemma~54]{reis2023subspace}.  The exact
quotient matrices and the lifted descriptions of $K_i$ have the same type
of bound.  Enforce this determinant cap and the corresponding bit caps on
every branch; no good branch is removed.  The internal descent caps proved
above then apply uniformly at every quotient.  Failed radius checks,
arithmetic, or caps return failure.

At each of at most $k$ stages there are a barycenter call, a covariance call,
a rank-weighted descent, and a volume estimate.  Give each failure budget
$\delta/(8k)$, using the uniform finite interfaces and the confidence
argument in Lemma~\ref{lem:rank-weighted-smooth-sublattice}; the union of
these failures has probability at most $\delta/2$.  Adaptivity is
harmless because each guarantee holds uniformly for the current bounded
input and uses fresh random bits.  All exact algebra and radius-root approximations have fixed
polynomial costs, and
\begin{equation*}
    \sum_i2^{O(m_i)}\operatorname{poly}(L,\log(1/\delta))
       \leq2^{O(k)}\operatorname{poly}(L,\log(1/\delta)).
\end{equation*}
This proves the stated confidence and all-branch resource guarantees.
\end{proof}

\begin{proof}[Proof of Proposition~\ref{prop:finite-reversible-flatness}]
We first make the routine finite and checked on every tape, then treat
geometrically valid outputs.  Apply Theorem~\ref{thm:constructive-flatness} with failure budget $\beta$.
Enlarge $C_\Phi$ in \eqref{eq:flatness-interface} to absorb its universal
constant and the radius approximation.  The theorem's deterministic loop,
norm, determinant, and bit caps bound every tape by
$2^{O(k)}\operatorname{poly}(L,\log(1/\beta))$ resources.  Pad the random tape and retain
timeout and arithmetic-failure flags.  Exact rational algebra checks the
encoding, positivity of $R$, and the projected and fiber ranks before a
subspace is used.  A rejected output produces a \textsc{fail} leaf.

With probability at least $1-\beta$, the geometric inequalities hold, so
the shrinking argument of Section~\ref{subsec:classical-recursion} and Lemma~\ref{lem:branch-bound} prove
\eqref{eq:shrinkage-equivalence} and \eqref{eq:branch-bound}.  The exact
structural checks do not certify those geometric inequalities on every tape.
A well-formed output that is not geometrically valid still removes at least
one dimension, and it cannot produce an invalid accepted point because the
leaf is verified exactly.
Bennett simulation gives the capped reversible implementation with the
stated resource bound.
\end{proof}

\subsection{Finite geometric primitives}

The barycenter and moment estimators below run the technical cube walk of
\cite{dyer1991random} on a mesh finer than the one fixed in that source.
The following lemma extracts the parameter dependence of its conductance
argument, so that the mixing time of the refined walk is explicit.

\begin{lemma}[Common refinement of the finite cube walk]
\label{lem:finite-grid-refinement}
Let $d\geq3$, and let $C_0\subseteq\mathbb R^d$ be a compact convex
body represented by a rational lifted linear program of encoding length
$L$.  Suppose that, for a rational $q\geq1$,
\begin{equation*}
    \mathbb{B}_2^d\subseteq C_0\subseteq q\mathbb{B}_2^d.
\end{equation*}
Write $(C_0)_{+a}:=C_0+a\mathbb{B}_2^d$.  Choose a dyadic rational
\begin{equation*}
    \frac1{200d^3}\leq h_0\leq\frac1{100d^3},
    \qquad
    \alpha_0:=24d^2h_0,
    \qquad
    \alpha'_0:=\frac{h_0}{4d}.
\end{equation*}
For a dyadic $\nu\in(0,1]$, put
$(h,\alpha,\alpha'):=\nu(h_0,\alpha_0,\alpha'_0)$.

There is a deterministic predicate on the closed grid cubes
$Q_z:=h(z+[0,1]^d)$, $z\in\mathbb Z^d$, which accepts every cube meeting
$(C_0)_{+\alpha}$ and accepts only cubes meeting
$(C_0)_{+\alpha+\alpha'}$.  Let $\mathcal C_h$ be its accepted cubes and
let $U_h$ be their half-open union.  With
$\beta:=\alpha+\alpha'+\sqrt d\,h$, one has $\beta\leq\nu/(4d)\leq1$ and
\begin{equation*}
    C_0\subseteq U_h\subseteq(C_0)_{+\beta}
       \subseteq(1+\beta)C_0.
\end{equation*}

Set $k:=\lceil\log_2(4d)\rceil$.  Consider the walk which draws $k$ fair
bits at each step, assigns exactly $2d$ strings to the moves
$z\mapsto z\pm e_i$, assigns all other strings to holding, and rejects
moves to unaccepted cubes.  This walk is connected, symmetric, and lazy,
and its stationary law $\pi$ is uniform on $\mathcal C_h$.  Its
worst-start total-variation mixing time satisfies, for $0<\theta<1/2$,
\begin{equation*}
    t_{\mathrm{mix}}(\theta)
    \leq
    \left\lceil
    10^{22}d^{16}(q+1)^4\nu^{-4}
    \left[
       d\log\frac{8(q+1)}{\nu h_0}+\log\frac1\theta
    \right]
    \right\rceil .
\end{equation*}
Accepted cubes and their neighbors have indices of
$O(d\log(8(q+1)/h))$ bits.  On such indices the predicate costs
$\operatorname{poly}(d,L,\log(q+1),\log(1/\nu))$ time.
Neither the predicate nor the walk lists the state set.
\end{lemma}

\begin{proof}
Put $R:=q+1$.  We first construct the predicate, then extract the
parameter dependence of the conductance argument.

\proofstep{The deterministic predicate.}
Let $r:=\alpha+\alpha'$ and $a:=\alpha'/(2R)$.  For a closed cube $Q$,
apply deterministic rational ellipsoid feasibility to
\begin{equation*}
    A:=C_0\cap Q_{+r},
    \qquad Q_{+r}:=Q+r\mathbb{B}_2^d,
\end{equation*}
starting from the enclosing ball $q\mathbb{B}_2^d$.  Accept if a rational
feasible point is found, and reject after a fixed iteration cap sufficient
to reduce the enclosing volume below $(a/d)^d$.

If $Q$ meets $(C_0)_{+\alpha}$, choose $x\in C_0$ with
$\operatorname{dist}(x,Q)\leq\alpha$.  Convexity and
$\mathbb{B}_2^d\subseteq C_0$ give
\begin{equation*}
    (1-a)x+a\mathbb{B}_2^d\subseteq C_0.
\end{equation*}
Every point of this ball has distance at most
$\alpha+a(q+1)=\alpha+\alpha'/2$ from $Q$.  Thus $A$ contains a ball of
radius $a$, whose volume is at least $(a/d)^d$, and the predicate must
accept.  Conversely, an accepted feasible point certifies that $Q$ meets
$(C_0)_{+\alpha+\alpha'}$.

All separation and stopping tests can be rational.  Separation from
$C_0$ follows from exact linear programming and a Farkas certificate for
its lifted description.  For separation from $Q_{+r}$, let $p$ be the
coordinatewise projection of a rational test point $y$ onto $Q$, and put
$v:=y-p$ and $s:=v^\top v$.  If $s>r^2$, the rational inequality
\begin{equation*}
    v^\top(z-p)\leq\frac{s+r^2}{2}
\end{equation*}
contains $Q_{+r}$ and excludes $y$.  Indeed, for $z=p'+w$ with $p'\in Q$
and $\|w\|_2\leq r$, the projection property and
$2r\sqrt s\leq s+r^2$ prove validity.  Membership is tested by the exact
comparison $s\leq r^2$.

Use a standard bit-capped rational ellipsoid implementation, with
outward rounding and fixed slack in its volume decrease.  Its iteration
cap is $O(d^2\log(2dq/a))$, with a universal implied constant.
No irrational volume comparison is required: for an ellipsoid with
shape matrix $H$, the test
$4^d\det H<(a/d)^{2d}$ certifies the required volume bound, since the
unit-ball volume is at most $2^d$.  The usual rounded-ellipsoid encoding
bounds, together with the rational separation procedures above, give
the asserted polynomial predicate cost.  This is the weak-feasibility
construction of \cite[Section~1]{dyer1991random}, with an explicit
inscribed-ball bound in place of its cone-volume threshold.

The predicate inclusions imply the stated sandwich, because a cube has
diameter $\sqrt d\,h$ and
$C_0+\beta\mathbb{B}_2^d\subseteq C_0+\beta C_0=(1+\beta)C_0$.
Moreover,
\begin{equation*}
    \beta
       =h\left(24d^2+\frac1{4d}+\sqrt d\right)
       \leq25d^2h
       \leq\frac{\nu}{4d}<1.
\end{equation*}
An accepted cube meets $R\mathbb{B}_2^d$.  Counting its possible
coordinate indices therefore gives
\begin{equation*}
    |\mathcal C_h|\leq(8R/h)^d
\end{equation*}
and the claimed index bound.

\proofstep{Local geometric estimates.}
We record the parameter checks needed in the cube-counting arguments,
rather than applying the fixed-parameter mixing theorem of
\cite{dyer1991random}.  Our choices satisfy
\begin{equation*}
    h\leq1,\qquad
    \beta\leq1,\qquad
    \alpha\geq12\sqrt2\,d^{3/2}h,\qquad
    0<\alpha'\leq\frac{h}{2\sqrt d}.
\end{equation*}
More explicitly, the proofs of Propositions~8--11 of that source use
the following local estimates:
\begin{align*}
    \alpha^2-\frac{2\alpha\ell h}{\sqrt{2d}}+\ell^2h^2
       &\leq(\alpha-\sqrt d\,h)^2
          &&(2d\leq\ell\leq20d),\\
    \frac{8dh}{\alpha}
       &\leq\frac{2}{3\sqrt{2d}},\\
    (\alpha+\alpha')^2
       -\frac{2h(\alpha+\alpha')}{\sqrt{2d}}+h^2
       &\leq\alpha^2,\\
    \frac{6\sqrt d\,h}{\alpha}
       &\leq\frac{2}{3\sqrt{2d}}.
\end{align*}
They follow from $\alpha/h=24d^2$ and $\alpha'/h=1/(4d)$.
For the first inequality, the difference between the right and left
sides is concave in $\ell$, so it suffices to check the two endpoints.
The other inequalities follow directly by substitution.  Common
refinement leaves these homogeneous inequalities unchanged and only
decreases $h$ and $\beta$.

Let $\mathcal N_h$ be the natural state set, consisting of all closed
cubes meeting $(C_0)_{+\alpha}$.  The first two local estimates, together
with Propositions~7 and~9 of \cite{dyer1991random}, give exactly the
charging argument of its Proposition~10:
\begin{equation}
    |S_B|\leq2d\,e(S)+18|S_I|,
    \qquad S\subseteq\mathcal N_h.
    \label{eq:fgr-natural-count}
\end{equation}
Here $e(S)$ is the number of natural graph edges crossing the cut,
$S_I$ consists of cubes wholly contained in $(C_0)_{+\alpha}$, and
$S_B:=S\setminus S_I$.
For completeness, we prove the technical charging estimate with the
stronger receiving bound available for our parameters.
Fix $S\subseteq\mathcal C_h$, put $S'=S\cap\mathcal N_h$, and let
$e(S)$ count technical graph edges crossing this cut.
For each weak cube $Q\in\mathcal C_h\setminus\mathcal N_h$, choose
$x_Q\in Q$ with
$r_Q:=\operatorname{dist}(x_Q,C_0)\leq\alpha+\alpha'$.
Such a witness exists by the predicate's outer inclusion, and
$r_Q>\alpha$ because $Q$ is not natural.  Set
\begin{equation*}
 u_Q=\frac{x_Q-\Pi_{C_0}x_Q}{r_Q},\qquad
 t=\frac1{\sqrt{2d}},\qquad
 I_Q=\{i:|(u_Q)_i|>t\}.
\end{equation*}
The denominator
$D_Q:=\sum_{i\in I_Q}(u_Q)_i^2$ is at least $1/2$,
since the other $d$ coordinates contribute at most $dt^2=1/2$.
Send a fraction
$\phi_i(u_Q):=(u_Q)_i^2/D_Q$ of one unit of charge to the
coordinate neighbor
$Q-h\operatorname{sign}((u_Q)_i)e_i$, for each $i\in I_Q$.
Every such neighbor is natural.  Indeed, with
$p_Q=\Pi_{C_0}x_Q$, the shifted witness satisfies
\begin{equation*}
 \|x_Q-h\operatorname{sign}((u_Q)_i)e_i-p_Q\|_2^2
 \leq r_Q^2-2hr_Qt+h^2
 \leq(\alpha+\alpha')^2-2h(\alpha+\alpha')t+h^2
 \leq\alpha^2.
\end{equation*}
The middle inequality holds because $r\mapsto r^2-2hrt+h^2$
is increasing for $r\geq\alpha>ht$; the last is the third local
estimate displayed above.

We bound charge received by one natural cube.
For $x,y\notin C_0$, nonexpansiveness of $I-\Pi_{C_0}$ and
normalization give
\begin{equation*}
 \left\|
 \frac{x-\Pi_{C_0}x}{\operatorname{dist}(x,C_0)}
 -
 \frac{y-\Pi_{C_0}y}{\operatorname{dist}(y,C_0)}
 \right\|_2
 \leq
 \frac{2\|x-y\|_2}
 {\min\{\operatorname{dist}(x,C_0),
        \operatorname{dist}(y,C_0)\}}.
\end{equation*}
Witnesses in coordinate neighbors of the same receiving cube
are at pairwise distance at most
$2h+\sqrt d\,h\leq3\sqrt d\,h$.
Their normals therefore differ by at most
\begin{equation*}
 \varepsilon_0:=\frac{6\sqrt d\,h}{\alpha}
       =\frac1{4d^{3/2}}<\frac{t}{2}.
\end{equation*}
Opposite neighbors in one coordinate cannot both send positive
charge to this cube: their active normal coordinates would have
opposite signs and absolute values greater than $t$, contradicting
this bound.  There is consequently at most one sender per coordinate.
If there is any sender, fix its normal $u_0$.
For the sender with normal $u^{(i)}$ in coordinate $i$,
$|u_i^{(i)}|>t$ and
$|u_i^{(i)}-(u_0)_i|\leq\varepsilon_0<t/2$, so
$|u_i^{(i)}|\leq2|(u_0)_i|$.  Thus
\begin{equation*}
 \phi_i(u^{(i)})
 \leq2|u_i^{(i)}|^2
 \leq8|(u_0)_i|^2.
\end{equation*}
Summing over the distinct sender coordinates bounds the total
received charge by $8\|u_0\|_2^2=8$.
This also covers the zero-receiver case trivially.

Now distribute charge only from the weak cubes in $S\setminus S'$.
Charge going to natural cubes outside $S$ crosses cut edges, each
of which carries at most one unit.  Charge remaining in $S'$
is at most $8|S'|$.  Hence
$|S\setminus S'|\leq e(S)+8|S'|$, and in particular
\begin{equation}
 |S|\leq e(S)+9|S'|\leq e(S)+18|S'|,
 \qquad S\subseteq\mathcal C_h,\quad S'=S\cap\mathcal N_h.
 \label{eq:fgr-technical-count}
\end{equation}
These arguments require only the two predicate inclusions and the
displayed local estimates, not the source's particular mesh or outer
radius.
The argument allows witness distances throughout
$(\alpha,\alpha+\alpha']$, rather than requiring a fixed offset.

\proofstep{Natural expansion with explicit radius dependence.}
We claim that every nonempty $S\subseteq\mathcal N_h$ with
$|S|\leq|\mathcal N_h|/2$ satisfies
\begin{equation}
    \frac{e(S)}{|S|}
       \geq\kappa_{\mathrm{nat}}
       :=\frac{h^2}{2000dR^2}.
    \label{eq:fgr-kappa-nat}
\end{equation}
Following the proof of \cite[Lemma~1]{dyer1991random}, choose a smooth
strictly convex outer approximation $H$ to $(C_0)_{+\alpha}$ which meets
exactly the same closed cubes and is contained in $R\mathbb{B}_2^d$.
Such an approximation exists because the excluded cubes in a bounded
neighborhood have a positive distance from the compact offset body.
It may also be chosen transverse to the relevant grid faces.  Put
$M:=\partial H$.

Write $e:=e(S)$ and $S^c:=\mathcal N_h\setminus S$.
If $e/|S|\geq1/(4d)$, the claim follows.  Otherwise
\eqref{eq:fgr-natural-count}, applied to both sides of the cut, gives
\begin{equation*}
    |S_I|\geq|S|/38,\qquad |S^c_I|\geq|S^c|/38.
\end{equation*}
Let $T$ be the intersection of $H$ with the union of the cubes in $S$,
and let $T^c:=H\setminus T$.  Ignoring their measure-zero boundaries,
\begin{equation*}
    \operatorname{Vol}_d(T)\geq h^d|S|/38,\qquad
    \operatorname{Vol}_d(T^c)\geq h^d|S^c|/38.
\end{equation*}
The Euclidean isoperimetric inequality and
$\operatorname{Vol}_d(H)\leq
\operatorname{Vol}_d(\mathbb{B}_2^d)R^d$ imply
\begin{equation*}
    \operatorname{Per}(T)\geq
       \frac dR\operatorname{Vol}_d(T),
    \qquad
    \operatorname{Per}(T^c)\geq
       \frac dR\operatorname{Vol}_d(T^c).
\end{equation*}
The shared interior interface has area at most $eh^{d-1}$.
If it accounts for at least half the perimeter of either set, then
$e/|S|\geq dh/(76R)$.

Otherwise the portions of $M$ belonging to $T$ and $T^c$ have areas at
least $dh^d|S|/(76R)$ and $dh^d|S^c|/(76R)$, respectively.
Choose the one, denoted by $E$, having at most half the area of $M$.
Then
\begin{equation*}
    \operatorname{area}_{d-1}(E)\geq\frac{dh^d|S|}{76R}.
\end{equation*}
The inequality derived in the proof of
\cite[Proposition~6, pp.~20--21]{dyer1991random}, before specializing
the radius, states that a smooth closed manifold of dimension $d-1$
with nonnegative Ricci curvature and intrinsic diameter $D_M$ satisfies
\begin{equation*}
    \operatorname{area}_{d-2}(\partial_M E)
       \geq
       \frac{2}{\pi D_M\sqrt{d-1}}
       \operatorname{area}_{d-1}(E)
\end{equation*}
for such a half-area set $E$.  It applies to the smooth convex
hypersurface $M$.  Moreover, $D_M\leq\pi R$: a planar section through
the origin and two boundary points has perimeter at most $2\pi R$,
and its shorter boundary arc joins those points within $M$.
Consequently,
\begin{equation*}
    \operatorname{area}_{d-2}(\partial_M E)
       \geq\frac{\operatorname{area}_{d-1}(E)}
                    {10R\sqrt d}.
\end{equation*}

Every point of $\partial_M E$ lies on a cut facet, apart from
lower-dimensional intersections.  For each such facet $F$,
$M\cap F$ is contained in the relative boundary of the convex set
$H\cap F$.  Monotonicity of the surface area of convex sets under
inclusion therefore gives
\begin{equation*}
    \operatorname{area}_{d-2}(M\cap F)
       \leq\operatorname{area}_{d-2}(\partial F)
       \leq2d\,h^{d-2}.
\end{equation*}
This is the facet estimate used in the proof of
\cite[Lemma~1]{dyer1991random}.  Hence
\begin{equation*}
    2d\,h^{d-2}e
       \geq\frac1{10R\sqrt d}\frac{dh^d|S|}{76R},
    \qquad
    \frac e{|S|}\geq\frac{h^2}{1520R^2\sqrt d}.
\end{equation*}
Each of the three possible lower bounds
$1/(4d)$, $dh/(76R)$, and $h^2/(1520R^2\sqrt d)$ is at least
$\kappa_{\mathrm{nat}}$, since $h\leq1\leq R$.
This proves \eqref{eq:fgr-kappa-nat}.

\proofstep{Technical expansion and mixing.}
Apply the transfer argument of \cite[Lemma~2]{dyer1991random}.
For a technical cut $S$ with $|S|\leq|\mathcal C_h|/2$, either
$e(S)\geq|S|/2$, or \eqref{eq:fgr-technical-count} on both sides gives
\begin{equation*}
    |S'|\geq|S|/36,\qquad
    |(S^c)'|\geq|S^c|/36\geq|S|/36.
\end{equation*}
Applying \eqref{eq:fgr-kappa-nat} to the smaller natural side yields,
in either case,
\begin{equation*}
    \frac{e(S)}{|S|}
       \geq\frac{\kappa_{\mathrm{nat}}}{36}
       =\frac{h^2}{72000dR^2}.
\end{equation*}
In particular the technical graph is connected.  It contains the
origin's cube.

The fair-bit walk is symmetric and holds with probability at least
$1/2$.  Each accepted coordinate edge has transition probability
$2^{-k}\geq1/(8d)$.  Its conductance therefore satisfies
\begin{equation*}
    \Phi\geq\frac{h^2}{576000d^2R^2}
       \geq\frac{h^2}{10^6d^2R^2}.
\end{equation*}
By \cite[Theorem~2, p.~13]{dyer1991random}, its pointwise transition
error is at most $\exp(-t\Phi^2/2)$.  Summing over the states gives,
from every accepted start,
\begin{equation*}
    d_{\mathrm{TV}}(P^t(x,\cdot),\pi)
       \leq\frac12(8R/h)^d
          \exp\left(-\frac{th^4}{2\cdot10^{12}d^4R^4}\right).
\end{equation*}
It suffices to take
\begin{equation*}
    t\geq
    2\cdot10^{12}d^4R^4h^{-4}
    \left[d\log\frac{8R}{h}+\log\frac1\theta\right].
\end{equation*}
Finally,
\begin{equation*}
    2\cdot10^{12}d^4R^4h^{-4}
       \leq3.2\cdot10^{21}d^{16}R^4\nu^{-4},
\end{equation*}
because $h_0^{-1}\leq200d^3$.  This proves the stated conservative
mixing bound.  Integer caps may use dyadic upper bounds for the
logarithms within the slack in its universal prefactor.
Each step uses $k$ fair bits, one bounded-length index update, and one
predicate evaluation.
\end{proof}

\begin{lemma}[Finite-bit geometric estimators]
\label{lem:finite-bit-geometric-estimators}
Let $C\subseteq\mathbb R^d$ be a rational compact convex body represented by a
rational lifted linear program of encoding length $L$, with explicit rational
inner and outer ellipsoids.  For rational
$\varepsilon,\tau\in(0,1/2)$, finite randomized Boolean algorithms output
rational $\zeta>0$ and $\widetilde b\in\mathbb Q^d$ such that, jointly with
probability at least $1-\tau$,
\begin{equation*}
    \frac{\operatorname{Vol}_d(C)}{\zeta}
       \in[1-\varepsilon,1+\varepsilon],
    \qquad
    \lVert \widetilde b-b(C)\rVert_{C-C}\leq\varepsilon.
\end{equation*}
Here $b(C):=\operatorname{Vol}_d(C)^{-1}\int_Cx\,dx$ is the barycenter.
Their running time, random-tape length, output encoding, and workspace are
polynomial in
\begin{equation*}
    d,L,\frac1\varepsilon,\log\frac1\tau.
\end{equation*}
The algorithms can be capped and reversibly simulated with polynomial
overhead.
\end{lemma}

\begin{proof}
We first reduce to $d\geq3$.  For $d=0$, output $\zeta=1$ and the unique
vector in $\mathbb Q^0$.  For $d\in\{1,2\}$, apply the construction below
in dimension $3$ to $C\times[-1,1]^{3-d}$, project its barycenter estimate
onto the first $d$ coordinates, and divide its volume estimate by
$2^{3-d}$.  The volume ratio is unchanged, and projection maps the lifted
difference body onto $C-C$, so the barycenter guarantee and polynomial
bounds are preserved.

Now let $d\geq3$.  The proof rounds $C$, refines a cube walk until its
stationary grid-center mean is close to the barycenter, and averages
independent chains.  Rational affine rounding gives a body $C_0$ with
\begin{equation*}
    \mathbb{B}_2^d\subseteq C_0\subseteq q\mathbb{B}_2^d,
    \qquad q=\operatorname{poly}(d).
\end{equation*}
Volume ratios and barycenter errors in the difference-body norm are affine
invariant.  For volume, run the estimator of \cite{dyer1991random} with
failure budget $\tau/2$.  For the barycenter, use the common-refinement
construction of Lemma~\ref{lem:finite-grid-refinement}.

Let $h_0,\alpha_0,\alpha'_0$ be the rational base parameters in that
lemma.  For a dyadic $\nu\in(0,1]$ fixed below, put
$(h,\alpha,\alpha'):=\nu(h_0,\alpha_0,\alpha'_0)$.
Let $\mathcal C_h$ be the accepted cubes and $U_h$ their half-open
union.  With $\beta:=\alpha+\alpha'+\sqrt d\,h$, the lemma gives
\begin{equation}
    C_0\subseteq U_h\subseteq(C_0)_{+\beta}
       \subseteq(1+\beta)C_0.
    \label{eq:dfk-cube-union-sandwich}
\end{equation}
In particular, $\beta\leq\nu/(4d)$.  Choose the largest dyadic
$\nu\leq\min\{1,4\varepsilon/c_0\}$, where $c_0$ is a sufficiently
large universal constant.  Then $\nu^{-1}=O(1/\varepsilon)$ and
\begin{equation}
    \beta\leq\frac{\varepsilon}{c_0d}.
    \label{eq:dfk-barycenter-parameter-choice}
\end{equation}
All mesh and predicate parameters are rational and have polynomial
encoding lengths.

Run the fair-bit walk of Lemma~\ref{lem:finite-grid-refinement}, starting
from the cube with index zero.  Each step draws exactly
$k=\lceil\log_2(4d)\rceil$ fair bits, with $2d$ strings assigned to signed
coordinate moves and all remaining strings assigned to holding;
proposals to unaccepted cubes are rejected.  The lemma supplies
connectivity, uniform stationarity on $\mathcal C_h$, and a fixed
mixing cap polynomial in $d,q,\nu^{-1}$ and $\log(1/\theta)$ for
total-variation error $\theta$.  It also supplies a deterministic
polynomial-cost predicate and polynomial-length index registers.
The state set is never listed.

Next we bound the bias of the stationary grid-center mean.  By
\eqref{eq:dfk-cube-union-sandwich},
\begin{equation*}
    d_{\mathrm{TV}}(\operatorname{Unif}(U_h),
                    \operatorname{Unif}(C_0))
       \leq(1+\beta)^d-1=O(d\beta).
\end{equation*}
Both supports lie in $(1+\beta)C_0$, so maximal coupling bounds the
first-moment error by $O(d\beta)$ in the difference-body norm.  Replacing a
uniform point in a cube by its rational center adds at most $\sqrt d\,h/2$.
Thus the stationary grid-center mean is within $\varepsilon/3$ of $b(C_0)$.

Finally, we control the sampling error.  Take
$N=\lceil C_Hq^2d\varepsilon^{-2}\log(4d/\tau)\rceil$
independent chains, each run for the mixing cap of
Lemma~\ref{lem:finite-grid-refinement} with $\theta=\tau/(4N)$.  Their joint law is
within $\tau/4$ of independent stationary samples.  Coordinatewise Hoeffding
and $2\mathbb{B}_2^d\subseteq C_0-C_0$ bound the sampling error of the
average by $\varepsilon/3$, except with probability $\tau/4$.  The mesh
fixes the bias deterministically, so only mixing and finite-sample errors
enter this probability budget.  This proves the barycenter claim with
budget $\tau/2$, in addition to the volume budget $\tau/2$.  Grid centers
share a polynomial-bit denominator, and averaging adds only $O(\log N)$
bits.
All chain lengths, sample counts, and index bounds are fixed from
$d,L,1/\varepsilon,\log(1/\tau)$ before the tape is drawn.  Applying the
inverse affine map and Bennett simulation gives the reversible algorithms.
\end{proof}

\begin{lemma}[Finite-bit log-affine moments]
\label{lem:finite-bit-log-affine-moments}
Let $C\subseteq\mathbb R^d$ be a full-dimensional compact rational
lifted polytope with explicit rational inner and outer ellipsoids.
Let $s\in\mathbb Q^d$ and $B\geq1$ be rational, with
$|\langle s,x\rangle|\leq B$ for every $x\in C$.  Write $\mu_s$ and
$\Sigma_s\succ0$ for the mean and covariance of the probability law
on $C$ with density proportional to $e^{\langle s,x\rangle}$.
For rational $\varepsilon,\tau\in(0,1/2)$, a capped finite randomized
Boolean algorithm returns rational $\widetilde\mu$ and
$\widetilde\Sigma\succ0$ such that, with probability at least $1-\tau$,
\begin{align*}
    |v^\top(\widetilde\mu-\mu_s)|
       &\leq\varepsilon\sqrt{v^\top\Sigma_s v}
       \qquad(v\in\mathbb R^d),\\
    (1-\varepsilon)\Sigma_s
       &\preceq\widetilde\Sigma\preceq(1+\varepsilon)\Sigma_s.
\end{align*}
Its time, workspace, random-tape length, and output encoding are
\begin{equation*}
    \operatorname{poly}(d,L,B,1/\varepsilon,\log(1/\tau)),
\end{equation*}
where $L$ includes all supplied rational data.  The same bound holds for its
reversible implementation.
\end{lemma}

\begin{proof}
We reduce log-affine moments to uniform moments on a rational polytope and
estimate the latter with the finite walk of
Lemma~\ref{lem:finite-bit-geometric-estimators}.  We bound the
density-approximation error first and the moment-estimation error second.
Put $\eta:=\varepsilon/100$ and
$m:=\lceil2B+B^2/\eta\rceil$.  The rational polytope
\begin{equation*}
    \mathcal U:=\left\{(x,y):x\in C,\ y\in\mathbb R_{\geq0}^m,
        \sum_{i=1}^m y_i\leq1+\frac{\langle s,x\rangle}{m}\right\}
\end{equation*}
has dimension $D:=d+m$.  Its fibers are simplices, so the $x$-marginal
of its uniform law has density proportional to
$(1+\langle s,x\rangle/m)^m$.  For $|t|\leq B$, the choice of $m$ gives
\begin{equation*}
    e^{t-\eta}\leq(1+t/m)^m\leq e^t.
\end{equation*}
Thus the normalized marginal density differs pointwise from the
target density by a factor in $[e^{-\eta},e^\eta]$.  If $\mu_m$ and
$\Sigma_m$ are its moments, then
\begin{align*}
    |v^\top(\mu_m-\mu_s)|
       &\leq(e^\eta-1)\sqrt{v^\top\Sigma_s v},\\
    e^{-\eta}\Sigma_s&\preceq\Sigma_m\preceq e^\eta\Sigma_s.
\end{align*}
The first inequality is Cauchy--Schwarz applied to the density ratio
minus one.  The second follows by writing each directional variance
as the minimum, over centers, of the corresponding second moment.

For the uniform sampler, the fiber height is between
$1/2$ and $3/2$, so the given radius bounds and an interior point of
the simplex give inner and outer radii for $\mathcal U$ with polynomial
bit lengths.  Rational affine rounding gives a body $V$ satisfying
$\mathbb{B}_2^D\subseteq V\subseteq q\mathbb{B}_2^D$, with
$q=\operatorname{poly}(D)$.  For $Z$ uniform on $V$, its covariance
satisfies
\begin{equation*}
    \operatorname{Cov}(Z)\succeq\frac{c}{(D+1)^2}I_D
\end{equation*}
for a universal $c>0$.  To see this, fix a unit vector $v$ and choose
$p\in\{v,-v\}\subseteq V$ with
$|v^\top(p-\mathbb EZ)|\geq1$.  Integrating the squared centered
functional over $(1-a)V+ap\subseteq V$, with $a:=1/(D+1)$, bounds
its variance below by $(1-a)^Da^2$, which is at least
$\lambda_0:=1/(4(D+1)^2)$.

We now bound the bias.  Apply
Lemma~\ref{lem:finite-grid-refinement} in dimension $D\geq3$ to $V$.
For a common dyadic scale $\nu\in(0,1]$ fixed below, use that lemma's
parameters $(h,\alpha,\alpha')=\nu(h_0,\alpha_0,\alpha'_0)$ and put
$\beta:=\alpha+\alpha'+\sqrt D\,h$.  Its accepted cubes have half-open
union $U_h$ satisfying
\begin{equation*}
    V\subseteq U_h\subseteq V_{+\beta}\subseteq(1+\beta)V,
    \qquad \beta\leq\frac{\nu}{4D}.
\end{equation*}
This sandwich and maximal coupling bound the bias of every first and
second coordinate moment of grid centers by
$O(D\beta q^2+\sqrt D\,hq)$.
Fix a rational $\bar\beta\in[\beta,2\beta]$ of polynomial encoding.
Contracting centers inward by the rational factor $1/(1+\bar\beta)$ puts
them in $V$, keeps a common rational denominator, and adds at most
$O(\beta q^2)$ to this bias.  To turn coordinate-moment
errors into relative covariance error, write $\mu_V:=\mathbb EZ$ and
$M_V:=\mathbb EZZ^\top$.  For estimates $\widehat\mu_V,\widehat M_V$, put
$\widehat\Sigma_V:=\widehat M_V-\widehat\mu_V\widehat\mu_V^\top$.
Since $\|\mu_V\|_2\leq q$,
\begin{align*}
    \|\widehat\Sigma_V-\operatorname{Cov}(Z)\|_2
    &\leq\|\widehat M_V-M_V\|_2\\
    &\quad+(2q+\|\widehat\mu_V-\mu_V\|_2)
                     \|\widehat\mu_V-\mu_V\|_2.
\end{align*}
It suffices to show that
$\|\widehat M_V-M_V\|_2\leq\eta\lambda_0/4$ and
$\|\widehat\mu_V-\mu_V\|_2\leq\eta\lambda_0/(8(q+1))$.  The display
then gives relative covariance error at most $\eta$, and the mean error
is at most $\eta$ in covariance norm.  Entrywise tolerances incur only
factors $\sqrt D$ for the mean and $D$ for the second-moment matrix.
Since $\sqrt D\,h\leq\beta$, it suffices, for a sufficiently large
universal $c_1$, to require
\begin{equation*}
    \beta\leq
       \frac{\eta\lambda_0}{c_1D^2(q+1)^3}.
\end{equation*}
Choose the largest dyadic
\begin{equation*}
    \nu\leq
       \min\left\{1,\frac{4\eta\lambda_0}
                            {c_1D(q+1)^3}\right\}.
\end{equation*}
Then the required bias bound holds, and
$\nu^{-1}=\operatorname{poly}(D,q,1/\eta)$ because
$\lambda_0=1/[4(D+1)^2]$.
Lemma~\ref{lem:finite-grid-refinement} therefore gives polynomial mixing
and predicate costs for these moment tolerances.  The comparison concerns
moments of coupled points, not total variation between the discrete and
continuous laws.

For the sampling error, run $N=\operatorname{poly}(D,q,1/\eta)\log(2/\tau)$
independent chains from the cube with index zero, each for the mixing cap
of Lemma~\ref{lem:finite-grid-refinement} with $\theta=\tau/(4N)$.  Hoeffding's inequality for coordinates and coordinate
products, with a union bound over $D+D^2$ entries, gives the same moment
tolerances for the empirical estimates, except with probability $\tau/2$.
Give the mesh bias and the empirical error fixed shares of these
tolerances.  The mixing errors add at most $\tau/4$, and the combined mean
and covariance errors are at most $4\eta$.

Finally, undo the affine rounding and project onto $x$.  Linear maps
preserve covariance-order and directional mean-error bounds; combining
them with the displayed density-ratio bounds proves the claim with
room in $\varepsilon=100\eta$.  Before the fixed affine map, the samples
have a common rational grid denominator, so their empirical moments have
polynomial encodings.  Reject failed arithmetic or
non-positive-definite outputs and use fixed rational defaults on those
branches; the good event is never rejected.  Since $D$ is polynomial in $d,B,1/\varepsilon$, all
walk, sample, and register caps have the claimed size, and retaining their
finite tapes and histories gives the reversible implementation.  The lift
is used only by a polynomial-time moment estimator and introduces no
additional integer variables.
\end{proof}

\begin{lemma}[Finite-bit $\ell$-position]
\label{lem:finite-bit-ell-position}
Let $d\geq1$, and let $Q\subseteq\mathbb R^d$ be a rational symmetric convex
body represented by a rational lifted linear program of encoding length $L$, with explicit rational
inner and outer ellipsoids.  For every $\tau\in(0,1/2)$, a finite randomized
Boolean algorithm outputs a rational invertible map $T$ such that, with
probability at least $1-\tau$,
\begin{equation}
    \ell_{TQ}\,\ell_{(TQ)^\circ}
    \leq C_\ell d\log\bigl(2d_{\mathrm{BM}}(Q,\mathbb{B}_2^d)\bigr)
    \leq C_\ell d\log(2d),
    \label{eq:finite-bit-ell-position}
\end{equation}
where $C_\ell$ is universal and
\begin{equation*}
    \ell_C:=
    \left(\mathbb E_{g\sim N(0,I_d)}\lVert g\rVert_C^2\right)^{1/2}.
\end{equation*}
Its running time, random-tape length, output encoding, and workspace are
$\operatorname{poly}(d,L,\log(1/\tau))$.  It can be capped and reversibly
simulated with polynomial overhead.
\end{lemma}

\begin{proof}
The geometric source is the Gaussian polar comparison of
\cite[Lemma~4.4.2]{dadush2012integer}.  We compute an approximate
determinant maximizer for the true Gaussian matrix norm by optimizing
against a finite empirical norm, and right-orthogonal invariance of the
true norm then gives the duality bound that the comparison needs.

Apply rational ellipsoidal rounding and let $S\in\mathbb Q^{d\times d}$ be
invertible such that, for $Q_0:=SQ$,
\begin{equation}
    \mathbb{B}_2^d\subseteq Q_0\subseteq d\mathbb{B}_2^d.
    \label{eq:ell-position-rounding}
\end{equation}
Constant rational slack in the two inclusions absorbs the rounding error.  For
a real $d\times d$ matrix $M$, define the Gaussian matrix norm and its trace-dual
by
\begin{equation*}
    \alpha(M):=
    \left(\mathbb E_g\lVert Mg\rVert_{Q_0}^2\right)^{1/2},
    \qquad
    \alpha^*(Y):=
    \sup_{\alpha(M)\leq1}\operatorname{tr}(Y^\top M).
\end{equation*}
The rounding in~\eqref{eq:ell-position-rounding} gives, for every $M$,
\begin{equation}
    \frac1d\lVert M\rVert_F\leq\alpha(M)\leq\lVert M\rVert_F.
    \label{eq:gaussian-matrix-norm-frobenius}
\end{equation}

For a simultaneous empirical approximation, fix $\gamma:=1/(100d)$ and
\begin{equation*}
    R_c=C_g\left(d+\left\lceil\log_2(1/\tau)\right\rceil+1\right),
    \qquad R_g:=\sqrt d\,R_c,
    \qquad \eta_g:=\frac{\gamma}{16d^2},
\end{equation*}
where $C_g$ is a sufficiently large universal integer.  Choose
\begin{equation}
    N:=\left\lceil
        C_N R_g^4\eta_g^{-2}
        \left[
            d^2\log\left(\frac{C_NR_g^2}{\eta_g}\right)
            +\log\frac4\tau
        \right]
    \right\rceil
    \label{eq:empirical-gaussian-sample-count}
\end{equation}
for a sufficiently large universal $C_N$.
Put $q_g:=\lceil\log_2(16d(R_c+1)/\eta_g)\rceil$ and
$\delta:=2^{-q_g}$, so $\delta\leq\eta_g/(16d(R_c+1))$.
Set $a:=2^{-\lceil2R_c^2\rceil-4}$ and
$b_g:=q_g+\lceil2R_c^2\rceil+4$, so $2^{-b_g}=a\delta$.
Write $\Phi_{\mathrm N}$ for the standard normal CDF.
For each coordinate draw exactly $b_g$ fair bits, obtaining a uniform
integer $J\in\{0,\ldots,2^{b_g}-1\}$, and put
$u_{b_g}:=(J+1/2)2^{-b_g}$.
Approximate the clipped Gaussian quantile
$f(u_{b_g}):=\max\{-R_c,\min\{R_c,\Phi_{\mathrm N}^{-1}(u_{b_g})\}\}$
by at most $\lceil\log_2(2R_c/\delta)\rceil+1$ bisections on
$[-R_c,R_c]$.  At each midpoint use a certified rational enclosure of
$\Phi_{\mathrm N}$ of width at most $a\delta/4$.  If the enclosure
contains $u_{b_g}$, stop at that midpoint; otherwise its comparison
with $u_{b_g}$ determines the next half-interval.  At the cap use the
bracket midpoint, and round the result toward zero to the $\delta$-grid.
Such CDF enclosures are obtained in polynomial time by integrating the
Taylor series for $e^{-x^2/2}$ with certified remainder bounds and
approximating the normalizing constant with certified rational bounds;
their required precision is $O(R_c^2+\log(1/\delta))$.
Use independent bit blocks for all coordinates and samples.

For analysis only, couple $J=\lfloor2^{b_g}U\rfloor$ to a continuous
uniform $U$, and set $g=\Phi_{\mathrm N}^{-1}(U)$.
Since $\Phi_{\mathrm N}'(x)\geq a$ on $[-R_c,R_c]$, the clipped
quantile is $a^{-1}$-Lipschitz.  Each resulting coordinate differs from
$f(U)$ by at most $3\delta$.  Thus $\|\widetilde g_i\|_2\leq R_g$,
and, uniformly for $\|M\|_F=1$, the change in the squared gauge caused
by finite-input quantization, bisection, and inward rounding is at most
$6dR_c\delta\leq\eta_g$.  All these caps are fixed before sampling.
Define
\begin{equation*}
    \widehat\alpha(M):=
    \left(\frac1N\sum_{i=1}^N
        \lVert M\widetilde g_i\rVert_{Q_0}^2
        +\eta_g\lVert M\rVert_F^2
    \right)^{1/2}.
\end{equation*}
On the Frobenius unit sphere, summands lie in $[0,R_g^2]$ and are
$2R_g^2$-Lipschitz.  A net of radius $\eta_g/(8R_g^2)$ has logarithmic size
$O(d^2\log(CR_g^2/\eta_g))$.  Hoeffding, the union bound, and Lipschitz
extension therefore give uniform sampling error at most $\eta_g$, except
with probability $\tau$.  Gaussian clipping, the finite quantile approximation above,
and the regularizer each add at most $\eta_g$.  Thus
\begin{equation*}
    \left|\widehat\alpha(M)^2-\alpha(M)^2\right|
       \leq4\eta_g=\frac{\gamma}{4d^2}
       \qquad(\lVert M\rVert_F=1).
\end{equation*}
Since
\eqref{eq:gaussian-matrix-norm-frobenius} gives
$\alpha(M)^2\geq1/d^2$ on that sphere, homogeneity yields, jointly with
probability at least $1-\tau$,
\begin{equation}
    (1-\gamma)\alpha(M)
    \leq\widehat\alpha(M)
    \leq(1+\gamma)\alpha(M)
    \qquad\text{for every }M\in\mathbb R^{d\times d}.
    \label{eq:empirical-gaussian-matrix-norm}
\end{equation}
The regularizer makes $\widehat\alpha$ a norm on every tape.  We apply
duality only to the true Gaussian norm, not to this empirical approximation.

Next, solve the finite convex program
\begin{equation}
    \max\ \det(\widehat A)^{1/d}
    \quad\text{subject to}\quad
    \widehat A\succeq0,
    \quad \widehat\alpha(\widehat A)\leq1
    \label{eq:finite-ell-position-program}
\end{equation}
to relative objective accuracy $1-\varepsilon_{\mathrm{num}}$, where
$\varepsilon_{\mathrm{num}}:=1/(100d)$, and set
\begin{equation}
    A:=(1-\gamma)\widehat A.
    \label{eq:true-gaussian-program-scaling}
\end{equation}
Let $\mathrm{OPT}_\alpha$ be the true optimal determinant root.
On~\eqref{eq:empirical-gaussian-matrix-norm}, $\alpha(A)\leq1$, and scaling
a true optimizer by $1/(1+\gamma)$ gives
\begin{equation}
    \det(A)^{1/d}
    \geq
    \frac{(1-\gamma)(1-\varepsilon_{\mathrm{num}})}{1+\gamma}
       \mathrm{OPT}_\alpha
    \geq(1-\varepsilon_L)\mathrm{OPT}_\alpha,
    \qquad
    \varepsilon_L\leq\frac{1}{20d}.
    \label{eq:true-gaussian-approximate-optimizer}
\end{equation}

For the duality bound, Gaussian right-orthogonal invariance and polar
decomposition $M=PU$ show
that the maximum of $|\det M|$ over $\alpha(M)\leq1$ equals
$D_\alpha:=\mathrm{OPT}_\alpha^d$, even without the PSD restriction.
Equation~\eqref{eq:true-gaussian-approximate-optimizer} gives
$\det A\geq(1-\varepsilon_L)^dD_\alpha\geq9D_\alpha/10$.
For any $H$ with $\alpha(H)\leq1$, write
$A^{-1}H=PU$ with $P\succeq0$ and $U$ orthogonal, and put
$H':=HU^\top$.  Then $\alpha(H')\leq1$ and $A^{-1}H'=P$.
For $t:=1/(2d)$, feasibility after scaling by $1+t$ gives
\begin{equation*}
    1+t\operatorname{tr}P
    \leq\det(I+tP)
    =\frac{\det(A+tH')}{\det A}
    \leq\frac{10}{9}(1+t)^d
    \leq\frac{10}{9}e^{1/2}<2.
\end{equation*}
Thus $\operatorname{tr}(A^{-1}H)\leq\operatorname{tr}P<2d$.
Taking the supremum over $H$ and using $\alpha(A)\leq1$ yields
\begin{equation}
    \alpha(A)\alpha^*(A^{-1})\leq2d.
    \label{eq:true-gaussian-lewis-duality}
\end{equation}

It remains to check that the program~\eqref{eq:finite-ell-position-program}
has exact rational separation on every tape.  Put $F(M):=\widehat\alpha(M)^2/2$.  For rational symmetric $M$, each
$q_i:=\lVert M\widetilde g_i\rVert_{Q_0}$ is the optimum of a rational
lifted LP with a rational dual optimizer $y_i\in Q_0^\circ$ satisfying
$\langle y_i,M\widetilde g_i\rangle=q_i$; take $y_i=0$ when $q_i=0$.
The chain rule for subgradients gives
\begin{equation}
    G_M:=\operatorname{sym}\!\left(
        \frac1N\sum_{i=1}^N q_i y_i\widetilde g_i^\top
    \right)+\eta_gM,
    \qquad
    \operatorname{sym}(H):=\frac{H+H^\top}{2},
    \label{eq:empirical-norm-subgradient}
\end{equation}
as a rational subgradient on symmetric matrices.  When $F(M)>1/2$, $G_M$
separates, since $\langle G_M,Z-M\rangle_F\leq1/2-F(M)<0$ for every feasible
$Z$.  Rational $LDL^\top$ elimination separates the PSD cone by $vv^\top$
when $v^\top Mv<0$; determinants and inverses supply the determinant-root
oracle on positive definite matrices.  The regularizer bounds feasible
Frobenius norms by $\operatorname{poly}(d)$, while a small rational multiple
of $I_d$ has objective at least $1/\operatorname{poly}(d,\log(1/\tau))$.
In the half-optimal region this gives minimum eigenvalue
$2^{-\operatorname{poly}(d,\log(1/\tau))}$.  Thus all radius logarithms are
polynomial, and rational ellipsoid optimization gives the required
polynomial-bit approximate optimizer
\cite{grotschel2012geometric,schrijver1998theory}.

Finally, we convert~\eqref{eq:true-gaussian-lewis-duality} into the
required geometric position.  The Gaussian polar comparison
\cite[Lemma~4.4.2]{dadush2012integer}, with the trace-dual convention above,
states that
\begin{equation*}
    \left(\mathbb E_g\|Yg\|_{Q_0^\circ}^2\right)^{1/2}
    \leq4\left(1+\log_2d_{\mathrm{BM}}(Q_0,\mathbb{B}_2^d)\right)\alpha^*(Y),
\end{equation*}
where $g$ is a standard Gaussian vector in $\mathbb R^d$.  Apply it to
$Y=A^{-1}$.  For $T:=A^{-1}S$, the two factors on the left
of~\eqref{eq:finite-bit-ell-position} are $\alpha(A)$ and
$\left(\mathbb E_g\|A^{-1}g\|_{Q_0^\circ}^2\right)^{1/2}$.  Affine invariance
of $d_{\mathrm{BM}}$ and \eqref{eq:true-gaussian-lewis-duality} give
\eqref{eq:finite-bit-ell-position}.  All sampling and optimization caps are
fixed in advance.  The empirical approximation is the only probabilistic
event; on it, rational optimization and the true-norm comparison give the
claimed bound.  On every tape, failed arithmetic, a nonpositive
determinant, or an exhausted cap yields a fixed rational invertible default.  The
bounded samples, regularizer, and eigenvalue bounds make all registers
polynomial in $d,L,\log(1/\tau)$.  Bennett simulation gives the finite
reversible implementation.
\end{proof}

\begin{lemma}[Finite-bit discrete Gaussian sampling]
\label{lem:finite-bit-dgs}
Let $\Lambda$ be a rational lattice of rank $d$, and let $L$ denote the
encoding length of the supplied lattice and scale data.
Let $k\geq d$ be an integer and let $s>0$ be rational, with
$|\log s|\leq\operatorname{poly}(L,k)$.  For every fixed constant $C>0$, a
capped randomized Boolean algorithm outputs a lattice vector whose
distribution has statistical distance at most $2^{-Ck}$ from
$D_{\Lambda,s}$.  It uses
\begin{equation*}
    2^{O(d)+\operatorname{polylog}k}\operatorname{poly}(L)
\end{equation*}
time, space, and finite random bits, and hence $2^{O(k)}\operatorname{poly}(L)$
of each.  Its reversible simulation has the corresponding gate and qubit
bounds.
\end{lemma}

\begin{proof}
We first treat ranks $d\leq1$.  If $d=0$, output $0$ exactly.  If $d=1$,
write $\Lambda=\mathbb Zb_1$ and apply directly the capped one-dimensional
inverse-transform construction below, with shift $c=0$ and width
$r=s/\|b_1\|_2$.  For this single draw, take $N_{\mathrm{step}}:=1$ and
$B_{\mathrm{tail}}:=\lceil(C+12)k\rceil+16$ in that construction.
Choose the truncation and CDF budgets so that their
sum is at most $2^{-Ck}$, and multiply the sampled integer by $b_1$.
The width satisfies $r^2=s^2/(b_1^\top b_1)$, which is
rational, and the construction uses $\operatorname{poly}(L,k)$ resources,
within the claimed bound.

From now on assume $k\geq d\geq2$.  We run the general sampler of
\cite[Theorem~3.7]{aggarwal2015svp} (ADRS below)
with confidence parameter $\kappa_{\mathrm A}=C_1k$ and retain one sample.
Let $\ell_{\mathrm A}$ be its number of combiner stages; the source's
parameter choice gives
$\ell_{\mathrm A}=O(\log(2\kappa_{\mathrm A})+\log^2(2d))$.
Fix an integer $C_1\geq2$, depending only on $C$, large enough that the
ideal statistical error is at most $2^{-(C+5)k}$ and
$\ell_{\mathrm A}e^{-\kappa_{\mathrm A}/2}\leq2^{-(C+6)k}$.
The rest of the proof couples a capped finite computation to this ideal
sampler and charges the errors of all its random primitives to one
transcript.  We fix all caps before any random bit
is read, discretize each random primitive and total the transcript error,
and then bound the registers, truncate the output, and count the cost.

\proofstep{Scale.}
The scale $s$ is rational, so no scale approximation enters the error
budget.  This is the case needed in
Lemma~\ref{lem:rank-weighted-smooth-sublattice}, where the actual Gaussian
scales are the rational reciprocals of \eqref{eq:rank-weighted-sigma}.
There, determinant computation and root extraction give a constant-factor
rational upper approximation to the original normalized determinant, which
stays fixed throughout the descent.

\proofstep{Primitives.}
Recall that the ADRS proof consists of basis reduction and a sublattice
computation (Proposition~2.17),
$M_0=(32\kappa_{\mathrm A})^{\ell_{\mathrm A}+2}2^d$ high-parameter
samples (Theorem~2.16), its $\ell_{\mathrm A}$ combiner stages
(Proposition~3.5), and truncation to its output list.  In the Gaussian
sampler of \cite[Lemma~2.3 and Section~5]{brakerski2013classical} (BLP+
below), we treat the output of each complete one-dimensional discrete
Gaussian sampler as a primitive; its internal continuous half-line
proposals are not primitives.  The other primitives are the BLP+
outer-correction Bernoulli choices, Poisson counts in the square sampler
(Theorem~3.3 of ADRS), and finite uniform or Bernoulli choices.

\proofstep{Estimator cap.}
Before fixing the transcript caps, we cap the maximum-probability
estimator of ADRS Proposition~3.1, which each square-sampler call uses
once.  For a call on $N=2^h$ categories, $h\leq d$, permit only stages
$j=0,\ldots,h+1$, with $p_{\mathrm{est}}=2^{-j}$.  On a successful count
test return the current $p_{\mathrm{est}}$, as in the source, and retain the source's
input-exhaustion failure rule.  If the last permitted stage ends without
an estimate, abort the sampler and return $0$ with a failure flag.
Conditionally on reaching that stage, its fresh Poisson draw has mean
$\mu=2\kappa_{\mathrm A}N$.  If there is no input-exhaustion failure
and all category counts are below $\kappa_{\mathrm A}/3$, their sum
$R_{\mathrm P}$ satisfies $R_{\mathrm P}<\kappa_{\mathrm A}N/3=\mu/6$.
The lower-tail Chernoff bound therefore gives
\begin{equation*}
    \Pr[R_{\mathrm P}<\mu/6]\leq e^{-25\mu/72}
       \leq e^{-\kappa_{\mathrm A}/2}.
\end{equation*}
This bounds the additional truncation event for every input label
stream, with no independence assumption on its labels.
Since there are at most $\ell_{\mathrm A}$ estimator calls, their total
additional error is at most $2^{-(C+6)k}$, which we charge to the
transcript budget below.  Every returned estimate satisfies
$\widetilde p_{\max}=2^{-j}\geq1/(2N)$.  The actual Poisson means in
the estimator and the square sampler are, respectively,
$\kappa_{\mathrm A}/p_{\mathrm{est}}$ and $1/\widetilde p_{\max}$.  On every path they
are therefore exact rationals, bounded by $2\kappa_{\mathrm A}N$ and $2N$.
Unknown category probabilities occur only in the source's analysis.

\proofstep{Transcript caps.}
Let $N_{\mathrm{base}}$ be a deterministic upper bound on the operations
and primitive-call sites in this level-capped computation, not counting
repetitions of the outer-correction and finite-uniform rejection loops;
the finite cost of a one-dimensional draw is charged below.  The displayed ADRS parameters
give fixed universal $c_4,c_5$ with
\begin{equation*}
    N_{\mathrm{base}}
    \leq
    2^{c_4d+(\log_2(2k))^{c_5}}\operatorname{poly}(L).
\end{equation*}
Choose a sufficiently large integer $C_{\mathrm{tail}}$, set
$B_{\mathrm{tail}}:=C_{\mathrm{tail}}
(k+\lceil\log_2(2N_{\mathrm{base}})\rceil)$, and cap every remaining
rejection loop at $R_{\mathrm{rej}}:=8B_{\mathrm{tail}}$.  A conservative
bound on the fixed nesting depth of the remaining BLP+ outer-correction
and finite-uniform rejection loops gives a universal integer
$c_{\mathrm{loop}}$ such that
\begin{equation}
    N_{\mathrm{step}}:=N_{\mathrm{base}}(1+R_{\mathrm{rej}})^{c_{\mathrm{loop}}}
    \leq2^{O(d)+\operatorname{polylog}k}\operatorname{poly}(L).
    \label{eq:dgs-capped-transcript-length}
\end{equation}
Taking $C_{\mathrm{tail}}$ sufficiently large in terms of $C$ makes
\begin{equation}
    N_{\mathrm{step}}2^{-B_{\mathrm{tail}}}\leq2^{-(C+8)k}.
    \label{eq:dgs-transcript-tail-budget}
\end{equation}
These choices are noncircular: the confidence constant and estimator
level cap precede $N_{\mathrm{base}}$, which precedes the rejection caps
and $N_{\mathrm{step}}$; the remaining tail constants follow
$C_{\mathrm{tail}}$.  The estimator's at most $h+2$ levels are included
in $N_{\mathrm{base}}$, not in the fixed rejection-nesting exponent
$c_{\mathrm{loop}}$.  Uniform upper bounds on every truncation radius,
support length, register size, and probability precision are determined
by $d,k,L,C$ before any random bit is read.  Individual call parameters
may depend on earlier outcomes, but their work and tape usage are bounded
by these predetermined maxima.

\proofstep{One-dimensional draws.}
We now discretize the primitives.  Write $\rho_r(x):=e^{-\pi x^2/r^2}$ and
$\rho_r(A):=\sum_{x\in A}\rho_r(x)$.  Fix one complete one-dimensional
discrete Gaussian draw inside BLP+.  After scaling its line lattice, it
samples $D_{\mathbb Z+c,r}$.  Let $u$ be the corresponding rational
Gram--Schmidt vector, let $t$ be the current rational coset shift, and
let $s_{\mathrm A}$ be the current lattice Gaussian scale.  Store
\begin{equation*}
    c\equiv\frac{\langle t,u\rangle}{\langle u,u\rangle}\pmod{\mathbb Z},
    \qquad r^2=\frac{s_{\mathrm A}^2}{\langle u,u\rangle}.
\end{equation*}
Since $s$ is rational and the ADRS scales differ from it by powers of
$\sqrt2$, $s_{\mathrm A}^2$ is rational.
Thus $c$ and $r^2$ are rational.  We track the integer coordinate choices
rather than the physical one-dimensional samples, so each updated coset
shift remains rational.  Interval endpoints come from outward rational
square-root bounds, and all weights are evaluated using $r^2$; no
irrational lattice coordinate is stored.

After an integer shift we may assume $c\in[-1/2,1/2]$.  We replace the
whole one-dimensional sampler, including its continuous half-line
proposals and inner rejection, by a draw from $D_{\mathbb Z+c,r}$
conditioned on
\begin{equation*}
    [-X_r,X_r],\qquad X_r:=1+C_2r\sqrt{B_{\mathrm{tail}}},
\end{equation*}
with outward rational rounding.  No continuous half-line proposal is
ever generated, so the ideal object being approximated is the complete
output law $D_{\mathbb Z+c,r}$.  To bound the truncation, put
$T:=C_2r\sqrt{B_{\mathrm{tail}}}$.
On each side of $0$, list the points of $\mathbb Z+c$ of absolute value
greater than $1+T$ as $y_0,y_1,\ldots$ in increasing absolute value, so
$|y_j|\geq1+T+j$.  If $r<1$, some $x_0\in\mathbb Z+c$ has
$|x_0|\leq1/2$, so the normalizer is at least $\rho_r(x_0)$; since
$y_j^2-x_0^2\geq T^2+2j$, the omitted probability is at most
$2\sum_{j\geq0}e^{-\pi(T^2+2j)/r^2}\leq3e^{-\pi T^2/r^2}$.
If $r\geq1$, the interval $[-r/2,r/2]$ contains at least
$\lfloor r\rfloor\geq r/2$ points of $\mathbb Z+c$, so the normalizer is
at least $(r/2)e^{-\pi/4}$, while the unnormalized omitted mass is at
most $2\int_T^\infty e^{-\pi t^2/r^2}\,dt\leq re^{-\pi T^2/r^2}$.
In both cases the omitted probability is at most
$2e^{\pi/4}e^{-\pi C_2^2B_{\mathrm{tail}}}$, uniformly in $c$ and $r$;
outward rounding only enlarges the interval.  Equivalently, truncating
the normalizer $\rho_r(\mathbb Z+c)$ to the rounded interval loses at
most this fraction of it.

It remains to choose $C_2$.  Put
\begin{equation*}
    \alpha:=\frac{2^{-B_{\mathrm{tail}}}}{64(d+1)N_{\mathrm{step}}}.
\end{equation*}
By \eqref{eq:dgs-capped-transcript-length},
$\log_2N_{\mathrm{step}}\leq\log_2N_{\mathrm{base}}
+c_{\mathrm{loop}}\log_2(1+R_{\mathrm{rej}})=O(B_{\mathrm{tail}})$.
Generating the $M_0\geq2^d$ initial samples gives
$N_{\mathrm{base}}\geq2^d\geq d+1$, so
$\log_2(d+1)\leq B_{\mathrm{tail}}$.  Hence a universal $C_2$ makes
$2e^{\pi/4}e^{-\pi C_2^2B_{\mathrm{tail}}}\leq\alpha$.
All scales have polynomial-bit logarithms, so $X_r$ has polynomial
encoding; we take its outward rounding with additive slack at most $1$.
The conditioned law lives on the at most $M\leq2X_r+3$ consecutive points
of $\mathbb Z+c$ in the rounded interval, so $\log M$ is polynomial.  We
sample it by the finite inverse transform, using the Gaussian CDF
evaluation described below, with polynomial work per draw, and couple
each finite draw to the ideal complete output $D_{\mathbb Z+c,r}$.  With
the per-primitive budget
$\delta_{\mathrm{prim}}=2^{-B_{\mathrm{tail}}}/N_{\mathrm{step}}$ fixed in
the inverse-transform step below, truncation and CDF error together cost
at most $\alpha+\delta_{\mathrm{prim}}/16\leq2\delta_{\mathrm{prim}}$ per
draw.

\proofstep{Outer correction.}
Here $b,p,a,w_x$ have the support-dependent meanings specified in the
finite inverse-transform and Gaussian CDF steps below.
The BLP+ outer correction \cite[Section~5.2]{brakerski2013classical}
accepts with probability
\begin{equation*}
    A=\prod_j\frac{\rho_{r_j}(\mathbb Z+c_j)}{\rho_{r_j}(\mathbb Z)}
       \in(e^{-2},1],
\end{equation*}
a product of at most $d$ ratios and hence a quotient of at most $2d$
one-dimensional normalizers.  Here $r_j$ and $c_j$ are the normalized
parameters of the one-dimensional draws, and integer shifts give
$c_j\in[-1/2,1/2]$ without changing $A$.  The bound $A>e^{-2}$ uses the
full, dimension-dependent width hypothesis of
\cite[Lemma~2.3]{brakerski2013classical}, which the ADRS scales satisfy
once their universal constant is taken sufficiently large; this
hypothesis also gives $r_j\geq\sqrt{\ln6/\pi}>1/2$.  By the choice of $C_2$, truncating each
normalizer to $[-X_{r_j},X_{r_j}]$ loses at most a fraction $\alpha$ of
it.  We evaluate each truncated normalizer to relative error at most
$\alpha$ with the Gaussian sum routine of the CDF step below, applied to
its support with $p$ replaced by $p+\lceil\log_2(d+1)\rceil+8$; the
displayed $b$ and $p$ remain the precision rules for CDF comparisons.
This computes $\sum_xw_x\geq1$ to additive, hence relative, error at most
$\alpha/2^{16}$.  The truncated normalizer is
$e^{-\pi a^2/r_j^2}\sum_xw_x$, where $a$ is as in that step.  Since
$a\leq1/2<r_j$, the factor $e^{-\pi a^2/r_j^2}$ is at least $e^{-\pi}$
and has exponent in $[0,\pi]$; evaluating it to additive error
$e^{-\pi}\alpha/4$ adds relative error at most $\alpha/4$.

Each computed normalizer therefore has relative error at most $2\alpha$
from the corresponding full normalizer, so the computed $\widetilde A$
satisfies $|\log\widetilde A-\log A|\leq2d\cdot4\alpha\leq1$ and
$|\widetilde A/A-1|\leq16d\alpha\leq\delta_{\mathrm{prim}}/4$.  As
$A\leq1$, this is also an additive bound.  Round the product with
additive error at most $\delta_{\mathrm{prim}}/4$ and clip it to $[0,1]$.
Clipping does not increase the distance to $A$, so the computed
probability is within $\delta_{\mathrm{prim}}/2$ of $A$.  Draw the
correction Bernoulli by comparison with a uniform dyadic rational of
$\lceil\log_2(16/\delta_{\mathrm{prim}})\rceil$ bits, which adds at most
$\delta_{\mathrm{prim}}/16$.  Each correction decision therefore couples
to the ideal one with error at most $\delta_{\mathrm{prim}}$.  These
precisions use only $O(\log(d+1))$ guard bits beyond the displayed rules,
so all bit lengths remain polynomial.

\proofstep{Finite inverse transform.}
For each finite inverse-transform draw with support size $M$, put
\begin{equation*}
    \delta_{\mathrm{prim}}
       :=\frac{2^{-B_{\mathrm{tail}}}}{N_{\mathrm{step}}},
    \qquad
    b:=B_{\mathrm{tail}}
       +\lceil\log_2N_{\mathrm{step}}\rceil
       +\lceil\log_2M\rceil+6.
\end{equation*}
The draw uses the midpoint of a uniformly chosen dyadic interval of
length $2^{-b}$, and every CDF comparison is evaluated to additive error
at most $2^{-b}$.  Binary search locates the output index with at most
$\lceil\log_2M\rceil+1$ CDF evaluations.  For the analysis, couple this
midpoint to a continuous uniform variable in its interval.
A comparison can disagree with the ideal inverse transform only within
distance $2^{1-b}$ of a CDF boundary.  A union bound over the at most
$M-1$ boundaries gives coupling error at most
$4M2^{-b}\leq\delta_{\mathrm{prim}}/16$.

\proofstep{Gaussian CDF evaluation.}
The Gaussian CDF evaluations needed here have an explicit polynomial-time
implementation.  Each conditioned one-dimensional discrete Gaussian has
support on a consecutive segment of a translate of $\mathbb Z$.
Let $a$ be the smallest absolute value of a support
point and use the equivalent unnormalized weights
\begin{equation*}
    w_x:=\exp\!\left(-\pi\frac{x^2-a^2}{r^2}\right).
\end{equation*}
Then $0<w_x\leq1$ and the normalizer is at least $1$.
Put $p:=b+\lceil\log_2M\rceil+10$.
If $r<1$, the truncated support has
$M=O(1+\sqrt{B_{\mathrm{tail}}})$ points, so we evaluate its weights
individually to additive error at most $2^{-p}$.
Exponentials with exponent larger than the requested precision can be
replaced by zero, and the remaining ones are evaluated by their Taylor
series with rational interval bounds.

If $r\geq1$, all exponents lie in $[0,U]$, where the rational upper bound
\begin{equation*}
    U:=4(X_r+1)^2/r^2=O(B_{\mathrm{tail}}+1)
\end{equation*}
suffices.  Approximate $e^{-u}$ on this interval by
$\sum_{j=0}^J(-u)^j/j!$, with
$J:=\lceil8(U+p+1)\rceil$.
The remainder is at most
$U^{J+1}/(J+1)!\leq2^{-p}$.
After substituting $u=\pi(x^2-a^2)/r^2$, we sum the polynomial over any
initial segment of the support using rational power-sum recurrences,
without enumerating the support.
Approximating $\pi$ with rational interval bounds adds at most another
$2^{-p}$ error per weight.
Thus the numerator and denominator of each CDF have additive errors at
most $2M2^{-p}$.  Since the exact denominator is at least $1$, rational
division with suitable guard bits evaluates the CDF to error at most
$2^{-b}$.
The polynomial degree, arithmetic precision, and coefficient encodings
are polynomial in
$b,B_{\mathrm{tail}},\log M,L,k$.
Bernoulli probabilities are evaluated to additive error at most
$\delta_{\mathrm{prim}}$.

\proofstep{Poisson draws.}
We truncate each $\operatorname{Pois}(\lambda)$ draw at
\begin{equation*}
    R_\lambda
    :=\left\lceil\max\left\{8\lambda,
    2(B_{\mathrm{tail}}+\lceil\log_2N_{\mathrm{step}}\rceil+2)\right\}\right\rceil.
\end{equation*}
The Poisson bound \cite[Lemma~2.19]{aggarwal2015svp} gives tail at most
$2e^{-R_\lambda}\leq2^{-B_{\mathrm{tail}}}/N_{\mathrm{step}}$.
We sample the law conditioned on $\{0,\ldots,R_\lambda\}$ by the finite
inverse transform above, with $M:=R_\lambda+1$, from a table of relative
weights.  For $\lambda=0$ return zero.
Otherwise put $j_\lambda:=\lfloor\lambda\rfloor$, set
$\psi_{j_\lambda}:=1$, and generate the relative weights in both directions by
\begin{equation*}
    \psi_{j+1}=\psi_j\frac{\lambda}{j+1}\quad(j\geq j_\lambda),
    \qquad
    \psi_{j-1}=\psi_j\frac{j}{\lambda}\quad(1\leq j\leq j_\lambda).
\end{equation*}
Every multiplier in these recurrences is at most $1$, and the exact
normalizer is at least $1$.
Round each update to a multiple of $2^{-P}$, where
$P:=b+2\lceil\log_2M\rceil+10$.
Each weight error is at most $M2^{-P}$, and the sum of all weight errors
is at most $M^22^{-P}$.
Consequently, normalized partial sums, followed by rational division
with guard bits, give CDF error at most $2^{-b}$.
The table takes $O(M\operatorname{poly}(L,k,b,\log M))$ operations and
bits of workspace.

\proofstep{Rejection loops and transcript error.}
The complete one-dimensional sampler is now a capped inverse-transform
computation and has no inner rejection loop.  The ideal BLP+ outer
correction accepts with probability at least $e^{-2}>1/8$, so its capped
loop times out with probability at most
$e^{-B_{\mathrm{tail}}}\leq2^{-B_{\mathrm{tail}}}$.  Deviations caused by
finite one-dimensional draws and approximate correction probabilities
are charged separately through the transcript coupling.  Finite-set
uniforms use binary rejection, with acceptance at least $1/2$ and the
same cap; random permutations are sequences of these choices.  All
remaining finite probabilities are approximated to additive error
$2^{-B_{\mathrm{tail}}}/N_{\mathrm{step}}$.

We now total the transcript error.  Put
$E:=N_{\mathrm{step}}2^{-B_{\mathrm{tail}}}$.
The capped outer-correction and finite-uniform timeouts contribute at
most $E$, and all remaining primitive tail and coupling errors contribute
at most $4E$, conservatively.  Adding the estimator-level truncations
gives, by \eqref{eq:dgs-transcript-tail-budget}, total transcript error
at most
\begin{equation*}
    5E+2^{-(C+6)k}
    \leq5\cdot2^{-(C+8)k}+2^{-(C+6)k}
    \leq2^{-(C+4)k}.
\end{equation*}
Every failure returns $0\in\Lambda$ with a flag.  The distance guarantee
applies to this unconditional vector output; callers do not discard or
resample flagged zeros.

\proofstep{Registers.}
We bound the registers directly in terms of the original input, before
using any final-output tail.  Let $B=(b_1,\ldots,b_d)$ be the input
basis, and let $B^+=(B^\top B)^{-1}B^\top$.
The initialization in \cite[Theorem~2.15, Proposition~2.17, and the proof of
Theorem~3.7]{aggarwal2015svp} computes one recursively reduced basis
$\widehat B=(\widehat b_1,\ldots,\widehat b_d)$ and selects one prefix
lattice $\Lambda'=\widehat B_{d'}\mathbb Z^{d'}$.  Size-reduce this basis
once, so that its Gram--Schmidt vectors $u_i$ and coefficients
$\mu_{ji}$ satisfy $|\mu_{ji}|\leq1/2$ for $i<j$.  Size reduction
preserves every prefix lattice and every Gram--Schmidt length, hence
also the selected prefix and its high-parameter sampling condition.
The initial reduction is charged once at the standard bit-operation
cost of the basis-reduction algorithm used in ADRS Theorem~2.15.

Here is a bound for this fixed basis in the original input length.
Put $R_0=\max_j\|b_j\|_2$, and let $\gamma$ be the approximation factor
of the source's recursive basis reduction; its parameter choice has
$\log\gamma=\operatorname{poly}(L,k)$.
In every nonzero orthogonal quotient used by the reduction, some
projected input vector $b_j$ is nonzero and has length at most $R_0$.
The recursive reduction guarantee therefore gives
$\|u_i\|_2\leq\gamma R_0$.
Orthogonality and size reduction give
$\|\widehat b_i\|_2\leq\sqrt d\,\gamma R_0$.
Writing $\widehat b_i=Bz^{(i)}$, with $z^{(i)}\in\mathbb Z^d$, yields
\begin{equation*}
 \|z^{(i)}\|_2
 \leq\|B^+\|_2\sqrt d\,\gamma R_0.
\end{equation*}
The original rational matrix $B^+$ has polynomial encoding.
Consequently the integers $z^{(i)}$, and hence the entries of this
single fixed rational basis, have $\operatorname{poly}(L,k)$ bits.
Its rational Gram--Schmidt data have polynomial encoding by
determinant bounds applied once to this fixed matrix.

If $d'=0$, all vectors sampled in $\Lambda'$ are zero.
Otherwise put
$b_*:=\min\{1,u_1^\top u_1,\ldots,u_{d'}^\top u_{d'}\}$.
This is a positive rational lower bound for all $\|u_i\|_2$ with
$i\leq d'$, with polynomial encoding.
ADRS Theorem~3.7 draws its initial samples once, at scale
$\widehat s=2^{\ell_{\mathrm A}/2}s$.  Put
$S_*:=\max\{1,2^{\ell_{\mathrm A}}s^2\}$.  This rational number bounds
$\widehat s$ from above, and it has polynomial encoding and polynomial
$|\log S_*|$.
Choose the tail constant $C_2$ to be an integer, as is permitted
by its earlier lower-bound requirement, and set
\begin{equation*}
 X_{\mathrm{cap}}:=2+C_2S_*B_{\mathrm{tail}}/b_*
       \geq2+C_2S_*\sqrt{B_{\mathrm{tail}}}/b_*.
\end{equation*}
This rational number has polynomial encoding.
The one-dimensional support truncation above, including its
at-most-one-unit outward rounding, bounds every normalized
Gram--Schmidt component of every initial draw by $X_{\mathrm{cap}}$,
on every tape.

For a centered initial draw $\widehat B_{d'}w$, denote these components
by $\eta_i$.  The backward recursion chooses the integer coordinates in
descending order, with
\begin{equation*}
 w_i=\eta_i-\sum_{j>i}\mu_{ji}w_j.
\end{equation*}
Descending induction gives
$|w_i|\leq(3/2)^{d'-i}X_{\mathrm{cap}}\leq2^{d'}X_{\mathrm{cap}}$:
the inductive sum is bounded by the geometric series with ratio
$3/2$.  This bounds coordinates as they are chosen, not just after
a complete vector has been produced.
In rational units the BLP+ coset-shift update is
\begin{equation*}
 t_{d'}=0,\qquad t_{i-1}=t_i+w_i\widehat b_i-\eta_iu_i.
\end{equation*}
Thus
$t_i=\sum_{j>i}(w_j\widehat b_j-\eta_ju_j)$, and its next scalar shift is
\begin{equation*}
 \frac{\langle t_i,u_i\rangle}{\langle u_i,u_i\rangle}
 =\sum_{j>i}\mu_{ji}w_j.
\end{equation*}
The terms involving $u_j$ disappear by orthogonality.
Compute these scalar shifts directly modulo $\mathbb Z$ from the
fixed Gram coefficients and the chosen integers.
Their denominators divide a common denominator of those
coefficients; the same is true of
$\eta_i=w_i+\sum_{j>i}\mu_{ji}w_j$.
If the vector $t_i$ is needed, evaluate its displayed fixed-basis
sum directly.  Its denominators divide a common multiple of the
denominators of $\widehat B$ and of the fixed Gram vectors multiplied by
the common denominator for the $\eta_i$.
There are at most $d'$ terms, with the coefficient bounds just proved.
Reducing fractions or retaining these sums therefore bounds every
shift, center, and arithmetic intermediate by
$\operatorname{poly}(L,k)$ bits, with no iterated projection or
changing-lattice denominator argument.

Encode every subsequent list vector in the same prefix basis
$\widehat B_{d'}$.  Each combiner pairs coefficient vectors of equal
parity and replaces $w,w'$ by $(w+w')/2$.  The result is integral and
has infinity norm at most $\max\{\|w\|_\infty,\|w'\|_\infty\}$; forming
the numerator requires only one additional bit.  No new lattice basis or
growing chain of normal forms is used.  Indices, counters, scales, CDF
computations, and rounded probability tables have the caps already
specified above.  Thus every individual register has polynomial
bit length on every tape.  The list and transcript caps bound the
number of such registers; their single-exponential number does not
increase the bit length of one entry.

\proofstep{Output truncation.}
Map to zero every output outside a rational outward rounding of
the centered ball of radius $C_3s\sqrt k$; the rounded ball has radius at
most $2C_3s\sqrt k$.  The centered Gaussian tail bound
\cite[Lemma~2.4]{aggarwal2015svp} makes the removed ideal mass at most
$2^{-(C+5)k}$.  Let $B$ be the full-column-rank rational basis matrix of
$\Lambda$, and put $B^+:=(B^\top B)^{-1}B^\top$.
This rational left inverse has polynomial encoding length.
If a retained vector is $Bz$, then
\begin{equation*}
    \|z\|_2
       \leq\|B^+\|_2\|Bz\|_2
       \leq2\|B^+\|_2C_3s\sqrt k.
\end{equation*}
The input encoding and the bound on $|\log s|$ therefore give polynomial
output encoding, even when the lattice lies in a proper ambient subspace.
The ideal ADRS error, transcript coupling, and final tail
sum to at most $2^{-Ck}$.

\proofstep{Cost.}
Set $\lambda_*:=4\kappa_{\mathrm A}2^d$ and fix
\begin{equation*}
    R_{\max}:=\left\lceil\max\left\{1,8\lambda_*,
       2(B_{\mathrm{tail}}
          +\lceil\log_2N_{\mathrm{step}}\rceil+2)\right\}\right\rceil.
\end{equation*}
This is a deterministic bound for every allowed Poisson table on every
tape, not a maximum over the tables encountered on one execution.
All list sizes are bounded by $N_{\mathrm{step}}$, and
\begin{equation*}
    R_{\max}\leq
    2^{O(d)+\operatorname{polylog}k}\operatorname{poly}(L).
\end{equation*}
For reversible simulation, reserve the global maximum register and tape
lengths and pad each bounded computation to its predetermined work cap;
inactive steps and unused tape bits have no effect on the vector output.
The Gaussian CDF evaluations and scalar probability evaluations take
$\operatorname{poly}(L,k,B_{\mathrm{tail}})$ work per call, and Poisson
tabulation adds work at most
$R_{\max}\operatorname{poly}(L,k,B_{\mathrm{tail}})$ per call.
The number of stochastic primitive calls is bounded by
$N_{\mathrm{step}}$; the additional table operations are deterministic.
Thus total work is at most
\begin{equation*}
    N_{\mathrm{step}}(R_{\max}+1)
       \operatorname{poly}(L,k,B_{\mathrm{tail}})
    =
    2^{O(d)+\operatorname{polylog}k}\operatorname{poly}(L).
\end{equation*}
Table entries have polynomial encoding length, and their number has the
same single-exponential bound, so the workspace bound is preserved.
The finite random-tape bound is unchanged up to polynomial factors.  All
loops and registers are now capped, so Bennett simulation gives the
reversible implementation.
\end{proof}

\subsection{A variant from the estimates of Reis and Rothvoss}
\label{subsec:published-variant}

This subsection derives a variant of
Theorem~\ref{thm:constructive-flatness} from earlier estimates, with a factor
$O(\log^3(2k))$ in place of $\Phi(k)$.  It shows what the argument gives
without the recent estimates of Facts~\ref{fact:reverse-minkowski}
and~\ref{fact:isotropic-mean-norm}.  In their place it uses the
$O(\log(2m))$ reverse Minkowski threshold and $\ell$-position, which
replaces approximately isotropic position.

For a full-dimensional symmetric convex body $C$ with the origin in its
interior, its gauge and polar are
\begin{equation*}
    \|x\|_C:=\inf\{t>0:x\in tC\},
    \qquad
    C^\circ:=\{y:\langle x,y\rangle\leq1\text{ for every }x\in C\}.
\end{equation*}
The gauge is a norm, and $\|y\|_{C^\circ}=\max_{x\in C}\langle x,y\rangle$
is the support function of $C$.  Define
\begin{equation*}
    \ell_C:=\left(\mathbb E_{Y\sim N(0,I_d)}\|Y\|_C^2\right)^{1/2}.
\end{equation*}
For example, a Euclidean ball of radius $r>0$ satisfies
\begin{equation*}
    \ell_{rB_2^d}=\frac{\sqrt d}{r},
    \qquad \ell_{(rB_2^d)^\circ}=r\sqrt d.
\end{equation*}
Thus enlarging the ball decreases its $\ell$-value and increases that
of its polar; their product is $d$, independently of the radius.
For general $C$, the quantity $\ell_C$ measures the typical $C$-norm of a
Gaussian vector, and $\ell_{C^\circ}$ measures the body's width in Gaussian
directions.  For bodies in a proper subspace, we use a standard Gaussian in
that subspace.

\begin{fact}[Urysohn's inequality and $\ell$-position]
\label{fact:urysohn-ell-position}
For a symmetric convex body $C\subseteq\mathbb R^d$,
\begin{equation}
    \operatorname{Vol}_d(C)^{1/d}
    \leq\frac{C_U}{d}\ell_{C^\circ},
    \label{eq:urysohn-gaussian-form}
\end{equation}
where $C_U$ is universal.  Moreover, an invertible linear map $T$ can be
chosen so that
\begin{equation}
    \ell_{TC}\ell_{(TC)^\circ}\leq C_{\mathrm{pos}}d\log(2d),
    \label{eq:ell-position-preliminary}
\end{equation}
for a universal $C_{\mathrm{pos}}$.  We call this an $\ell$-position.
These are \cite[Corollary~34 and Theorem~30]{reis2023subspace}; the finite
randomized computation of $T$ is established in Lemma~\ref{lem:finite-bit-ell-position}.
\end{fact}

Urysohn's inequality bounds volume by average width.  With $T$ as in
\eqref{eq:ell-position-preliminary} it gives $\operatorname{Vol}_d(TC)^{1/d}\ell_{TC}=O(\log(2d))$,
which turns a covering bound expressed through $\ell_{TC}$ into a
determinant--volume bound for the original body.

The proof of Theorem~\ref{thm:constructive-flatness} in
Appendix~\ref{app:flatness-proofs} adapts as follows.
Lemma~\ref{lem:rank-weighted-smooth-sublattice} and its
proof hold for every nondecreasing integer-valued $\xi\geq3$
satisfying~\eqref{eq:reverse-minkowski-threshold} for which an integer in
$[\xi(m),\xi(m)+1]$ is computable in polynomial time, in particular for the
$O(\log(2m))$ threshold of
\cite[Definition~2.12 and Theorem~2.13]{dadush2019approximating}.  Replace
the covariance map by an $\ell$-position $T$ of the core $C$, computed by
Lemma~\ref{lem:finite-bit-ell-position}; on the barycenter event the core
has polynomially bounded radius logarithms.  The full-dimensional case of
\cite[Proposition~37]{reis2023subspace}, including its
approximate-barycenter clause, gives
$\operatorname{Vol}_m(K_{\mathcal L})^{1/m}\leq C_1\operatorname{Vol}_m(C)^{1/m}$,
and Urysohn's inequality~\eqref{eq:urysohn-gaussian-form} gives
\begin{equation*}
    \operatorname{Vol}_m(TC)^{1/m}\ell_{TC}
       \leq\frac{C_2}{m}\ell_{(TC)^\circ}\ell_{TC}
       \leq C_3'\log(2m).
\end{equation*}
Since $\mathbb E\|G\|_{TC}\leq\ell_{TC}$ and
$\operatorname{Vol}_m(TK_{\mathcal L})^{1/m}\leq C_1\operatorname{Vol}_m(TC)^{1/m}$,
the first line of the per-stage bound~\eqref{eq:rank-weighted-cover} in
that proof is then at most
\begin{equation*}
    C_5'\log^2(2m)\frac qm
    \left(\frac{\det\mathcal L}{\operatorname{Vol}_mK_{\mathcal L}}\right)^{1/m},
\end{equation*}
and the harmonic sum gives $O(\log^3(2k))$.

\section{Encoding, compression, and reductions}
\label{app:recursion-details}

The quantum algorithm replays classical computations inside a circuit, so
every classical step must have a fixed finite cost on every input.  This
appendix fixes the encoding conventions that make this possible, confines
the root body to a bounded box, and proves the rank-preserving
Frank--Tardos lemma that compresses every recursive affine subspace back to
polynomial encoding length.  The resulting invariant bounds the encodings
of nodes and witnesses in the classical recursion of
Section~\ref{subsec:classical-recursion}, of the rational lifts in the
local access of Section~\ref{sec:search}, and of the circuits and setup
of the quantum algorithm in Section~\ref{sec:quantum-ip}.

The encoding length $L$ includes all dimensions and rational data;
projections and fibers append only rational variables and equalities, so
their lifts remain polynomial-size.  In recursive use, compressed nodes
and witnesses have $\operatorname{poly}(L)$ bits, while auxiliary
geometric data may have $\operatorname{poly}(L,\log(1/\delta))$ bits.
Each local call is charged for its actual input encoding.  Exact
membership, optimization, separation, lattice normal form, and rational
arithmetic cost $\operatorname{poly}(L)$.  Over a fixed finite
inverse-closed universal gate set, we make finite classical routines
reversible with polynomial overhead, cap and pad all loops, and retain
random tapes and work registers until uncomputation
\cite{bennett1989time}.  We synthesize approximate phase rotations to an
explicit total operator-norm budget (Appendix~\ref{subsec:phase-synthesis}).

In coefficient coordinates of the root affine lattice, write
$K^\sharp=\{z:x_0+Bz\in K\}$.  Standard small-solution bounds
\cite[Theorem~17.1]{schrijver1998theory} compute
$M\leq2^{\operatorname{poly}(L)}$ such that feasibility is unchanged after
intersecting $K^\sharp$ with $[-M,M]^n$:
\begin{equation}
    K^\sharp\cap\mathbb Z^n\neq\varnothing
    \Longleftrightarrow
    K^\sharp\cap[-M,M]^n\cap\mathbb Z^n\neq\varnothing.
    \label{eq:root-compactification}
\end{equation}
We keep this box in every lifted recursive description.  Compression
replaces an affine subspace $H$ by $H'$ with
\begin{equation}
    H\cap\{-M,\ldots,M\}^n=H'\cap\{-M,\ldots,M\}^n.
    \label{eq:ipco-fiber-compression}
\end{equation}
The following lemma constructs $H'$ satisfying
\eqref{eq:ipco-fiber-compression} with $\dim H'=\dim H$.

\begin{lemma}[Rank-preserving bounded-box compression]
\label{lem:rank-preserving-frank-tardos}
Let $H\subseteq\mathbb R^k$ be a nonempty rational affine subspace and let
$M\geq1$ be an integer.  In time polynomial in the encoding of $H$ and
$\log M$, one can compute a rational affine subspace $H'$ such that
\begin{equation*}
    H\cap\{-M,\ldots,M\}^k
    =H'\cap\{-M,\ldots,M\}^k,
    \qquad
    \dim H'=\dim H,
\end{equation*}
and $H'$ has a basis of encoding length $\operatorname{poly}(k,\log M)$.
\end{lemma}

\begin{proof}
We compress the rows one at a time, then use the pivot pattern to check the
dimension.  Write $H=\{x:Fx=f\}$ using a rational reduced-row-echelon system with
$q=k-\dim H$ independent rows.  For row $i$, apply the sign-preserving
Frank--Tardos vector compression theorem \cite{frank1987application} to
\begin{equation*}
    w_i:=(F_{i,*},-f_i)\in\mathbb Q^{k+1}
\end{equation*}
with norm parameter $N:=(k+1)M+1$.  It returns an integer vector
\begin{equation*}
    \overline w_i=(\overline F_{i,*},-\overline f_i)
\end{equation*}
of $\operatorname{poly}(k,\log M)$ encoding such that
\begin{equation}
    \operatorname{sign}(w_i^\top z)
    =\operatorname{sign}(\overline w_i^\top z)
    \quad
    \text{for all }z\in\mathbb Z^{k+1}
    \text{ with }\lVert z\rVert_1\leq N-1.
    \label{eq:frank-tardos-row-signs}
\end{equation}
For $x\in\{-M,\ldots,M\}^k$, the vector $(x,1)$ has $\ell_1$-norm at most
$kM+1\leq N-1$, so it is admissible in
\eqref{eq:frank-tardos-row-signs}.  Hence
$F_{i,*}x=f_i$ if and only if
$\overline F_{i,*}x=\overline f_i$, simultaneously for every row, and
$H':=\{x:\overline Fx=\overline f\}$ has exactly the required bounded
integer points.

It remains to check the dimension, which the bounded-point identity alone
does not determine.  Let $p_i$ be the pivot column of row $i$ in the original
reduced-row-echelon system.  Apply
\eqref{eq:frank-tardos-row-signs} to the unit vectors $e_j$, including the
augmented coordinate.  Every zero coefficient remains zero and every nonzero
coefficient retains its sign.  In particular,
\begin{equation*}
    \overline F_{i,p_i}>0,
    \qquad
    \overline F_{h,p_i}=0\quad(h\neq i).
\end{equation*}
The compressed rows are therefore independent.  Since
$\overline F$ has full row rank $q$, its map onto $\mathbb R^q$ is
surjective, so $H'$ is nonempty and
$\dim H'=k-q=\dim H$.  Rational Gaussian elimination produces the asserted
basis within the same polynomial encoding bound.
\end{proof}

Since the root box remains in every node description, applying the lemma
after each hyperplane or fiber step preserves the node's feasible integer
points exactly, together with its dimension, and bounds the encoding of its
affine subspace independently of the incoming coefficients.  Rational lattice
normal form then gives $\operatorname{poly}(L)$-bit local coordinates and
lattice bases.  Projecting or taking the next fiber appends polynomially many
variables and equalities.  The auxiliary projection parameters, position
maps, and cover descriptions have $\operatorname{poly}(L,\log(1/\delta))$
bits, and each local call is charged for them.
Compression returns each child node to the original $\operatorname{poly}(L)$
encoding bound, so induction over the at most $n$ levels gives the invariant
for nodes and witnesses stated at the start of this appendix
\cite[Theorem~62 and the proof of Theorem~61]{reis2023subspace}.  The
reversible node computations include these deterministic arithmetic
operations.

\paragraph{Optimization and mixed-integer feasibility.}
\label{app:reductions}

We now state and prove the two standard reductions to the
feasibility algorithm of Theorem~\ref{thm:main} that give
Corollary~\ref{cor:worst-case-ip}.

\begin{corollary}[Mixed-integer linear feasibility]
\label{cor:mixed-integer-linear-feasibility}
Let $A,A',b$ be rational and $\delta\in(0,1/2)$.  Within the worst-case bounds of
Theorem~\ref{thm:main}, a quantum algorithm returns a
rational pair $(x,y)$ with $x\in\mathbb Z^n$, $y\in\mathbb R^{n_y}$, and
$Ax+A'y\leq b$ with probability at least $1-\delta$ if such a pair exists,
and returns no pair otherwise.  The exponential dependence is only on $n$;
$n_y$ enters through $L$.
\end{corollary}

\begin{proof}[Proof of Corollary~\ref{cor:mixed-integer-linear-feasibility}]
Project out the continuous variables while retaining the rational lift
\begin{equation*}
    D:=\{x\in\mathbb R^n:\exists y\in\mathbb R^{n_y},
    Ax+A'y\leq b\}.
\end{equation*}
Standard small-solution bounds for rational mixed-integer systems compute
$M\leq2^{\operatorname{poly}(L)}$ such that feasibility is unchanged after
intersecting $D$ with $[-M,M]^n$
\cite[Chapter~17]{schrijver1998theory}.  This compact rational polytope is presented
by the original inequalities together with the box, so membership,
optimization, and separation are rational linear programs of encoding length
$\operatorname{poly}(L)$.  Apply
Theorem~\ref{thm:main} to
$D\cap[-M,M]^n$ and $\mathbb Z^n$.  Given a returned integer $x$, rational
linear programming produces a rational $y$ satisfying $Ax+A'y\leq b$, and the
pair is checked exactly.  Thus continuous variables affect only polynomial
factors and one-sided witness soundness is preserved.
\end{proof}

\begin{corollary}[Integer linear optimization]
\label{cor:integer-linear-optimization}
For rational $A,b,c$ and $\delta\in(0,1/2)$, a quantum algorithm within the
worst-case gate and qubit bounds of Theorem~\ref{thm:main}
classifies~\eqref{eq:ilo} as infeasible, finite, or unbounded with
probability at least $1-\delta$.  In the finite case it returns an
optimizer.  In the unbounded case it returns integer vectors $x,r$ with
$Ax\leq b$, $Ar\leq0$, and $c^\top r>0$.  Every returned point and
unboundedness certificate is checked exactly.
\end{corollary}

\begin{proof}[Proof of Corollary~\ref{cor:integer-linear-optimization}]
Given a verified integer feasible point, one linear program detects
unboundedness; in the bounded case we binary-search the objective value.
Let $P=\{x:Ax\leq b\}$.  Scale the rows so that $A$ and $b$ are integral,
let $R\geq1$ be an integer upper bound on the Euclidean norms of the rows of
$[A\ b]$, and put $M:=(n+1)R^{n+1}$.  By Hadamard's inequality every
subdeterminant of $[A\ b]$ is at most $R^{n+1}$ in absolute value, so
\cite[Theorem~17.1]{schrijver1998theory} writes the integer hull of $P$ as
$\operatorname{conv}\{x_1,\ldots,x_t\}+\operatorname{cone}\{y_1,\ldots,y_s\}$
with integral $x_i,y_j\in[-M,M]^n$.  If $P\cap\mathbb Z^n$ is nonempty,
then $t\geq1$ and $x_1\in P\cap[-M,M]^n\cap\mathbb Z^n$.  If the integer
optimum is finite, then $c^\top y_j\leq0$ for every $j$, so the optimum over
the integer hull, which equals the integer optimum, is attained at some
$x_i$; thus some optimizer lies in $[-M,M]^n$.  Moreover
$\log M=\operatorname{poly}(L)$.  Clear the denominators of $c$ to
obtain an integral vector $\bar c$ with the same optimizers, and put
\begin{equation*}
    B_{\mathrm{obj}}:=M\lVert\bar c\rVert_1,
    \qquad
    J:=2+\left\lceil\log_2(2B_{\mathrm{obj}}+1)\right\rceil.
\end{equation*}
The numbers $M,B_{\mathrm{obj}},J$ have polynomial encoding, and
$J=\operatorname{poly}(L)$.

First run feasibility on $P\cap[-M,M]^n$.  Given a verified feasible point
$x$, we solve the rational linear system
\begin{equation*}
    Ar\leq0,\qquad c^\top r\geq1.
\end{equation*}
If it is feasible, clear denominators in $r$ and exactly verify the resulting
integer pair $(x,r)$; then $x+tr$ is feasible and improving for every integer
$t\geq0$.  If it is infeasible, $c$ is nonpositive on
$\operatorname{rec}(P)=\{r:Ar\leq0\}$, since any direction with positive
objective could be positively rescaled to satisfy the system, and the
objective is therefore bounded above.

In the bounded case,
$\bar c^\top x\in[-B_{\mathrm{obj}},B_{\mathrm{obj}}]\cap\mathbb Z$
for every $x\in P\cap[-M,M]^n\cap\mathbb Z^n$.
Binary-search this interval using feasibility of
\begin{equation*}
    P\cap[-M,M]^n\cap\{x:\bar c^\top x\geq\gamma\}
\end{equation*}
and retain the witness at the largest successful threshold $\gamma$.  Including the
initial call, at most $J$ feasibility calls are made.  Run each with error at
most $\delta/J$; a union bound makes every answer simultaneously correct with
probability at least $1-\delta$.  Since
$\log(J/\delta)=O(\log L+\log(1/\delta))$, these sequential calls preserve the
stated asymptotic gate and qubit bounds.
\end{proof}

\begin{proof}[Proof of Corollary~\ref{cor:worst-case-ip}]
Corollaries~\ref{cor:integer-linear-optimization}
and~\ref{cor:mixed-integer-linear-feasibility} prove, respectively,
the general-ILP and mixed-integer assertions within the worst-case
bounds of Theorem~\ref{thm:main}.
\end{proof}

The classical reference of Section~\ref{subsec:reference}, together with
the reduction in the proof of
Corollary~\ref{cor:integer-linear-optimization}, also proves the
classical bound stated in the introduction.

\begin{proof}[Proof of Corollary~\ref{cor:classical-ip}]
Algorithm~\ref{alg:classical-reference}
implements the recursion with the flatness routine of
Proposition~\ref{prop:finite-reversible-flatness}.  Generate its setup as
in the proof of Theorem~\ref{thm:main}.  The setup is good with
probability at least $1-\delta/8$, and on this event
Lemma~\ref{lem:classical-reference} shows that the algorithm decides
integer feasibility correctly.  The same lemma bounds its running time
by~\eqref{eq:classical-bound} on every tape.  The recurrence~\eqref{eq:classical-recurrence} for $T(k)$ in
Section~\ref{subsec:classical-complexity} is written after suppressing one
common $\operatorname{poly}(L,\log(1/\delta))$ factor, with the maximum
over $1\leq d\leq k$ and $T(0)=1$.  This every-tape bound uses
the explicit $D_{\max}$ stop of Algorithm~\ref{alg:classical-reference},
under which all node processings and address decodings have capped costs
independently of geometric validity.  Generating the primes and the seed
takes $\operatorname{poly}(L,\log(1/\delta))$ time.  For optimization, we
run the reduction in the proof of
Corollary~\ref{cor:integer-linear-optimization} with this algorithm in
place of the quantum one, with error $\delta/J$ per call.
\end{proof}

\section{Bounded equation form}
\label{app:der-cell-test}

This appendix proves Lemma~\ref{lem:der-cell-test}, the capped cell test
that the instance-wise algorithm for bounded equation form in
Section~\ref{subsec:bounded-equality} runs on each reflection cell, and
then proves Theorem~\ref{thm:der-instance-wise}.
Each DER cell, doubled about any of its points, stays inside the slightly
enlarged box of~\eqref{eq:der-cell-reflection}, and approximate integer
programming needs exactly this property.  If a cell's slice of
$\{x:Ax=b\}$ contains an integer point, preprocessing produces a body $P$
inside that slice that keeps all of its integer points.  In positive
dimension, on a good cover seed and a successful barycenter estimate
$\widetilde c$ of $P$, a $2$-approximate closest-vector computation with
unit ball $P-\widetilde c$ returns an integer point in
$\widetilde c+2(P-\widetilde c)$.
This set lies in the cell doubled about $\widetilde c\in P$, so
integrality places the returned point in the original box.
The zero-dimensional case is handled directly.  The cell test is the
classical rounding step of \cite[Section~6]{dadush2013sparse}, with a
finite barycenter estimate and the deterministic near-symmetric
closest-vector algorithm of that paper.  We record a finite
implementation so that the comparison in Section~\ref{sec:der}
accounts for the cost of the test.

We first list the cells.  For each coordinate $i$, sort the distinct
breakpoints
$0,u_i,2^{j-1},u_i-2^{j-1}$ for
$0\leq j\leq\lceil\log_2u_i\rceil$.  The number $m_i$ of consecutive
closed intervals they determine satisfies
$m_i\leq2\lceil\log_2u_i\rceil+3\leq7\log_2(u_i+1)$.
The cells $\mathcal R_0,\ldots,\mathcal R_{N-1}$ are the products of
these intervals in lexicographic order, where
\begin{equation}
    N:=\prod_i m_i\leq7^n\Psi(u),
    \qquad \log N=\operatorname{poly}(L).
    \label{eq:der-cell-count}
\end{equation}
Mixed-radix arithmetic computes $\mathcal R_j$ from $j$ in polynomial
time.  By \cite[Lemma~21]{dadush2025approximate}, every cell has the
reflection property
\begin{equation}
    v+2(\mathcal R_j-v)
    \subseteq\prod_{i=1}^n[-1/2,u_i+1/2]
    \quad\text{for every }v\in\mathcal R_j.
    \label{eq:der-cell-reflection}
\end{equation}

\begin{lemma}[Finite DER cell test]
\label{lem:der-cell-test}
Let $I=(A,b,u)$ be a bounded equation-form instance of encoding length $L$,
and let $\beta,\tau\in(0,1/2)$.  There is a computable encoding bound
$L'=\operatorname{poly}(L,\log(1/\beta),\log(1/\tau))$ for which the
following holds.  Draw a cover seed $\eta$ from
Lemma~\ref{lem:rational-indexed-cover} with failure budget $\tau$.  For
each reflection cell $\mathcal R$, a capped test
$V_\eta(\mathcal R;\omega_{\rm est})$ uses a uniform
random tape $\omega_{\rm est}$ of polynomial length and returns either no
point or an exactly verified point of $\mathcal P_I\cap\mathbb Z^n$.
Conditional on a good $\eta$, if $\{x:Ax=b\}\cap\mathcal R$ contains an
integer point, the test returns a point with probability at least
$1-\beta$ over $\omega_{\rm est}$.  Its running
time and the gate and qubit costs of its reversible implementation are
\begin{equation*}
    2^{O(n)}\operatorname{poly}
    \bigl(L,\log(1/\beta),\log(1/\tau)\bigr).
\end{equation*}
With probability at least $1-\tau$ over $\eta$, the seed is good
simultaneously for all cells and all estimator tapes.
\end{lemma}

\begin{proof}
Fix a cell $\mathcal R$ and put $P_{\mathcal R}:=\{x:Ax=b\}\cap\mathcal R$.

\proofstep{The test.}
Given $\mathcal R$, $\eta$, and $\omega_{\rm est}$, the test is the
following deterministic computation.
\begin{enumerate}
    \item Reject if rational linear programming shows
    $P_{\mathcal R}=\varnothing$, or if lattice normal form shows that its
    affine hull has no integer point.  Otherwise pass to affine-hull
    coordinates and apply the ellipsoid-or-hyperplane preprocessing of
    Section~\ref{subsec:classical-recursion}.  After each restriction, compress the affine
    subspace as in Appendix~\ref{app:recursion-details} and repeat the two
    tests.  For these compressions take the root lattice $\mathbb Z^n$ in
    the original coordinates and $M:=\max\{1,u_1,\ldots,u_n\}$ in
    \eqref{eq:ipco-fiber-compression}; since
    $P_{\mathcal R}\subseteq[0,u]\subseteq[-M,M]^n$, compression then
    preserves every integer point of $P_{\mathcal R}$, not only
    feasibility.  Retain the original equations and cell inequalities.  Stop
    either at a rounded body $P\subseteq P_{\mathcal R}$ containing every
    integer point of $P_{\mathcal R}$, or at a point, which goes directly
    to step~4.  In positive dimension $k\leq n$, use the rounding
    ellipsoid and coordinate linear programs to compute rational
    $r_0,R_0>0$ such that $P$ contains a Euclidean ball of radius $r_0$
    and $P\subseteq R_0B_2^k$.
    \item Apply Lemma~\ref{lem:finite-bit-geometric-estimators} to $P$,
    reading $\omega_{\rm est}$, with accuracy $1/(2(k+1))$ and failure budget
    $\beta$.  Write its barycenter estimate as $\widetilde c$ and put
    $r_1:=r_0/(2(k+1))$.  Reject unless $\widetilde c\in P$ and all $2k$
    exact membership tests $\widetilde c\pm r_1e_i\in P$ pass.
    \item Write the local affine lattice as $a+\Lambda$.
    Run the fixed-seed $2$-approximate CVP implementation described
    below, with unit ball $P-\widetilde c$, lattice $\Lambda$, and target
    $\widetilde c-a$.  Supply inner and outer radii
    $r_2:=r_1/k$ and $R_1:=4R_0$, and cap its work at the bound for
    $1/4$-symmetric gauges.  If its output is $v\in\Lambda$, put $y:=a+v$.
    \item Map the candidate back to the original coordinates and accept
    only if it is integral and satisfies $Ay=b$ and $0\leq y\leq u$.
\end{enumerate}
Every timeout, register overflow, or undefined arithmetic operation
rejects, and all loop bounds and register sizes are fixed in advance.  In step~1 the body and its coordinates are deterministic
functions of $\mathcal R$ and do not depend on $\eta$ or
$\omega_{\rm est}$.  These restrictions preserve exactly the feasible
integer points, and the retained box bounds and compression keep every
encoding polynomial.

\proofstep{The closest-vector call.}
The guard in step~2 makes the gauge well defined on every continuing tape.  Convexity gives $\widetilde c+r_1B_1^k\subseteq P$, so
\begin{equation*}
    r_2B_2^k\subseteq P-\widetilde c\subseteq2R_0B_2^k.
\end{equation*}
Exact rational linear programming supplies membership and gauge
evaluation, and hence the weak membership and gauge-distance oracles
required by \cite[Theorem~3.3]{dadush2013sparse}.  With approximation
parameter $\varepsilon=1$ and symmetry parameter $\gamma=1/4$, that
theorem gives a deterministic $2$-approximate CVP algorithm in
time $2^{O(k)}$ times a polynomial in the encodings of the body, lattice,
target, and oracle parameters.  To initialize its scalar distance search,
compute $d_2:=\operatorname{dist}_2(\widetilde c-a,\Lambda)$ by exact
ellipsoidal CVP.  If $d_2=0$, send $y=\widetilde c$ to step~4.
Otherwise rational bisection on the exact squared distance gives
$d_-\in[d_2/2,d_2]$.  With $l:=d_-/(2R_0)$ and
\begin{equation*}
    d_P:=\min_{v\in\Lambda}
          \|v-(\widetilde c-a)\|_{P-\widetilde c},
\end{equation*}
the radius bounds imply
\begin{equation*}
    l\leq\frac{d_2}{2R_0}\leq d_P
    \leq\frac{d_2}{r_2}\leq\frac{2d_-}{r_2}=l\frac{R_1}{r_2}.
\end{equation*}
These bounds initialize the search as in the proof of that theorem.

\proofstep{Enumeration through the cover.}
Put $K:=P-\widetilde c$.  We implement the lattice-enumeration calls,
including those inside the shortest-vector and sparsifier routines, with
Lemma~\ref{lem:rational-indexed-cover} and exact ellipsoidal enumeration.
For a requested body $D$ and lattice $\Gamma$, enumerate each valid
$s_j+\widehat E$, retain only points belonging to $D$, and remove
duplicates.  Conditional on a good cover seed,
\begin{equation*}
    \sum_{j\ {\rm valid}} |(s_j+\widehat E)\cap\Gamma|
    \leq C_{\rm cov}^k C_{\rm pt}^k
          \nu_{\max}(\Gamma,D),
\end{equation*}
so enumeration takes $2^{O(k)}(\nu_{\max}(\Gamma,D)+1)$ times a
polynomial in the relevant encodings and $\log(1/\tau)$.
The packing bounds in the proofs of
\cite[Theorems~3.2 and~3.3]{dadush2013sparse} give the required
single-exponential bounds for these calls when
$\varepsilon=1$ and $\gamma=1/4$.
The requested bodies are rational translates and positive scalings of
$K$ or $K\cap(-K)$.

\proofstep{Encoding bounds.}
The caps rest on a seed-independent encoding bound.
Choose a computable $h=\operatorname{poly}(L,\log(1/\beta))$ large enough
to bound the encodings of the preprocessed body, estimator output,
radii, affine shift $a$, and the original local lattice basis
$B$, and increase $h$ if necessary so that
\begin{equation*}
    r_2\geq2^{-h},\qquad R_1\leq2^h,\qquad
    \|B\|_2,\ \|B^{-1}\|_2,\ \|\widetilde c-a\|_2\leq2^h.
\end{equation*}
These bounds hold uniformly over all cells and all continuing estimator
tapes.  The outer distance search has
$O(1+\log(R_1/r_2))$ stages, with scalar scales bounded by $2^{O(h)}$.
For the original local lattice,
\begin{equation*}
    \lambda_1(K\cap(-K),\Lambda)
       \geq\frac{1}{R_1\|B^{-1}\|_2}\geq2^{-2h}.
\end{equation*}
Each sparsifier call uses this original lattice and a parameter
$t\leq2^{O(h)}$.  Its number of stages therefore satisfies
\begin{equation*}
    s=O\!\left(1+\log\!\left(
       1+\frac{t}{\lambda_1(K\cap(-K),\Lambda)}\right)\right)
      =O(h)
\end{equation*}
by \cite[Section~5, proof of Theorem~3.2]{dadush2013sparse}.
Each stage has sublattice index at most $2^{O(k)}$, so the cumulative
index is at most $2^{O(ks)}$.  We represent every intermediate sublattice
as $BH_s\mathbb Z^k$, where $H_s$ is an integer Hermite normal form matrix
relative to the original basis $B$.  Its entries are bounded by the
cumulative index, so these bases have polynomial encoding.

Exact membership filtering bounds the coordinates of retained
enumeration outputs by the radii of their requested bodies.
These outputs lie in sublattices of the original rational lattice, so
a common denominator of $B$ bounds their coordinate denominators.
The rational linear-programming bit bounds then give polynomial
encodings for the gauge values and scalar scales computed from them.
Hence one computable polynomial bound $L'$ covers the rational lift of
every requested body.  We choose register and loop caps above these
bounds, and reject a nonpositive scale or an overflow before invoking a
cover call.  The cover matrices may affect enumeration choices, but they
do not enter the defining lifts of later requested bodies.  The cover
construction and ellipsoidal enumeration use separate work registers,
with the bounds supplied by their implementations.

Fix this $L'$ before drawing $\eta$.
Lemma~\ref{lem:rational-indexed-cover} then applies simultaneously to
every requested body, for every cell and estimator tape, so its failure
budget is paid once.  On a good cover seed and an accurate estimator
tape, the packing, stage, and encoding bounds above ensure that the caps
do not reject.  On inaccurate estimator tapes the symmetry guarantee can
fail, but the guards and caps still keep the test well defined and within
the stated resources.

\proofstep{Completeness.}
Assume $P_{\mathcal R}$ has an integer point, the cover seed is
good, and the estimator succeeds.  Let $b_P$ be the barycenter of $P$
and put $w:=\widetilde c-b_P$.  The centroid inequality
$-(P-b_P)\subseteq k(P-b_P)$
\cite[Section~5]{reis2023subspace} gives
\begin{equation*}
    \frac{P-P}{k+1}
    \subseteq S:=(P-b_P)\cap(b_P-P).
\end{equation*}
The estimator's error bound puts $w$ in $S/2$, so by convexity
\begin{equation*}
    \frac12S\subseteq
    (P-\widetilde c)\cap(\widetilde c-P).
\end{equation*}
The Milman--Pajor inequality
$\operatorname{Vol}_k(S)\geq2^{-k}\operatorname{Vol}_k(P)$
\cite{milman2000entropy}, also stated in
\cite[Theorem~5]{dadush2025approximate}, implies
\begin{equation*}
    \operatorname{Vol}_k\bigl((P-\widetilde c)
                      \cap(\widetilde c-P)\bigr)
    \geq4^{-k}\operatorname{Vol}_k(P).
\end{equation*}
Thus the gauge is $1/4$-symmetric.  Moreover,
$P-P\supseteq2r_0B_2^k$ implies
$\widetilde c+(r_0/(k+1))B_2^k\subseteq P$, so the membership guards
pass.  Since $P$ contains an integer point, its lattice distance from
$\widetilde c$ in this gauge is at most one, and the approximate CVP
output therefore lies in $\widetilde c+2(P-\widetilde c)$.  It satisfies
the original affine equations and, since $P\subseteq\mathcal R$ and
$\widetilde c\in\mathcal R$, lies in
$\widetilde c+2(\mathcal R-\widetilde c)$, so by
\eqref{eq:der-cell-reflection} and integrality it satisfies
$0\leq y\leq u$ and step~4 accepts.  The zero-dimensional case accepts
directly.  Because the estimator succeeds with probability at least
$1-\beta$, this proves the conditional acceptance guarantee.

\proofstep{Resources and soundness.}
Once the cover seed is fixed, the only random part of the test is the
finite barycenter estimator, whose random tape has polynomial length by
Lemma~\ref{lem:finite-bit-geometric-estimators}.  The deterministic CVP
implementation, exact arithmetic, and caps use the stated
single-exponential resources.  Retaining the computation history gives the
reversible test and its inverse by Bennett simulation
\cite{bennett1989time}.  The final exact verification makes the test
sound independently of all geometric and probabilistic guarantees.
\end{proof}

\paragraph{The instance-wise algorithm.}
\label{app:der-instance-wise}

The classical DER reference tests the cells of Section~\ref{subsec:bounded-equality} in order and stops at the
first verified point of $\mathcal P_I\cap\mathbb Z^n$, or reports
infeasibility after the last cell.  It uses the capped cell test of
Lemma~\ref{lem:der-cell-test} with failure parameter
$\beta:=\delta/(8N)$.  The shared setup $\omega$ consists of one cover
seed with failure budget $\tau:=\delta/8=N\beta$ and one independent
uniform random tape for the barycenter estimator, reused in every cell;
both have length polynomial in $L$ and $\log(1/\delta)$.
Since $\log(1/\tau)\leq\log(1/\beta)$, each cell test costs
$2^{O(n)}\operatorname{poly}(L,\log(1/\beta))$.  Let
$T_{\mathrm{DER}}(I,\omega)$ be the reference's bit running time,
including setup.  In this classical execution, cheap rejections need not
be padded.

\begin{proof}[Proof of Theorem~\ref{thm:der-instance-wise}]
We bound the setup failure, compare depth-first search on an interval
tree over the cells with the reference, and apply
Theorem~\ref{thm:tree-search}.

Once $\omega$ is fixed, Lemma~\ref{lem:der-cell-test} gives a
deterministic test.  Its coherent implementation costs
$A_{\mathrm{cell}}=2^{O(n)}\operatorname{poly}(L,\log(1/\delta))$ gates and
qubits, since $\log(1/\beta)=O(\log N+\log(1/\delta))$.  Except with
probability $\delta/8$, the cover seed is good for all bodies arising in
all cells and on all estimator tapes.  Conditional on such a seed, each
fixed cell containing an integer point accepts except with probability
$\beta$.  Reusing the estimator tape correlates these events but leaves
their marginal bounds unchanged.  A union bound over the $N$ cells shows
that, except with total probability $\delta/8+N\beta=\delta/4$, every
feasible cell accepts and hence the reference is correct.  No exponential
table of independent per-cell tapes is generated or loaded.

For a fixed $\omega$, take the interval tree $\mathcal I(N)$ of
Section~\ref{subsec:binary-address-tree} and mark leaf $j$ exactly when
its cell test accepts.  The tree has $2N-1$ vertices, depth
$\Delta:=\lceil\log_2N\rceil$, and leaves in the reference's order.
Each vertex is encoded by its binary path from the root.  Interval
splitting and mixed-radix decoding give the local child and depth
circuits in polynomial time, and a leaf label uses the reversible cell
test.  Up to polynomial factors, all local circuits therefore cost at
most $A_{\mathrm{cell}}$ on every tape.  The case $N=1$ is a single cell
test and satisfies the same bounds directly.

Let $N'$ be the number of cells tested by the classical reference.  Every
vertex that depth-first search visits before the first marked leaf lies
on the path to one of these $N'$ leaves, and testing each cell takes at
least one classical bit operation.  Consequently,
\begin{equation}
    T_{\mathrm{dfs}}(\mathcal I(N))
    \leq(\Delta+1)N'
    \leq(\Delta+1)T_{\mathrm{DER}}(I,\omega).
    \label{eq:der-dfs-comparison}
\end{equation}
Apply Theorem~\ref{thm:tree-search} with $T_1:=2N-1$ and
$\epsilon:=\delta/4$.  Its instance-wise query bound is
\begin{equation*}
    c_{\mathrm{tree}}\Delta^{5/2}
    \sqrt{T_{\mathrm{dfs}}(\mathcal I(N))}\,
    \log^2\frac{16\Delta^2\log(2T_1)}{\delta}.
\end{equation*}
Together with~\eqref{eq:der-dfs-comparison}, this gives
\eqref{eq:der-instance-wise}.  Replacing the square root by
$\sqrt{T_1}$ gives the every-tape cap, and \eqref{eq:der-cell-count}
turns it into the stated worst-case bound.  The extra registers have
polynomial size, and the cell-test workspace is reused.  As in
Appendix~\ref{subsec:phase-synthesis}, we synthesize the rotations to
total operator-norm error at most $\delta/8$ across the adaptive run,
which changes its transcript distribution by at most $\delta/4$.
The setup, tree-search, and synthesis failure probabilities sum to at most
$\delta/4+\delta/4+\delta/4<\delta$.  Recompute the returned cell's point
and verify it against the original input; this also proves soundness on
every tape.
\end{proof}

\section{Priority-driven cover enumeration}
\label{subsec:priority-cover}

This appendix proves Corollary~\ref{cor:cover-enumeration}, which extends
the comparison of Theorem~\ref{thm:main} in Section~\ref{sec:quantum-ip}
from the depth-first reference to classical implementations that
enumerate whole covers and select nodes by a monotone priority.
The depth-first reference decodes one address at a time.  Cover
enumeration instead lists all ellipsoid points before
recursion~\cite[Theorem~58]{reis2023subspace}, and an implementation may
then select pending nodes by a priority.  The comparison of
Theorem~\ref{thm:main} extends to any monotone priority that depends only
on a node's own path.  Write $\ell_H$ for a node's canonical-address path,
ordered lexicographically with proper prefixes first.  An admissible
priority $h(H,\ell_H)$ is a nonnegative polynomial-bit integer, computable
from that path and the shared setup in capped
$2^{O(n)}\operatorname{poly}(L,\log(1/\delta))$ time, and nondecreasing
along paths; it cannot depend on the search frontier.  Depth and running
maxima of path-local scores are admissible.
The proof of Corollary~\ref{cor:cover-enumeration} ranks the binary
address tree by integers that encode the priority, checks that least-rank
search follows the comparator, and runs a branch-local quantum search
interleaved with the solver of Theorem~\ref{thm:main}.

\begin{corollary}[Monotone priorities]
\label{cor:cover-enumeration}
Fix an admissible priority $h$.  Run the recursion of
Section~\ref{subsec:classical-recursion}, selecting pending nodes by $(h(H,\ell_H),\ell_H)$.
At each selected internal node, enumerate every valid cover ellipsoid,
filter and deduplicate its points, and insert all unshrunken fiber children
before selecting another node.  Return a point only when a marked leaf is
selected.  Let $T_{\rm cov,h}(I,\omega)$ be its bit running time.
With probability at least $1-\delta$, a quantum algorithm sharing this
setup is correct and uses at most
\begin{equation}
 2^{O(n)}\operatorname{poly}(L,\log(1/\delta))
 \sqrt{T_{\rm cov,h}(I,\omega)}
 \label{eq:cover-enumeration-comparison}
\end{equation}
gates.  It retains the every-tape bounds of
Theorem~\ref{thm:main} and returns only verified points.  Taking $h=0$
recovers the canonical-order comparison.
\end{corollary}

\begin{proof}[Proof of Corollary~\ref{cor:cover-enumeration}]
Dimension zero is checked directly.  Otherwise, use the same setup and
binary address tree $\mathsf{T}_\omega$ as in the
proof of Theorem~\ref{thm:main}.  We first fix a good setup.  Node occurrences
are distinguished by their canonical-address paths, even when their
compressed affine subspaces agree.  The comparator selects a pending node
by $(h(H,\ell_H),\ell_H)$, generates every child of a selected internal
node, and tests for a solution only on selecting a leaf.  Thus a point
encountered during child generation is not returned early.

Next we give $\mathsf{T}_\omega$ distinct positive integer priorities
without enumerating its vertices.  The deterministic address
cap~\eqref{eq:address-cap} supplies a known
integer $A_\star\geq\max_H|\mathcal A_H|$; for example, take
\begin{equation*}
    A_\star:=1+2C_{\mathrm{cov}}^n2^{g_{\mathrm{top}}+3},
    \qquad B_\star:=2A_\star+1,\qquad J_\star:=2A_\star.
\end{equation*}
Write a path as $\ell=(\iota_1,\ldots,\iota_t)$, where $\iota_s$ is the
zero-based raw index of its $s$th address, and put
\begin{equation*}
    \operatorname{code}(\ell)
       :=\sum_{s=1}^t(\iota_s+1)B_\star^{n-s}\in\{0,\ldots,B_\star^n-1\}.
\end{equation*}
Zero padding preserves lexicographic order and places a proper prefix
before its extensions.  For an auxiliary interval vertex or dead singleton
at $H$, let $\operatorname{pre}(v)$ be its zero-based preorder index in
$\mathcal I(|\mathcal A_H|)$; at a recursion leaf put
$\operatorname{pre}(v)=0$.  Every such index is less than $J_\star$.  If an
interval $[a,b)$ splits at $c$, its two child indices are its own index
plus $1$ and plus $2(c-a)$, respectively.  These indices are therefore
locally computable.  A canonical singleton is identified with its child's
root and uses that child's path and priority, as in
Definition~\ref{def:binary-address-tree}.

For a vertex associated with $H$ and path $\ell$, set
\begin{equation}
    \mathfrak r(v):=1+J_\star\bigl(B_\star^nh(H,\ell)
                       +\operatorname{code}(\ell)\bigr)
                       +\operatorname{pre}(v).
    \label{eq:priority-integer-ranking}
\end{equation}
The ranks are distinct.  They increase along every edge: preorder indices
increase within one interval tree, while a canonical child has
nondecreasing $h$ and a strictly larger path code.  If $h$ has at most
$b_h$ bits, the ranks have
$O(b_h+n\log B_\star+\log J_\star)=\operatorname{poly}(L,\log(1/\delta))$
bits.
Lemma~\ref{lem:address-tree-access} and the assumed capped circuit for $h$
compute them with the same single-exponential local gate and space bounds.
The quantum algorithm stores no frontier or priority queue.

Consider classical least-rank search on $\mathsf{T}_\omega$.  When it
selects the root associated with an internal node $H$, all of $H$'s
auxiliary vertices precede every previously pending recursion root and
every newly generated child root.  The first claim follows from
minimality of $(h(H,\ell_H),\ell_H)$, and the second from monotonicity
of $h$ and the prefix convention.  Consequently the auxiliary vertices
at $H$ are exhausted before another recursion root is selected.  Induction
shows that the selected recursion roots occur in exactly the comparator's
order, through its first marked leaf or until exhaustion.

Let $\mathscr V_h$ be these selected recursion nodes and let $Q_h$ be the
number of selected binary vertices.  Each selected internal node
contributes at most $2|\mathcal A_H|-1$ auxiliary vertices.  Thus
\begin{align}
    Q_h
    &\leq2\sum_{H\in\mathscr V_h\text{ internal}}|\mathcal A_H|
           +\#\{H\in\mathscr V_h\text{ leaves}\}\notag\\
    &\leq2C_{\mathrm{addr}}^n
           \sum_{H\in\mathscr V_h}\chi(H)
     \leq2C_{\mathrm{addr}}^nT_{\mathrm{cov},h}(I,\omega).
    \label{eq:priority-cover-charge}
\end{align}
The last inequality charges one bit operation per selected node and per
point emitted from its cover ellipsoids, regardless of how their occupancy
compares with the number of genuine branches.

On a feasible instance the ranked binary tree has a marked leaf.  Apply
the branch-local heuristic-search result of the arXiv preprint
\cite[Definition~1.2 and Theorem~1.5]{chakrabarti2022universal}.
The ranking in~\eqref{eq:priority-integer-ranking} is branch-local: it is
a function of the vertex encoding and fixed setup, with no dependence on
the explored frontier.  Its evaluation, including replay of node data,
is charged as a local computation.  The imported theorem finds a marked
vertex using $\sqrt{Q_h}$ queries times a polynomial in the depth, the
rank bit bound, $\log T_{\max}$, and $\log(1/\delta)$, with error at most
$\delta/4$.  Constant-degree child, mark, and ranking queries each cost
$2^{O(n)}\operatorname{poly}(L,\log(1/\delta))$ gates and qubits.
Its classical control consists of polynomially many threshold and path
registers; its walk and size-estimation subroutines have the same finite
register implementation as in Appendix~\ref{subsec:tree-search-proof}.  Hence
the remaining work adds only polynomial factors, not a materialized list
of vertices.

Cap this priority routine before any subroutine that would exceed a fixed
\begin{equation*}
    \sqrt{T_{\max}}\operatorname{poly}
       (\Delta_{\max},b_h+n\log B_\star+\log J_\star,\log T_{\max},
        \log(1/\delta))
\end{equation*}
query budget, with effective constants from the imported construction.
On a good setup this does not truncate its success event, since
$Q_h\leq|\mathsf{T}_\omega|\leq T_{\max}$.  Check every measured candidate
exactly.  On a cap, failure, or unverified candidate, the routine supplies
no answer.  Synthesize its rotations to total operator-norm error
$\delta/8$ using this every-tape cap, so synthesis changes the transcript
distribution by at most $\delta/4$ and adds only polynomial gate factors.

The imported theorem assumes that a marked node exists.  To obtain a
decision algorithm, interleave this marked-only routine with the DFS
quantum solver of Theorem~\ref{thm:main}, alternating one elementary
operation of each.  Return the first verified point, or report
infeasibility when the DFS solver does so.  A failed or capped priority
routine is suspended and never decides infeasibility.  The joint good-setup and DFS event fails with
probability at most $\delta/4$, as shown in that theorem's proof.  On a
feasible instance, the additional ideal priority-search and synthesis
errors total at most $\delta/2$.  Combining these events and
\eqref{eq:priority-cover-charge} proves the stated bound with probability
at least $1-\delta$.

Finally consider an infeasible instance.  Every node is then selected by the classical
comparator.  Its occupancy charge gives
\begin{equation*}
    |\mathsf{T}_\omega|
       \leq2C_{\mathrm{addr}}^nT_{\mathrm{cov},h}(I,\omega).
\end{equation*}
There are no marks, so the DFS solver has
$T_{\mathrm{dfs}}(\mathsf{T}_\omega)=|\mathsf{T}_\omega|$ and supplies
the required square-root bound.  Exact verification prevents a false
point on every tape.  The interleaving stops no later than twice the DFS
solver's every-tape cap, and the two workspaces together retain its
asymptotic qubit bound.  Setup and final verification are absorbed since
$T_{\mathrm{cov},h}\geq1$.  For $h=0$, prefix-lexicographic selection
is canonical depth-first order, recovering the former comparator.
\end{proof}

Both comparisons concern the specified implementations with the same
geometric choices and canonical branch order; they do not cover arbitrary
IP solvers or sibling priorities.  They charge cover occupancy as well
as the number of branches, and the reference depends on the input
length and confidence parameter through the shared-setup construction.

\section{Finite and reversible implementation of the quantum recursion}
\label{app:finite-quantum-recursion}

\subsection{Prime ladder, certified hashes, and geometric indexing}
\label{app:hash-proofs}

\begin{proof}[Proof of Lemma~\ref{lem:ellipsoid-difference-enlargement}]
We prove the inclusion with the parallelogram identity and then bound
each parity slice through a midpoint minimizer.
Take $z,w\in Z$ and let
$e\equiv z-w\pmod2$.  Their midpoint lies in $e/2+\mathbb Z^d$, and the
parallelogram identity gives
\begin{equation*}
    q^\star_e+\frac14(z-w)^\top Q_B(z-w)
    \leq q\!\left(\frac{z+w}{2}\right)
       +\frac14(z-w)^\top Q_B(z-w)
    =\frac{q(z)+q(w)}2\leq1.
\end{equation*}
Hence $Z-Z\subseteq S$.

For the count, fix one minimizer $m_e$ attaining $q^\star_e$.
For $v\in S_e$, both $m_e+v/2$ and $m_e-v/2$ are integer vectors, and
the average of their $q$-values is $q^\star_e+v^\top Q_Bv/4\leq1$, so at
least one of them belongs to $Z$.  Thus
\begin{equation*}
    S_e\subseteq2(Z-m_e)\cup-2(Z-m_e),\qquad |S_e|\leq2\nu.
\end{equation*}
For $e=0$, nonemptiness gives $m_0\in Z$, so the two sets share zero
and $|S_0|\leq2\nu-1$.  The parity slices are disjoint, which gives
\eqref{eq:ellipsoid-difference-bound}.  Exact CVP in
$e/2+\mathbb Z^d$ under $Q_B$ computes each $m_e$ and $q^\star_e$ with
the claimed resources, and every distance and threshold is rational.
\end{proof}

The hashes draw their moduli from primes
\begin{equation}
    2^{g+2}<p_g<2^{g+3},
    \qquad g=0,1,\ldots,g_{\mathrm{top}},
    \label{eq:prime-ladder}
\end{equation}
which exist by Bertrand's postulate.  Appendix~\ref{app:finite-quantum-recursion} fixes
their common cap for the IP recursion.  For general input bodies, an
all-point coordinate bound fixes a polynomial-bit count bound and hence a
finite ladder.  The prime number theorem, in the explicit form of
\cite{rosser1962approximate}, implies that, for an effective
universal threshold $x_*$,
\begin{equation*}
    \pi(2x)-\pi(x)\geq\frac{x}{3\log x}\qquad(x\geq x_*),
\end{equation*}
where $\pi(x)$ counts the primes at most $x$.
Hence a uniformly sampled integer in the level-$g$ interval is prime
with probability $\Omega(1/(g+1))$; a fixed table handles the finitely
many smaller levels.  For any allocated failure budget
$\varepsilon\in(0,1/2)$, use
$O((g+1)\log((g_{\mathrm{top}}+1)/\varepsilon))$ independent candidates
at level $g$, with deterministic polynomial-time primality testing
\cite{agrawal2004primes}.  A union bound shows that the entire ladder is
valid except with probability $\varepsilon$, using
$\operatorname{poly}(g_{\mathrm{top}}+1,\log(1/\varepsilon))$ time and
random bits and $O(g_{\mathrm{top}}+1)$-bit candidates.
A missing entry makes setup fail.

\begin{proof}[Proof of Proposition~\ref{prop:certified-global-hash}]
We give the exact test, show that a random normal passes once $p$ is
large, and then search over primes.

For a proposed pair let
$\mathcal K_a:=\{z\in\mathbb Z^d:a^\top z\equiv0\pmod p\}$, with the explicit
basis of Lemma~\ref{lem:bucket-basis}.  For $e\neq0$, choose an integer
$z_e$ satisfying
\begin{equation*}
    a^\top z_e\equiv-2^{-1}a^\top e\pmod p,
\end{equation*}
which is possible since $a\neq0$ and $p$ is odd, and compute
\begin{equation*}
    \gamma_e:=\min_{w\in e/2+z_e+\mathcal K_a}w^\top Q_Bw,
    \qquad
    \gamma_0:=\min_{w\in\mathcal K_a\setminus\{0\}}w^\top Q_Bw.
\end{equation*}
For nonzero parity, $\gamma_e$ is a CVP to the origin in the displayed
affine lattice, a coset that cannot contain zero.  For zero parity it is
an SVP, not a CVP at zero.  Since
$\mathcal K_a\cap2\mathbb Z^d=2\mathcal K_a$, these quantities give the
exact test
\begin{equation}
    \mathcal K_a\cap(S\setminus\{0\})=\varnothing
    \quad\Longleftrightarrow\quad
    q^\star_e+\gamma_e>1\quad\text{for every }e.
    \label{eq:certified-hash-test}
\end{equation}
Equality rejects because the ellipsoid is closed.  By $Z-Z\subseteq S$,
passing proves global injectivity.  A rejection need not exhibit an actual
collision, since $S$ can contain artificial differences.  Exact rational
ellipsoidal SVP and CVP have the stated single-exponential costs
\cite{micciancio2010deterministic}, \cite[Theorem~5.2.2]{dadush2012integer}.

Suppose now that $p>2^{d+2}\nu$.  Then no nonzero element of $S$ vanishes
modulo $p$: if $pw\in S_e$ with $w\neq0$, the $p+1$ vectors
$\pm w,\pm3w,\ldots,\pm pw$ all lie in $S_e$, because odd multiples
preserve parity and their squared norms do not exceed that of $pw$, and
this contradicts $|S_e|\leq2\nu$.  For a uniform nonzero $a$, each nonzero
$v\in S$ is therefore annihilated with probability less than $1/p$.  The
vectors $v$ and $-v$ give the same event, and there are at most
$2^d\nu-1$ nonzero antipodal pairs.  Hence
\begin{equation*}
    \Pr\bigl[\mathcal K_a\cap(S\setminus\{0\})\neq\varnothing\bigr]
    \leq\frac{2^d\nu-1}{p}<\frac14,
\end{equation*}
and one random normal passes the exact test with probability greater
than $3/4$ at this modulus scale.

When $\nu$ is unknown, an all-point rational coordinate box gives a count
bound $U$ with $\log U=\operatorname{poly}(L)$.  Use the prime ladder
\eqref{eq:prime-ladder} of Appendix~\ref{app:finite-quantum-recursion} through $O(d+\log U)$ levels.  At each level, try
$O(\log(1/\tau))$ independent normals and stop at the first certificate
that passes; if every capped trial fails, return failure.  The first level
whose lower prime bound exceeds $2^{d+2}\nu$ has $p\leq C2^d\nu$.
Conditional on reaching that level, all its trials reject with probability
at most their allocated $\tau/2$.  Capped prime and finite uniform-normal
generation have total failure at most $\tau/2$.  This proves the
probability claim without a counting oracle or enumeration.  Accepted
overshoots remain injective, although their small-modulus bound is not
guaranteed, and all have polynomial-bit encodings.  The number of stages
and tests is polynomial in $L,\log(1/\tau)$.
\end{proof}

\begin{proof}[Proof of Theorem~\ref{thm:certified-geometric-indexing}]
Every nonempty tile has a checked
injective hash, so each point of the tile is the unique point of its
bucket within the tile, and exact CVP from the center recovers it.  The
first-tile rule leaves exactly the address
\eqref{eq:certified-cover-inverse} for each point of $X$, which proves
the bijection on every completed setup.

For the second item, apply Proposition~\ref{prop:certified-global-hash}
to each nonempty tile with allocated error $O(\tau/q_D)$; empty tiles are
handled exactly.  All point-encoding and ladder bounds follow from the
rational tile data, before the proposals are drawn.  A union bound over
tiles, not points, gives simultaneous completion and
\eqref{eq:tile-prime}.  Certified injectivity gives the lower bound in
\eqref{eq:address-count}.  For the upper bound, use
$|V|\leq q_D<2C_{\mathrm{cov}}^d$:
\begin{equation*}
    1+\sum_{j\in V}P_j
    \leq1+C_{\mathrm{hash}}^d
       \left(2C_{\mathrm{cov}}^d+\sum_{j\in V}\nu_j\right)
    \leq C_{\mathrm{addr}}^d\left(1+\sum_{j\in V}\nu_j\right).
\end{equation*}

For the circuits, prefix sums of $P_j$ give the raw index conversions and
identify the final sentinel.  Decoding uses one bucket CVP and exact
membership tests.  The canonical-address test checks only membership in
earlier tiles.  Inverse recovery scans the tiles for $j_x$ and computes
one modular inner product; it does not scan hash proposals.  Replaying
the capped setup and these deterministic operations costs
single-exponential time and space, and Bennett simulation
\cite{bennett1989time} gives the stated reversible circuits.  Off-domain
arithmetic and missing certificates reject.
\end{proof}

\begin{proof}[Proof of Theorem~\ref{thm:coherent-access}]
If $d=0$, then $D$ and the affine lattice each consist of the sole point
of $\mathbb{R}^0$. Take the trivial one-slot cover, a raw domain of one
live canonical address with no sentinel, and the trivial decoder and
canonicity test. These satisfy the bounds of the theorem because
$\kappa\ge 1$. Assume from now on that $d\ge 1$. We build the cover,
truncate the ladder and split the failure budget,
and then check the deterministic guarantees.

Generate the cover seed of Lemma~\ref{lem:rational-indexed-cover}
with failure budget $\tau/2$ and an encoding bound covering $D$.
Run the capped cover construction, and return failure if it fails, times
out, or has no valid slot.  The deterministic covering tests
in Appendix~\ref{app:indexable-cover-proof} certify every accepted
candidate, on every seed.  Consequently every retained cover satisfies
both \eqref{eq:rational-indexed-cover} and
\eqref{eq:rational-cover-point-bound}; the good-seed event is needed
only to guarantee completion with probability at least $1-\tau/2$.

Put $\kappa_E:=\lceil C_{\rm pt}^d\kappa\rceil$; every retained
tile has $\nu_j\leq\kappa_E$.  Truncate the prime ladder after
\begin{equation*}
    g_{\max}:=d+\lceil\log_2\kappa_E\rceil+1,
\end{equation*}
and generate this shared ladder with failure budget $\tau/4$.
Conditional on a valid ladder, allocate failure budget $\tau/(4q_D)$
to the capped normal generation and hash certification of each nonempty
tile; empty tiles are handled exactly.  The proof of
Proposition~\ref{prop:certified-global-hash}, with the primes fixed,
gives these conditional bounds, and the sufficient level
$d+\lceil\log_2\nu_j\rceil$ lies within the cap.  Since $|V|\leq q_D$,
a union bound over cover generation, prime generation, and tiles gives
completion and simultaneous small-modulus bounds with probability
at least $1-\tau$.  Equation~\eqref{eq:address-count} then gives
$|\mathcal A|\leq2^{O(d)}\chi$.

On every completed setup, even an accepted overshoot has
$P_j<2^{g_{\max}+3}=2^{O(d)}\kappa_E$, so
$|\mathcal A|\leq2^{O(d)}\kappa$ deterministically.
Certified coverage puts every point of $D\cap(x_0+\Lambda)$ in a valid
tile, certified hash injectivity and exact bucket CVP recover it uniquely
within that tile, and the first-tile rule leaves exactly one canonical
address.  Thus live decoding is a bijection on every completed setup,
unconditionally.  Theorem~\ref{thm:certified-geometric-indexing}
supplies the reversible query costs; cover generation and all failure
checks have the stated capped preprocessing cost.  No point-count
estimate is needed, and testing canonicity requires no scan over
multiple seeds.
\end{proof}

\subsection{Quantum tree search}
\label{subsec:tree-search-proof}
\label{subsec:tree-search}

Theorem~\ref{thm:tree-search} combines a detection result of Ambainis
and Kokainis with the usual reduction of search to detection by descent,
as in \cite{montanaro2018backtracking}.  We give the reduction in full,
because \cite[Theorem~7]{ambainis2017quantum} is stated for detection
only.  The proof caps the detection algorithm as a subroutine
\textsc{Detect}, searches by descent with it, and then bounds the cost.

\begin{proof}[Proof of Theorem~\ref{thm:tree-search}]
We begin with detection.  Ambainis and Kokainis treat binary trees in this model,
with a marking predicate that takes the values \emph{true} or \emph{false}
exactly at leaves \cite[Section~3.4]{ambainis2017quantum}.  Let a tree
$\mathcal T$ be given, together with upper bounds $n'$ on its depth and
$T_1'$ on its size, and an error parameter $\epsilon'$.  Their algorithm
outputs $1$ if $\mathcal T$ contains a marked vertex and $0$ otherwise, with
probability at least $1-\epsilon'$ \cite[Theorem~7]{ambainis2017quantum}.  It
makes
\begin{equation*}
    O(n'^{3/2}\sqrt T\log^2(2n'\log(2T_1')/\epsilon'))
\end{equation*}
queries, with logarithms regularized at the endpoints, where
$T=T_{\mathrm{dfs}}(\mathcal T)$ is the number of vertices explored by
depth-first search.  Their proof bounds the number of queries
on the event that all its subroutines succeed.  That event has probability at
least $1-\epsilon'$, and it implies the correct output.  The algorithm runs
subroutines whose query counts are fixed classically before they start.

To implement its walk in the stated local-access model, represent a vertex
coherently by its root-to-vertex path.  This path is unique because the
input is a tree.  Under the depth promise it has at most $n'$ edges, so it
stores at most $n'+1$ vertex encodings.  A child is obtained by appending
the output of the supplied child circuit, and a parent by truncating the
path; compute--copy--uncompute gives the incident neighbourhood operations
reversibly with $O(bn')$ additional qubits and polynomial overhead per
query.  At path length $n'$ (measured in edges), preserve the labels of
genuine leaves, but replace any genuinely internal boundary vertex by an
unmarked leaf with no children.  This changes no tree satisfying the depth
promise and keeps the implementation bounded otherwise.  The remaining
registers hold phase estimates, counters, and the classical paths used to describe depth-first prefixes.
All non-query operations therefore cost
$\operatorname{poly}(b,n',\log(T_1'/\epsilon'))$ gates per query and use
$\operatorname{poly}(b,n',\log(T_1'/\epsilon'))$ qubits.

Fix an effective constant $c_0\geq1$ for which this query bound holds in the
form
\begin{equation*}
    c_0n'^{3/2}\sqrt T
    \log^2\frac{2n'\log(2T_1')}{\epsilon'}.
\end{equation*}
For a vertex $w$,
let $\textsc{Detect}(w)$ run this algorithm on $\mathsf{T}(w)$ with
$n':=\Delta$, $T_1':=T_1$, and $\epsilon':=\epsilon/(2\Delta)$.  It stops with
output $0$ before any subroutine that would make the number of queries exceed
\begin{equation*}
    N_{\mathrm{det}}:=
    c_0\Delta^{3/2}\sqrt{T_1}
    \log^2\frac{4\Delta^2\log(2T_1)}{\epsilon}.
\end{equation*}
Suppose that $\Delta\geq\operatorname{depth}(\mathsf{T})$ and
$T_1\geq|\mathsf{T}|$.  Then $\Delta$ and $T_1$ bound the depth and the size
of every $\mathsf{T}(w)$, and
$T_{\mathrm{dfs}}(\mathsf{T}(w))\leq|\mathsf{T}(w)|\leq T_1$.  Therefore
the stopping rule never triggers on the success event of
\cite[Theorem~7]{ambainis2017quantum}.  Hence, with probability at least
$1-\epsilon/(2\Delta)$, $\textsc{Detect}(w)$ is correct and makes at most
\begin{equation*}
    c_0\Delta^{3/2}\sqrt{T_{\mathrm{dfs}}(\mathsf{T}(w))}
    \log^2\frac{4\Delta^2\log(2T_1)}{\epsilon}
\end{equation*}
queries.  This holds conditionally on any history, since every call uses
fresh randomness.

Search proceeds by descent.
\begin{enumerate}
    \item Set $v$ to the root.  If $\textsc{Detect}(v)=0$, return
    \textsc{none}.
    \item Repeat at most $\Delta$ times: if $v$ is a leaf, leave the loop;
    otherwise, let $v_1,v_2$ be its children in order, and set $v:=v_1$ if
    $\textsc{Detect}(v_1)=1$ and $v:=v_2$ otherwise.
    \item If $\operatorname{mark}(v)=\textsc{marked}$, return $v$; otherwise
    return \textsc{none}.
\end{enumerate}
Claim~(a) holds by step~3.  At most $\Delta+1\leq2\Delta$ calls of
\textsc{Detect} are made, each with at most $N_{\mathrm{det}}$ queries, and
the descent itself makes $O(\Delta)$ queries.  This proves~(b).

For~(c), assume that $\Delta\geq\operatorname{depth}(\mathsf{T})$ and
$T_1\geq|\mathsf{T}|$.  Let $\mathcal E$ be the event
that every call of \textsc{Detect} is correct and within its query bound.
Since at most $2\Delta$ calls are made, $\Pr[\mathcal E]\geq1-\epsilon$.
Assume $\mathcal E$.  If $\mathsf{T}$ has no marked vertex, the algorithm
returns \textsc{none} in step~1.  Otherwise, let $m^\star$ be the first marked
vertex in depth-first order.  We show by induction that every vertex $v$
reached in step~2 is $m^\star$ or an ancestor of $m^\star$.  This holds for the
root.  Let $v\neq m^\star$ be such an ancestor, with children $v_1,v_2$.  Then
$m^\star$ lies in
$\mathsf{T}(v_1)$ or in $\mathsf{T}(v_2)$.  Depth-first search visits all of
$\mathsf{T}(v_1)$ before $\mathsf{T}(v_2)$, so if
$m^\star\in\mathsf{T}(v_2)$, then $\mathsf{T}(v_1)$ has no marked vertex.  In
both cases the correct answer of $\textsc{Detect}(v_1)$ leads to the child
whose subtree contains $m^\star$.  The loop therefore ends at $m^\star$ after
$\operatorname{depth}(m^\star)\leq\Delta$ iterations, and $m^\star$ is
returned.

It remains to bound the cost.  We show that every call is made on a subtree
$\mathsf{T}(w)$ with
$T_{\mathrm{dfs}}(\mathsf{T}(w))\leq T_{\mathrm{dfs}}(\mathsf{T})$.  If
$\mathsf{T}$ has no marked vertex, the only call is on $\mathsf{T}$ itself.
Otherwise, if $m^\star\in\mathsf{T}(w)$, the inequality holds because
depth-first search on $\mathsf{T}$ visits the vertices of $\mathsf{T}(w)$
consecutively and in their own depth-first order.  If
$m^\star\notin\mathsf{T}(w)$, then $w=v_1$ for an ancestor $v$ of $m^\star$
with $m^\star\in\mathsf{T}(v_2)$.  So $\mathsf{T}(w)$ has no marked vertex
and is visited entirely before $m^\star$, which gives
$T_{\mathrm{dfs}}(\mathsf{T}(w))=|\mathsf{T}(w)|<T_{\mathrm{dfs}}(\mathsf{T})$.
Summing over at most $2\Delta$ calls and adding $O(\Delta)$ queries for the
descent gives~\eqref{eq:tree-search-cost}.  The other operations are those of
the detection algorithm and $O(\Delta)$ classical steps of the descent.
\end{proof}

\subsection{Shared setup, node processing, and the classical reference}

\subsubsection{Shared setup}
\label{app:shared-setup}

The \emph{shared setup} $\omega$ consists of the prime
table~\eqref{eq:prime-ladder}, the cover seed of
Lemma~\ref{lem:rational-indexed-cover}, and a recursion tape with two
independent parts: a flatness segment for
Proposition~\ref{prop:finite-reversible-flatness}, and a list of
proposed hash normals for every scale and dimension, shared by all tiles
of that dimension.  Node processing is indexed by the set $\mathscr N$ of
bit strings of length at most $L_{\mathrm{node}}=\operatorname{poly}(L)$
that encode the affine subspaces the recursion can produce; this set
depends only on the instance.  This subsection lays out the tape and fixes the parameters used
below, including the
per-tile point bound $\vartheta_k(d)$ and the security parameter.

Let $L_{\mathrm{node}}=\operatorname{poly}(L)$ bound the encoding length of
the compactified root $\mathbb R^n$ and of the affine subspaces produced by
Lemma~\ref{lem:rank-preserving-frank-tardos} for the root box $[-M,M]^n$.  Let
$\mathscr N$ be the set of bit strings of length at most
$L_{\mathrm{node}}$ that encode such subspaces, so
$|\mathscr N|\leq2^{L_{\mathrm{node}}+1}$.  The set $\mathscr N$ depends
only on the instance and is fixed before any randomness is drawn.  For every
tape, each node of the recursion below is identified by an element of
$\mathscr N$.  For
$1\leq d\leq k\leq n$, put
\begin{align}
    \kappa_k(d)&:=(C_{\mathrm{br}}\Phi(k))^d,
    \qquad \vartheta_k(d):=C_{\mathrm{pt}}^d\kappa_k(d),\notag\\
    g_{\max}(k,d)&:=d+\bigl\lceil\log_2\vartheta_k(d)\bigr\rceil+1,
    \qquad
    g_{\mathrm{top}}:=\max_{1\leq d\leq k\leq n}g_{\max}(k,d).
    \label{eq:g-top}
\end{align}
Thus $g_{\mathrm{top}}=O(n\log\log(4n))$.  Let $\lambda_{\mathrm{sec}}$ be the
least integer at least $2$ such that
\begin{equation}
    2^{L_{\mathrm{node}}+1}\Bigl[2^{-\lambda_{\mathrm{sec}}}
    +2C_{\mathrm{cov}}^n4^{-\lambda_{\mathrm{sec}}}\Bigr]
    +n(g_{\mathrm{top}}+1)\lambda_{\mathrm{sec}}2^{-\lambda_{\mathrm{sec}}}
    \leq\frac\delta{16},
    \label{eq:security-parameter}
\end{equation}
and set
\begin{equation*}
    \beta_{\mathrm{flat}}:=2^{-\lambda_{\mathrm{sec}}},
    \qquad
    N_{\mathrm{hash}}:=\lambda_{\mathrm{sec}}.
\end{equation*}
Since $\log\vartheta_n(n)=O(n\log\log(4n))$, we have
$\lambda_{\mathrm{sec}}=\operatorname{poly}(L,\log(1/\delta))$.  The union
bound over $\mathscr N$ in Lemma~\ref{lem:good-tapes} below forces
$\lambda_{\mathrm{sec}}>L_{\mathrm{node}}$.

The shared setup $\omega$ records the prime table, the cover seed, and
the recursion tape, and every quantity indexed by $\omega$ depends on all
these data.  Conditional on the table and seed, the recursion tape
consists of two independent uniform parts.  The first, the flatness
segment, serves
Proposition~\ref{prop:finite-reversible-flatness} with failure budget
$\beta_{\mathrm{flat}}$ and is padded to the largest length needed in
dimension at most $n$; each node reads the prefix it needs.  The second,
the hash-proposal part, holds, for every $g\leq g_{\mathrm{top}}$,
$1\leq d\leq n$, and $t\in[N_{\mathrm{hash}}]$, a normal
$a_{g,d,t}\in\mathbb F_{p_g}^d\setminus\{0\}$.  Every tile of
dimension $d$ uses this same proposal sequence at scale $g$.
In the bit model, each such normal is one capped draw.  Identify
$\mathbb F_p$ with $\{0,\ldots,p-1\}$, and $\mathbb F_p^d\setminus\{0\}$ with
$\{0,\ldots,p^d-2\}$ by writing $m+1$ in base $p$.  A uniform element of a set
$\{0,\ldots,N-1\}$ is drawn by rejection from blocks of
$\lceil\log_2N\rceil$ uniform bits, with at most $\lambda_{\mathrm{sec}}$
blocks.  Each block is accepted with probability greater than $1/2$, so a cap
is exceeded with probability less than $2^{-\lambda_{\mathrm{sec}}}$.
There are $n(g_{\mathrm{top}}+1)\lambda_{\mathrm{sec}}$ capped normal
draws.  Conditioned on no overflow, the two parts are exactly uniform and
independent.  The tape has
$2^{O(n)}\operatorname{poly}(L,\log(1/\delta))$ bits.

\subsubsection{Node-processing instructions}
\label{app:node-processing}

\begin{definition}[Node processing with a fixed tape]
\label{def:node-processing}
Fix a shared setup $\omega$, including the prime
table~\eqref{eq:prime-ladder} and the cover seed.  For
$H\in\mathscr N$, the processing $\operatorname{Proc}_\omega(H)$ is the
capped deterministic computation specified by the steps below.  Any timeout, register overflow,
or undefined arithmetic produces a leaf of type \textnormal{\textsc{fail}}.
\end{definition}
Processing follows the classical recursion of
Section~\ref{subsec:classical-recursion}, with every random choice read
from $\omega$.  It forms the node, tests it for emptiness, and applies
ellipsoid-or-hyperplane preprocessing until the node is rounded, empty,
or a point; a point is marked exactly when it passes exact verification
against the input.  At a rounded node of dimension $k\geq1$ it runs
Proposition~\ref{prop:finite-reversible-flatness} on the flatness segment
of $\omega$, forms the projected body $D=K_W$, covers it by
Lemma~\ref{lem:rational-indexed-cover}, and certifies one hash per
nonempty tile by trying the stored normals at increasing prime scales.
If all of this succeeds, $H$ is internal, with the ordered address set
$\mathcal A_H$ of~\eqref{eq:output-sensitive-addresses}, and its children
are the compressed fibers $H_y\in\mathscr N$ of its canonical addresses,
in address order.  Every child has dimension at most $k-d\leq k-1$, where
$d\geq1$ is the dimension of the subspace $W$ found by flatness, so
the resulting recursion tree $\mathcal T_\omega$ is finite and has at
most $n$ internal levels.  The step-by-step instructions follow.

\begin{enumerate}
    \item Form the node $K^\sharp\cap[-M,M]^n\cap H$.  If
    $\mathbb Z^n\cap H=\emptyset$, which Hermite normal form decides, or if
    linear programming shows the node to be empty, $H$ is a \emph{leaf} of
    type \textnormal{\textsc{empty}}.  Otherwise, use the lattice
    $\mathbb Z^n\cap H$ in local coordinates.
    \item Apply ellipsoid-or-hyperplane preprocessing
    \cite[Lemma~60]{reis2023subspace}.  On every dimension reduction, replace
    $H$ by the compressed affine subspace and repeat the two emptiness tests
    of step~1.  Stop when the node is rounded, is a leaf of type
    \textnormal{\textsc{empty}}, or has dimension $0$.  In dimension $0$,
    $H$ is a leaf of type \textnormal{\textsc{point}}.  Write $z$ for its
    unique candidate in root coefficient coordinates.  Using the original
    root data $x_0,B$, it is \emph{marked} exactly when the exact checks
    $z\in\mathbb Z^n$ and $x_0+Bz\in K$ both pass.
    \item Let $k\geq1$ be the rounded dimension.  Run
    Proposition~\ref{prop:finite-reversible-flatness} with failure budget
    $\beta_{\mathrm{flat}}$ and the flatness segment of $\omega$.
    On \textnormal{\textsc{fail}}, $H$ is a leaf of type \textnormal{\textsc{fail}}.  Otherwise, form
    $\theta$, $\widetilde K$, $K_W$, and $\Pi_W(x_0+\Lambda)$ as in
    Section~\ref{subsec:classical-recursion}, and put $D:=K_W$.
    \item Apply Lemma~\ref{lem:rational-indexed-cover} to $D$.  If no
    cover slot is valid, return a \textnormal{\textsc{fail}} leaf.
    For each valid tile, test emptiness by exact CVP.  On an empty tile
    use the all-dead default $(P_j,a_j)=(p_0,e_1)$.  Otherwise, for
    $g=0,\ldots,g_{\max}(k,d)$ and
    $t=1,\ldots,N_{\mathrm{hash}}$ in that order, apply
    \eqref{eq:certified-hash-test} to $(p_g,a_{g,d,t})$ and keep the
    first pair that passes.  If none passes for a nonempty tile, return
    a \textnormal{\textsc{fail}} leaf.
    \item The node is \emph{internal}, with ordered address set
    $\mathcal A_H$ given by~\eqref{eq:output-sensitive-addresses}.  Its
    children are taken one for each canonical address, in increasing address
    order.  The child of a canonical address decoding to $y$ is the
    compressed global affine subspace $H_y\in\mathscr N$ of the fiber
    $K\cap\Pi_W^{-1}(y)$, as in~\eqref{eq:ipco-fiber-compression}.
\end{enumerate}
The \emph{recursion tree} $\mathcal T_\omega$ has the compactified root
$\mathbb R^n$ as root and the children defined in step~5.  Two nodes reached
by different sequences of canonical addresses are distinct, even if they
determine the same subspace.  Every child has dimension at most
$k-d\leq k-1$, and every internal node has finitely many children, so
$\mathcal T_\omega$ is finite and has at most $n$ internal levels.

No step consults the path from the root to $H$, so
$\operatorname{Proc}_\omega(H)$ depends only on $H$ and $\omega$, and one
tape serves every node.

\subsubsection{Good setups and the classical reference}
\label{app:good-setups}

The shared setup $\omega$ is \emph{good} if its prime table is valid,
its cover seed is good, no draw overflows, and the following hold for every
$H\in\mathscr N$ whose processing reaches step~3 of the node-processing instructions in
Appendix~\ref{app:node-processing}.
\begin{enumerate}[label=(G\arabic*)]
    \item The output of Proposition~\ref{prop:finite-reversible-flatness}
    is geometrically valid.
    \item At step~4 every nonempty tile receives a passing certificate,
    and its selected modulus satisfies
    $P_j\leq C_{\mathrm{hash}}^d(\nu_j+1)$.
\end{enumerate}

\begin{lemma}[Good setups]
\label{lem:good-tapes}
Conditional on any fixed valid prime table and good cover seed, a
uniformly random recursion tape makes $\omega$ good with probability at
least $1-\delta/16$.  On a good setup, at every internal node of
$\mathcal T_\omega$, the following hold for every valid slot $j$:
\begin{equation*}
    \nu_j\leq\vartheta_k(d),
    \qquad
    \nu_j\leq P_j\leq C_{\mathrm{hash}}^d(\nu_j+1),
\end{equation*}
and the canonical addresses of $\mathcal A_H$ are in bijection with the
branch set $X=K_W\cap\Pi_W(x_0+\Lambda)$.
\end{lemma}

\begin{proof}[Proof of Lemma~\ref{lem:good-tapes}]
We bound the failure probability of (G1) and (G2) at one fixed
$H\in\mathscr N$, take a union bound over $\mathscr N$, and then derive the
bijection.  Fix the table and seed as in the lemma, and condition on no
overflow.  The flatness segment is then uniform, and the proposed normals
are mutually independent uniform elements, independent of the flatness
segment.  Fix $H\in\mathscr N$ whose processing reaches step~3.  Its
steps~1--2 are deterministic.  Step~3 reads only the flatness segment, so by
Proposition~\ref{prop:finite-reversible-flatness}, (G1) fails with
probability at most $\beta_{\mathrm{flat}}=2^{-\lambda_{\mathrm{sec}}}$.

Condition on the flatness segment and assume (G1).  By
\eqref{eq:branch-bound} and the point bound
\eqref{eq:rational-cover-point-bound}, which holds for every translate of
$\widehat E$, every tile satisfies $\nu_j\leq\vartheta_k(d)$.
The tiles are now fixed independently of the proposed normals.  For a
nonempty tile, set
\begin{equation*}
    g_j^\star:=d+\lceil\log_2\nu_j\rceil\leq g_{\max}(k,d).
\end{equation*}
Then $p_{g_j^\star}>2^{d+2}\nu_j$ and
$p_{g_j^\star}<2^{d+4}\nu_j\leq C_{\mathrm{hash}}^d\nu_j$, since
$C_{\mathrm{hash}}\geq32$ and $d\geq1$.  Each of the
$N_{\mathrm{hash}}=\lambda_{\mathrm{sec}}$ independent proposals at
that level passes the certificate with probability greater than $3/4$.
The event that none passes has probability at most
$4^{-\lambda_{\mathrm{sec}}}$.  A pass at an earlier level only makes
the selected prime smaller, so this tile fails to complete with
$P_j\leq p_{g_j^\star}<C_{\mathrm{hash}}^d\nu_j$ with probability at most
$4^{-\lambda_{\mathrm{sec}}}$.  On completion, certified injectivity of
the tile hash on its $\nu_j$ points gives $P_j\geq\nu_j$.  Empty tiles
have $\nu_j=0$ and the default $P_j=p_0<8\leq C_{\mathrm{hash}}^d$.  A union
bound over fewer than $2C_{\mathrm{cov}}^n$ slots bounds failure of (G2) by
$2C_{\mathrm{cov}}^n4^{-\lambda_{\mathrm{sec}}}$.

The selected normals can be correlated between tiles and nodes.  This is
harmless: we conditioned on the geometry of a fixed encoded node, not on
the event that the recursion reaches it or on which proposals were
selected, and the argument uses only that the proposals at each sufficient
level are uniform and independent.  No union bound over lattice points is
taken.

The events at different nodes need not be independent either, so we take a
union bound over the fixed set $\mathscr N$ and add the overflow
probability.  This bounds the probability of a bad tape by the left side
of~\eqref{eq:security-parameter}, which is at most $\delta/16$.

For the last claim, let $H$ be an internal node on a good setup.  Its
nonempty tiles have checked injective hashes and the cover seed is good,
so Theorem~\ref{thm:certified-geometric-indexing} gives the bijection
directly.  The same bijection holds for any completed node setup with a
good cover seed, even if a selected prime overshoots the actual-occupancy
bound.
\end{proof}

On a good setup, $\mathcal T_\omega$ is therefore the recursion tree of
Section~\ref{subsec:classical-recursion}, with the flatness choices
of Proposition~\ref{prop:finite-reversible-flatness} and with children in
canonical-address order, that is, the first-tile/residue order induced by
the checked hashes.

The quantum algorithm will be compared with one classical implementation
of this recursion, Algorithm~\ref{alg:classical-reference}.  It runs the
recursion depth first on $\omega$ and uses the same order.  The call
\textsc{Explore}$(H)$ processes $H$ and returns its point if $H$ is a
marked leaf.  At an internal node it decodes the addresses of
$\mathcal A_H$ in increasing order, calls itself on the child of each
canonical address, and returns the first point found.  The top-level
call stops with \textsc{infeasible} once more than $D_{\max}$ node
processings and address decodings have been made, a cap fixed in
Lemma~\ref{lem:classical-reference}(c), and verifies any returned point
in the original coordinates.

For a setup $\omega$, call a node $H$ of $\mathcal T_\omega$ \emph{visited} if
the uncapped procedure \textsc{Explore} of
Algorithm~\ref{alg:classical-reference}, started at the root, calls
\textsc{Explore}$(H)$.  This procedure halts on every tape, because
$\mathcal T_\omega$ is finite.  For a visited internal node $H$, let
$t_H\geq1$ be the number of addresses of $\mathcal A_H$ that the procedure
decodes at $H$.  For
the instance $I=(K,x_0+\Lambda)$, put
\begin{equation}
    \operatorname{Dec}(I,\omega):=
    \#\{\text{visited nodes}\}
    +\sum_{H\text{ visited, internal}}t_H,
    \label{eq:decodings}
\end{equation}
the number of node processings and address decodings.  The visited nodes, the
numbers $t_H$, and $\operatorname{Dec}(I,\omega)$ are determined by $I$ and
$\omega$.

For an internal node $H$ of $\mathcal T_\omega$ with valid slots $V_H$ and
tile counts $\nu_j(H)$, define its \emph{ellipsoid work} as
\begin{equation*}
    \chi(H):=1+\sum_{j\in V_H}\nu_j(H),
\end{equation*}
and put $\chi(H):=1$ for a leaf.  Define
\begin{equation}
    \operatorname{Work}(I,\omega):=\sum_{H\in\mathcal T_\omega}\chi(H),
    \qquad
    \operatorname{Work}^{\mathrm{first}}(I,\omega):=
    \sum_{H\text{ visited}}\chi(H).
    \label{eq:reference-work}
\end{equation}
On an infeasible instance every node is visited, so the two are equal.  The
quantity $\chi(H)$ counts the lattice points of all covering ellipsoids at
$H$, which is the occupancy of the cover.  It can be much larger than the
number $|X|$ of branches when $K_W$ is thin.  An implementation that
enumerates the lattice points of these ellipsoids inspects all of them, and
the $M$-ellipsoid enumeration of Dadush~\cite{dadush2012integer} does the
same for its own cover.

\begin{lemma}[The classical reference]
\label{lem:classical-reference}
Let the prime table be valid and $\omega$ be good.
\begin{enumerate}[label=(\alph*)]
    \item Algorithm~\ref{alg:classical-reference} returns a point of
    $K\cap(x_0+\Lambda)$ if the instance is feasible, and \textnormal{\textsc{infeasible}}
    otherwise.  On every tape it returns only feasible points.
    \item Its running time $T_{\mathrm C}(I,\omega)$ satisfies
    \begin{equation}
        \operatorname{Dec}(I,\omega)
        \leq T_{\mathrm C}(I,\omega)
        \leq2^{O(n)}\operatorname{poly}\!\left(L,\log\frac1\delta\right)
        \operatorname{Dec}(I,\omega).
        \label{eq:reference-time}
    \end{equation}
    If the instance is infeasible, then
    $\operatorname{Work}(I,\omega)\leq\operatorname{Dec}(I,\omega)$.
    \item With $C_W:=2C_{\mathrm{br}}C_{\mathrm{cov}}C_{\mathrm{pt}}$,
    \begin{align}
        \operatorname{Dec}(I,\omega)
        &\leq2C_{\mathrm{addr}}^n
        \operatorname{Work}^{\mathrm{first}}(I,\omega),\notag\\
        \operatorname{Work}^{\mathrm{first}}(I,\omega)
        &\leq\operatorname{Work}(I,\omega)\leq(3C_W\Phi(n))^n.
        \label{eq:reference-worst-case}
    \end{align}
    Hence
    \begin{gather*}
        \operatorname{Dec}(I,\omega)\leq D_{\max}:=
        2C_{\mathrm{addr}}^n(3C_W\Phi(n))^n,
        \\
        T_{\mathrm C}(I,\omega)
        \leq2^{O(n)}\Phi(n)^n\operatorname{poly}(L,\log(1/\delta)).
    \end{gather*}
    The bound on $T_{\mathrm C}(I,\omega)$ holds on every tape.
\end{enumerate}
\end{lemma}

\begin{algorithm}[t]
\caption{Classical reference recursion with shared setup $\omega$}
\label{alg:classical-reference}
\begin{algorithmic}[1]
\Require the compactified root and a shared setup $\omega$, including
the prime table~\eqref{eq:prime-ladder} and cover seed
\Function{Explore}{$H$}
    \State run $\operatorname{Proc}_\omega(H)$
    \If{$H$ is a leaf}
        \State \Return its point if $H$ is marked, and $\bot$ otherwise
    \EndIf
    \For{each $\alpha\in\mathcal A_H$, in increasing order}
        \State decode $\alpha$ and decide whether it is canonical
        \If{$\alpha$ is canonical, with child $H_y$}
            \State $x\gets$ \Call{Explore}{$H_y$}
            \If{$x\neq\bot$}
                \State \Return $x$
            \EndIf
        \EndIf
    \EndFor
    \State \Return $\bot$
\EndFunction
\State run \Call{Explore}{$\mathbb R^n$}, stopping with \textsc{infeasible}
once more than $D_{\max}$ node processings and address decodings have been
made (Lemma~\ref{lem:classical-reference}(c)); return its point, mapped back
to the original coordinates and verified, or \textsc{infeasible}
\end{algorithmic}
\end{algorithm}

\begin{proof}[Proof of Lemma~\ref{lem:classical-reference}]
We first compare the number of addresses with the occupancy.  Parts (a)
and (b) use the bound $\operatorname{Dec}(I,\omega)\leq D_{\max}$ of (c),
whose proof does not depend on them.  Let $H$ be an internal node.
Its body $D=K_W$ is nonempty, so $V_H\neq\emptyset$.  For every $j\in V_H$,
certified injectivity gives $P_j\geq\nu_j$.  Including the final sentinel,
\eqref{eq:address-count} gives
\begin{equation}
    \chi(H)\leq1+\sum_{j\in V_H}P_j=|\mathcal A_H|
    \leq C_{\mathrm{addr}}^d\chi(H).
    \label{eq:address-vs-work}
\end{equation}

(a) Only marked \textsc{point} leaves, which are verified exactly, are
returned.  This proves the last claim on every tape.  By (c), the cap
$D_{\max}$ is not reached on a good setup, so the algorithm returns the output
of the uncapped \textsc{Explore}.  We show by induction on the dimension
that \textsc{Explore}, called on a feasible node, returns a point.
\begin{itemize}
    \item Preprocessing preserves the feasible lattice points, so a feasible
    node is not of type \textsc{empty}.
    \item In dimension $0$, the node is a marked \textsc{point} leaf.
    \item Otherwise, by (G1) and~\eqref{eq:shrinkage-equivalence}, there is
    a point $x\in\widetilde K\cap(x_0+\Lambda)$; fix one and put
    $y^\star:=\Pi_Wx\in X$.  By
    Lemma~\ref{lem:good-tapes}, $y^\star$ has a canonical address
    $\alpha^\star$.  Its compressed fiber still contains $x$, because
    compression preserves every integer point of that fiber in the root box.
    \item \textsc{Explore} decodes the addresses in increasing order.  It
    either returns a point from the child of a canonical address before
    $\alpha^\star$, or it decodes $\alpha^\star$ and calls itself on
    $H_{y^\star}$.  There the induction hypothesis applies.
\end{itemize}

(b) By (c), the cap is not reached on a good setup, so the algorithm
performs exactly $\operatorname{Dec}(I,\omega)$ node processings and address
decodings.  Each takes at least one step, which gives the lower bound.

For the upper bound, the cost of one node processing is at most
\begin{equation*}
    2^{O(k)}\operatorname{poly}(L,\lambda_{\mathrm{sec}})
    +2^{O(d)}(g_{\mathrm{top}}+1)N_{\mathrm{hash}}
        \operatorname{poly}(L,\lambda_{\mathrm{sec}}).
\end{equation*}
One decoding costs $2^{O(d)}\operatorname{poly}(L,\lambda_{\mathrm{sec}})$,
including the test whether the address is canonical and, for a canonical
address, the formation and compression of the child's subspace.  Both costs
are
$2^{O(n)}\operatorname{poly}(L,\log(1/\delta))$, and the final verification
costs $\operatorname{poly}(L)$.  Since $\operatorname{Dec}\geq1$, this gives
the upper bound in~\eqref{eq:reference-time}.

On infeasible instances every node is visited and every address is
decoded, so
$\operatorname{Dec}=|\mathcal T_\omega|+\sum_{H\text{ internal}}|\mathcal A_H|$.
By~\eqref{eq:address-vs-work} and $\chi=1$ at leaves, this is at least
$\operatorname{Work}$.

(c) For the first bound, a visited internal node contributes
$1+t_H\leq1+|\mathcal A_H|$ to $\operatorname{Dec}$, and a visited leaf
contributes $1=\chi$.  By~\eqref{eq:address-vs-work},
$1+|\mathcal A_H|\leq2C_{\mathrm{addr}}^n\chi(H)$.

To bound the work, let $\mathcal W(k)$ bound the work of the subtree
below any node of dimension at most $k$, so $\mathcal W(0)=1$.  We show by
induction that
$\mathcal W(k)\leq(3C_W\Phi(k))^k$ for $k\geq1$.  Consider an internal node
of dimension at most $k$, with dimension $k'\leq k$ after preprocessing and
projected dimension $d$.  Put
\begin{equation*}
    \psi:=C_W\Phi(k)\geq1,
    \qquad
    \gamma:=2C_{\mathrm{cov}}C_{\mathrm{pt}}\geq8.
\end{equation*}
There are fewer than $2C_{\mathrm{cov}}^d$ tiles, each with at most
$\vartheta_{k'}(d)\leq(C_{\mathrm{br}}C_{\mathrm{pt}}\Phi(k))^d$
points, so $\chi\leq1+\psi^d$.  By~\eqref{eq:branch-bound}, there
are at most $(C_{\mathrm{br}}\Phi(k))^d=(\psi/\gamma)^d$ children, each of
dimension at most $k-d$.  Since $\Phi$ is nondecreasing, the induction
hypothesis gives $\mathcal W(k-d)\leq(3\psi)^{k-d}$, also when $k-d=0$.  Hence
the work below the node is at most
\begin{equation*}
    1+\psi^d+(\psi/\gamma)^d(3\psi)^{k-d}
    =\psi^k\bigl(\psi^{-k}+\psi^{d-k}+3^{k-d}\gamma^{-d}\bigr)
    \leq\psi^k\bigl(2+3^{k-1}/8\bigr)
    \leq(3\psi)^k.
\end{equation*}
Here we used $\psi\geq1$, $1\leq d\leq k$, and $\gamma\geq8$.  A leaf has work
$1$.  Thus $\operatorname{Work}\leq(3C_W\Phi(n))^n$, and the remaining bounds
follow from~\eqref{eq:reference-time}.  On an arbitrary tape, the explicit
$D_{\max}$ stop still bounds the number of node processings and decodings,
and their costs are capped independently of geometric validity.  Hence the
stated worst-case time holds on every tape.
\end{proof}

\subsection{The binary address tree and the proof of the main theorem}

\subsubsection{The binary address tree and vertex encoding}
\label{app:binary-tree}

This subsection defines the binary address tree sketched in
Section~\ref{subsec:binary-address-tree}, using the interval trees
defined there, and proves its structure.

\begin{definition}[Binary address tree]
\label{def:binary-address-tree}
For a setup $\omega$, the tree $\mathsf{T}_\omega$ has three kinds of vertices:
\begin{enumerate}[label=(\roman*)]
    \item pairs $(H,J)$, where $H$ is an internal node of
    $\mathcal T_\omega$ and $J$ is an internal vertex of
    $\mathcal I(|\mathcal A_H|)$;
    \item pairs $(H,\{\alpha\})$ with $\alpha\in\mathcal A_H$ dead;
    \item pairs $(H,\star)$ with $H$ a leaf of $\mathcal T_\omega$.
\end{enumerate}
A canonical address $\alpha\in\mathcal A_H$ with child $H_y$ is identified
with the root of $H_y$.  That root is $(H_y,[0,|\mathcal A_{H_y}|))$ if $H_y$
is internal, and $(H_y,\star)$ otherwise.  The root of $\mathsf{T}_\omega$ is
the root of $\mathcal T_\omega$, identified in the same way.  The children of
$(H,J)$, in order, are the images of the two children of $J$ in
$\mathcal I(|\mathcal A_H|)$ under these identifications.  Vertices of kinds
(ii) and (iii) are leaves.  Define
\begin{equation*}
    \operatorname{mark}(v):=
    \begin{cases}
        \textnormal{\textsc{internal}},
            &v\text{ is of kind (i)},\\
        \textnormal{\textsc{marked}},
            &v=(H,\star)\text{ and }H\text{ is a marked leaf of }\mathcal T_\omega,\\
        \textnormal{\textsc{unmarked}},
            &\text{otherwise}.
    \end{cases}
\end{equation*}
\end{definition}

\begin{lemma}[Structure of the binary address tree]
\label{lem:address-tree-structure}
For every setup $\omega$, the following hold.
\begin{enumerate}[label=(\alph*)]
    \item Every internal vertex of $\mathsf{T}_\omega$ has exactly two
    children.  Every marked vertex carries an exactly verified feasible
    point.  Moreover,
    \begin{equation}
        \operatorname{depth}(\mathsf{T}_\omega)\leq\Delta_{\max}
        :=n\bigl(\lceil\log_2C_{\mathrm{cov}}\rceil
        +\lceil\log_2(C_{\mathrm{br}}C_{\mathrm{pt}}\Phi(n))\rceil+12\bigr).
        \label{eq:binary-depth}
    \end{equation}
    \item The size satisfies
    \begin{equation*}
        |\mathsf{T}_\omega|
        \leq2\sum_{H\text{ internal}}|\mathcal A_H|+1.
    \end{equation*}
    \item Depth-first search on $\mathsf{T}_\omega$ meets the nodes of
    $\mathcal T_\omega$ in the order in which the uncapped procedure
    \textnormal{\textsc{Explore}} of Algorithm~\ref{alg:classical-reference}
    calls them, and it stops at the same marked leaf.  Consequently,
    \begin{align}
        T_{\mathrm{dfs}}(\mathsf{T}_\omega)
        &\leq(\Delta_{\max}+2)\operatorname{Dec}(I,\omega),\notag\\
        T_{\mathrm{dfs}}(\mathsf{T}_\omega)
        &\leq2\sum_{H\text{ visited, internal}}|\mathcal A_H|
        +\#\{\text{visited leaves}\}.
        \label{eq:dfs-vs-reference}
    \end{align}
\end{enumerate}
If moreover the prime table is valid and $\omega$ is good:
\begin{enumerate}[resume*]
    \item $\mathsf{T}_\omega$ has a marked vertex if and only if the instance
    is feasible.
    \item We have
    \begin{equation}
        T_{\mathrm{dfs}}(\mathsf{T}_\omega)
        \leq2C_{\mathrm{addr}}^n
        \operatorname{Work}^{\mathrm{first}}(I,\omega),
        \qquad
        |\mathsf{T}_\omega|\leq T_{\max}
        :=4C_{\mathrm{addr}}^n(3C_W\Phi(n))^n.
        \label{eq:binary-size}
    \end{equation}
\end{enumerate}
\end{lemma}

The node $H$ of a vertex is encoded by its \emph{label}, the sequence of
canonical addresses on the path from the root of $\mathcal T_\omega$ to $H$.  The
label determines $H$ through $\operatorname{Proc}_\omega$.  A vertex is
encoded by its label together with $J$, $\{\alpha\}$, or $\star$.  The
address set satisfies the deterministic cap
\begin{equation}
    |\mathcal A_H|
    \leq1+q_D\,2^{g_{\max}(k,d)+3}
    <1+2C_{\mathrm{cov}}^d2^{g_{\max}(k,d)+3},
    \label{eq:address-cap}
\end{equation}
because every selected prime is at a level at most $g_{\max}(k,d)$,
whether or not the small-modulus event holds.  Hence encodings have
$\operatorname{poly}(L,\log(1/\delta))$ bits.

\begin{proof}[Proof of Lemma~\ref{lem:address-tree-structure}, parts~(a), (b), (d), and~(e)]
(a) The interval tree splits every interval of length at least two into two
nonempty halves.  Marked vertices are verified \textsc{point} leaves.  A
root-to-leaf path meets at most $n$ internal nodes of $\mathcal T_\omega$,
whose projected dimensions sum to at most $n$, and inside the interval tree
of such a node $H$ it uses at most $\lceil\log_2|\mathcal A_H|\rceil$
edges.  By~\eqref{eq:address-cap} and
\begin{equation*}
    g_{\max}(k,d)\leq
    d+d\log_2(C_{\mathrm{br}}C_{\mathrm{pt}}\Phi(n))+2,
\end{equation*}
this number of edges is at most
\begin{equation*}
    d\log_2C_{\mathrm{cov}}
    +d\log_2(C_{\mathrm{br}}C_{\mathrm{pt}}\Phi(n))+d+7.
\end{equation*}
Summing over the path gives~\eqref{eq:binary-depth}.

(b) Each internal $H$ contributes the $|\mathcal A_H|-1$ internal vertices of
its interval tree.  Every other vertex except the root is a leaf of exactly
one interval tree: a dead address, a canonical address whose child is a
leaf of $\mathcal T_\omega$, or a canonical address whose child is
internal, in which case the vertex was already counted as that child's
root.  There are $\sum_H|\mathcal A_H|$ interval-tree leaves.

(d) This follows from Lemma~\ref{lem:classical-reference}(a) and part (a).

(e) Combine (b), the second bound in~(c), \eqref{eq:address-vs-work}, and
Lemma~\ref{lem:classical-reference}(c), using that leaves have $\chi=1$ and
that $\operatorname{Work}\geq1$.
\end{proof}

\begin{proof}[Proof of Lemma~\ref{lem:address-tree-structure}, part~(c)]
We first show that both searches meet the nodes of $\mathcal T_\omega$
in the same order, and then count the vertices visited at each node.  In
$\mathcal I(m)$, depth-first search meets the singletons in increasing
order.  A singleton identified with the root of a child node $H_y$ is
followed by the search below that root, before the next singleton.
\textsc{Explore} decodes the addresses of $\mathcal A_H$ in increasing order
and calls itself on each canonical one at once.  Both procedures stop at the
first marked leaf.  By induction on the height of $H$, they meet the nodes
below $H$ in the same order and stop at the same marked leaf, if any.  In
particular, at a visited internal node $H$, depth-first search meets exactly
the first $t_H$ singletons of $\mathcal I(|\mathcal A_H|)$.

Depth-first search visits an internal vertex of
$\mathcal I(|\mathcal A_H|)$ exactly when the leftmost singleton below it is among these.  Such vertices
lie on the root paths of the first $t_H$ singletons.  Consider the union of
these paths.  Its vertices with two children in the union number $t_H-1$.
Its vertices with one child in the union are ancestors of the $t_H$th
singleton, so there are at most $\lceil\log_2|\mathcal A_H|\rceil$ of them.
Hence, at $H$, depth-first search visits at most
$t_H-1+\lceil\log_2|\mathcal A_H|\rceil$ internal vertices of
$\mathcal I(|\mathcal A_H|)$ and at most $t_H$ dead addresses.  A canonical
singleton is the root of a child node and is counted there.  The visited
vertices $(H,\star)$ are those of the visited leaves of $\mathcal T_\omega$.
The interval tree of $H$ is embedded in $\mathsf{T}_\omega$, so~(a) gives
$\lceil\log_2|\mathcal A_H|\rceil\leq\Delta_{\max}$.  Hence each summand
below is at most $2t_H+\Delta_{\max}\leq(\Delta_{\max}+2)t_H$, because
$t_H\geq1$.  By~\eqref{eq:decodings},
$\operatorname{Dec}(I,\omega)\geq\sum_Ht_H+\#\{\text{visited leaves}\}$.
Therefore
\begin{align*}
    T_{\mathrm{dfs}}(\mathsf{T}_\omega)
    &\leq\sum_{H\text{ visited, internal}}
    \bigl(2t_H-1+\lceil\log_2|\mathcal A_H|\rceil\bigr)\\
    &\phantom{\leq{}}+\#\{\text{visited leaves}\}\\
    &\leq(\Delta_{\max}+2)\operatorname{Dec}(I,\omega).
\end{align*}
The second bound in~\eqref{eq:dfs-vs-reference} holds because the interval
tree of $H$ has $2|\mathcal A_H|-1$ vertices.
\end{proof}

\subsubsection{Gate synthesis}
\label{subsec:phase-synthesis}

The tree search and the other quantum algorithms of Section~\ref{sec:quantum-ip} use
non-native phase rotations, which we account for once.
Let $N_{\mathrm{rot}}$ bound the number of non-native phase rotations on
every branch of a capped computation.  Synthesizing each rotation to
operator-norm accuracy $\varepsilon_{\mathrm{syn}}/N_{\mathrm{rot}}$ gives
total error at most $\varepsilon_{\mathrm{syn}}$ by a hybrid argument.  This
changes output probabilities, or the transcript distribution of an adaptive
computation, by at most $2\varepsilon_{\mathrm{syn}}$.  Each search algorithm
specifies its synthesis budget as a fixed fraction of $\delta$.
Solovay--Kitaev synthesis costs
$\operatorname{poly}(\log(N_{\mathrm{rot}}/\delta))$ gates per rotation
\cite{dawson2006solovay}.  For the geometric integer programming algorithm,
\begin{equation*}
    \log N_{\mathrm{rot}}
    =O\bigl(n\log\log(4n)+\log L+\log(1/\delta)\bigr).
\end{equation*}
Since $n\leq L$, this adds only polynomial gate factors and does not change
the dimension exponent.
We charge the synthesized phases separately from local-access circuit
calls.  After measurement we run the exact rational verifier again, so
invalid witnesses are rejected regardless of the synthesis error.

\subsubsection{Local access, the main theorem, and operation counts}

\begin{lemma}[Local access]
\label{lem:address-tree-access}
Fix a shared setup $\omega$, including the prime table and cover seed,
as classical data.  There are
reversible circuits that, on the encoding of a vertex $v$ of
$\mathsf{T}_\omega$, compute $\operatorname{depth}(v)$, $\deg(v)$, the $i$th
child of $v$, and $\operatorname{mark}(v)$, and return all work registers to
zero.  Each uses
$2^{O(n)}\operatorname{poly}(L,\log(1/\delta))$ gates and qubits.
\end{lemma}

\begin{proof}[Proof of Lemma~\ref{lem:address-tree-access}]
Each circuit first replays $\operatorname{Proc}_\omega$ along the label of
$v$.  At each of the at
most $n$ levels of the replay, the circuit runs steps~1--4 of
Definition~\ref{def:node-processing}, computes $|\mathcal A_H|$, decodes
the next address of the label and checks whether it is canonical, using
Theorem~\ref{thm:certified-geometric-indexing}, and forms and compresses
the fiber.  Steps~1--4 cost
\begin{equation*}
    2^{O(k)}\operatorname{poly}(L,\lambda_{\mathrm{sec}})
    +2^{O(d)}(g_{\mathrm{top}}+1)N_{\mathrm{hash}}
        \operatorname{poly}(L,\lambda_{\mathrm{sec}}).
\end{equation*}
The depth is the sum over levels of the interval-tree depth of the chosen
address, plus the depth of $J$, which is integer arithmetic.  For a vertex
$(H,J)$, the circuit computes the two children by splitting $J$.  For a
singleton half $\{\alpha\}$, it decodes $\alpha$ and decides whether
$\alpha$ is canonical; if so, it processes the child node to determine
whether that node is internal and to compute its address count.  The mark
reads the leaf type and, for a \textsc{point} leaf, runs the exact
verification.
All of these are capped deterministic computations with
$2^{O(n)}\operatorname{poly}(L,\log(1/\delta))$ steps.  Bennett's simulation
gives reversible circuits with the stated cost.  The tape is read but never
modified.
\end{proof}

\begin{proof}[Proof of Theorem~\ref{thm:main}]
Handle dimension zero directly and otherwise compactify the root by
\eqref{eq:root-compactification}.  Fix effective upper bounds on the
universal constants in the constructive geometric routines and in
Theorem~\ref{thm:tree-search}, including $c_{\mathrm{tree}}$, and hard-wire
them into the algorithm, so that all tape, depth, and work caps are
computable.
Generate the primes of~\eqref{eq:prime-ladder} with failure budget
$\delta/32$, choose the seed of
Lemma~\ref{lem:rational-indexed-cover} with encoding bound
$L'=\operatorname{poly}(L,\log(1/\delta))$ and failure budget $\delta/32$,
and draw the flatness/hash recursion tape.  Denote these data together
by $\omega$.  The encoding invariant bounds every internal body
on every tape by $L'$.  Missing data or an overflowing draw makes both
algorithms return \textsc{infeasible}.  Conditional on a valid table and
a good cover seed, Lemma~\ref{lem:good-tapes} bounds tape failure by
$\delta/16$, so the setup is good with probability at least
$1-\delta/8$.  On this event Lemma~\ref{lem:classical-reference} gives
classical correctness, \eqref{eq:reference-time}, and the classical
worst-case bound.

The quantum algorithm applies Theorem~\ref{thm:tree-search} to
$\mathsf{T}_\omega$, using the
circuits of Lemma~\ref{lem:address-tree-access}, depth cap
$\Delta=\Delta_{\max}$, size cap $T_1=T_{\max}$, and error $\delta/16$.
It synthesizes the rotations to total operator-norm error $\delta/32$,
using the every-tape query cap~\eqref{eq:tree-search-worst}; the adaptive
hybrid argument then changes the transcript distribution by at most
$\delta/16$.  It recomputes any returned coefficient vector $z$, forms
$x=x_0+Bz$ using the original root data, and exactly verifies
$z\in\mathbb Z^n$ and $x\in K$ against the original input.
It returns $x$ if both checks pass, and \textsc{infeasible} if no
candidate is returned or a check fails.  Exact verification proves
soundness on every tape and certain rejection on infeasible instances.

Now fix a good setup.  Lemma~\ref{lem:address-tree-structure} supplies valid
depth and size bounds and a marked vertex exactly when the instance is
feasible.  Except with probability $\delta/8$ over search and synthesis,
Theorem~\ref{thm:tree-search}(c) returns the correct answer using
\begin{equation*}
    c_{\mathrm{tree}}\Delta_{\max}^{5/2}
    \sqrt{T_{\mathrm{dfs}}(\mathsf{T}_\omega)}\,
    \log^2\frac{64\Delta_{\max}^2\log(2T_{\max})}{\delta}
\end{equation*}
local queries.  Each costs
$2^{O(n)}\operatorname{poly}(L,\log(1/\delta))$ gates; the other
operations and synthesized rotations add polynomial factors.  Since
$\Delta_{\max}$ and $\log T_{\max}$ are polynomially bounded, the gate
count is at most
\begin{equation}
    2^{O(n)}\operatorname{poly}\!\left(L,\log\frac1\delta\right)
    \sqrt{T_{\mathrm{dfs}}(\mathsf{T}_\omega)}
    \leq
    2^{O(n)}\operatorname{poly}\!\left(L,\log\frac1\delta\right)
    \sqrt{T_{\mathrm C}(I,\omega)}.
    \label{eq:universal-bound}
\end{equation}
The second inequality uses
$T_{\mathrm{dfs}}(\mathsf{T}_\omega)\leq
(\Delta_{\max}+2)\operatorname{Dec}(I,\omega)
\leq(\Delta_{\max}+2)T_{\mathrm C}(I,\omega)$.
Shared setup and final verification are absorbed because
$T_{\mathrm{dfs}}\geq1$.  Combining the setup and search events gives
joint probability at least $1-\delta/4$, stronger than the claimed
$1-\delta$.

On every tape, Theorem~\ref{thm:tree-search}(b) caps the query count by
\eqref{eq:tree-search-worst}, and
\begin{equation*}
    \sqrt{T_{\max}}=2^{O(n)}\Phi(n)^{n/2}.
\end{equation*}
The local-access bound also holds on every tape, which proves the
every-tape gate bound.
Each local-access circuit uses at most
$2^{O(n)}\operatorname{poly}(L,\log(1/\delta))$ qubits, and the shared
setup has the same bound in bits.  The tree-search registers use
$\operatorname{poly}(b,\Delta_{\max},\log(T_{\max}/\delta))$ qubits.
The encoding, depth, and size bounds make this polynomial in
$L$ and $\log(1/\delta)$.  Reusing the local work registers and
including the setup registers therefore gives, on every input and tape,
the bound
\begin{equation*}
    2^{O(n)}\operatorname{poly}(L,\log(1/\delta))
    \qquad\text{qubits}.
\end{equation*}
\end{proof}

We now justify the constants $A$ and $Q_{\max}$ of
Section~\ref{subsec:never-slower}, for the operation counts $S$,
$T_{\rm C}(I,\omega)$, and $Q(I,\omega;\rho)$ defined there.  The setup
counted in $S$ consists of the prime table, the cover seed, and the
flatness and hash recursion tape, together with the failure flags of the
capped draws.  The time $T_{\rm C}(I,\omega)$ includes root node
processing and the final verification.  The count $Q$ exceeds the bare
count of unitary gates, and even for fixed $\omega$ it is random, because
it depends on~$\rho$.  Every run uses at least one operation, so
$T_{\rm C},Q\geq1$; this includes the branch in which a missing setup or
an overflowed draw makes both algorithms return infeasibility at once.
The controller of Theorem~\ref{thm:tree-search} spends only polynomially
many bit operations per local query, so the bounds in the proof of
Theorem~\ref{thm:main} also hold for the fully charged count, with larger
polynomial factors.  Hence there are
$A=2^{O(n)}\operatorname{poly}(L,\log(1/\delta))\geq1$ and
$Q_{\max}=2^{O(n)}\Phi(n)^{n/2}\operatorname{poly}(L,\log(1/\delta))$
such that $S\leq A$ on every run, $Q(I,\omega;\rho)\leq Q_{\max}$ for
every setup and transcript, and, with probability at least $1-\delta$,
both algorithms are correct and
$Q(I,\omega;\rho)\leq A\sqrt{T_{\rm C}(I,\omega)}$.

\begin{proof}[Proof of Corollary~\ref{cor:never-slower}]
Fix $I$. Common initialization is performed once, at cost $S$. Both algorithms
read the same $\omega$, neither writes to it, and their workspaces are
disjoint. Once $\omega$ is fixed, Algorithm~\ref{alg:classical-reference} is
deterministic. The only remaining randomness is in the quantum measurement
outcomes, and these are not part of~$\omega$.

\proofstep{Coupling.} Pausing the quantum computation between two of its own
operations adds no measurement and does not act on its registers. The
classical steps act only on the classical workspace and on the read-only
$\omega$. So for fixed $\omega$, the quantum operations and measurement
outcomes in the combined run have the same distribution as the corresponding
prefix of a standalone quantum run with setup~$\omega$. We couple the two
runs. Draw a complete standalone transcript~$\rho$, and let the combined run
use its outcomes until it stops. Similarly, the classical part of the
combined run is a prefix of the standalone classical run.

\proofstep{Time.} Both standalone runs halt on every setup and transcript. The
classical run halts because of the cap $D_{\max}$ in
Algorithm~\ref{alg:classical-reference} and the caps on its constituent
computations. The quantum run halts because of its caps. Only top-level final
outputs stop the race. A top-level final output is either a verified feasible
point or a report of infeasibility. Internal subroutine returns, candidate
measurements and failing leaves of the quantum search do not stop it. Let
$m=\min\{T_{\rm C}(I,\omega),Q(I,\omega;\rho)\}\ge 1$. After $m$ complete
rounds, the faster algorithm has executed all of its operations, so it has
produced its final output. The race therefore stops within the first $m$
rounds. It may stop earlier, possibly in the middle of a round, and that only
lowers the count. Each round has at most two operations, which proves the
first inequality in~\eqref{eq:never-slower-bound}. The second inequality
follows from $Q(I,\omega;\rho)\le Q_{\max}$. Common initialization is charged separately and
is not part of the factor~$2$.

\proofstep{Correctness and the high-probability bound.} Let $E$ be the event that
three things hold: $\omega$ is good, the quantum algorithm returns a correct
answer, and $Q(I,\omega;\rho)\le A\sqrt{T_{\rm C}(I,\omega)}$. The proof of
Theorem~\ref{thm:main} shows that $\omega$ is good with probability at least
$1-\delta/8$. It also shows that, on a good setup, the second or third
condition fails with probability at most $\delta/8$. Hence
$\Pr[E]\ge 1-\delta/4\ge 1-\delta$. Both algorithms share one setup, so no
union bound over two setups is needed. On a good setup,
Algorithm~\ref{alg:classical-reference} is correct by
Lemma~\ref{lem:classical-reference}(a). So on~$E$, both final outputs are
correct, and whichever one arrives first is correct. No vote and no
resolution of disagreements is needed. The first bound in~(ii) follows from
the first inequality in~\eqref{eq:never-slower-bound} on~$E$. The second
bound in~(ii) follows from $A\ge 1$. The sharper
bound with $\min\{T_{\rm C},A\sqrt{T_{\rm C}}\}$ is the one that keeps the
property of never being slower than the reference.

\proofstep{Soundness.} By Lemma~\ref{lem:classical-reference}(a), the classical
reference returns only verified feasible points on every setup. The quantum
algorithm recomputes the returned coefficient vector~$z$ and maps it to
$x=x_0+Bz$ using the original root data. It checks exactly that $z$ is
integral and $x\in K$, and otherwise reports infeasibility. So every point
returned by the combined algorithm is feasible, which proves~(iii). On an
infeasible input no verified point exists, so every run reports
infeasibility. A report of infeasibility is not a certificate, however. It
can come from a bad setup, from the cap $D_{\max}$, or from a failed quantum
search. On a feasible input this is the only possible error, and it occurs
with probability at most $1-\Pr[E]\le\delta$.
\end{proof}

\subsection{Random recursion paths}

\subsubsection{A direct proof of the worst-case bound}
\label{subsec:random-path}

For the worst-case bound of Corollary~\ref{cor:worst-case-ip}, it suffices
to follow one random root-to-leaf path of the recursion of
Section~\ref{subsec:classical-recursion}, with fresh random choices at
every node, and to amplify the probability that the path ends at a
feasible point.  The path reaches a fixed feasible child if it picks a
cover ellipsoid that contains the child, a hash that isolates the child
in that ellipsoid, and the child's residue; one closest-vector
computation then decodes the child.  From Section~\ref{sec:search}, the argument uses
Lemmas~\ref{lem:rational-indexed-cover} and~\ref{lem:bucket-basis}.
Its success analysis conditions on the cover seed being good
simultaneously for all projected bodies of the prescribed encoding
length, as arranged below.  It needs neither certified hashes nor
the address-tree construction.  We
fix the primes, describe one path, prove
Proposition~\ref{prop:random-path-worst-case}, and then deduce the
general-ILP and mixed-integer assertions of
Corollary~\ref{cor:worst-case-ip}.

For $1\leq d\leq k\leq n$ put $\vartheta_k(d):=C_{\rm pt}^d\kappa_k(d)$,
with $\kappa_k(d)$ as in~\eqref{eq:branch-bound}, and fix primes
\begin{equation}
    2^{d+3}\vartheta_k(d)<p_{k,d}\leq2^{d+4}\vartheta_k(d),
    \label{eq:random-path-primes}
\end{equation}
which exist by Bertrand's postulate.  At a rounded node of dimension $k$
with a geometrically valid flatness output, a translate of $\widehat E$
contains at most
$\nu_{\max}(\Gamma,\widehat E)\leq C_{\rm pt}^d\nu_{\max}(\Gamma,K_W)\leq\vartheta_k(d)$
points of $\Pi_W(x_0+\Lambda)$, by Lemma~\ref{lem:rational-indexed-cover}
and~\eqref{eq:branch-bound}.  Hence $p_{k,d}\geq2\nu$ whenever $\nu$
counts the points of $\Pi_W(x_0+\Lambda)$ in a cover ellipsoid.

A path draws a uniform element of a set of size $N$ from
$\lceil\log_2N\rceil$ coin bits and fails if these bits encode a number
at least $N$; this happens with probability less than $1/2$, and
otherwise the element is uniform.  The coins form one segment for each
rounded node on the path, so every node reads fresh coins.

Starting at
the compactified root~\eqref{eq:root-compactification}, a path processes
a node as follows.  It first applies the emptiness tests, by linear
programming and lattice normal form, and the ellipsoid-or-hyperplane
preprocessing of
Section~\ref{subsec:classical-recursion}, compressing the affine subspace
by Lemma~\ref{lem:rank-preserving-frank-tardos} after every restriction.
An empty node ends the path with failure.  In dimension zero the path
verifies the unique candidate exactly against the input and ends.  At
a rounded node of dimension $k\geq1$, the path calls
Proposition~\ref{prop:finite-reversible-flatness} with failure budget
$\beta:=1/(4n)$, fails on a rejected output, and otherwise forms $\theta$,
$\widetilde K$, $\Gamma$, and $D:=K_W$ as in
Section~\ref{subsec:classical-recursion}, in rational coordinates of $W$.
Write $\Pi_W(x_0+\Lambda)=y_0+B_\Gamma\mathbb Z^d$ with a basis $B_\Gamma$
of $\Gamma$, put $p:=p_{k,d}$, and hash by
$r_a(y_0+B_\Gamma z):=a^\top z\bmod p$.  The path draws a uniform slot
$j\in[C_{\rm cov}^d]$ of the cover of $D$ from
Lemma~\ref{lem:rational-indexed-cover}, a uniform nonzero
$a\in\mathbb F_p^d$ (an integer $t<p^d-1$ encodes the base-$p$ digits of
$t+1$), and a uniform residue $r\in\mathbb F_p$.  It fails if slot $j$ is
invalid.  Otherwise an exact closest-vector computation in the norm
$\lVert\cdot\rVert_{\widehat Q}$ returns a point $y$ of the bucket
of residue $r$ of Lemma~\ref{lem:bucket-basis} closest to the center $s_j$
\cite{micciancio2010deterministic}, \cite[Theorem~5.2.2]{dadush2012integer}.
The path keeps $y$ only if $y\in s_j+\widehat E$ and $y\in D$, and then
moves to the compressed fiber $H_y$ of $K\cap\Pi_W^{-1}(y)$, as
in~\eqref{eq:ipco-fiber-compression}.

\begin{proof}[Proof of Proposition~\ref{prop:random-path-worst-case}]
Handle dimension zero directly.  After a classical setup, we bound the
probability that one path ends at a verified point and then amplify it.  The classical setup fixes effective upper bounds on the
universal constants $C_{\rm cov}$, $C_{\rm pt}$, $C_{\rm br}$, and
$C_\Phi$, which are used throughout.  It draws the seed of
Lemma~\ref{lem:rational-indexed-cover} with failure budget $\delta/8$
and with $L'=\operatorname{poly}(L)$
bounding every projected body that a path can meet; the encoding
invariant of Appendix~\ref{app:recursion-details} supplies such a bound,
since $\log(1/\beta)=O(\log n)$.  It generates the
primes~\eqref{eq:random-path-primes} by capped uniform sampling with
deterministic primality testing.  Since
$\log\vartheta_n(n)=O(n\log\log(4n))$ and a uniform integer in the range
of~\eqref{eq:random-path-primes} is prime with probability
$\Omega(1/\log\vartheta_k(d))$, this takes
$\operatorname{poly}(L,\log(1/\delta))$ time and fails with probability
at most $\delta/8$; on failure, the algorithm reports infeasibility.

We first show that a random hash isolates a fixed point with good
probability.  Let $T$ be a convex set that contains exactly $\nu\geq1$
points of $y_0+B_\Gamma\mathbb Z^d$, let $p\geq2\nu$ be prime, and fix
one of these points, $y^\star=y_0+B_\Gamma z^\star$.  If another point
$y_0+B_\Gamma z$ of $T$ had $z-z^\star=pw$ with $w\in\mathbb Z^d$, the
$p+1$ distinct lattice points $y^\star+iB_\Gamma w$, $0\leq i\leq p$,
would lie on the segment between the two points and hence in $T$,
contradicting $\nu\leq p$.  Thus $z-z^\star\not\equiv0\pmod p$, and the
nonzero $a\in\mathbb F_p^d$ with $a^\top(z-z^\star)\equiv0$ form a
hyperplane minus the origin, a fraction less than $1/p$ of all nonzero
$a$.  A union bound over the other $\nu-1$ points shows that, for a
uniform nonzero $a$, another point of $T$ has the residue of $y^\star$
with probability less than $\nu/p\leq1/2$.

Next we bound the success probability of a single level.  Let a feasible node end its preprocessing at a
rounded node of dimension $k$, and condition on a geometrically valid
flatness output $(W,d,R)$.  By~\eqref{eq:shrinkage-equivalence} there is a
point $x\in\widetilde K\cap(x_0+\Lambda)$; fix one, put
$y^\star:=\Pi_Wx\in X\subseteq D$, and fix a valid slot $j^\star$ with
$y^\star\in s_{j^\star}+\widehat E$, which exists
by~\eqref{eq:rational-indexed-cover}.  This ellipsoid contains $\nu$
points of $\Pi_W(x_0+\Lambda)$ with $p_{k,d}\geq2\nu$.  The path draws
$j=j^\star$ with probability at least $\frac12C_{\rm cov}^{-d}$; then a
nonzero $a$ that isolates $y^\star$ in $s_{j^\star}+\widehat E$ with
probability at least $\frac14$; and then the residue $r=r_a(y^\star)$ with
probability at least $1/(2p_{k,d})$.  In this event every other point of
the bucket lies outside the ellipsoid, at $\widehat Q$-distance greater
than one from $s_{j^\star}$, while $y^\star$ lies within distance one.
The closest-vector computation therefore returns $y^\star$, which passes
both membership tests.  The fiber $K\cap\Pi_W^{-1}(y^\star)$ contains $x$,
so the path moves to a feasible child of dimension at most $k-d$ with
probability at least
\begin{equation*}
    \frac1{16C_{\rm cov}^dp_{k,d}}
    \geq\frac1{2^{d+8}(C_{\rm cov}C_{\rm pt})^d\kappa_k(d)}
    \geq\bigl(c_{\rm path}\Phi(k)\bigr)^{-d},
\end{equation*}
where $c_{\rm path}:=2^9C_{\rm cov}C_{\rm pt}C_{\rm br}$ and the last step
uses $d\geq1$.

Now put $\lambda_0:=1$ and
$\lambda_k:=(1-\beta)^k(c_{\rm path}\Phi(k))^{-k}$ for $k\geq1$; these
numbers do not increase with $k$, because $\Phi$ is nondecreasing and
$c_{\rm path}\Phi\geq1$.  We show by induction on $k$ that a path started
at a feasible node of dimension at most $k$ ends at a verified point with
probability at least $\lambda_k$.  Preprocessing keeps every feasible
lattice point, so the path meets no empty node, and in dimension zero the
candidate passes verification.  Otherwise preprocessing ends at a rounded
node of dimension $k'$ with $1\leq k'\leq k$.  Its flatness output is
geometrically valid with probability at least $1-\beta$.  Given that, the
path reaches a feasible node of dimension at most $k-d$ with probability
at least
$(c_{\rm path}\Phi(k'))^{-d}\geq(c_{\rm path}\Phi(k))^{-d}$, and from that
node, with fresh coins, it succeeds with probability at least
$\lambda_{k-d}\geq(1-\beta)^{k-d}(c_{\rm path}\Phi(k))^{-(k-d)}$.  The
product of these three probabilities is at least
$(1-\beta)^{k-d+1}(c_{\rm path}\Phi(k))^{-k}\geq\lambda_k$.  At the root,
$(1-\beta)^n\geq3/4$, so one path ends at a verified point, flatness
failures included, with probability
\begin{equation*}
    P_{\rm path}\geq\lambda:=\tfrac34\bigl(c_{\rm path}\Phi(n)\bigr)^{-n}.
\end{equation*}

Every step of a path is a capped deterministic computation of its node
and its coins: linear programming, lattice normal form, the preprocessing
of \cite[Lemma~60]{reis2023subspace}, compression, the capped flatness
routine, the cover circuit, the bucket basis, one closest-vector
computation, and exact membership tests.  A path visits at most $n$
rounded nodes and makes at most $n$ restrictions, and each step costs
$2^{O(n)}\operatorname{poly}(L,\log(1/\delta))$.  Bennett simulation
\cite{bennett1989time} and Hadamard gates on the coin register therefore
give a unitary $\mathcal B$ with
$2^{O(n)}\operatorname{poly}(L,\log(1/\delta))$ gates and qubits that
prepares a superposition of path histories with a success flag; the flag
of $\mathcal B|0\rangle$ is one with probability $P_{\rm path}$.

The probability $P_{\rm path}$ is unknown, so we amplify it as in the
exponential search of Brassard, H{\o}yer, Mosca, and Tapp
\cite{brassard2002quantum}, but with a single fixed range of iteration
counts, which suffices because $P_{\rm path}\geq\lambda$.  Let
$\mathcal Q:=-\mathcal BS_0\mathcal B^{-1}S_1$, where $S_1$ flips the sign
of the states with flag one and $S_0$ that of the all-zero state, and
write $P_{\rm path}=\sin^2\phi$ with $0\leq\phi\leq\pi/2$.  After $\ell$
applications of $\mathcal Q$ to $\mathcal B|0\rangle$, the flag is one
with probability $\sin^2((2\ell+1)\phi)$
\cite[Section~2]{brassard2002quantum}.  Let $m$ be the least power of two
with $m\geq\lambda^{-1/2}$.  One round measures $\mathcal B|0\rangle$,
then measures $\mathcal Q^\ell\mathcal B|0\rangle$ for a uniform
$\ell\in\{0,\ldots,m-1\}$ given by $\log_2m$ fresh random bits, and
verifies each candidate exactly.  If
$P_{\rm path}\geq3/4$, the first measurement succeeds with probability at
least $3/4$.  If $\lambda\leq P_{\rm path}<3/4$, then
$\sin(2\phi)=2(P_{\rm path}(1-P_{\rm path}))^{1/2}\geq P_{\rm path}^{1/2}$
and $mP_{\rm path}^{1/2}\geq1$, so the second measurement succeeds with
probability
\begin{equation*}
    \frac1m\sum_{\ell=0}^{m-1}\sin^2((2\ell+1)\phi)
    =\frac12-\frac{\sin(4m\phi)}{4m\sin(2\phi)}
    \geq\frac12-\frac1{4mP_{\rm path}^{1/2}}\geq\frac14.
\end{equation*}
On a feasible instance with a good setup, each round therefore finds a
verified point with probability at least $1/4$, and
$N_{\rm round}:=\lceil\log_{4/3}(4/\delta)\rceil$ rounds all fail with
probability at most $\delta/4$.

A round applies $\mathcal B$ or its inverse at most $2m$ times.  Each
of the reflections $S_0$ and $S_1$ costs
$2^{O(n)}\operatorname{poly}(L,\log(1/\delta))$ gates.
Since $m=2^{O(n)}\Phi(n)^{n/2}$, the $N_{\rm round}$ rounds use
$2^{O(n)}\Phi(n)^{n/2}\operatorname{poly}(L,\log(1/\delta))$ gates on
every input.  Synthesizing the non-native gates with total operator-norm
budget $\delta/8$, as in Appendix~\ref{subsec:phase-synthesis}, changes the
distribution of the adaptive transcript by at most $\delta/4$ and
multiplies the gate count by $\operatorname{poly}(L,\log(1/\delta))$.  The
rounds reuse one set of $2^{O(n)}\operatorname{poly}(L,\log(1/\delta))$
qubits.  On feasible instances, setup, rounds, and synthesis fail with
total probability at most $\delta/8+\delta/8+\delta/4+\delta/4$, which
is less than $\delta$.
A returned point has passed exact verification against the original
input, so infeasible instances are reported infeasible on every run.
\end{proof}

The proofs of Corollaries~\ref{cor:mixed-integer-linear-feasibility}
and~\ref{cor:integer-linear-optimization} use the feasibility algorithm
only through its gate and qubit bounds, its verified witnesses, and its
one-sided error, with error at most $\delta/J$ per call.  With
Proposition~\ref{prop:random-path-worst-case} in place of
Theorem~\ref{thm:main}, they prove the general-ILP and mixed-integer
assertions of Corollary~\ref{cor:worst-case-ip} with the same bounds.

\end{document}